\documentclass[aps,pra,nofootinbib,twocolumn]{revtex4}
\usepackage{graphicx,epstopdf}
\usepackage{amsmath}
\usepackage{amssymb}
\usepackage{braket}
\usepackage{color}
\usepackage{xcolor}
\usepackage{float}
\usepackage{epstopdf}
\DeclareMathOperator{\tr}{tr}
\usepackage{enumitem}
\newcommand{\PRLsep}{\noindent\makebox[\linewidth]{\resizebox{0.667\linewidth}{1pt}{$\bullet$}}\bigskip}
\let\svitem\item%
\def\mybox#1{\makebox[2.5cm][l]{\bfseries#1}}
\newenvironment{leftitemize}
{\renewcommand\item[1][$\bullet$]{\svitem[\mybox{##1}]}%
  \begin{itemize}[leftmargin=\dimexpr2.5cm+\labelsep]}{\end{itemize}}

\usepackage[colorlinks=true, citecolor=blue, urlcolor=blue ]{hyperref}
\input{epsf}
\begin{document}

\title{Quantum discord and its allies: a review}
\author{Anindita Bera\(^{1,2,3}\), Tamoghna Das\(^{2,3}\), Debasis Sadhukhan\(^{2,3}\), Sudipto Singha Roy\(^{2,3}\),\\
 Aditi Sen(De)\(^{2,3}\), and Ujjwal Sen\(^{2,3}\)}
\affiliation{\(^1\)Department of Applied Mathematics, University of Calcutta, 92 Acharya Prafulla Chandra Road, Kolkata 700 009, India\\
\(^2\)Harish-Chandra Research Institute, Chhatnag Road, Jhunsi, Allahabad 211 019, India\\
\(^3\)Homi Bhabha National Institute,
Training School Complex, Anushakti Nagar, Mumbai 400 085, India}
%\pacs{}
%\date{today}

%\pacs{}

\begin{abstract}
We review concepts and methods associated with quantum discord and related topics. We also describe their possible connections with other aspects of quantum information and beyond, including quantum communication, quantum computation, many-body physics, and open quantum dynamics. Quantum discord in the multiparty regime and its applications are also discussed.
%and the current experimental status of these concepts. 
\end{abstract}

\maketitle

{\hypersetup{linkcolor=black}
% or \hypersetup{linkcolor=black}, if the colorlinks=true option of hyperref is used
\tableofcontents
}
\section{Introduction}
\label{intro}
The quantum theory of nature, formalized in the first few decades of the $20^{\text{\tiny th}}$ century, contains elements that are fundamentally different from those  required in the classical physics description of nature.
One of the most prominent features in quantum physics is the existence of quantum correlations between different quantum systems. In a classical world,  if a system in a pure state can be divided into two subsystems, then the sum of the information of the subsystems makes up the complete information of the whole system.
This is no longer true in the quantum formalism. In particular, there exists quantum states consisting of two (or more) physical systems for which  complete information of the whole is available, even when the subsystems are completely random.
Erwin Schr\"odinger~\cite{ErwinSchroedinger} coined the term ``quantum entanglement"~\cite{Horo_RMP} to describe this quantum feature.

About three decades ago, with the spectacular discoveries of quantum communication and computational 
schemes~\cite{Nielsen, MWildebook, Gisin-rmp-crypto, Horo_RMP, qcom},
it has been realized  that apart from its fundamental importance, entanglement can be used as a resource to efficiently achieve certain  information processing tasks  which  cannot be performed by using unentangled states.  Several of these phenomena and protocols have already been realized in the laboratories by using different  physical substrates
%  making the field of 
% quantum information and computation more attractive
(see e.g.~\cite{lab11,lab12,lab2,lab31, lab32,lab4}).

%At the same time, counter-intuitive inventions are reported  ranging from  fundamental to applicational nature.
However, a thin but steady stream of developments keep being reported which challenge the belief that entanglement is the only form of quantum correlation in shared quantum systems. For example,  
Knill and Laflamme~\cite{Knill1}
discovered the protocol of deterministic quantum computation with one quantum bit where the
natural bipartite split of the system is unentangled, even though the phenomenon demonstrated is nonclassical, under a plausible assumption. This naturally leads to the quest for quantum correlations beyond entanglement in the same split.  

Distinguishability of quantum states lies at the heart of 
physics~\cite{gordon-enrico,levitin-inf,Holevo,yuen-measurement,Schumacher-channel,helstrom-book}. And herein we get another whiff of evidence in the same direction, viz. quantum correlation beyond entanglement. For a single-party quantum system, a set of mutually orthogonal states can always be discriminated with certainty.
% which is, in general, not true for two-party systems with restricted operations like local operations and classical communication (LOCC).
%The intuitive reason behind such findings in the two-party  case is the presence of entanglement which can not be created by LOCC.
For quantum systems of two (or more) parties, there is a practical and useful restriction on the set of allowed operations to consider only local quantum operations supplemented by classical communication, which has been acronymized as
 ``LOCC"~\cite{Horo_RMP}.
In this case, even orthogonal states may not be distinguishable. It may seem that the reason behind such indistinguishability is that entangled states cannot be created by LOCC. In sharp disagreement to such intuition,  Bennett
\emph{et al.}~\cite{Bennett3} (see also~\cite{UPB1, UPB2}) presented a set of pure  states of two quantum spin-1 particles, that despite being product and orthogonal, cannot be locally distinguished, i.e., distinguished by LOCC-based measurement strategies.  
On the other hand,
it was demonstrated that two pure quantum states can always be locally distinguished if they are orthogonal, irrespective of their entanglement content, and irrespective of the number of parties and their dimensions~\cite{WHSV} (see also~\cite{Virmani,Chen1,Chen2}). It was moreover exposed that local indistinguishability of certain ensembles
of quantum states can be increased by decreasing its average entanglement~\cite{Horodecki2003a}. These results indicate that the physical quantity or quantities responsible for the nonclassical behavior of local indistinguishability of orthogonal states is clearly of a different nature than entanglement. Indeed, the seminal paper of Bennett \emph{et al.}~\cite{Bennett3} was titled ``Quantum nonlocality without entanglement". Such a correlation quantity beyond entanglement can be the property of equal or unequal mixtures of the ensembles discussed above\footnote{This review considers the question of defining  quantum correlation beyond entanglement for quantum states. One may however go further and ask whether it is possible to measure this ``nonlocality" in multiparty quantum ensembles. See~\cite{Sir_maam_new_new,horo_new_new}, and compare with~\cite{fuchs_new,englart_new}.}.

Peres and Wootters~\cite{peres_new_new}  provided a plausible reasoning that one would require to utilize non-LOCC measurement strategies to optimally distinguish between elements of a two-party quantum ensemble, where the elements are identically prepared pure qubits at the two locations (parties). Such an ensemble is therefore built of ``parallel" states\footnote{Parallel states are product states of the form $|\uparrow_{\hat{n}}\rangle\otimes |\uparrow_{\hat{n}}\rangle$, where $|\uparrow_{\hat{n}}\rangle$ can, e.g., be the spin-up state  in the $\hat{n}$-direction of a quantum spin-up  system.}. See also~\cite{popescu_new_new} in this regard. Furthermore, it was discovered by Gisin and Popescu~\cite{rsp_new_new1} that ``antiparallel" states\footnote{ Antiparallel states are product states of the form $|\uparrow_{\hat{n}}\rangle\otimes |\downarrow_{\hat{n}}\rangle$, where $|\uparrow_{\hat{n}}\rangle$ and $|\downarrow_{\hat{n}}\rangle$ can, e.g., be the spin-up and spin-down states
  in the $\hat{n}$-direction of a quantum spin-$\frac{1}{2}$  system.} can contain more information about the spin-direction than the parallel ones. Cf.~\cite{rsp_new_new2,rsp_new_new3,mani_new}.

 In another direction, a non-maximally entangled state was found to provide the best resolution for frequency  measurements in presence of decoherence~\cite{huelga_new}. Furthermore, it was discovered that a non-maximally entangled state  furnishes the highest violation of a certain Bell inequality~\cite{Acin-Durt}. It was also observed that  maximally entangled states do not have a special status when considering asymptotic local transformations between
two-party entangled quantum states~\cite{Horo_asym_ent}.
%(see also \cite{Dutta-Shaji, Vedral-against-Dutta-Shaji,shraaban})

\begin{figure}[t]
\begin{center}
\includegraphics[scale=0.14 ,angle =0]{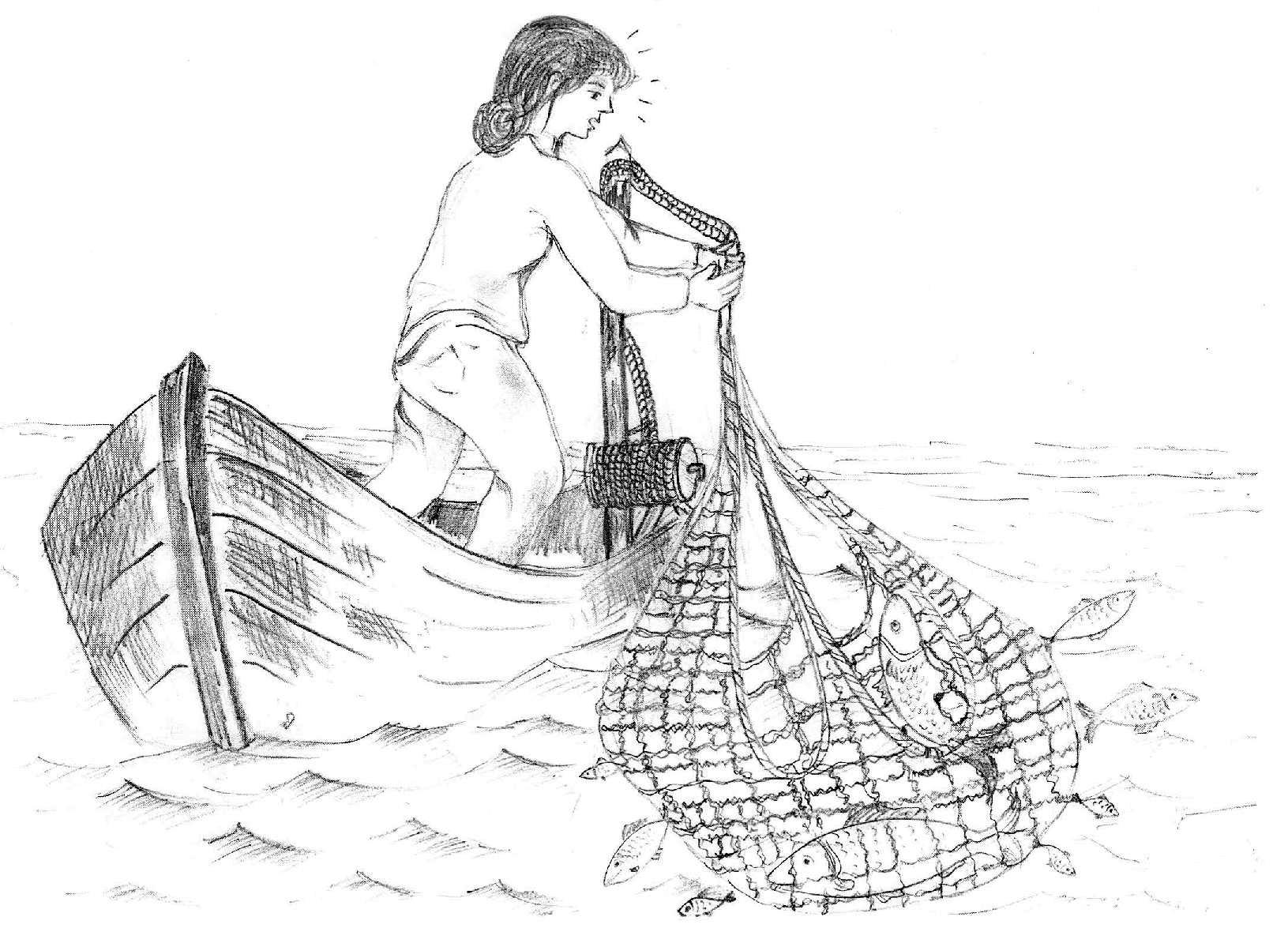}
\end{center}
\caption{About fishes, large and little. In any catch of fishes, the mesh would net
 large fishes and let out the little ones with the water. Likewise, the
net for non-separable states lets out some quantum correlated states along
with states that are deemed as having only ``classical'' correlations.
``Little'' fishes are by no means unimportant, as any fisherperson would
swear. And, for example, one remembers that in the 1940s and later, a
theater group evolved in India that was named ``Little Theatre'': they
certainly weren't staging insignificant pieces. [The sketch is by Mahasweta
Pandit.]
}
\label{fig:schematic_discord}
\end{figure}

These developments are some of the potential ones that have led researchers to address the  question whether entanglement is the only way to quantify quantum correlations present  in a shared quantum state, and whether there are
% Is entanglement the only resource for such tasks?
resources independent of entanglement
that can be used to implement quantum protocols with nonclassical efficiencies.  It turns out  that the non-separability sieve  can indeed be seen as leaving out some states that are quantum correlated in a different way. See figure~\ref{fig:schematic_discord}. One can fine-grain the sieve, via several approaches, and 
conceptualize disparate 
measures of quantum correlations beyond the entanglement-separability paradigm~\cite{Kavan-rmp,review-qd1,review-qd2,review-qd3,review-qd4}.
%It turns out that fine-grained quantification of quantum correlations leading to different quantum correlation measures, independent of entanglement,  is indeed possible  (for review, see~\cite{Kavan-rmp,review-qd1,review-qd2,review-qd3,review-qd4}). See figure \ref{fig:schematic_discord}.
Reviewing such quantum correlation measures and the ensuing  implications  is the main objective of this survey.

One of the first among such approaches was discovered around 2000, when Ollivier and Zurek~\cite{Oliver1,zurek-conditional} and Henderson and Vedral~\cite{Henderson1} proposed a measure of quantum correlations, known as quantum discord (QD), by quantizing concepts from
classical information theory~\cite{nou-do-igyara}. Around the same time, several other measures were introduced including quantum work deficit~\cite{Oppenheim1, Sir-mam2, Devetak1, Sir-mam3},  quantum deficit~\cite{QDefi1, QDefi2}, measurement-induced nonlocality~\cite{MIN}, etc.
Interestingly, there appeared in this way, quantum states of two or more parties that are not entangled, and yet quantum correlated.  The non-vanishing of quantum  discord for separable states may be contrasted with the fact that there exists an entanglement measure called distillable entanglement~\cite{Bennet_EoF,rains_new}, which is vanishing for certain entangled states, viz. the bound entangled states\footnote{ There exists a physical quantity called shared purity that can be zero for certain entangled states and non-zero for certain separable states~\cite{shared-anindyada}.}~\cite{bound-ent}. 

Sec.~\ref{general measures} reviews  definitions  of quantum correlation beyond entanglement and some general properties.
% In this review, we mainly focus on characterizations and applications of original definition of QD and  mention important results on other discord-like measures.
This is followed by strategies for  detection and the computational complexities of these measures which we briefly review in Secs.~\ref{sec:Computability} and~\ref{sec:witness_QD} respectively. Some attention is given to the class of states having vanishing QD in Sec.~\ref{sec:volume_QD}. Understanding  this set is useful for classifying the set of bipartite quantum states according to these quantum correlation measures.  

Quantum information processing tasks in which QD or discord-like measures are expected to be important are discussed in Sec.~\ref{Sec:Applications}.
%The works presented in this section is  debatable and hence  require further careful analysis to settle this issue and to establish QD as resource.
%However, it has already been confirmed that
QD can be an interesting  tool to detect cooperative phenomena like quantum phase transition and  disorder-induced-order in  quantum spin systems.   This is discussed in Sec.~\ref{Sec:manybody}. The relation  of QD with open quantum system is taken up in Sec.~\ref{sec:open_QD}.
%Moreover, the dynamical pattern of QD in the evolved state, both in closed (Sec.~\ref{Sec:manybody}) and in open systems  ( Sec.~\ref{sec:open_QD}) are important for realizing quantum information protocols. Implementation of any protocols in physical systems are always disturbed by the interaction of  the system-environment duo. Studying the interplay of quantum  and classical correlations between different parts of  the system and the environment is the key focus in Sec.~\ref{sec:open_QD}.

In Secs.~\ref{general measures} to~\ref{sec:open_QD}, investigations  are  restricted, in the main,  to bipartite states. We  move on to discuss QD for multipartite states in the succeeding sections. The  constraints  on the sharability of quantum correlations  between different parts of a multiparty quantum system has been referred to as the monogamy  of quantum correlations. Different aspects of this concept are considered in Secs.~\ref{sec:monogamy},~\ref{Sec:Connection} and~\ref{sec:app_discord monogamy score}. 
Definitions of a few multiparty quantum correlation measures are considered in Sec.~\ref{Disccord_other_multiparty}. Some miscellaneous items are collected in Sec.~\ref{sec:miss-QD}.
A short conclusion is presented in Sec.~\ref{end}.

\section{Measures of quantum correlations}
\label{general measures}
Quantification of quantum correlation (QC) present in any quantum state is one of the primary tasks related to the understanding and efficient utilization of the state for various quantum information processing schemes. In this review, we are mainly interested in  QC measures which are different from the ones conceptualized within the entanglement-separability paradigm. 
% Towards that aim, in this section, first we will discuss QC measures . Subsequently, in the forthcoming section we will provide definition of different discord and discord-like measures, in detail.
Quantum discord (QD) is a prominent example of such a measure. In this section, we first provide definitions of these QC measures in three categories: A. Measurement-based QD (Subsec.~\ref{meas}), B. Distance-based QD (Subsec.~\ref{dis}) and C. Other QC measures (quantum discord-like measures) (Subsec.~\ref{oth}). 
%The entire review is devoted on the discussion of the results, based on original measurement-based QD. The other direction in which substantional work has been carried out is geometric QD which we will define, briefly discuss some of its properties in Subsec.~\ref{Geometric_quantum_discord} and some of its applications in relevent subsections with QD.

\subsection{\bf Measurement-based quantum discord}
\label{meas}
There are several ways that lead to the concept of QD  of a bipartite quantum system. These can be classified into two broad categories of which one is based on measurement in any one of the subsystems, which we will discuss now. The other category consists of the distance-based measures which is discussed in the succeeding subsection.

%\subsubsection*{\bf Measurement based}
\subsubsection{Quantum discord}
Consider two classical random variables $X$ and $Y,$ for which the joint probability distribution of getting outcome $X=x$ and $Y=y$ is $p_{x,y}$. 
A measure of mutual interdependence of any of the variables on the other one is the classical mutual information~\cite{nou-do-igyara} between the variables, which 
% between two random variables \(X\) and \(Y\) 
can be written as 
\begin{equation}
\label{mut}
 I(X:Y) = H(X) + H(Y) - H(X,Y),
\end{equation}
where \(H(X)\) and \(H(Y)\) are the Shannon entropies\footnote{
%Let the classical variable  $A$, having probability distribution of obtaining $A=a$ be $p_a$. 
Let $A$ be a classical random variable, which takes the value $a$ with probability $p_a$. The Shannon entropy of $A$ is then given by
\begin{equation}
H(A)=-\sum_{a} p_a \log_2 p_a.
\end{equation}
}
 of the marginal distributions \(p_{x,.}\) and \(p_{.,y}\), with dots indicating variables that have been 
summed over and \(H(X,Y)\) is the Shannon entropy of the joint distribution \(p_{x,y}\).
The same quantity in Eq.~(\ref{mut}) can be expressed as 
\begin{equation}
\label{mute1}
 I(X:Y) = H(X) - H(X|Y),
\end{equation}
where the conditional entropy, \(H(X|Y)\), is defined as 
\begin{equation}
\label{eq:con1}
H(X|Y) = \sum_{y \in Y} p_y H(X|Y=y) = H(X,Y) - H(Y).
\end{equation}
A sleight-of-hand equivalence of these two definitions of mutual information can also be observed from Venn diagram representations of the entropic quantities.
 
These definitions of classical mutual information can be taken over to the quantum domain~\cite{Henderson1, zurek-conditional, Oliver1}. It was proposed that the quantum version of the first definition, the quantum mutual information, can be obtained by replacing Shannon entropies by von Neumann entropies\footnote{The von Neumann entropy~\cite{concave1} of a density matrix \(\sigma\) is given by \(S(\sigma) = -\mbox{tr}(\sigma \log_2 \sigma)\), which reduces to $-\sum_i \lambda_i \text{log}_2 \lambda_i$, where 
\(\lambda_i\) are the eigenvalues of \(\sigma\).}
\cite{Nielsen} in Eq.~(\ref{mut}). For a bipartite quantum state \(\rho_{AB}\), shared between two parties, $A$ and $B$, usually referred to as Alice and Bob, possibly situated in two distant locations, the quantum mutual information is defined as
\begin{equation}
\label{Eq:quant1}
 I_{AB} = S(\rho_A)+S(\rho_B)-S(\rho_{AB}),
\end{equation}
where $\rho_i=\mbox{tr}_j(\rho_{AB})$ 
$(\{i,j\} \in \{A, B\}, i \neq j)$ are local density matrices of $\rho_{AB}$. One may similarly try to quantize the concept of conditional entropy, which would then lead us to a quantization of classical mutual information, as defined via Eq.~(\ref{mute1}). However, 
% But the problem in this case is that 
 replacing Shannon entropy to von Neumann~\cite{Nielsen} in Eq.~(\ref{eq:con1}) leads to a quantity which can be positive as well as negative~\cite{cerf-negative,HOW05,HOW07,Li07}. The quantum conditional entropy of \(\rho_{AB}\) was argued to be given by 
\begin{equation}
\label{cond_entropy}
S_{A|B} =  \min_{\{\Pi^B_k\} \in \mathcal{M}^B} \sum_k p_k S(\rho_{A|k}),
\end{equation}
where the minimization is taken over all quantum measurements, \(\{\Pi^B_k\}\), performed on the system  $B$, and \(\mathcal{M}^B\) forms the set of all such measurements. 
Here, \(\{p_k,\rho_{A|k}\}\) is the post-measurement ensemble that is formed at Alice's side, where
 \(\rho_{A|k} = \mbox{tr}_B (\mathbb{I}^A_m \otimes \Pi^B_k \rho_{AB} \mathbb{I}^A_m \otimes \Pi^{B\dagger}_k)/p_k\), with \(p_k = \mbox{tr}_{AB} (\mathbb{I}^A_m \otimes \Pi^B_k \rho_{AB} \mathbb{I}^A_m \otimes \Pi^{B\dagger}_k)\), and with \(\mathbb{I}^A_m\) being the identity operator on the 
Hilbert space of Alice's subsystem\footnote{Throughout the review, we consider bipartite states on \(\mathbb{C}^m \otimes \mathbb{C}^n\), except when considering continuous variable systems.} with dimension $m$. Therefore, the second form of the classical mutual information, as given in Eq.~(\ref{mute1}), when quantized in the way mentioned above, gives us the quantity
%another form of classical mutual information in the quantum domain can be expressed as 
\begin{equation}
\label{Eq:quant2}
 J_{A|B} = S(\rho_A) - S_{A|B}. 
\end{equation}
It can be shown that in general, \(I_{AB} \geq J_{A|B}\). However, the inequality can be strict, and indeed it was noticed that they are unequal for almost all two-party quantum states~\cite{Vedral-against-Dutta-Shaji}. Moreover, $I_{AB}$ and $J_{A|B}$ are argued to quantify total correlations~\cite{goRar-katha} and classical correlations~\cite{Henderson1} respectively of a bipartite state $\rho_{AB}$. Therefore, for a given two-party quantum state, \(\rho_{AB}\),
the difference between these two quantities, given in Eqs.~(\ref{Eq:quant1}) and~(\ref{Eq:quant2}) was proposed to be a measure of QC and was called as quantum discord (QD) \cite{Oliver1,zurek-conditional}, given by
\begin{equation}
\label{Eq:discord1}
\mathcal{D}^{\leftarrow}(\rho_{AB}) = I_{AB} - J_{A|B}.
% = S(\rho_A) - S_{A|B}. 
\end{equation}
The notation ``$\leftarrow$" in the superscript of QD denotes that the measurement has been performed in the subsystem `$B$' while $\mathcal{D}^{\rightarrow}$ denotes QD for the measurement in the first subsystem, i.e. in `$A$'. Unless defined otherwise, we will henceforth consider the quantum discord $\mathcal{D}^{\leftarrow}$, and denote it for convenience\footnote{Since we are using $2$ as the base of the logarithm, in the definition of the von Neumann entropy, 
the unit of QD will be ``bits".} as $\mathcal{D}$.
%For our convenience, from now onwards we use $\mathcal{D}^{\leftarrow}$ and denote it, unless further specified as $\mathcal{D}$.
The definition of QD also provides a justification for 
%natural way to motivate 
considering a maximization in the definition of \(J_{A|B}\), since  to obtain the amount of 
QC present in the state, one must pump out all the classical correlations from the total correlations, assuming that total correlations contain only classical and quantum correlations, and that the constituents are additive. Although we will predominantly be dealing with the case when the measurement in the definition is a projective-valued (PV) 
one, positive-operator valued measurements\footnote{A positive operator valued measure (POVM)~\cite{Nielsen} is a set of
 generalized measurement operators $\{{\cal A}_i\}$, which are positive semidefinite, and acts on a quantum state $\rho$ in the following way:
 \begin{equation}
 \rho\rightarrow\rho_i = {\cal A}_i \rho {\cal A}^{\dagger}_i/p_i, ~\text{with}~ p_i = \text{tr}({\cal A}_i \rho {\cal A}^{\dagger}_i),
\end{equation}  
where $\sum_i{\cal A}^{\dagger}_i{\cal A}_i = \mathbb{I}$, and $p_i$ is the probability  of obtaining the  post-measurement  state $\rho_i $.
 } (POVMs) have also been considered for defining QD. Indeed, POVMs are already present in the definition of classical correlation, $J_{A|B}$ in Ref.~\cite{Henderson1}. 
In general, a definition of QD that utilizes POVMs is useful in relating the quantity to other information-theoretic quantities like accessible (classical) information~\cite{nou-do-igyara} through the Holevo bound~\cite{Holevo} and the entanglement of formation (see Appendix \ref{Entanglement of formation}) through the Koashi-Winter relations~\cite{koashi_winter}.
 Performing a POVM, however, may render a physical system open and, therefore, has to be cautiously used while providing thermodynamic interpretation of QD and related quantitites~\cite{benn-demon1,benn-demon2,landauer1,szilard,zurek2003,Oppenheim1,Sir-mam2,Sir-mam3}. It is interesting to note here that
  projective measurements are shown  to be optimal among all POVMs for rank-2 bipartite quantum states~\cite{Galve2}. On the other hand, there exist states already for two-qubits, for which projective measurements are not optimal~\cite{povm-qd1,Chen3,Galve2,povm-qd2,lang-caves-povm}. See also Ref.~\cite{DAriano}.

\begin{figure}[!ht]
\centering
\includegraphics[width=0.95\columnwidth,keepaspectratio]{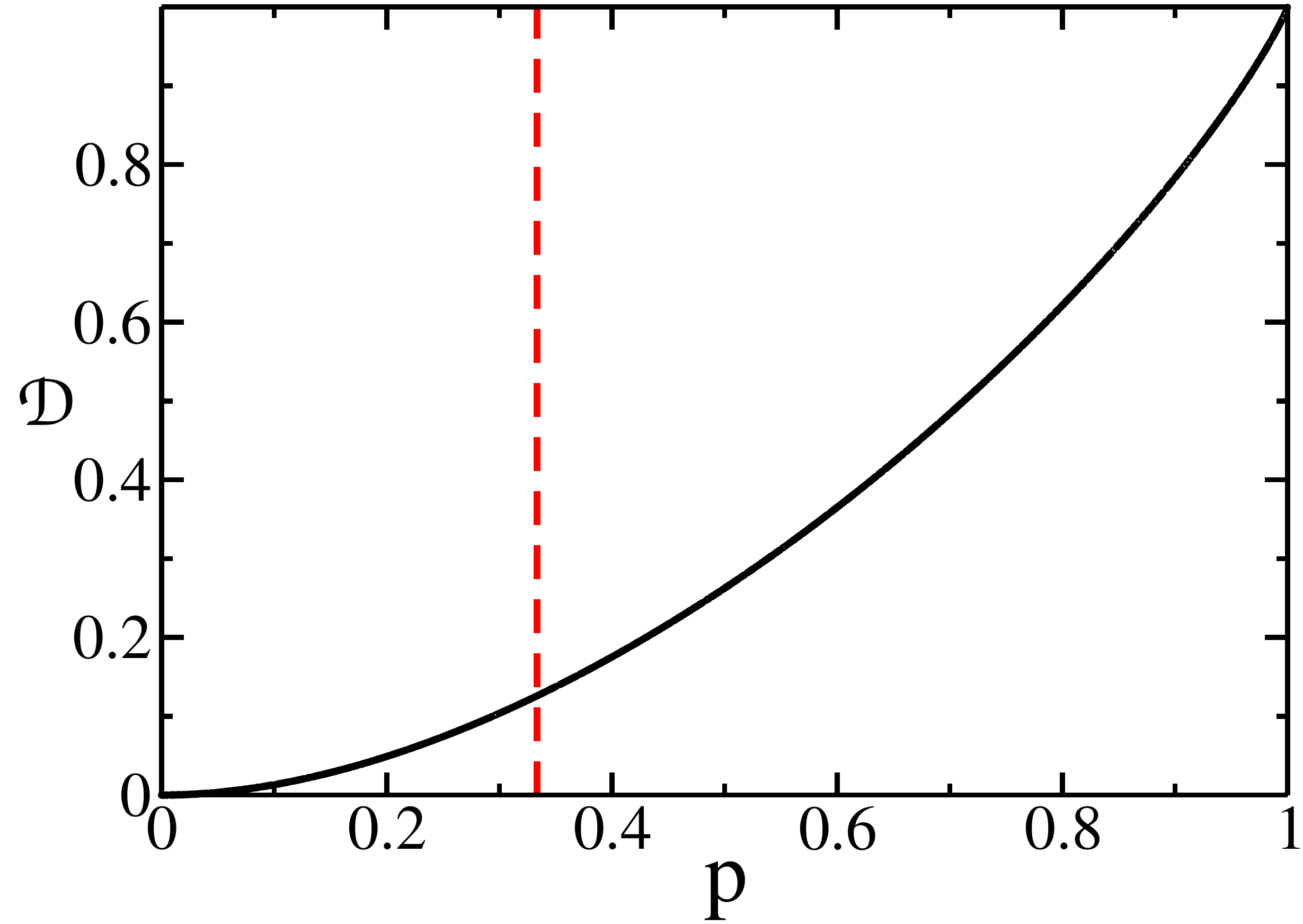}
\caption{Quantum discord for the Werner states $\rho_W = \frac{1-p}{4} \mathbb{I}_2\otimes\mathbb{I}_2 +p |\psi^-\rangle \langle \psi^-|$. The red dashed line corresponds to $p=1/3$ below which the state is unentangled.  [Adapted from Ref.~\cite{Oliver1} with permission. Copyright 2001 American Physical Society.]}
\label{werner_plot} 
\end{figure}

%A major portion of this review is about properties of the QC just defined. 
Let us begin here by enumerating some properties of QD, which come to the mind rather immediately, or which are used more frequently later in the review. 
%After the definition, we now discuss the properties of QD which are given as follows:  
\begin{itemize}
\item[a)] $\mathcal{D}(\rho_{AB}) \geq 0$, since $I_{AB} \geq J_{A|B}$. 
%To obtain this, one uses the fact that conditional entropy is a concave function~\cite{concave1,zurek-conditional}, and that quantum mutual information decreases under local measurement~\cite{mutual-dekh}.
\item[b)] QD is not symmetric, i.e., in general, $\mathcal{D}^{\leftarrow}(\rho_{AB}) \neq \mathcal{D}^{\rightarrow}(\rho_{AB})$.
% unless $\rho_{AB}$ is symmetric. 
This is clearly visible, as conditional entropy is not symmetric for all states. They, of course, coincide for states which are symmetric under interchange of the two parties (cf.~\cite{Diff_discord}).
%\textcolor{orange}{For any arbitrary bipartite state, the difference between these two kind of discord i.e., between $\mathcal{D}^{\rightarrow}$ and $\mathcal{D}^{\leftarrow}$  has been quantified in  and a tight upper bound has been provided.}
\item[c)] QD is invariant under local unitary transformations, i.e., $\mathcal{D}(\rho_{AB})=\mathcal{D}[(U_A \otimes U_B) \rho_{AB} (U_A \otimes U_B)^\dagger]$, for arbitrary unitaries $U_A$ and $U_B$ on the subsystems $A$ and $B$. One of the important characteristics of von Neumann entropy is that it is invariant under unitary transformations, and hence $I_{AB}$ is invariant under local unitaries. Consider now the effect of a local unitary transformation $U_A \otimes U_B$, acting upon the state $\rho_{AB}$, on the quantum conditional 
entropy. Suppose that the minimum in Eq.~(\ref{cond_entropy}) is reached in the measurement $\{\tilde{\Pi}_k^B\}$, for the state $\rho_{AB}$. For the local unitarily transformed state $(U_A \otimes U_B) \rho_{AB} (U_A \otimes U_B)^\dagger$, a measurement $\{\Pi_k^B\}$ leads to the ensemble $\{p'_k,\rho'_{A|k}\}$, 
where $\rho'_{A|k}=U_A \mbox{tr}_B (\mathbb{I}^A_m \otimes \Pi{'}^{B}_k~\rho_{AB}~\mathbb{I}^A_m \otimes \Pi{'}^{B \dagger}_k) U_A^\dagger/{p'_k}$, $p'_k=\mbox{tr}_A [U_A \mbox{tr}_B  (\mathbb{I}^A_m \otimes \Pi{'}^{B}_k~\rho_{AB}~\mathbb{I}^A_m \otimes \Pi{'}^{B \dagger}_k) U_A^\dagger]$, $\Pi{'}^{B}_k=U^\dagger_B \Pi_k^B U_B$. Thereby, the optimization of the local unitarily transformed state is reached in the measurement $\{\tilde{\Pi}{'}^{B}_k=U^\dagger_B \tilde{\Pi}^B_k U_B\}$, and leads to the same value of the quantum conditional entropy as of $\rho_{AB}$.
% Under the local unitary transformation, $U_A \otimes U_B$,  the term $S(\rho_{A|k})$, in the conditional entropy  can be expressed as
% $U_A \mbox{tr}_B(\mathbb{I}^A \otimes \pi^B{'} \rho_{AB} \mathbb{I}^A \otimes \pi^B{'}^\dagger){U_A}^\dagger$, where $\pi^B{'}=\pi^B U^B$. Thus the optimization with respect to $\pi^B{'}$, end up with a same result as above, prove the property.
\item[d)] QD is zero if and only if their exists a local measurement on $B$ that does not disturb the quantum system~\cite{Oliver1,Vedral-against-Dutta-Shaji,Dutta_thesis}. 
\item[e)] For a bipartite pure state, QD reduces to entanglement, i.e., von Neumann entropy of the local density matrices.
\item[f)] QD is upper bounded by the von Neumann  entropy of the measured subsystem\footnote{The Authors in Ref.~\cite{Wang-monogamydekh} have given a necessary and
sufficient condition for saturation of this upper bound of QD, by using the conditions for equality
of the Araki-Lieb inequality~\cite{Lieb1,Lieb2,Lieb3}.} $B$ i.e. 
$\mathcal{D}^{\leftarrow}(\rho_{AB}) \leq S(\rho_B)$~\cite{mutual-dekh,continuity-discord}, while $J_{A|B} \leq \min \{S(\rho_A), S(\rho_B)\}$~\cite{luo-li1}.
\end{itemize}
While QD and entanglement coincide for pure states, by considering mixed states, it can be shown that QD is different than entanglement. Specifically, it is non-vanishing for some separable states. An example which illustrates this is the class of Werner state~\cite{Werner}, given by 
\begin{equation}
\rho_W = \frac{1-p}{4} \mathbb{I}_2\otimes\mathbb{I}_2 +p |\psi^-\rangle \langle \psi^-|
\label{eqn:werner}
\end{equation}
 with $|\psi^-\rangle=\frac{1}{\sqrt{2}} (|01\rangle-|10\rangle)$ being the singlet state. Here $\mathbb{I}_n$ denotes the identity operator on the $n$-dimensional complex Hilbert space.  It is separable when $p \leq \frac{1}{3}$. However, $\mathcal{D}(\rho_W)>0$ in the entire range of $p$ except at $p=0$~\cite{Oliver1, Luo} (see figure~\ref{werner_plot}).

\subsubsection{Gaussian quantum discord}
\label{GaussianQD}
The concept of QD has been extended to continuous-variable systems, specifically to the case of two-mode Gaussian states~\cite{pirandola1,duan-gauss,Braunstein-gauss,ragy-gauss,arun-book-continuous,Giedke-thesis,cirac1, mista1,wenger-gauss,bowen-gauss,adesso-illu,simon-gauss}.
When the measurement involved in QD  is restricted only to the set of Gaussian measurements \cite{paris1, adesso1},  it is  called Gaussian QD, which is an upper bound of QD of continuous-variable systems.
%i.e., measurements as well as states are Gaussian. 
%In particular,  can be defined by following the original definition of QD given in Eq.~(\ref{Eq:discord1}) where the measurement required can now be restricted to the 
% under the restriction that the  measurements on $B$ are the two-mode 
%Gaussian POVMs~\cite{paris1, adesso1}. 
Note that these measurements can be implemented by 
% are all viable using 
 linear optics and homodyne detection.
% is defined for bipartite Gaussian states exactly like the ordinary QD, except that the measurement involved in QD is restricted to generalized Gaussian positive operator valued  measurements (POVMs),that can be implemented by  

Gaussian QD was first evaluated for %\cite{paris1} where initially the anaytical form was computed for 
squeezed thermal states  \cite{paris1, adesso1} and then further extended to arbitrary two-mode Gaussian states~\cite{adesso1}. %Here one should note that any mapping from G. S. to G. S. is known as Gaussian operations \cite{mukunda1,dutta1} and unlike in Eq.~[\ref{Eq:discord1}], the optimization involve in computing  Gaussian QD is generalized local Gaussian positive operator valued measure (POVM). \cite{adesso1,paris1}.
 It has been shown~\cite{paris1, adesso1} that almost all two-mode Gaussian states have non-zero Gaussian QD. Moreover, squeezed thermal states are entangled if 
 $\mathcal{D}_{Gauss} (\rho_{AB})>1$. However, if 
% they may be entangled or separable if
  $\mathcal{D}_{Gauss} (\rho_{AB}) \leq 1$, conclusive identification of entangled state is as yet not possible.  
 %For two-mode separable Gaussian states, Gaussian QD possesses a value less than or equal to unity. % of two-mode separable Gaussian states is less than or equal to 1. 
%Hence, Gaussian QD can detect entangled Gaussian states whenever it is more than unity.
%\textcolor{red}{Bujhte parchina r ki ki deoa uchit}.

Quantum correlation in continuous variable systems beyond Gaussian states has also been investigated. 
% in Ref.~\cite{tatham-gauss}.  
For a particular type of non-Gaussian two-mode 
 Werner states~\cite{filip-gauss}, which is obtained by mixing a two-mode squeezed state with the vacuum, 
 QD has been  computed analytically. See also~\cite{tatham-gauss}. Giorda \emph{et al.}~\cite{giorda-gaussdekho} asked whether non-Gaussian measurements can be optimal for obtaining QD for Gaussian states. 
%can lead to a stronger minimization in Gaussian QD than the Gaussian measurements. 
To address this query, two-mode squeezed thermal states and mixed thermal states have been studied
 by considering a range of experimentally feasible non-Gaussian measurements. It is observed that  Gaussian measurements always provide the optimal value of Gaussian QD~\cite{giorda-gaussdekho}.  Moreover, there are numerical evidences which also reveal that QD for Gaussian  states require only Gaussian measurements~\cite{giorda-gaussdekho,Olivares-gauss}. 
 %Hence, both theoretical and experimental observations indicate that QD for Gaussian  state may not require to include non-Gaussian measurements.
%  is the original QD for Gaussian states
Pirandola \emph{et al.}~\cite{pirandola-gaussdekho} connected the Gaussian QD with the 
% solved conjecture~\cite{guha-gauss} on the bosonic minimum output entropy~\cite{CERF-GAUSS} (see also~\cite{mari-gauss}) which states 
result that the minimum von Neumann entropy at the  output of a bosonic Gaussian channel is achieved by Gaussian input states~\cite{guha-gauss,CERF-GAUSS} (see also~\cite{mari-gauss}). The Authors
 showed that the solution of the minimization problem for the bosonic Gaussian channel implies the optimality of QD by using Gaussian measurements for a large family of Gaussian states. It is important to note that
 %the minimization of the bosonic output entropy implies the optimality of Gaussian QD for a class of  Gaussian states including squeezed thermal states.In subsequent years, 
 several experiments have been performed and proposals for the same given to detect and measure Gaussian QD~\cite{madsen-gauss,Chille-gauss,qars-gauss,meda-gauss}.

\subsubsection{Symmetric quantum discord}
\label{bipartite global qd}
The original QD~\cite{Henderson1,Oliver1} in Eq.~(\ref{Eq:discord1}) is not symmetric under the exchange  of $A$ and $B$ \cite{Diff_discord}. However, by performing von Neumann measurements $\{\Pi_i^A \otimes \Pi_j^B\}$ on the entire system, a symmetric version of QD~\cite{rulli-sarandy1} can be defined. Before presenting the definition of the symmetric version of QD, it is useful to rewrite the original QD in the following way. Note first that the quantum mutual information of a bipartite quantum state can be expressed in the following way: % as a relative entropy distance between $\rho_{AB}$ and $\rho_A \otimes \rho_B$:
\begin{equation}
\label{mutual-distance}
I(\rho_{AB})=S(\rho_{AB}||\rho_A \otimes \rho_B),
\end{equation}
 where $\rho_A$ and $\rho_B$ are local density matrices of $\rho_{AB}$. The relative entropy between the two quantum states  $\sigma$ and $\xi$ is given by
\begin{equation}
\label{rel-main}
S(\sigma||\xi)=\tr(\sigma \log \sigma-\sigma \log \xi).
\end{equation}
Clearly, it is not a symmetric function of its arguments, and therefore does not conform to the usual notion of a distance. However, time and again, this ``non-standard" distance turns up in different formulae and notions in many areas including in quantum information. Furthermore, 
% Reverting back to quantum mutual information, it is easy to see that $I(\rho_{AB})=S(\rho_{AB}||\rho\otimes \rho_B)$.
one can see that for rank-1 PV measurements,  $J_{A|B} (\rho_{AB})=\max_{\{\Pi_k^B\} \in \mathcal{M}^B} S(\phi_{B}(\rho_{AB})||\rho_A \otimes \phi_B(\rho_B))$. 
%where  and $\phi_{B}(\rho_{B})=\sum_{k} \Pi_k^B \rho_{B} \Pi_k^B$. 
A symmetric version of QD can now be defined as
\begin{eqnarray}
\label{sym-qd1}
\mathcal{D}_{sym} (\rho_{AB})=\min_{\{\Pi_i^A \otimes \Pi_j^B\}} [S(\rho_{AB}||\rho_A \otimes \rho_B)-\nonumber\\
S(\phi_{AB}(\rho_{AB})||\phi_A(\rho_A) \otimes \phi_B(\rho_B))].
\end{eqnarray}
Here
\begin{eqnarray}
\label{new-somikoron}
\phi_{AB}(\rho_{AB})=\sum_{i,j} (\Pi_i^A \otimes \Pi_j^B) \rho_{AB} (\Pi_i^A \otimes \Pi_j^B),\nonumber\\
\phi_{B}(\rho_{AB})=\sum_{k} \mathbb{I}_m^A \otimes \Pi_k^B \rho_{AB} \mathbb{I}_m^A \otimes \Pi_k^B,  \nonumber\\
\phi_\alpha(\rho_\alpha)=\sum_{k} \Pi_k^\alpha \rho_\alpha \Pi_k^\alpha,~\alpha=A, B.
\end{eqnarray}
One can rewrite  Eq.~(\ref{sym-qd1}) in terms of quantum mutual information $I$ as~\cite{symmqd3}
\begin{equation}
\label{sym-qd2}
\mathcal{D}_{sym}^{mutual} (\rho_{AB})=\min_{\{\Pi_i^A \otimes \Pi_j^B\}} [I(\rho_{AB})-I(\phi_{AB} (\rho_{AB}))].
\end{equation}
It expresses the minimal amount of correlations which are lost due to the measurements~\cite{okrasa1}. 
A similar interpretation is possible for the original QD, but for measurement performed only on one party~\cite{Geom_discord1,okrasa1}. The symmetric version of QD has also been considered for the case when POVMs are used for the measurements at the two parties~\cite{Piani_nolocalbroad,symmqd4}.
%It is important to mention here that Eq.~(\ref{sym-qd2}) is equivalent to the The symmetric version of QD becomes 
$\mathcal{D}_{sym}$ is equivalent to what has been termed as measurement-induced disturbance
(MID)~\cite{QDefi2}, if instead of considering the minimization, the measurement is performed in the eigenvectors of the reduced density matrices of each part. 
%Clearly, MID is not a good measure of quantum correlation~\cite{symmqd5} 
Since MID does not consist of any optimization over the local measurements, it usually returns an overestimation of the  amount of nonclassical correlations compared to the symmetrized version of QD~\cite{symmqd5}.
% However, Eq.~(\ref{sym-qd2}) is known as an ameliorated version of the measurement-induced disturbance (or AMID)~\cite{symmqd3,symmqd4} where optimization is achieved over arbitrary complete local projective measurements~\cite{symmqd3} or POVMs~\cite{symmqd4} which can be performed on both the subsystems $A$ and $B$.
An analytical formula of the symmetric version of QD has been discussed in Ref.~\cite{symmqd1} for the Bell-diagonal (BD) states, given by 
\begin{equation}
\label{bell-diagonal}
\rho_{AB}=\frac{1}{4}\left(\mathbb{I}_2 \otimes \mathbb{I}_2+\sum_{i=x,y,z} T_{ii} \sigma^i \otimes \sigma^i\right),
\end{equation}
where $T_{ii}$ are reals with $-1 \leq T_{ii}\leq 1$ $\forall i$. 
The \(T_{ii}\)'s must also satisfy the additional condition coming from the constraint that the state is
positive semidefinite. \(\sigma^i\), \(i=x,y,z\), are the Pauli spin-1/2 operators.
% with $\lambda_1 \geq \lambda_2 \geq \lambda_3 \geq \lambda_4$  and $|B_{1,2} \rangle=\frac{|00 \rangle \pm |11 \rangle}{\sqrt{2}} \equiv |\phi^ \pm \rangle$, $|B_{3,4} \rangle=\frac{|01 \rangle \pm |10 \rangle}{\sqrt{2}} \equiv |\psi^ \pm \rangle$.
 The Authors in Ref.~\cite{symmqd2} presented an experimental implementation of a witness operator for the symmetric version of QD in a composite system. %In this regard, Eq.~(\ref{sym-qd1}) can be expressed in the following form:

The symmetric version of QD can also be expressed as
\begin{eqnarray}
\label{sym-qd3}
\mathcal{D}_{sym}^f (\rho_{AB})=\min_{\{\Pi_i^A \otimes \Pi_j^B\}} [S(\rho_{AB}||\phi_{AB}(\rho_{AB}))-\nonumber\\
S(\rho_{A}||\phi_A(\rho_A))-S(\rho_{B}||\phi_B(\rho_B))].
\end{eqnarray}
We will discuss the extensions of some of these forms 
to the multiparty domain. One of these extensions will give rise to the concept of global QD, discussed in Sec.~\ref{Sec:Global_QD}.
Note that Eq.~(\ref{sym-qd3}) can be seen as the difference between two terms, one of which is $S(\rho_{AB}||\phi_{AB}(\rho_{AB}))$, that can be interpreted as the ``global distance" of the state $\rho_{AB}$ from the resultant state after local measurements have been carried out. This global distance can be seen as the quantum correlations in $\rho_{AB}$ except that there can be local contributions to this distance in the form of $S(\rho_{A}||\phi_{A}(\rho_{A}))$ and $S(\rho_{B}||\phi_{B}(\rho_{B}))$, and the sum of these two expressions forms the second term, which is subtracted from the global distance to obtain the symmetric version of QD.
%the first term, which is global on A and B, and the second term, which is the sum of two local contributions.

Another symmetrized version of QD is the ``two-way quantum discord", defined as~\cite{symmqd6}
\begin{equation}
\mathcal{D}^\leftrightarrow (\rho_{AB})=\max\{\mathcal{D}^\leftarrow (\rho_{AB}),\mathcal{D}^\rightarrow (\rho_{AB})\}.
\end{equation} 
%Where `$\rightarrow$' and `$\leftarrow$' indicate that the measurements performed on the subsystem $A$ and $B$ respectively. 
Other versions of symmetric QD are discussed in  Refs.~\cite{symmqd7,symmqd8}.

\subsection{\bf Distance-based quantum discord}
\label{dis}
%Upto now, we have discussed about QD which is based on the concept of classical information theory.
We have until now tried to conceptualize QD by quantizing certain concepts in classical information theory. Since such definitions of QD involve optimization over sets of local measurements, the computation of which is in general a challenging task. Moreover, while dealing with the theory of entanglement, we have realized that the quantifications of entanglement originating from different concepts lead to new insights in quantum information.

%One of the challenging task in computing QD using Eq.~(\ref{cond_entropy}) is that it  involves optimization over  all possible local measurements performed on any one of the subsystems. Though presence of symmetry in some cases may reduce the computational complexity to some extent \cite{Luo,Ali,Chen2,Computational_complexity,Huang2},  in general, for an arbitrary quantum state it is quite difficult to end up with an analytical form of QD. 

In this subsection, we are  going to discuss the distance-based formulations of QD. The minimization involved in this definition, can often be performed explicitly and hence it becomes a convenient tool for analyzing QC associated with the system. In general, the distance between two quantum states can be defined in several ways~\cite{distance_based}. Here, we consider two broad directions by which distance measures are defined, namely, the relative entropy and the norm distance.
%, and discuss various applications of the QD, defined using them.

In the preceding subsection, we have seen that the original information-theoretic version of QD can be written as the difference between two relative entropy distances. The relative entropy-based QD that is considered in this subsection is a qualitatively different one, and is akin to the relative entropy of entanglement\footnote{See Appendix~\ref{Sec:ReoE} for a definition of the relative entropy of entanglement.}
 and geometric measures of entanglement. The idea here is to consider a set of states that are devoid of quantum correlations in some sense. Quantum correlation of a given state is then defined as the minimal distance of the state from that set.

\subsubsection{Relative entropy-based discord}
\label{relative_QD}
The concept of entanglement has led to the realization that there is a class of states, the separable states\footnote{A separable state is of the form
\begin{equation}
\label{Eq:seperable-defn}
\eta_{AB} = \sum_{i} p_{i} \eta^{i}_A \otimes \eta^{i}_B,
\end{equation}
where $\eta^{i}_j$ is a density matrix of the $j^{th}$ site and \(p_{i} \geq 0\) with \(\sum_{i} p_{i} =1\). These are exactly those states that can be prepared by LOCC between the sites. %However, the difference between Eq.~(\ref{Eq:classical_state}) and~ (\ref{Eq:seperable-defn}) is that $\sigma^{i}_A$ or $\sigma^{i}_B$ may not always orthonormal.
}, which are ``useless" for certain tasks and have zero entanglement. This in turn has been used to quantify entanglement by measuring the shortest distance of entangled state to the set of separable states~\cite{knight1, vedral-plenio1, bravyi1}. In similar vein, one is led to the set of ``quantum-classical"(q-c) states having the form

%Clearly it is not a symmetric measure. A unified view of total, classical, quantum correlations in both the bipartite  as well as in multipartite  scenarios is developed by Modi \emph{et al.}~\cite{relative1} with the help of relative entropy. The approach~\cite{relative1} is based on the idea that ``a distance from a given state to the closest state without the desired property (e.g., entanglement or discord) is a measure of that property". For example, the relative entropy of entanglement, 
%$\mathcal{E}_R$, ~\cite{knight1, vedral-plenio1, bravyi1} (see Appendix \ref{Sec:ReoE}) is defined as the shortest distance  of a quantum state $\rho_{AB}$  from a separable state $\sigma_{AB}$\footnote{A separable state is
%\begin{equation}
%\label{Eq:seperable-defn}
%\sigma_{AB} = \sum_{i_A, i_B} p_{i_A i_B} \sigma^{i}_A \otimes \sigma^{i}_B,
%\end{equation}
%where $\sigma^{i}_j$ is the density matrix in the $j^{th}$ site and \(p_{i_A i_B} \geq 0\) with \(\sum_{i_A, i_B} p_{i_A i_B} =1\). These states can also be prepared by LOCC. However, the difference between Eq.~(\ref{Eq:classical_state}) and~ (\ref{Eq:seperable-defn}) is that $\sigma^{i}_A$ or $\sigma^{i}_B$ may not always orthonormal.}. 
%
%Now before giving the definition of QD based on the relative entropy, let us first define the quantum-classical (QC) state $\chi_{AB}^{QC}$ in the following form
\begin{equation}
\label{defn-qc}
\chi_{AB}^{\text{q-c}}=\sum_i p_i \rho_i \otimes |\phi_i\rangle \langle \phi_i|
\end{equation}
with $p_i \geq 0$, $\sum_i p_i=1$, $\langle\phi_i|\phi_j\rangle=\delta_{ij}$, and $\rho_i$'s belonging to the subsystem $A$. Clearly, for the q-c state, there exists a von Neumann measurement on the subsystem $B$ that does not perturb the state. They form the class of ``useless" states for tasks where $\mathcal{D}^\leftarrow$ is predicted to be a resource. The relative entropy-based QD with distance being considered from the q-c states (with the set of q-c states being denoted below as ``q-c''), for a state $\rho_{AB}$  is given by~\cite{relative1}
\begin{equation}
\label{Eq:Rel-qc}
\mathcal{D}_{rel}^{\text{q-c}} (\rho_{AB})=\min_{\chi_{AB}^{\text{q-c}}\in \text{q-c}} S(\rho_{AB}||\chi_{AB}^{\text{q-c}}).
\end{equation}
%Here $QC$ denotes the set of all quantum-classical states over which the minimization is considered. 
Note that the role of $A$ and $B$ will be exchanged for ``classical-quantum" (c-q) states, which are of the form
\begin{equation}
\label{Eq:defn-cq}
\chi_{AB}^{\text{c-q}}=\sum_i p_i |\phi_i\rangle \langle \phi_i| \otimes \rho_i,
\end{equation}
with the same conditions stated above, except that  
$\rho_i$'s belong to the subsystem $B$. These are exactly the set of ``useless" states for tasks for which $\mathcal{D}^\rightarrow$ is considered to be useful. One can now define a $\mathcal{D}_{rel}^{\text{c-q}}$ using the c-q states.
% One can similarly define the relative entropy-based QD in the above manner. 

We will now briefly discuss the relative entropy-based QD where the distance is taken from the bipartite ``classical-classical" (c-c) states $\chi_{AB}^{\text{c-c}}$, given by
\begin{equation}
\label{Eq:classical_state}
\chi_{AB}^{\text{c-c}} = \sum_i p_i |\phi_i\rangle \langle \phi_i| \otimes |\psi_i\rangle \langle \psi_i|,
\end{equation}
with $\langle\phi_i|\phi_j\rangle=\delta_{ij}$ and 
$\langle\psi_i|\psi_j\rangle=\delta_{ij}$.
% Note that all QC, CQ, CC states can be prepared by local operations and classical communications (LOCC)~\cite{Werner}. The relative entropy-based QD of a bipartite quantum state $\rho_{AB}$ is then defined as the shortest distance of a given state from CC state $\chi_{AB}^{CC}$. 
Mathematically, the measure can be expressed as 
\begin{equation}
\label{Eq:ReD_def}
\mathcal{D}_{rel} (\rho_{AB})=\min_{\chi_{AB}^{\text{c-c}}\in C} S(\rho_{AB}||\chi_{AB}^{\text{c-c}}).
\end{equation}
Here $C$ denotes the set of all c-c states. 
%Following the same spirit, in this case, the shortest distance is taken from the states $\chi_{1 2 \ldots N}$ which contains only classical correlations, while $\rho_{12\ldots N}$ contains all possible correlations.

It is important to mention here that the set of q-c, c-q and c-c states are all subsets of the set of separable states, which form a convex set, while the formers do not. 

One may now define a classical correlation based on relative entropy distance of the state $\rho_{AB}$, as 
\begin{equation}
J_{rel}^{\text{q-c}} (\rho_{AB})=S(\chi^{\text{q-c}}_{\rho_{AB}} || \Pi_{\chi^{\text{q-c}}_{\rho_{AB}}}),
\end{equation}
where $\chi^{\text{q-c}}_{\rho_{AB}}$ and $\Pi_{\chi^{\text{q-c}}_{\rho_{AB}}}$ are respectively the closest q-c state of $\rho_{AB}$ 
and the closest product state\footnote{A product state of two parties is of the form 
\(\tilde{\eta}_A \otimes \tilde{\eta}{'}_B\).} of  $\chi^{\text{q-c}}_{\rho_{AB}}$, with respect to the relative entropy distance. % Clearly the quantum mutual information in Eq.~(\ref{mutual-distance}) is nothing but the total correlation $T_{rel}$.  
Interestingly, it was found that $I-(\mathcal{D}^{\text{q-c}}_{rel}+ J^{\text{q-c}}_{rel})=-\mathcal{L}$ where $\mathcal{L}=S(\Pi_{\rho_{AB}} || \Pi_{\chi^{\text{q-c}}_{\rho_{AB}}})$~\cite{relative1} and $I$ is the %total correlation which is nothing but the 
quantum mutual information given in Eq.~(\ref{mutual-distance}). Here $\Pi_{\rho_{AB}}$ is the closest product state of $\rho_{AB}$. A similar relation among the same quantities, but by using a linearized variant of relative entropy has also been addressed in Ref.~\cite{rel-ref2}. %Additionally the Authors~\cite{rel-ref3} has established the relation between relative entropy and QD for projecive measurements. 
Note here that the above concept of relative entropy-based QD can also be extended to the multipartite domain~\cite{relative1}. See Sec.~\ref{dissonance}.
%, known as quantum dissonance (see Sec.~\ref{dissonance}).

\subsubsection{Geometric quantum discord}
\label{Geometric_quantum_discord}
In this subsection, we consider the quantification of quantum correlation again by using a distance to the set of q-c, c-q, or c-c states, but here the distance is defined via a norm on the relevant space of quantum states. Such quantifications are generally referred to as geometric quantum discord (GQD).

%We present here the definition and various analytical bounds of geometric quantum discord (GQD)~\cite{Vedral-against-Dutta-Shaji} and some of its variations. 
%along with its  range of possible applications in different quantum information theoretic schemes is now discussed  in detail. 
Let us begin with the definition of GQD of a bipartite state $\rho_{AB}$, as proposed  by Daki\'c \emph{et al.}~\cite{Vedral-against-Dutta-Shaji}, based on  the Hilbert-Schmidt distance\footnote{
The Hilbert-Schmidt norm and the trace norm are special cases of the Schatten p-norm, which, for an arbitrary operator $X$, is defined as
\begin{equation*}
|| X ||_p = \Big[\text{tr}\big((X^{\dagger}X)^{\frac{p}{2}}\big)\Big]^{\frac{1}{p}}.
\end{equation*}
The Hilbert-Schmidt norm and the trace norm are obtained for $p=2$ and $p=1$ respectively. 
%The spectral norm is obtained for $p=\infty$.
}. It is given by
\begin{eqnarray}
\mathcal{D}_G(\rho_{AB})= \min_{\chi^{\text{q-c}}_{AB}\in {\text{q-c}}}||\rho_{AB}-\chi^{\text{q-c}}_{AB}||^2,
\label{eqn:geomt_discord}
\end{eqnarray}
where the minimization is performed over the set of all quantum states with vanishing $\mathcal{D}^\leftarrow$. See Eq.~(\ref{defn-qc}). %QD, $\mathcal{D}^\leftarrow (\rho_{AB})=0$, i.e., $\xi_{AB}$ are QC states given in Eq.~(\ref{defn-qc}).
% quantum states of the form $\xi_{AB}=\sum_i p_i \rho_i \otimes |\phi_i\rangle \langle \phi_i|$, with $\sum_ip_i=1$, $\langle\phi_i|\phi_j\rangle=\delta_{ij}$ and $\rho_i$'s belonging to the subsystem $B$. These states  are known as quantum-classical state.
 One of the utilities of the above definition lies in the fact that for a general two-qubit quantum state, one can show that  Eq.~(\ref{eqn:geomt_discord}) has the closed analytical form given by  
\begin{eqnarray}
\mathcal{D}_G(\rho_{AB})=\frac{1}{4}\sum_{i} \sum_{j} (||x_i||^2+||T_{ij}||^2-k_{\max}),
\label{eqn:geomt_discord_two_qubit}
\end{eqnarray}
where $\rho_{AB}$ is expressed using one- and two-point classical correlators as   
\begin{eqnarray}
\label{eq:gen1}
\rho_{AB}&=&\frac{1}{4}\big(\mathbb{I}_2 \otimes \mathbb{I}_2+\sum_i~x_i \sigma^i \otimes \mathbb{I}_2+\sum_i~y_i \mathbb{I}_2 \otimes \sigma^i\nonumber \\&+&\sum_{ij} T_{ij} \sigma^i\otimes \sigma^j\big),
\end{eqnarray}
with  $T_{ij}=\text{tr}(\rho_{AB} ~\sigma^i\otimes \sigma^j)$ being the two-point classical correlators forming a $3 \times 3$ correlation matrix  $T$, while  $x_i=\text{tr}(\rho_{AB} ~\sigma^i\otimes \mathbb{I}_2)$ and $y_i=\text{tr}(\rho_{AB} ~\mathbb{I}_2\otimes \sigma^i)$,  $i,j\in \{x,y,z\}$.
%  and $\sigma^i$'s are the Pauli matrices. 
$k_{\max}$ is the largest eigenvalue of the  matrix $K=TT^T +xx^T$, where 
\(x\) is a column vector of the magnetizations \(x_i\). Using this, one can show that for two-qubits, the states with maximal $\mathcal{D}_G$ are the singlet state and states connected to it by local unitaries. Among separable states, the states exhibiting maximum  GQD are given by
\begin{eqnarray}
\sigma_{i_1 i_2 i_3}=\frac{1}{4}\left(\mathbb{I}_2 \otimes \mathbb{I}_2+\frac{1}{3}\sum_{k=1}^3 (-1)^{i_k} \sigma^k \otimes \sigma^k\right), i_k=\pm1.\nonumber \\
\end{eqnarray}

Further generalizations in this direction have been carried out  by Luo and Fu~\cite{Geom_discord1}. An arbitrary bipartite  quantum state $\rho_{AB}$ on  
$\mathbb{C}^m \otimes \mathbb{C}^n$ can be expressed as
\begin{equation}
\label{Eq:rho_mXn}
\rho_{AB}=\sum_{ij} c_{ij}~X_i\otimes Y_j,
\end{equation}  where $\{X_i,i=1,2,\ldots, m^2\}$ and $\{Y_j,j=1,2,\ldots, n^2\}$ are  sets of Hermitian operators, forming orthonormal bases in the space of Hermitian operators on $\mathbb{C}^m$ and $\mathbb{C}^n$ respectively, with the inner product and so does the operator $X_i\otimes Y_j$  on $\mathbb{C}^m\otimes\mathbb{C}^n$. By using Eq.~(\ref{Eq:rho_mXn}), another form of GQD in terms of state parameters can be obtained and is given in the following theorem:\\  
{\bf Theorem 1}~\cite{Geom_discord1}: {\it The analytical form of GQD for an arbitrary bipartite state $\rho_{AB}$  on $\mathbb{C}^m \otimes \mathbb{C}^n$ can be expressed as}
\begin{eqnarray}
\mathcal{D}_G(\rho_{AB})= \text{tr} (\tilde{C}\tilde{C}^T)-\max_{A} \text{tr}(A\tilde{C} \tilde{C}^T A^T),
\label{eqn:geomt_discord_general}
\end{eqnarray}
{\it where $\tilde{C}=(c_{ij})$, }$c_{ij}=\mbox{tr}(\rho_{AB}~ X_i \otimes Y_j)$, {\it and the maximization is performed  over  $A=(a_{kl})$, with } $a_{kl}=\text{tr}(|k\rangle \langle k|~X_l^{\dagger})$. {\it Here $k=1,2, \ldots,m$, $l=1,2, \ldots,m^2$ and $\{|k\rangle\}$ is an orthonormal basis in $\mathbb{C}^m$.} \\
From the above form of $\mathcal{D}_G(\rho_{AB})$, one obtains a lower bound on $\mathcal{D}_G(\rho_{AB})$, namely
\begin{eqnarray}
\mathcal{D}_G(\rho_{AB})\geq \text{tr}(\tilde{C}\tilde{C}^T)-\sum_{i=1}^m \lambda_i=\sum_{i=m+1}^{m^2}\lambda_i,
\label{gqd_lb_1}
\end{eqnarray}  
where $\lambda_i$'s are the eigenvalues of $\tilde{C}\tilde{C}^T$ in a non-increasing order. 

Motivated by the original definition of QD~(see Eq.~(\ref{Eq:discord1})), a ``modified" GQD can be defined as~\cite{Geom_discord1}
% Moreover, it was shown~\cite{Geom_discord1} that motivated by the original definition by QD given in Eq.~(\ref{Eq:discord1}), GQD can be modified as 
\begin{eqnarray}
\tilde{\mathcal{D}}_G(\rho_{AB})= \min_{\{\Pi^B_k\}}||\rho_{AB}-\phi_B(\rho_{AB})||^2,
\label{eqn:geomt_discord2}
\end{eqnarray}
where the minimization is performed over the set of projective measurements $\{\Pi^B_k\}$ performed by $B$ and the definition of $\phi_B(\rho_{AB})$ is given in Eq.~(\ref{new-somikoron}), 
and it turns out that 
$\mathcal{D}_{G}(\rho_{AB})=\tilde{\mathcal{D}}_G(\rho_{AB})$. Instead of the Hilbert-Schmidt distance, the above concept of the modified GQD has also been generalized by considering  the trace norm distance~\cite{MIN_GQD,GQD_new_extra1,GQD_extra5,GQD_extra4}.

%The exact expressions of $\mathcal{D}_G$ mentioned above are not easy to measure experimentally and 
In subsequent years,  several attempts have been made  to obtain  tighter lower bounds  of the above expression of GQD given in Eq.~(\ref{eqn:geomt_discord_two_qubit})  \cite{Hassan,Sai,Geom_discord_lower_bound_adesso1,
Geo_QD_decoherence,GQD_tomography,
Geom_discord_lower_bound2,Geo_addesso_analytic,Shi2,
GQD_measurement,GQD_two_sided,
Geom_discord_lower_bound_adesso2, GQD_square_norm,Rana1,wei1,luo_2_n,GQD_new_extra2,
GQD_new_extra3,GQD_bound_new1}. In particular,  Hassan \emph{et al.}~\cite{Hassan} showed that the lower bound on GQD obtained in Eq.~(\ref{gqd_lb_1}), can further be improved to
\begin{eqnarray}
\mathcal{D}_G(\rho_{AB}) \geq \frac{2}{m^2n} (||\vec{x'}||^2+\frac{2}{n} ||T'||^2-\sum_{j=1}^{m-1}\eta_j),\nonumber \\
\label{eqn:geomt_discord_general_improved}
\end{eqnarray}
where $\eta_j, j=1, 2,\ldots, m^2-1$ are the eigenvalues of the matrix $(\vec{x'}\vec{x'}^T+\frac{2 T' T'^T}{n})$,  arranged in non-increasing order. Here $x_i'=\mbox{tr}(\rho_{AB} \mathcal{L}_i \otimes \mathbb{I}_n)$ and $T'_{ij}=\mbox{tr}(\rho_{AB} \mathcal{L}_i \otimes \mathcal{L}_j')$ with $\mathcal{L}_i$ and $\mathcal{L}_j'$ being the generators of $SU(m)$ and $SU(n)$ respectively. % and $T$ is the correlation matrix. 
The lower bound in  Eq.~(\ref{eqn:geomt_discord_general_improved}) is tighter than that in Eq.~(\ref{gqd_lb_1}) and this can be illustrated by two examples, viz. $\rho_{AB}(p)=p |e_1\rangle\langle e_1|+\frac{1}{9}(1-p)\mathbb{I}_3 \otimes \mathbb{I}_3$,  and  
$\rho'_{AB}(p)=(1-p)|e_1\rangle\langle e_1|+
p|e_2\rangle \langle e_2|$, for any $p$, where  $|e_1\rangle=\frac{1}{\sqrt{6}}(|22\rangle+|33\rangle+|21\rangle+|12\rangle+|13\rangle+|31\rangle)$,  $|e_2\rangle=\frac{1}{2}(|11\rangle+|22\rangle+\sqrt{2}~|33\rangle)$ with $0\leq p \leq 1$ (see figures~\textcolor{red}{1} and \textcolor{red} {2} in Ref.~\cite{Hassan}). Additionally,  the lower bound obtained in Eq.~(\ref{eqn:geomt_discord_general_improved}) becomes exact for a 
$\mathbb{C}^2 \otimes \mathbb{C}^n$ system.\
 
%Furthermore, in Ref.~\cite{Shi2}, Shi \emph{et al.} gave an analytic formula of symmetric GQD for two-qubit system.
Furthermore, GQD for an arbitrary state $\rho_{AB}$ on $\mathbb{C}^2\otimes \mathbb{C}^n$ has also been derived  by  Vinjanampathy \emph{et al.}~\cite{Sai} and Luo \emph{et al.}~\cite{luo_2_n}. Moreover, in Ref.~\cite{GQD_measurement}, a tight measurement-based upper bound of GQD where the distance is calculated from the c-c states has been found for  two-qubit quantum states by considering the  Hilbert-Schmidt distance. % of the original state from the CC states only.
In addition to this, nonclassical correlations of some well known bipartite bound entangled states (on $\mathbb{C}^2 \otimes \mathbb{C}^4, \mathbb{C}^3 \otimes \mathbb{C}^3$ and 
$\mathbb{C}^4 \otimes \mathbb{C}^4$) have been calculated by using GQD \cite{Rana1,Bound_ent_GQD, volumehoro,UPB1,Rana2}.   \
A relation between QD and GQD for two-qubit systems has further been proposed in Ref.~\cite{Geom_discord_lower_bound_adesso2}. Moreover, in subsequent works, a relation between negativity (for defintion, see Appendix~\ref{LN}) and GQD has also been  conjectured~\cite{GQD_witness1,Rana_reply,GQD_witness2}. 
 %establishes  reveals that   $2\mathcal{D}_G\geq {\mathcal{D}}^2$~\cite{Geom_discord_lower_bound_adesso1}. 
 
 %that $\mathcal{D}_G\geq \mathcal{N}^2$ (See Appendix for the definition of negativity $\mathcal{N}$).
%This conjecture has been proved in 
%%Refs.~\cite{GQD_witness1,Rana_reply} by establishing a relation between GQD and entanglement witness
%\begin{eqnarray}
%\mathcal{D}_G(\rho_{AB})\geq \frac{\mathcal{E}_W^2(\rho_{AB})}{\text{tr}(W^2\rho_{AB})}.
%\label{ent_GQD}
%\end{eqnarray}
%Here, $W$ is the entanglement witness operator satisfying the matrix inequality $-\mathbb{I}\leq W\leq \mathbb{I}$ and $\mathcal{E}_W(\rho_{AB})$ is defined as 
%% the optimal entanglement witness operator, 
%$\mathcal{E}_W(\rho_{AB})=\max[0,-\text{tr}(W_{opt}\rho_{AB})]$. $W_{opt}$ is the 
%%Hermitian witness operator and $M$ being compact subset of the set $W$.  Here $W\rho_{AB}$ is the
% optimal entanglement witness operator of $\rho_{AB}$ given by $\text{tr}(W_{opt}\rho_{AB})=\text{min}_{W \in \mathcal{M}}\text{tr}(W\rho_{AB})$ with $M$ being the  compact subset of the set $W$. % Using Eq.~(\ref{ent_GQD}), the Authors established  a relation between squared negativity and GQD which turned out to be incorrect as shown by
% However, later on it was reported that there exist some counterexamples for which the above conjecture does not hold~\cite{GQD_witness2}.\

The original definition of QD involved one-sided measurements. Subsequently, certain symmetric versions of QD, involving measurements on both sides were defined. See Sec.~\ref{bipartite global qd}. Within the span of distance-based measures of QD, $\mathcal{D}_{rel}$ in Eq.~(\ref{Eq:ReD_def}) is such a symmetric version of QD. A symmetric version of GQD involving two-sided measurements was defined in Refs.~\cite{GQD_two_sided, Shi2,GQD_measurement} as
%As it was shown~\cite{Geom_discord1} that the definition of GQD given in Eq.~(\ref{eqn:geomt_discord}) is turned out to be equivalent to Eq.~(\ref{eqn:geomt_discord2}) based on the one-sided projective measurement performed by $B$. Such concept has been generalized to introduce GQD with two-sided projective measurement~\cite{GQD_two_sided, Shi2}. In particular, it is defined as
\begin{eqnarray}
\mathcal{D}^{sym}_G(\rho_{AB})=\min~ ||\rho_{AB}-\phi_{AB}(\rho_{AB})||^2,
\label{GQD_symm}
\end{eqnarray}
where $\phi_{AB}(\rho_{AB})$ is given in Eq.~(\ref{new-somikoron}) and the minimization is carried out over the set of all two-sided independent local measurements. A lower bound, similar to the one in Eq.~(\ref{gqd_lb_1}), can also be obtained in this case.  

Down the avenue, several works were reported where questions  have been raised regarding the validity of the above formulation of  GQD~\cite{geomet_discord_problem1,geomet_discord_problem2,geomet_discord_problem3}.  
In case of conventional  QD, as expressed in Eq.~(\ref{Eq:discord1}), it is  known that the value of  QD can be  increased by   applying some local operations on the  part on which measurement has to be performed, although it  can not be increased by performing any operations on the unmeasured part.  In contrast, it was shown~\cite{geomet_discord_problem1,geomet_discord_problem2,geomet_discord_problem5} that GQD in Eq.~(\ref{eqn:geomt_discord}) is not monotonic when the operations are performed even on the unmeasured subsystem of $\rho_{AB}$. In particular, if one considers the map on the unmeasured part say $A$, as $\tau:X\rightarrow \sigma \otimes X$, i.e. adding an ancilla at $A$, then the  Hilbert-Schmidt norm of the  state, after this action, is given by  
\begin{eqnarray}
||X||_2\rightarrow ||X||_2 \sqrt{\text{tr}(\sigma^2)},
\end{eqnarray}
using the property of the norm under tensor product.
In other words, the value of GQD becomes a function of the purity of the local ancilla upon addition of an ancillary system.

 In this regard, a possible remedy has also been suggested in Ref.~\cite{geomet_discord_problem1}. Specifically, the definition in Eq.~(\ref{eqn:geomt_discord}) can be modified  as
\begin{eqnarray}
\mathcal{D}^{mod}_G (\rho_{AB})=\max_{\Lambda_A} \mathcal{D}_G(\Lambda_A \rho_{AB}),
\end{eqnarray}
where the maximization is taken over all quantum channels acting on part $A$. However, such introduction of another maximization makes the computation of the resulting quantity difficult. It was also pointed out that the inherent non-monotonicity present in the GQD, in principle can still lead to unwanted results~\cite{dqc1_exp3,geomet_discord_problem3,geomet_discord_problem4,geomet_discord_problem5}.
%  Though this  fixes the nonmonotonicity problem of GQD which emerges under the operation on part $B$,  the advantage of  the close analytical form obtained earlier, is apparently lost.  Apart from this, additional shortcoming have also been reported~\cite{dqc1_exp3,geomet_discord_problem3,geomet_discord_problem4,geomet_discord_problem5}. 
It has been found that highly mixed states containing non-zero and even near-maximal quantum correlation as measured by QD may have negligible GQD.  
%One of the primary reason behind this contrary is the properties of 
This is at least partly due to the fact that
the Hilbert-Schmidt distance is highly sensitive to the purity of the state in its argument.
%  Eventually, in order to have a reliable  formulation of GQD, 
Attempts have also been made to define GQD by using other distance measures such as the trace norm~\cite{geomet_discord_problem3}, Bures distance~\cite{other_geometric_discord1, gqd-bures-abar},  Hellinger distance~\cite{other_geometric_discord2,GQD_new_extra1}, etc. See also~\cite{other_geometric_discord3}. In this regard, Bai \emph{et al.}~\cite{Bai2} have shown that for a class of symmetric two-qubit 
``$X$-states"\footnote{A bipartite state is called an 
\(X\)-state~\cite{Yu_Xstate,x-dekho} if in the computational basis, it has non-zero entries only in its diagonal and anti-diagonal positions, so that the state looks like the letter ``\(X\)''.},
 GQD using the trace norm~\cite{Ciccarello} serves as a lower bound for the same using the Hilbert-Schmidt distance. See also Ref.~\cite{Hilbert_trace_norm_comparision}. 
%there exists  a hierarchy relation for the GQDs, measured using trace-norm and Hilbert-Schdimt distance. In particular, in this case, the former serves as the lower bound of the later. Apart from this, 
%For a two-level atom evolving under a spontanuous emission process, it has been reported that the Hilbert-Schmidt norm GQD has nonphysical properties with respect to  local evolution% and thus trace norm discord serves as a better measure of QC
%~\cite{Hilbert_trace_norm_comparision}.

\subsection{\bf Other quantum correlation measures}
\label{oth}
Apart from QD, several other measures of quantum correlation
% different from classical one between a bipartite quantum state 
beyond entanglement have been introduced. Below, we briefly discuss some of them, specifically,  
%Some of them are the 
quantum work deficit (WD)~\cite{Oppenheim1,Sir-mam2,Sir-mam3}, quantum deficit~\cite{QDefi1}, and measurement-induced non-locality (MIN)~\cite{MIN}.

It is important to mention that there are further measures that have been put forward in the last fifteen years or so. These include the ones that have been proposed~\cite{renyi-dis1,Seshadreesan-SQUASH,sand-renyi-dis2,renyi-dis3,comment-gen1,tsallis-dis,rel-ref2,rel-ref3} based on  R{\'e}nyi and Tsallis entropies~\cite{renyi-main1,renyi-main2,renyi-main3, san-main,san-main2,tsallis-main1,tsallis-main2}. Further examples include Refs.~\cite{local-dis,oneway-dis,shared-anindyada,saitoh-dis,beggi-otherqc,li-other,uttam-other,marian-other,Tufarelli-other,yin-other,Sanders-classic,roga-response, doukas-discord,herbut-QD,xu-Generalizations}.

\subsubsection{Quantum work deficit}  
Quantum work deficit (WD)~\cite{Oppenheim1} was introduced to quantify quantum correlation by exploring the connection between thermodynamics and information~\cite{benn-demon1,landauer1,szilard, benn-demon2, seth-demon}. It is defined as the information, or work, that cannot be extracted from a bipartite quantum state when the two parties are in distant locations, as compared to the case when the same are together. Just as in any thermodynamical consideration for extracting work, one must be careful in setting up the stage with respect to the allowed operations for the work extraction. The set of allowed operations for work extraction for the bipartite quantum state when the two parties are at the same location is termed as ``closed operations (CO)". The same set in the distant laboratories paradigm is called ``closed local operations and classical communication (CLOCC)"~\cite{Oppenheim1, Sir-mam2, Devetak1, Sir-mam3,laws-horo,PT-Oppenheim,Reversible-horodecki,horoall-mutual}.
% concentrated in a bipartite quantum state $\rho_{AB}$ shared between two distant parties,
%of the amount of information concentrated in terms of the total work extractable under closed operation (CO) and closed local operation along with classical communication (CLOCC), when heat baths are attached~\cite{Oppenheim1, Sir-mam2}. 
 Here, closed operations are formed by $(i)$ global unitary operations, and $(ii)$ dephasing operation on the bipartite state by a projective measurement on the entire Hilbert space of the two-party system. On the other hand, CLOCC is constituted of  
$(i)$ local unitary operations, $(ii)$ dephasing by local measurements on either subsystem, and $(iii)$ communicating the dephased subsystem to the other one, by using a noiseless quantum channel. For a bipartite quantum state $\rho_{AB}$ on $\mathbb{C}^m \otimes \mathbb{C}^n$, it was shown that the works extractable by CO and CLOCC are respectively $I_{CO}$ and $I_{CLOCC}$, given by
% i.e., the communication consists of a single round from the measured party to the unmeasured one. Based on these operations, the physical quantity, that can be defined is known as the one-way WD~\cite{Sir-mam3,one-wayqd1}. The work extraction or the number of pure states that can be extraced from $\rho_{AB}$  under CLOCC ($I_{CLOCC}$) is less than or equal to that by CO $(I_{CO})$ and their  difference is the WD. For a bipartite quantum state $\rho_{AB} \in \mathbb{C}^A \otimes \mathbb{C}^B$, 
\begin{equation}
I_{CO}(\rho_{AB}) = \log_2 d - S(\rho_{AB}), 
\end{equation}
\begin{equation}
I_{CLOCC}(\rho_{AB}) = \log_2 d - \min_{\{\Pi^B_i\}} S(\rho'_{AB}),
\end{equation}
where $\rho'_{AB} = \sum_i \mathbb{I}^A_m \otimes \Pi_i^B~\rho_{AB}~\mathbb{I}^A_m \otimes \Pi_i^B$ is the locally dephased state, assuming that  CLOCC involved dephasing on $\mathbb{C}^n$, and $d =m n$. 
%\text{dim}(\mathcal{H}^A \otimes \mathcal{H}^B)$. % in the $B$ part and $d = \text{dim}(\mathbb{C}^A \otimes \mathbb{C}^B)$.
 Here, the minimization is performed over all projective measurements on the system at $B$. We have ignored here a multiplicative term, viz. $k_B T$, in the definitions of work, where $T$ represents the temperature of the heat bath involved, and $k_B$ is the Boltzmann constant. 
  The difference between  $I_{CO}$ and $I_{CLOCC}$ is defined as
 %over the set of projectors $\{\Pi_i\}$. Thus, 
 the ``one-way work deficit", given by 
\begin{equation}\label{Eq:deficit}
{\cal WD}^{\leftarrow}(\rho_{AB}) = I_{CO}(\rho_{AB}) - I_{CLOCC}(\rho_{AB}).
\end{equation}
 Note that like in the definition QD  in Eq.~(\ref{Eq:discord1}), $``\leftarrow"$ in the superscript 
indicates the subsystem $B$  as the 
dephased party. Moreover, WD also reduces to von Neumann entropy of the local density matrices for pure bipartite states. WD is similar to QD for states whose marginal states are maximally mixed~\cite{PrabhuErgo1}. See also~\cite{lian-1wdeficit} in this regard.
% Like QD, finding closed analytical form of WD for an arbitrary mixed state is also  difficult to obtain due to the optimization involved. 

If the dephasing process in CLOCC does not include any communication between the subsystems and both the parties completely dephase their subsystems by closed local operations, the corresponding work deficit is called zero-way work deficit.  On the 
other hand, if the dephasing protocol in CLOCC followed by the two parties involves
%does not commute with the previous one and include 
several communication rounds between them, the corresponding quantity is known as the two-way work deficit. Note that the relative entropy-based QD turns out to be zero-way work deficit when the distance is taken from the set of c-c states, whereas one-way work deficit is equal to relative entropy-based QD when the distance is considered from c-q or q-c states, whichever is relevant~\cite{Sir-mam3,one-wayqd1}. The definitions of extractable work are related to the concept of Maxwell's demon~\cite{benn-demon1,benn-demon2,landauer1,szilard,maxwell,law2,demon-dekhre1,demon-book,demon-dekhre2,seth-demon}. Indeed, for each bit of information obtained, an amount of work equal to $k_B T$ can be performed. This however does not violate the second law of thermodynamics, as an equal amount of work is needed to erase the memory corresponding to the information. For this and further discussions on this issue, see~\cite{Oppenheim1,Sir-mam2,Sir-mam3,Devetak1,zurek2003,lang-caves-povm,laws-horo,PT-Oppenheim,
Reversible-horodecki,horoall-mutual,alicki-information,hosoya-demon1,hosoya-demon2,peng-demon,demon-Brodutch}.

\subsubsection{Quantum deficit}
Another measure of quantum correlation, introduced by Rajagopal and Rendell, has been called quantum deficit~\cite{QDefi1, usha-raja}. %proposed by Rajagopal and Rendell \cite{QDefi1} in 2002.
 It is defined as the closeness of a given quantum state to its decohered classical counterpart. 
More precisely, for a bipartite quantum state $\rho_{AB}$, it is given  as the relative entropy distance between $\rho_{AB}$ and its decohered density operator $\rho^d_{AB}$:
\begin{eqnarray}\label{raja1}
\mathcal{R}(\rho_{AB})=S(\rho_{AB}||\rho^d_{AB}).
%&=& \text{tr}[\rho_{AB} \log_2 \rho_{AB}-\rho_{AB} \log_2 \rho^d_{AB}].
\end{eqnarray}
The quantum deficit uses the decohered density matrix %which possess the same information content as in the reduced density matrices $\rho_A$ and $\rho_B$ of $\rho_{AB}$. The form of this classically decohered counterpart $\rho^d_{AB}$ is given by
\begin{eqnarray}
\rho^d_{AB}=\sum_{a,b} p_{ab} |a\rangle \langle a| \otimes |b\rangle \langle b|,
%=\sum_{a,b}  p_{ab} |a,b \rangle \langle a,b|,
\end{eqnarray}
where $\{|a\rangle\}$ and $\{|b\rangle\}$ are eigenbases of the reduced density matrices $\rho_A$ and $\rho_B$ respectively of $\rho_{AB}$.
% Clearly, $\rho^d_{AB}$ is diagonal in this basis.
 Here, $p_{ab}=\langle a | \otimes \langle b |\rho_{AB}|a \rangle \otimes |b\rangle$ are the diagonal elements of $\rho_{AB}$. % and $\sum_{a,b} p_{ab}=1$. 
  Let $\lambda_i$ be the eigenvalues of $\rho_{AB}$. Eq.~(\ref{raja1}) then reduces to
\begin{equation}
\label{raja2}
\mathcal{R}(\rho_{AB})=\sum_i \lambda_i \log_2 \lambda_i-\sum_{a, b} p_{ab} \log_2 p_{ab}.
\end{equation}
It is important to observe that no optimization is required for the evaluation of quantum deficit. Also, unlike QD and one-way WD, this measure is symmetric with respect to the subsystems.

\subsubsection{Measurement-induced nonlocality}
%It is an well known fact that the  outcomes of a physical event obtained using quantum theory can not be 
% described by any local hidden variable theory \cite{Bell, CHSH, Horo_Bell, Horo_RMP} which try to make the quantum theory complete and compatible with the 
%local realistic model \cite{EPR_paper}. Violation of local realism  guarantees the existence of  nonlocality within a 
%bipartite system which can be  completely different from  the entanglement and QC present in the system. 
%As the QC  including entanglement, are the key resource  in many information theoretic and computational tasks \cite{Horo_RMP}, thus quantifying the non-locality has become an important area of research.

%Another geometric way of quantifying QC leads to the definition of measurement-induced nonlocality (MIN)~\cite{MIN}. The 
``Measurement-induced nonlocality (MIN)" is another measure of quantum correlation that is defined by using a distance to a set of states deemed as ``classical''~\cite{MIN}. It is to be noted that the ``nonlocality" in the name does not have any direct relation with the Einstein-Podolsky-Rosen argument~\cite{EPR_paper} or the Bell's theorem~\cite{Bell,CHSH}. For a bipartite quantum state $\rho_{AB}$, we first consider arbitrary projective measurements $\{\Pi_k^B\}$ on the party $B$, that keeps the reduced density matrix $\rho_B$ invariant if we forget the measurement outcome, i.e. $\sum_k \Pi_k^B \rho_B \Pi_k^B=\rho_B$. The MIN is then defined as the highest Hilbert-Schmidt distance between the pre- and post-measured states: 
%The MIN~\cite{MIN} of a shared bipartite quantum state $\rho_{AB}$, quantifies the effects of measurement on the state.
% nonlocality in terms of  the difference of the state before and after a local von Neumann  measurement which keeps the local subsystems invariant. 
%If $\{\Pi_k^A\}$ 
%If the local von Neumann measurements in  subsystem $B$ denoted by the set $\{\Pi^B(\rho_{AB}) = \sum_k \mathbb{I}^A \otimes \Pi_k^B~\rho_{AB}~\mathbb{I}^A \otimes \Pi_k^{B \dagger}\}$, the MIN is defined as the maximum of Hilbert-Schmidt distance between pre and post measured state which is given by
%
\begin{equation}
\label{Eq:MIN_def}
M_N(\rho_{AB}) = \max_{\{\Pi^B_k\}} || \rho_{AB} - \phi_B(\rho_{AB})||^2.
\end{equation} 
%where the maximization is taken over all von Neumann measurements  $\{\Pi_k^B\}$. 
%which keeps the reduced density matrix $\rho_B$ invariant, and $\Pi^B (\rho_{AB})=\sum_k \mathbb{I}^A \otimes \Pi_k^B \rho_{AB} \mathbb{I}^A \otimes \Pi_k^B$. %  $\sum_k \Pi_k^B \rho_B \Pi_k^B = \rho_B$.
%The Hilbert Schmidt norm is given by $|| X ||^2 = \text{tr}(X^{\dagger} X)$  \cite{Hilbert_schidt_norm, Nielson, Bhatia}.
The optimization over the $\{\Pi_k^B\}$ is required only when the spectrum of $\rho_B$ is degenerate and the definition of $\phi_B(\rho_{AB})$ is given in Eq.~(\ref{new-somikoron}). For the non-degenerate case, the only allowed measurement is on the eigenbasis of $\rho_B$. An analytical formula of  MIN for arbitrary-dimensional pure states has been found, and for $|\psi_{AB}\rangle = \sum_i \sqrt{\mu_i} |i_A\rangle |i_B\rangle$, the MIN is given by
\begin{equation}
M_N(|\psi_{AB}\rangle ) = 1 - \sum_i \mu_i^2,
\end{equation}
where $\sqrt{\mu_i}$ are the Schmidt coefficients.
Moreover, for mixed states on $\mathbb{C}^m\otimes \mathbb{C}^n$, there exists a tight upper bound of MIN, namely $M_N(\rho_{AB}) \leq \sum_{i=1}^{m^2-m} \lambda_i$ where $\{\lambda_i, i=1,2,\ldots,m^2-1\}$ are the eigenvalues of $TT^T$ in non-increasing order, with $T$ being the correlation matrix. 
% for arbitrary dimensional mixed state. 
%The motivation behind such measure is the presence of some kind of non-local behavior of quantum mechanics in 
%terms of its violation of Bell like inequalities  and  EPR paradox which  guarantees the 
%
Comparing the symmetric version of GQD as defined in Eq.~(\ref{GQD_symm}), with MIN, one notices that they are complementary.

\section{Computability of quantum discord}
\label{sec:Computability}
For a general quantum state, calculation of QD involves an optimization over measurements, which makes it difficult to obtain  a closed analytical expression.
% of it for a general quantum state. 
In particular, for  calculating $S_{A|B}$ in Eq.~(\ref{cond_entropy}), the minimum has to be taken over a certain set of measurements on the subsystem with \(B\). This set can, for example, be the set of all PV measurements or all generalized measurements described by POVMs. 
As shown in Refs.~\cite{DAriano, Davies78}, the number of elements in the extremal POVM need not be more than the square of the dimension of the system, and hence for states on $\mathbb{C}^2 \otimes \mathbb{C}^2$, the 
 optimization in the classical correlation does not need consideration of POVMs whose  elements number more than four~\cite{Hamieh}. 
Moreover, it was argued that on $\mathbb{C}^m \otimes  \mathbb{C}^m$, at most $\frac{m(m+1)}{2}$ POVM  elements are required for the optimization~\cite{Sasaki99}, implying that in $\mathbb{C}^2 \otimes  \mathbb{C}^2$, 
a $3$-element POVM is sufficient.  
%possibly the optimal set contain at most  i.e.,  as also estimated in the context of Shannon's mutual information for quantum ensemble .
%The definition of conditional entropy remains intact if number of POVM is restricted by $n^2$, where $n$ is the dimension of the subsystem $B$.  However, the optimal POVM must be an extremal \cite{Hamieh} which contains at most $n^2$ number of operators \cite{DAriano}. 
It is evident that for an arbitrary bipartite quantum state in arbitrary dimension, the optimized measurement setting over the set of PV measurements or POVMs  for classical correlation is generally hard to perform,  both analytically as well as numerically. 
%Hence, the exact  expression or numerical value of the quantity is far from reach.

In this direction,  Huang~\cite{Huang}
   showed that the time required to compute QD grows exponentially with the increase of the dimension of the Hilbert space, implying that computation of QD 
is NP-complete~\cite{Nielsen}. 
%The computational complexity of computing QD grows exponentially with the dimension of the Hilbert space, so that computability of QD in a quantum system of moderate size is practically intractable.  

In finite dimensions, a closed formula of QD is known only for specific classes of states.
However, an analytic expression of the  Gaussian QD can be obtained for continuous variable systems, as seen in  Sec.~\ref{GaussianQD}. 
%For two-mode squeezed thermal states, Giorda and Paris first explicitly obtained a analytic expression for Gaussian quantum discord and showed that almost  all squeezed-thermal states have nonzero Gaussian discord \cite{paris1}. They identified a threshold value of Gaussian discord, above which all the states are entangled. Below the thresholds, the states states can be either entangled or separable. Almost at the same time, Adesso and Datta came up with a similar definition of Gaussian quantum discord which reveals similar features for all two-mode Gaussian states \cite{adesso1}. {\color{red}( I'll include the formula, once I understand it.)} 
Furthermore, as  discussed in Sec.~\ref{Geometric_quantum_discord}, GQD can be evaluated analytically for arbitrary two-qubit systems~\cite{Rossignoli11}.
%As the general expressions for two-qubit systems are available only for the geometric discord \cite{Vedral-against-Dutta-Shaji} .

\subsection{Qubit systems}
\label{subsec:qubitsystem}
% However, for some special sets of states, it is possible to compute QD efficiently. 
 In the two-qubit scenario, POVMs with rank-1 elements are sufficient to optimize the QD~\cite{Hamieh}. A compact form of QD for arbitrary rank-2 states on \(\mathbb{C}^2\otimes \mathbb{C}^2\) is obtained in Ref.~\cite{Hamieh}, after performing the optimization over all POVMs where Koashi-Winter relation~\cite{koashi_winter} has been used. We will discuss the latter in Sec.~\ref{Sec:Connection}. 
It was shown~\cite{Shi} that a PV measurement is optimal for QD in this case while  it is  conjectured that 3-element POVM is required to obtain QD for states with rank more than 2. 
Let us for a while focus our attention on $X$-states. 
The reason for such a choice is partly because for such states, there has been some progress towards numerical and analytical tractability of a closed form of QD~\cite{Ali,Huang2}. 
Another reason is that $X$-states often appear in physical systems of interest. In particular, for a 
Hamiltonian, $H$, having $\mathbb{Z}_2$-symmetry on $\otimes_i \mathbb{C}_i^2$, the two-qubit 
  reduced density matrix, $\rho_{AB}$, of the ground state, $\rho$, boils down to a  $X$-state. %\cite{Yu_Xstate, Ciccarello}. 
  The argument runs as follows:
  %whose elements are zero except  the diagonal and offdiagonal terms. The $\mathbb{Z}_2$ symmetry enforces the  state to take  the following form of $\rho_{AB}$ in the computational basis :
\begin{align}
\label{eq:X}
[H,\otimes_i\sigma^z_i] = 0 &\implies [\rho,\otimes_i\sigma^z_i] = 0 \implies [\rho_{AB},\sigma_A^z \otimes \sigma_B^z] =0 \nonumber \\
&\implies \rho_{AB} = \begin{bmatrix}
a & 0 & 0 & e \\
0 & b & f & 0 \\
0 & f^{*} & c & 0 \\
e^{*} & 0 & 0 & d
\end{bmatrix},
\end{align}
where $a,b,c,d $ are real and non-negative with $a+b+c+d=1$. %We renamed $i^\text{th}$ and $j^\text{th}$ party to be $A$ and $B$. 
The positivity of $\rho_{AB}$ is ensured by $|e|^2\le a d$ and $|f|^2 \le b c$. In general, $e$ and $f$ may be complex numbers, although they can be made real and non-negative by local unitary transformations. Hence, without loss of generality, one can take $e,f\ge0$. To perform the optimization involved in  Eq.~(\ref{cond_entropy}), if we restrict ourselves to PV measurements, we can parametrize the measurement basis \{$ \Pi_k = |k'\rangle \langle k'|$\} by two angles $0 \leq \theta \leq \pi$ and $0\leq \phi < 2 \pi$:
\begin{align}
\label{parametric-form-qd}
|0'\rangle = \cos{\frac{\theta}{2}} |0\rangle + e^{i\phi}\sin{\frac{\theta}{2}} |1\rangle, \nonumber \\
|1'\rangle = \sin{\frac{\theta}{2}} |0\rangle - e^{i\phi}\cos{\frac{\theta}{2}} |1\rangle.
\end{align}
The original definition of QD in Eq.\ (\ref{Eq:discord1}) can be rewritten as the difference between the conditional entropy of the post- and pre-measured quantum states ${\rho_{AB}'}$ and $\rho_{AB}$, respectively and is given by
\begin{align}
\label{eqn:QDuCE}
\mathcal{D}(\rho_{AB})=S_{A|B} - S'_{A|B},
\end{align} 
where $S_{A|B}$ is the conditional entropy of the post-measured state, given in Eq.~(\ref{cond_entropy}) while   $S'_{A|B}=S(\rho_{AB})-S(\rho_B)$ is the  pre-measured conditional entropy which can be exactly obtained in closed form. 
It can be seen  that $S_{A|B} = \min_{\{\Pi_k^B\}}[S(\rho'_{AB})-S(\rho'_B)] $, $\rho'_{AB}=\sum_kp_k\rho_{A|k}\otimes\Pi_k^B$, where $\rho'_B$ is a reduced density matrix of the average post-measured state $\rho'_{AB}$, and $\{p_k, \rho_{A|k}\otimes \Pi_k^B\}$ is the post-measured ensemble. Therefore,
%  The measurement $\{\Pi_i\}$ has been performed on the subsystem $B$, with $p_i=\text{tr}(\mathbb{I}^A\otimes \Pi_i^B \rho_{AB})$ is the probability of the $i$th measurement outcome. The post-measurement state will be $\rho'_{AB}=\sum_ip_i\rho_A^i\otimes\Pi_i$ where $\rho_A^i=\text{tr}_B(\mathbb{I}^A\otimes \Pi_i^B \rho_{AB}\mathbb{I}^A\otimes \Pi_i^B)/p_i$.  
%So, one can also think QD to be the difference between the classical and quantum version of conditional entropies. The computation of $S_{A/B}$ does not require any optimization and can exactly be computed. But, for the post-measured state the quantum conditional entropy is given by In this case, 
\begin{align}
S_{A|B}=\min_{\theta, \phi}[\Lambda_+\log_2\Lambda_++\Lambda_-\log_2\Lambda_--\sum_{i=1}^4\lambda_i\log_2\lambda_i],
\label{eqn: qce}
\end{align}
where the eigenvalues of $\rho_B'$ are $\Lambda_{\pm}=(1\pm(a-b+c-d)\cos\theta)/2$ and the same of $\rho_{AB}'$ are given by 
\begin{eqnarray}
\lambda_{1,2}&=&\{1+(a-b+c-d)\cos\theta\nonumber\\
&&\pm[(a+b-c-d+(a-b-c+d)\cos\theta)^2\nonumber\\
&&+4(e^2+f^2+2 e f \cos2\phi)\sin^2\theta]^{1/2}\}/4,\nonumber\\
\lambda_{3,4}&=&\{1-(a-b+c-d)\cos\theta\nonumber\\
&&\pm[(a+b-c-d-(a-b-c+d)\cos\theta)^2\nonumber\\
&&+4(e^2+f^2+2 e f \cos2\phi)\sin^2\theta]^{1/2}\}/4.
\end{eqnarray}
To obtain $S_{A|B}$, 
we need to minimize the quantity over the parameters $\theta $ and $\phi$. 
The concavity of Shannon entropy ensures that 
minimization over $\phi$ happens at $\cos 2\phi =1$, although  the 
extremum points over $\theta$ has not be exactly located analytically.
Assuming that the optimal measurement basis is either the eigenstates of $\sigma^z$ or those of $\sigma^x$ has been found to provide a close estimate. 
%See Refs.~\cite{Ali,LuMa, Chen2,Huang2}. 
See Refs.~\cite{LuMa, Fanchini2, Celeri, Auyuanet, Galve, LiWang} in this regard. For the states satisfying the above assumption, we have
% even for a general ``$X$ state". In Ref. \cite{Ali}, it was claimed that the optimum measurement basis are the eigenstates of Pauli matrices $\sigma^z$ and $\sigma^x$ and hence the extremum value of $\theta$ can be either $ 0$ or $ \frac{\pi}{2}$. 
%If this is the case, then the QD for the ``$X$ state" state is given by 
%Hence the analytical form of the QD 
\begin{equation}
{\cal D}(\rho_{AB})\stackrel{?}{=}\min\{{\cal D}_{\{\sigma^x\}}(\rho_{AB}),{\cal D}_{\{\sigma^z\}}(\rho_{AB})\},
\label{discordW}
\end{equation}
where ${\cal D}_{\{\sigma^{\alpha}\}}(\rho_{AB})$ is the QD with the measurement basis being the eigenbasis of $\sigma^{\alpha}$ with $\alpha = x, z$. Here, the measurement in QD has been restricted to PV ones.
The question-mark is kept on the equality to indicate that the relation is not true in general.
%  in computing ${\cal D}(\rho_{AB})$. 
%Though obtained from a different approach, 
%Eq.~(\ref{discordW}) can also be obtained in Ref. \cite{Ali} by using a different approach.
%However, 
% it turns out that the result for the ``$X$ state" is in general not correct \cite{LuMa, Chen2} and hence it is still not possible to obtain  a closed  analytical form even  of QD
% even for the ``$X$ state". 
 For a subset of $X$-states, namely for Bell-diagonal states (for which  $a=d, b=c)$, Eq.~(\ref{discordW}) is valid~\cite{Luo, Lang}. % Luo \cite{Luo} analytically obtain a closed form expression of QD({\color{red} Check}). 
Even for symmetric $X$-states (i.e., with $b=c$)~\cite{Fanchini2}, 
 Eq.~(\ref{discordW})
 is   not always valid. % since not all the extrema are identified properly. 
% But the Authors in Ref.~\cite{Chen2, YuZhang}, identify two regions in the parameter space of `$X$' state, where Eq.\ (\ref{discordW}) holds. Without loss of generality, one can assume that 
%Assuming $| f + e | \ge | f - e |$ \footnote{ The positivity of $e$ can be ensured by local unitary transformation}.  
It was proven  that  optimal measurement for QD is $\{ \sigma^z \}$, if $(|e|+|f|)^2 \le (a-b)(d-c)$, while the optimal measurement will be $\{ \sigma^x \}$ when $|\sqrt{ad} - \sqrt{bc}| \le |e|+|f|$~\cite{Chen2, YuZhang}. 
Let us mention here that for
the two-parameter family of $X$-states within the specific regions mentioned in the previous sentence, QD (that contains an optimization  over all POVMs) is obtained from a POVM with 3 elements, confirming the conjecture of Ref.~\cite{Hamieh}.

Beyond $X$-states, 
recent studies in this direction reveal that for a large majority of two-qubit states, an optimal measurement is among the eigenstates of  $\sigma^x, \sigma^y$, and $\sigma^z$, and    
very small errors persist for the states which do not minimize on the aforementioned sets~\cite{Huang2, Namkung, Titas_freeze}. 
Motivated by these observations and the error analysis for the $X$-states, QD has been 
%Only for the Bell diagonal state, a full ranked state in the two-qubit system \cite{Luo}, optimize for those sets of measurements and an analytical form of discord took place. 
considered for different restricted classes of measurements,    
%which  is a bigger set than  the above mentioned class of measurements and the QD obtained under that measurement 
and the general term, ``constrained QD", has been used to identify them  \cite{Computational_complexity}.
%includes or in some sense generalizes the the measurement sets of the eigenstates of Pauli matrices . 
%It was shown that the nature of some physical systems remain unchanged even one allows some relaxation in the optimization process.
The differences of the original QD and such constrained QDs
%is quantified by the error $\epsilon = {\cal D}^c - {\cal D}$, and the error profile $\epsilon$,  
have been  investigated for Haar uniformly  generated 
bipartite two-qubit  as well as two-qutrit states with different ranks,  including some positive partial transpose (PPT\footnote{ A bipartite quantum state will be called PPT \cite{peres1996,Horodecki} if it remains positive under partial transposition.})  bound entangled states \cite{bound-ent}.  
In particular, for  the $X$-states, the maximal  absolute error is $0.0029$. 
It was found that the error decreases very rapidly with 
increase of size of the restricted measurement set. 
These restricted classes of projectors were chosen in several ways over the space of projection measurements. 
%However, the statistics of the error $\epsilon$ dictates the usefulness of ${\cal D}^c$ as it demands very limited 
%resources for realization in the laboratory. 
Moreover, it was also shown that for the quantum transverse $XY$ spin chain of finite length, constrained QD exactly matches with the actual QD and hence can detect the quantum phase transition (QPT) in that system  resulting the same scaling exponent~\cite{Computational_complexity}.
%For the other states this assumption allows the analytical form of discord with a very small error which can detect the quantum phase transition for XY model spin chain of finite length . 
%Thus in recent time to reduce the computational complexity, the discord has been calculated for a restricted class of PV measurements and the error $\epsilon = {\cal D}^c - {\cal D}$ was shown to be of the order of $10^{-3}$ which gradually diminishes with the increasing of the measurement set.
% Now the question is the discord for a restricted class of PV measurements the constrained discord ${\cal D}^C$ \cite{Computational_complexity} how much it differ from the actual discord. The error  has been checked to be of the order of 
Similar analysis has also been carried out for quantum WD for the same restricted classes of measurements.
%The quantum WD has also been calculated for the restricted class of measurements and a similar qualitative behavior of error profile has been reported in the same Ref.
%The same phenomena of error profile for the quantum work deficit has been also investigated and a same qualitative behavior has been reported \cite{Computational_complexity}.
%it was found that the error behaves qualitatively same way 

%\subsubsection*{Exact computability of  quantum discord}
%\label{subsubsec:gassian}
%\textcolor{blue}{No section, and put this before  the sec III A start}
%
%

\subsection{Higher dimensional systems}
\label{subsec:Higherdim}

Due to the optimization involved in computing QD, most of the studies therein are limited to 
$\mathbb{C}^2 \otimes \mathbb{C}^2$ 
 or $\mathbb{C}^2 \otimes \mathbb{C}^n$ systems, where measurements are considered in the qubit part, and only PV measurements are allowed, so that a relatively easy parametrization is possible,
 %is in the local qubit reduces to a standard spin 
 %measurement e.g. eigenstates of $\sigma^z$ and hence it is easy to parametrize 
 as discussed in the preceding subsection. 
This is no more true for higher dimensional systems.
 For example, to study QD of two spin-1 systems, one requires six 
% However, for computation of the QD between a qutrit (spin-1) and a complementary system, we need a convenient characterization of the PV measurements in the qutrit system since standard spin measurements are only a particular case of complete PV measurements and not sufficient to exhaust the entire qutrit subspace. In Ref.\ \cite{GQD_many_body4}, the Authors showed that though the standard spin measurements are not optimal in general for spins $s \ge 1$, the optimal measurements can  be thought of as the eigenstates of a generalized spin operator $S^z_U$, defined as $S^z_U = US^zU^{\dagger}$, where $U$ is a general unitary transformation which defines the generalized PV ($|m_U\rangle = U |m\rangle$) and $[S^{\mu}_U,S^{\nu}_U]=i\epsilon_{\mu \nu \sigma}S^{\sigma}_U$. Here, $U$ can be regarded as $U = e^{iH}$, where $H$ is a Hermitian operator. In general, $d^2$ (with $d=2s+1$) parameters are required to specify U. But, since phase is irrelevant in defining the projectors, we only need $d(d-1)$ number of parameters to write down the set of projectors, $\{ \Pi^m_U=|m_U\rangle\langle m_U |\}$. 
%
%linear combination of three standard spin operators (i.e. $S^x$, $S^y$, $S^z$ ) and the 
%six operators given below: $(S^{\mu}S^{\nu}+S^{\nu}S^{\mu})/2$ where $\mu , \nu~\in~\{ x,y,z \}$. Here spin $s=1$ implies the dimension $d = 3$. So, one needs $d(d-1)=6$ 
parameters to completely specify a general PV measurement. In general, for a spin-$s$ system, $n(n-1)-1$ parameters are required to define the complete set of PV measurements,
% required for QD, 
where $n=2s+1$. 
%This implies that with the increase of dimension, the optimization on the number of parameters in QD increase quadratically and hence finding the actual QD becomes harder. 
If the system possesses some special type of symmetry like parity symmetry, the number of free parameters can get reduced. For example, for two-qutrit states with \(S_z\)-parity symmetry~\cite{GQD_many_body4}, it is enough to consider the class of bases given by
%
%due to the parity symmetry reads as   
% given by
%. These 
%parameters can further be decomposed into 3 angles $(\alpha, \beta, \gamma)$, determining {\it `intrinsic' plot of vectors }$\langle {\bf S} \rangle_{m_U}$ and another 3 angles $(\psi, \theta, \phi)$, which specify the {\it orientation of the plot and of the ensuing states}. One can show that the most general PV measurement will contain the below mentioned orthonormal states
\begin{align}
&| 1_{\vec{r}} \rangle = \cos{\beta}(e^{-i\phi_0} \cos{\alpha} |1\rangle + e^{i\phi_0} \sin{\alpha} |-1\rangle) - \sin{\beta} ~e^{-i\gamma}|0\rangle, \nonumber \\
&| 0_{\vec{r}} \rangle = \sin{\beta}( e^{-i\phi_0} \cos{\alpha} |1\rangle + e^{i\phi_0} \sin{\alpha} |-1\rangle) + \cos{\beta}~ e^{-i\gamma}|0\rangle, \nonumber \\
&|\text{-}1_{\vec{r}}\rangle =-e^{-i\phi_0} \sin{\alpha} |1\rangle + e^{i\phi_0} \cos{\alpha} |-1\rangle,
\end{align}
where $\vec{r} = (\alpha, \beta, \gamma)$ and $\tan{\phi_0}=\tan{\gamma}~\tan{(\frac{\pi}{4}-\alpha)}$.

% In order to restrict the measurements in the $x,z$ plane, we need to fix $\phi_0$, s.t. $\tan{\phi_0}=\tan{\gamma}~\tan{(\frac{\pi}{4}-\alpha)}$. Now, for a $N$-party system with a qutrit and a complementary system, QD between the subsystem of $N-1$ arbitrary spins (subsystem A) and a spin-1 particle (subsystem B) is given by Eq.~(\ref{eqn:QDuCE}), where the PV measurements $M_B=\{\Pi_m^U\}$ are performed on the second subsystem. 

%For an $N$-party state with $S^z$ parity symmetry ($[\rho_{12\ldots N},\otimes_{k=1}^N P^z_k]=[\rho_{12\ldots N},\otimes_{k=1}^N(\otimes_{i=1}^{n-1}e^{i\pi (S^z_i-S_i)})]=0$), parity preserving measurements are optimal and can be described by just two or three parameters (or one in the real case) for $s=1$. However, there can exist some special type of measurements for which the number of free parameters get further reduced. Note that,  parity preserving measurements are not always guarantee the optimal measurements and hence does not always provide the actual QD. For example, in a degenerate case, the measurement set $\{\Pi_U^m\}$ and $\{P^z_B\Pi_U^mP^z_B\}$ results same value of QD. Therefore, in such cases, we need `parity breaking' measurements to get the actual QD. For a bound entangled states in $\mathbb{C}^3\otimes \mathbb{C}^3$ given in Ref. \cite{volumehoro, Horodecki_BE} the 

%In another work, QD has been evaluated numerically by considering  PV measurements in $\mathbb{C}^3$ \cite{Computational_complexity}. 
Interestingly, for bound entangled states in $\mathbb{C}^3\otimes \mathbb{C}^3$ given in Refs. \cite{volumehoro, Horodecki_BE}, it was observed that the error between the actual QD, 
obtained by considering arbitrary PV measurements in $\mathbb{C}^3\), and  the QD by using standard spin measurement bases corresponding to $S^x$, $S^y$, and $S^z$ is very low, thereby indicating the importance of  constrained QD.
% with the standard spins operators. % and an analytical form of QD for the later case has been found.

\section{Witnessing quantum discord}
\label{sec:witness_QD}
%In this section, we will discuss about experimental technique to witness QD. 
%Another important measure or rather a detection of entanglement is the entanglement witness. 
In entanglement theory, witness operators \cite{Horodecki, Maciej_entwitness,Bruss_entwitness} play an important role in detecting entangled states, especially in the laboratory. Its immense importance lies, at least partly, on the 
fact  that it is a tool to find out whether a state is entangled or not without state tomography \cite{Blatt_iontrap, Tomography, DAriano_tomo, White_tomo} and by performing 
a lower number of 
%much less 
local measurements than in other methods.
%a quantum state can be distinguished whether it is entangled or separable by a proper choice 
%of entanglement witness \cite{Horodecki, Maciej_entwitness,Bruss_entwitness} which is nothing but a Hermitian operator $W$ with at least one negative eigenvalue \cite{Terhal}.
%Hence for the measurement of a quantum state, the expectation value of the proper entanglement witness $W$ can experimentally identify the entanglement property of that state without any kind of state tomography .
%Although 

The concept of witness operators is based on the Hahn-Banach theorem. The Hahn-Banach theorem \cite{Simmons} in  functional analysis guarantees the existence of
a linear functional, $f : \mathbb{B} \rightarrow \mathbb{R}$, from a Banach space $\mathbb{B}$ to the set of real numbers $\mathbb{R}$,
such that for any convex and compact subspace ${\cal M} \subset \mathbb{B}$ and for any $x \in \mathbb{B}$ but $x \notin {\cal M}$, one has  
\begin{equation}
f({\cal M}) = 0, ~~ f(x) \neq 0.
\end{equation}
Since  the state space in quantum mechanics does form a Banach  space and the set of separable states, $\cal S$, is convex and compact \cite{sanpera11}, the existence of an operator which can distinguish an entangled state from the set of separable states is guaranteed by the Hahn-Banach theorem.
More precisely, 
%Mathematically, an entanglement witness operator (EW) is a Hermitian operator, denoted by   $W$ such that 
\begin{eqnarray}
\label{Eq:EW_def}
 \forall \rho \notin {\cal S}, ~ \exists ~W ~ \text{s.t. } \text{tr}(W\rho) < 0 \nonumber \\
 \text{while}~ \text{tr}(W\sigma) \geq 0 ~\forall \sigma \in {\cal S}, 
\end{eqnarray}
where $W$ is a Hermitian operator, and is referred to as an entanglement witness (EW).
It is important to note here that given an entangled state, finding an optimal witness operator is still a challenging task (see \cite{Maciej_entwitness,Guhne_entreview, ent_witness, ent_witness2}). Let us also mention here that
% for optimal EW operator).
% is guaranteed, whereas
%there is no universal witness operator which can distinguish all the entangled state from  $\cal S$. 
the Bell inequalities \cite{Bell,CHSH} can also be thought as  witnesses of quantum entanglement, albeit non-optimal. % due to its violation by some quantum states.
%If an arbitrary entangled state  of any number of parties,  $\rho \notin {\cal S}$, the entanglement witness $W$ which is an explicit function of $\rho$, represents a  hyper plane in the state space in between $\rho$ and $\cal S$, such that

In a similar spirit, one may wish to find a witness operator which can distinguish the set of zero discord states from a  discordant state. 
From the definition of an EW operator, in Eq. (\ref{Eq:EW_def}), one may be tempted to replace $\sigma$ by a zero discord state.
% which is classically correlated state, given in Eq. (\ref{Eq:classical_state}).
%One should remember that
 However, the set of states with vanishing discord do not form a  compact set and neither it is convex, and hence a direct use of the Hahn-Banach theorem in this case is not possible.
In this regard, it was shown \cite{Rahimi,Saitoh} that to detect discord-like nonclassical correlation, a non-linear witness operator,
%Precisely, it simply replace the $\sigma$ to a zero discord state, which is a classical state defined in Eq. (\ref{Eq:classical_state}). 
%\textcolor{red}{The witness of QD or any other measure of quantum correlations, should be a non linear operator as for any linear witness operator the expectation value over all non-zero discordant separable states which are the convex combination of the product states gives non-negative value and hence any linear operator will be unable to detect  QD or quantumness.}
%%A  has been proposed by 
%Rahimi \emph{et. al.} \cite{Rahimi} in 2010 and then SaiToh in Ref. \cite{Saitoh}, proposed a suitable witness of nonclassical correlation, which is a non-linear map 
${\cal W} : \mathcal{B}\left(\mathbb{C}^m \otimes \mathbb{C}^n\right) \rightarrow \mathbb{R} $,  can be defined\footnote{\(\mathcal{B}(\cdot)\) denotes the set of bounded linear operators 
on its argument.},  such that  
%from the Hilbert space of the multiparty quantum system to the set of real number 
for any c-c state, $\chi$ 
\begin{equation}
{\cal W}\chi \geq 0 ~\text{and}~ {\cal W}\rho < 0,
\end{equation}
where $\rho$ is any non-c-c  state, and 
%i.e., $\rho$ does not have any product eigenbasis and 
\begin{equation}\label{Eq:discord_witness1}
{\cal W}\tilde{\rho} = c - \text{tr}(\tilde{\rho} A_1)\text{tr}(\tilde{\rho} A_2)\ldots \text{tr}(\tilde{\rho} A_m),
\end{equation}
for an arbitrary quantum state $\tilde{\rho}$, with $c \geq 0$ and $A_1,A_2,\ldots,A_m $ being  positive Hermitian operators. 
%of a  means any state which is not classical i.e., 
%The chosen witness operator is a non-linear map as no separable state which has some non-classical property
%can not be detected by a linear map \cite{Rahimi}, as  stated by the theorem\\
Moreover, the following theorem can be proven. \\
{\bf Theorem 2}~\cite{Rahimi}: {\it A linear witness map cannot detect nonclassical correlation of a separable state.} \\
\textit{Proof:} Suppose ${\cal W}_{linear}$ is a  linear witness operator which can detect c-c states i.e. tr$(\chi {\cal W}_{linear}) \geq 0 $.
%  as well for any convex combination of $\chi$s, 
Therefore,  tr$(\sum_k p_k \chi_k{\cal W}_{linear}) \geq 0$, where $\{ p_k, \chi_k\}$ is any ensemble of c-c states. % with $\sum_k p_k = 1$.
 Now an arbitrary bipartite separable state $\sigma_{AB}$,
 % = \sum_{i} p_i |\psi_i\rangle\langle\psi_i|_A \otimes |\phi_i\rangle\langle\phi_i|_B $, 
 can always be written as a convex combination of product states \cite{Werner}, and hence convex combination of c-c states. 
 This implies 
%  Here each of the $|\psi_i\rangle\langle\psi_i|_A \otimes |\phi_i\rangle\langle\phi_i|_B$ is a product state so that these are individually classical states $\chi_i$.
 %tr$(\sum_k p_k\chi_k {\cal W}_{linear}) =  
 $\text{tr}(\sigma_{AB} {\cal W}_{linear})\geq 0 $.
%The above relation states that all separable states are non-CC. \textcolor{gray}{ The proof now follows from the fact that all CC states are separable}. 
 % which  implies.
\hfill $\blacksquare$  \\
 The proof of Theorem 2 considers  ``nonclassical" correlation as that in non-c-c state. However, the proof also goes through if one considers the same as that in non-q-c or non-c-q states.
%The Authors in Ref. \cite{Rahimi} define  is known to posses  non classical correlation.
It is worth mentioning that non-linear entanglement witness operators have also been investigated \cite{Nonl_entwitness, Nonlinear_entwit2, LUR_hoffmann}, and it was shown that non-linearities help to make the detection process 
more efficient.
%hence form local uncertainty relation \cite{LUR_hoffmann}, non linear EW operators can also be originated.

%for discrete system the  entanglement witness can be improved into a nonlinear map, with the help of local uncertainty relation \cite{LUR_hoffmann}.
%The nonlinear extension of the entanglement witness has also been studied by several Authors , in search of an improved version of witness and it was shown that for discrete symmetry it is possible.
The constant $c$ and the positive operators $A_i$ in Eq. (\ref{Eq:discord_witness1}) can be determined from the nonclassical state $\rho$.
%, whose nonclassical feature one wish to witness. 
The Hermitian operators   $A_1,A_2,\ldots, A_m $  are constructed by taking the projections of the eigenvectors of $\rho$~\cite{Rahimi} and 
$c = \sup_{\tilde{\rho}} \text{tr}(\tilde{\rho} A_1)\text{tr}(\tilde{\rho} A_2)\ldots \text{tr}(\tilde{\rho} A_m)$, where $\tilde{\rho}$ is any quantum state which has a bi-orthogonal product eigenbasis. 
For example, consider the mixed state
\begin{equation}
\tilde{\sigma}_{AB} = \frac{1}{2} \Big(|00\rangle \langle 00| + |1+\rangle \langle 1+|\Big),
\end{equation}
where $|+\rangle = \big(|0\rangle + |1\rangle \big)/{\sqrt{2}}$. The above state is a c-q state having non-zero QD if subsystem $B$ performs the measurement. 
%but does not possess any product eigenabasis. 
To successfully detect the state, it was found that one can assume 
%To obtained the witness operator corresponding to the above state it was shown that the form of the Hermitian operators 
$A_1 =  |00\rangle \langle 00|$ and $A_2 = |1+\rangle \langle 1+|$, and  $c = 0.182$. The above choice of course  leads to ${\cal W}\tilde{\sigma}_{AB} < 0$, while ${\cal W}\chi^\text{q-c}\ge 0 ~ \forall $ q-c states $\chi^\text{q-c}$.
One should note here that the witness operator proposed here can be implemented when multiple copies of the state are not available \cite{Guhne_entreview,Rahimi_witness}.
Another  non-linear witness operator has been proposed by Maziero \emph{et al.}~\cite{Maziero_witness}, to 
identify two-qubit states having non-vanishing QD. 
%capture the non classical phenomena for a restricted class of two-qubit states. 
The states for which the 
witness is provided is given in Eq. (\ref{eq:gen1}), with all off-diagonal elements   
of the correlation matrix, $T_{ij}~(i\neq j)$,  being zero. It was shown that for these two-qubit states, 
%those states, the witness is sufficient to detect classical correlation i.e., ${\cal W} \rho =0 \implies \rho$ is classical state,  it  became a necessary and sufficient if the state under consideration is Bell diagonal.  
%There is also other investigation regarding the witness of QD . A nonlinear witness operator $\cal W$ has been proposed, which   
%The two-qubit states given in 
%It means for any arbitrary two-qubit states $\rho$, the ${\cal W} \rho =0 $ is sufficient to detect the state as a classical state. It was also shown that the above condition  turns out to be necessary and sufficient for the Bell diagonal state. 
the proposed form of the witness operator is given by\footnote{For convenience of 
notation, 
we will interchangeably use 1, 2, 3 for \(x,y,z\).}
\begin{equation}\label{Eq:Witness_maziero}
{\cal W}\rho_{AB} = \sum_{i = 1}^3 \sum_{j = i + 1}^4 |\langle \hat{O}_i \rangle_{\rho} \langle \hat{O}_j \rangle_{\rho}|,
\end{equation}
where $\hat{O}_i = \sigma^i \otimes \sigma^i$ for $i = 1,2,3$ and $\hat{O}_4 = \sum_{i} x_i \sigma^i \otimes \mathbb{I}_2 + \sum_{i = 1}^3 y_i \mathbb{I}_2 \otimes  \sigma^i$, with $\sum_i x_i^2 = \sum_i y_i^2 = 1$ and $\langle \hat{O}_i \rangle_{\rho} = \text{tr}( \rho_{AB} \hat{O}_i)$.
It was shown that ${\cal W}\rho_{AB} = 0$ for states having  either (i)  all $T_{ii} $ are vanishing or (ii) all the $x_i=0=y_i$ $\forall i$ and all $T_{ii}$ except any one are vanishing. This implies that such states are c-c states.
%providing a sufficient condition to witness 
For the Bell-diagonal states, for which the magnetizations $x_i$ and $y_i$ are vanishing, the witness operator ${\cal W}\rho_{AB}$ turns out to be necessary and sufficient to detect the states with non-vanishing QD \cite{Maziero_witness}.
%The Authors also show that for the Werner state (given in Eq. (\ref{eqn:werner})), ${\cal W}\rho_W = 3p^2$, being a Bell diagonal state non zero witness guarantees the existence of quantumness in terms of discord of $\rho_W$ \cite{Maziero_witness}.
Moreover, it was proposed \cite{Auccaise_witnessexp}  that 
the witness operator can be implemented by the technique of nuclear magnetic resonance (NMR).  
%and the non classical feature of the above stated two-qubit states have been studied .

%The task of finding discord witness has been carried out in 2010 \cite{Bylicka_diswitness}, f
For arbitrary bipartite states on 
$\mathbb{C}^2 \otimes  \mathbb{C}^n$, another method to identify states with positive QD, based on the PPT criterion \cite{peres1996,Horodecki}  was proposed \cite{Bylicka_diswitness}.  
%``qubit - qudit" system, considered an important in quantum information theory \cite{Kraus_2XN}. The idea was to extend the Peres 
%Horodecki PPT separability criteria \cite{peres1996,Horodecki} into the detection of the zero discord state. A zero 
%discord state must be separable and hence PPT, a subclass of PPT states is defined as 
In particular, it was shown that all c-q states belong to 
 a new subclass of PPT states,  which was called strong PPT (SPPT) states\footnote{An arbitrary bipartite $\mathbb{C}^2 \otimes  \mathbb{C}^n$-dimensional quantum state $\rho_{AB} = {\bf X}^{\dagger}{\bf X}$ with ${\bf X} = \Bigl(\begin{smallmatrix} X_1 &SX_1\\ 0 & X_2 \end{smallmatrix}\Bigl)$ is SPPT iff there is a canonical conjugate ${\bf Y} = \Bigl(\begin{smallmatrix} X_1 &S^{\dagger}X_1\\ 0 & X_2 \end{smallmatrix}\Bigl)$, such that $\rho_{AB}^{T_{A}} = {\bf Y}^{\dagger}{\bf Y}$. Here $X_1, X_2$ and $S$ are $n \times n$ dimensional  matrices~\cite{SPPT}.} i.e., 
% These \textcolor{red}{states } positive  partial transposition with the canonical factorization
% of the required state \cite{SPPT}. 
${\cal D}^{\rightarrow}(\rho_{AB}) = 0 \implies $ the state is SPPT on  $\mathbb{C}^2 \otimes  \mathbb{C}^n$. 
%Note here that ${\cal D}^{\rightarrow}$ represents the QD with the measurement in the first party.
 %  classical quantum states of the above dimension are necessarily SPPT, and yield a natural witness.
% In the same time several other investigation has been made to provide a necessary and sufficient witness of QD.
%In 2011, Zhang and Yu with other Authors \cite{Zhang_witness,Yu} try to  introduce 

In Ref.~\cite{Zhang_witness,Yu}, unlike the witness operator described in Eq.~(\ref{Eq:Witness_maziero}), a single 
observable of QD witness was introduced which turns out to be invariant under local unitary (LU) operations. %\cite{Zhang_witness,Yu}. 
%for an unknown  quantum state when multiple copies are given. 
 Such witness operator  can detect  an arbitrary bipartite quantum state $\rho_{AB}$ of arbitrary dimensions (i.e., $ \mathbb{C}^m \otimes \mathbb{C}^n$) having positive QD, provided four copies of the state are available. %in the multicopy level. Moreover, a quantum circuit has been designed to evaluate witness of QD by using only local qubit measurements. Here, the witness operator is constructed by using the LU invariant operators and a polynomial LU invariant of degree k  is given by tr$(U_A \otimes U_B \rho_{AB}^{\otimes k})$, where $U_{A(B)}$ is some permutation operator acting on k copies of subsystem $A(B)$.
%For $k = 4$, 
The witness operator in this case is given by \cite{Zhang_witness}
\begin{align}
{\cal W} = u_1 - u_3 - \frac{2}{m}(u_2-u_4),
\end{align}
%where $d_A$ is the dimension of the subsystem $A$ 
where
\begin{align}
u_1 = V_{14}^AV_{23}^A V_{12}^BV_{34}^B, ~ u_2 = V_{14}^AV_{12}^BV_{34}^B, \\
u_3 =  V_{12}^AV_{34}^A V_{12}^BV_{34}^B, ~ u_4 = V_{12}^AV_{12}^BV_{34}^B. \nonumber
\end{align}
Here $V_{ij}^{A,B} = \sum_{k,l} | kl \rangle \langle lk |_{ij} $ is the swap operator on the $i^{\mbox{th}}$ and the $j^{\mbox{th}}$ copies of the subsystem $A$ or $B$. It was shown that $\text{tr}({\cal W}\rho_{AB}^{\otimes 4})=0 \implies {\cal D}^{\rightarrow}(\rho_{AB}) = 0$ in  $\mathbb{C}^{m} \otimes \mathbb{C}^{n}$. When $m=2$, the above theorem becomes necessary and sufficient. The treatment also provides a lower bound on GQD. Moreover, for two-qubit states, quantum circuit of the above witness operator by using local measurements have also been proposed.

In addition to this,  other attempts have  been made to detect the nonclassical correlations  in a quantum  state, as quantified by distance-based measures, without full-state tomography.  In Ref.~\cite{GQD_tomography}, Jin \emph{et al.} reported that  the exact GQD for an arbitrary unknown two-qubit state can be obtained by performing certain projective measurements, and was shown to be advantageous in comparison to tomography. In particular, it was shown that the method proposed requires measurements of  three parameters which are  three moments of the
 matrix $K=TT^T +xx^T$ 
 (see Sec.~\ref{Geometric_quantum_discord}), while in quantum state estimation~\cite{state_estimator}, 15 parameters have to be obtained. However, the former scheme needs more copies (not more than six~\cite{GQD_tomography}) of states in each round compared to the latter one.
 % The protocol was also generalized to  qubit-qunit states~\cite{Geom_discord_lower_bound2}. 
In the two-qubit case, it was found that a quantity, proposed to be related to GQD,  
% Specifically, it was shown that instead of the projective measuremnets on six copies of states~\cite{GQD_tomography}, GQD of two-qubit states 
 can be estimated by six or seven measurements on four copies of $\rho_{AB}$~\cite{Geom_discord_lower_bound2}. 
For further studies in the direction of discriminating quantum states with
non-zero QD from those with vanishing values of the same, see~\cite{Cialdi-discord,saguia1}.

% Moreover, 
%the eigenvalues of the matrix  $K$ are computed analytically~\cite{Geo_addesso_analytic} and it was found that
%the maximal eigenvalue of $K$ corresponding to a single state is $k_{\max}=\frac{\text{tr(K)}}{3}+\frac{\sqrt{6\text{tr}(K^2)-2\text{tr}(K)^2}}{3} \cos{\frac{\theta_{\max}}{3}}$,
% where $\theta_{\max}$ is the value of $\theta$ ($ 0\leq\theta\leq \pi$) for which $k_{\max}$ is being obtained. Hence, Eq.~(\ref{eqn:geomt_discord_two_qubit}) reduces to 
%\begin{eqnarray}
%\mathcal{D}_G(\rho_{AB})= \text{tr}(K)-k_{\max}.
%\end{eqnarray}
% Putting  $\theta=0$, a bound of GQD is obtained and given by
%\begin{eqnarray}
%\mathcal{D}_G (\rho_{AB}) \geq  \frac{1}{3}\big(2\text{tr}(K)-\sqrt{6~\text{tr}(K)^2-2 \text{tr}(K)^2}),\nonumber \\
%\end{eqnarray}
%which is same with $\mathcal{D}_G$ for pure states.

\section{Volume of states with vanishing quantum discord}
\label{sec:volume_QD}
With respect to the entanglement-separability problem, and for definiteness, considering the bipartite case,  
%In a bipartite domain, according to the entanglement content of a state, 
the entire state space can be divided into two sets, viz. those consisting of entangled and separable states. 
%The next question answered~\cite{sanpera11} was 
An important question in this regard is about the ``relative volume" of these two sets~\cite{sanpera11}. A similar question can be asked in the context of QD. Specifically, in this section, we will be discussing about the volume of set of states having vanishing QD.
%In this section, we are interested to highlight one of the fundamental question: How many entangled (separeble) states survive among all quantum states? More precisely, we want to estimate the volume of the space containing possible entangled (separable) states as a subspace of all possible quantum states spanning the entire complex Hilbert space.  

%Let us start this section by introducing the definition of a separable state. A state $\rho$ acting on a finite dimensional complex Hilbert space $\mathbb{C}=\mathbb{C}_1 \otimes \mathbb{C}_2$ is said to be separable if it can be expressed in the following way:
%\begin{equation}
%\label{separable1}
%\rho=\sum_{i=1}^k p_i \rho_i \otimes \tilde{\rho}_i,
%\end{equation}
%where $\rho_i$ and $\tilde{\rho}_i$ are states acting on $\mathbb{C}_1$ and $\mathbb{C}_2$ respectively.

Before discussing the division of the space of density operators into segments with zero and non-zero QD states, 
we notice that in a complex Hilbert space, separable pure states\footnote{An $N$-party separable pure state is defined as
\begin{equation}
|\psi\rangle_{1 2\ldots N} = |\chi_{A_1}\rangle \otimes  |\chi_{A_2}\rangle \ldots \otimes  |\chi_{A_l}\rangle,
\end{equation}
 where $2 \leq l \leq N$, $\cup_j A_j = \{1,2,\ldots, N \}$ and $A_i\cap A_j = \emptyset~\forall i,j$.
Note that such states can be fully separable \((l=N)\), bi-separable \((l=2)\), etc.} have vanishing volume
 % It is well-known that states with zero entropy have a set of measure zero. Hence, one can conclude that 
 in the subspace of all pure states. %the measure of separable states is equal to zero~\cite{volume111,volume211}. 
 QD coincides with entanglement in case of pure states and hence ``almost all"\footnote{The phrase ``almost all" is used to indicate that a certain property holds for all members of a space except for a set of measure zero~\cite{real-royden,math-gupta}.} pure states have non-vanishing QD. Therefore the question about volume of zero QD states remains non-trivial only for mixed states.

The Authors in Refs.~\cite{sanpera11, karol-volume} showed that the volume of separable states is non-zero. The result was independent of the dimension of the Hilbert spaces involved and the number of subsystems. This result initiated a series of  research works, among which is the work by Szarek~\cite{szarekvol}, where the radius of the separable ball was estimated for an arbitrary number of subsystems. An earlier result by  Braunstein \emph{et al.}~\cite{braunstein} showed that such estimates on the radius have implications for experiments using NMR.
We now try to see whether the set of all states having vanishing QD, which is a proper subset of the set of separable states, also has a non-zero volume. For a given state $\rho_{AB}$, 
$\mathcal{D}(\rho_{AB})=0$ iff the state is q-c.
%their exists a complete set of rank 1 projectors $\pi_a$ on $A$ such that 
%\begin{equation}
%\rho_{AB}=\sum_a p_a \pi^a \otimes \rho_{B|a},
%\end{equation}
%where $\sum_a \pi_a=\mathbb{I}$ and $\pi_a \pi_{a'}=\delta_{aa'} \pi_a$. 
%This implies that for this set of states, there exists a local measurement protocol by which the distant observers can extract maximal information about a bipartite system without perturbing it.
% This completeness of local measurements is featured by any classical state. 
%Therefore, set of states with vanishing QD has only classical correlations. 
\begin{figure}[t]
 \begin{center}
 \includegraphics[scale=0.62]{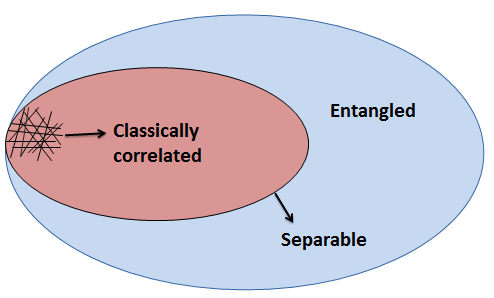}
 \end{center}
\caption{Classification of quantum states 
%Schematic diagram of the space of states 
of separated systems with respect to their entanglement and QD. Clearly, separable states contain the classically correlated states i.e. zero-discord states. Depending on the definition of QD utilized, the ``classically correlated" states can be q-c, c-q or c-c states.% \textcolor{red}{freezing part ta thakbe na.}
}
\label{volume-fig}
\end{figure}
%For the positive discord states, 
Ferraro \emph{et al.}~\cite{ferraro} 
%proposed a sufficient condition which is independent of the choice of basis, and is given by
proved that
\begin{eqnarray}
\label{cond}
\mathcal{D} (\rho_{AB})=0 \implies [\rho_{AB}, \mathbb{I}_A \otimes \rho_B]=0.
\end{eqnarray}
 Note that this is equivalent to $[\rho_{AB}, \mathbb{I}_A \otimes \rho_B] \neq 0 \implies \mathcal{D} (\rho_{AB})>0 $. 
% As an example, let us consider the following density operator $\rho_{AB}=\mu |+0\rangle\langle+0|+(1-\mu) |11\rangle\langle 11|$ where $|+\rangle=(|0\rangle+|1\rangle)/\sqrt{2}$ and $0<\mu<1$, which is a quantum classical state. One can check that for this state, the QD measured in the $B$ subsystem, the above relation holds.
The set of states with zero QD is surely a subset of the set which satisfies the equation on the right-hand-side of~(\ref{cond}). It was proven~\cite{ferraro} that the bigger set has measure zero.
 %However, there exist nonclassical correlations if we consider any Bell state as it commutes with its maximally mixed marginals. So, clearly Eq.~(\ref{cond}) is a necessary but not sufficient for vanishing QD. The Authors in Ref.~\cite{ferraro} also proved that typically any state picked at
 %It was shown that 
 %any randomly generated state contains positive QD and 
Furthermore, an arbitrarily small perturbation on a zero-discord state leads to a state having strictly positive QD.
 % implies that zero-discord states are negligible in the state space which is in a sharp contrast with the set of states having vanishing entanglement. 
 This is in sharp contrast to the situation of states having vanishing entanglement. Indeed, while there are separable states of non-full rank that can be made entangled by a small perturbation, full-rank separable states are submerged in the interior of the separable states and do not become entangled if the perturbation is sufficiently weak. For a schematic diagram, see figure~\ref{volume-fig}. 
 %To characterize the zero-discord states, a new necessary and sufficient criteria has been proposed~\cite{zero-QD} based on 
% A criterion to characterize zero-discord states have also been proposed~\cite{zero-QD} based on the commutativity of the normal matrices which are formed from the blocks of the given states (see also~\cite{animesh-nullity}).
 %zero-QD,animesh-nullity
 
 Instead of focusing on the volume of zero QD states, we can try to understand methods for knowing whether a state has zero QD. A method for this purpose was provided in Ref.~\cite{Vedral-against-Dutta-Shaji}, which works for bipartite states of arbitrary dimensions. 	To obtain this method, we consider the singular value decomposition $U\tilde{C}W^T=[c_1,c_2,\ldots,c_L]$ of a bipartite state $\rho_{AB}$ on $\mathbb{C}^m\otimes \mathbb{C}^n$ written out in Eq.~(\ref{Eq:rho_mXn}). Here,  $U$ and $W$ are $m^2 \times m^2$ and $n^2\times n^2$ orthogonal matrices, and $\mbox{diag}[c_1,c_2,\ldots,c_L]$ is an $m^2\times n^2$ matrix. In the new basis, $\rho_{AB}$ takes the form 
 $\rho_{AB}=\sum^L_{k=1} c_k S_k \otimes F_k$, where  $L=\text{rank}(\tilde{C})$. A necessary and sufficient condition for vanishing QD $(\mathcal{D}^\leftarrow)$ is then the mutual commutativity of the $F_k~(k=1,2,\ldots,L)$. It was also underlined in~\cite{Vedral-against-Dutta-Shaji} that this necessary and sufficient condition is more efficient than state tomography. 
Another necessary and sufficient criterion for zero QD of a bipartite quantum state was obtained in \cite{zero-QD} based on whether the corresponding density matrix can be written in a block form with the blocks being normal and mutually commuting.  See also~\cite{animesh-nullity}.

 The geometric pictures of the sets of states with zero and non-zero QD~\cite{geometry-cc,geometry-gen} and GQD~\cite{geometry-gen,Geo_QD_decoherence6,LiWang,GQD_CHSH1,Lang} have also been investigated, especially for two-qubit states. Note that mixing of two positive-discord states can lead to a zero-discord state, as also mixing two zero-discord  states can lead to a positive-discord state. 
 %Lang \emph{et al.}~\cite{Lang} showed that the geometrical representation of QD~\cite{geometry-cc,geometry-cc1,geometry-gen} for the  Bell-diagonal states of two-qubits is useful to show that QD is neither convex nor concave.

%Also it is shown that zero discord between a quantum system and its environment is necessary and sufficient for describing the evolution of the system through a completely positive map \cite{Shabani09,complete-positive} which we will discuss in Sec.~\ref{vanish-QD-CPTP}. However, Brodutch \emph{et al.}~\cite{Brodutch13} showed that non-vanishing QD can also lead to a completely-positive map.
The proposals that relate the vanishing of QD between a system and its environment with the complete positivity of the corresponding evolution of the system will be considered in Sec.~\ref{sec:open_QD}.

\section{Are quantum correlated states without entanglement useful?}
\label{Sec:Applications}
The early development of quantum information theory, especially in quantum communication~\cite{Bennett2,Bennett_teleport,Ekert91}, strongly suggests that entanglement shared between two or more parties is an important resource necessary  for achieving efficiencies  that cannot be reached by states without entanglement.
%emergence of nonclassicality  through entanglement led to a natural impression that the realm of nonclassicality  is perhaps entirely spanned by the quantum entanglement. Recognition of quantum entanglement  as a potential resource in the vast area of quantum information and
%%computation~\cite{Nielson} and especially in quantum communication with and without security~\cite{qcom}, can
%%be attributed as one of the primary reason behind this consideration.
However, about a decade ago, a prominent divergence from this line of thought has begun to emerge and researchers have started to ask: Is quantum entanglement the  only correlation-like resource for performing tasks with nonclassical efficiencies? 
%More specifically, Is entanglement the only way to quantify QC present in a shared quantum state? 
In this section, we are going to address this question. At the outset, let us mention that there exists, for example, the Bennett-Brassard 1984 quantum key distribution protocol~\cite{BB84}, that deals with only product states, at least in the ideal case, 
that is secure against even quantum adversaries,
%leading to sharing of identical keys between two parties without meeting, which can be shown to be secure against several, 
provided we do not require device-independent  security~\cite{device_ind1,device_ind2,device_ind3,device_ind4,device_ind5}. Let us also remember that there exists the Bernstein-Vazirani algorithm~\cite{bersten_vazirani,no_ent_algorithm,smolin_terhal} of quantum computation  that again uses product states at all stages of the protocol.  
However, the Bernstein-Vazirani algorithm requires a controlled-unitary operation that acts on a large number of qubits, which has product states as input and output. It is possible that the implementation of this unitary by breaking it up into single- and two-qubit unitaries~\cite{unitary_decomposition} will generate states having QC in the intermediate steps.  We will see that a similar situation appears in the deterministic quantum computation with a single qubit (DQC1)~\cite{Knill1} protocol.   Below we consider some QIP tasks in which, it is claimed that the states required  in the process possess non-zero amount of QD, while they do not have any or a significant amount of entanglement. We warn the potential readers that some parts of this section are currently contested in the community. The protocols that we are going to discuss about are deterministic quantum computation with a single qubit, remote state preparation, and local broadcasting, with  emphasis on whether  the resource involved can be identified as QD.
% start with some examples where signature of nonclassicality emerges even with the unentangled states and often led to  speedup in the efficiencies, over any classical counterparts.

\begin{figure}[t]
 \begin{center}
 \includegraphics[scale=0.3]{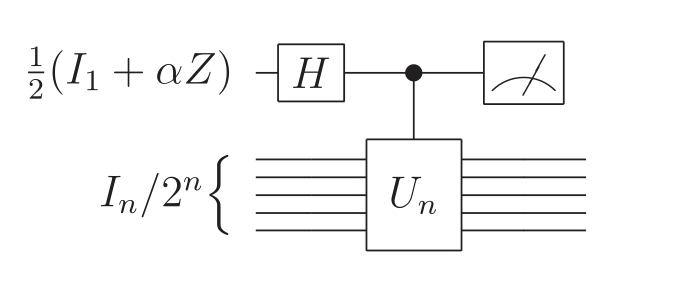}
 \end{center}
\caption{Schematic diagram of the DQC1 circuit, comprised  of a single-qubit system in a mixed state of polarization $\alpha$, together with a  N-qubit bath in which  each of the qubit is in a maximally mixed state $\mathbb{I}_2$.
%, where $\mathbb{I}_2$ is the identity operator acting on the two-dimensional Hilbert space. 
In order to compute the normalized trace of an arbitrary unitary operator,  the single-qubit system is subjected to a Hadamard gate, which is followed by a  controlled unitary $U_N$ on the qubits those belong to the $N$-party bath. $N$ is represented as $n$ in the figure. [Reprinted from Ref.~\cite{Dutta-Shaji} with permission. Copyright 2008 American Physical Society].}
 \label{fig:schematic_DQC1}
\end{figure}
 
Apart from the above mentioned schemes, QD has been claimed to be useful for several other QIP tasks. These include, e.g. the quantum state merging protocol~\cite{state_merging1,state_merging2, HOW05},  identification of unitaries and quantum channels~\cite{uni_discrimi,Channel_discrimi_QD}, and quantum metrology \cite{metrology_QD,QD_application2}. See also~\cite{QD_application1,QD_application2,QD_application3,QD_application4,
QD_application5,QD_application6}.
QD is also asserted to be useful in studying biological systems like 
%Computation of QD attracts much attention in several quantum mechanical systems  like 
photosynthesis in the light-harvesting pigment-protein complexes~\cite{Sai_QDFMO,Mahdian_QB, Saberi_QDFMO, Titas} and tunnelling through enzyme-catalysed reactions~\cite{Jevtic_enzyme}. For further claims on the usefulness of QD in QIP tasks, see~\cite{QD_application1,QD_application2,QD_application3,QD_application4,QD_application5,QD_application6}. Cf.~\cite{braunstein}.
%  In light-harvesting complexes of several biological models including cyanobacteria~\cite{cyanobacteria}, cryptophyte algae~\cite{Cryptophyte_algae, Cryptophyte_algae2} and higher plants~\cite{higher_plants}, a dimer state is claimed to form between two chromophores which are strongly coupled 
%%  QD has been found to be 
%  also been evaluated , in the photosynthesis process 
%   for a simple model of a dimer consist of two chromophore strongly coupled
   %% with a quasi-resonant undamped vibrational bosonic mode. A nonzero QD  between electronic excitations of the dimer has been reported where it is widely oscillatory over time \cite{Giorgi_QB, OReilly_QB}.
%The dimer state has been found to  exists   in the light harvesting antenna~\cite{OReilly_QB}. 
 %Quantifying QD can also be useful to identify the  and the environment.

\subsection{Deterministic quantum computation with single qubit}
\label{DQC1}
%Any illustration using mixed quantum state exhibiting signature of nonclassicality, was not known until Datta $\emph{et al.}$~\cite{Dutta-Shaji} gave a practical evidence based on the deterministic quantum computation with 1 qubit (DQC1) algorithm. Here Authors claimed that speed up in the efficiency is  entirely caused  due to the presence of QC in the system, which is quantified by QD.
Let us first briefly illustrate the task and circuit of DQC1~\cite{Knill1}. We then discuss the QC in different partitions of the set-up,  and ask whether the efficiency of the protocol is related to the QC. 
%The DQC1 is the model which is made of by a classical computer along with a single qubit, and hence is far from standard quantum computers. However, it is interesting to note that such a mere use of quantum information leads to an algorithm which cannot be solved by a classical computer efficiently till date.

The  task of the DQC1  algorithm, as proposed by Knill and Laflamme~\cite{Knill1}, is to assess the normalized trace\footnote{ The normalized trace of a matrix $M$ on $\mathbb{C}^m$ is defined as $\frac{1}{m} \text{tr}(M)$.} of a unitary matrix which cannot be solved efficiently by any known classical algorithms~\cite{classical_unitary_trace,dqc1_ent,Dutta_thesis,Dutta-Shaji}. The set-up consists of $N+1$ qubits, and the initial state is 
% where the first qubit which is the model or the system, is in a mixed state of polarization $\alpha$ and each of the remaining $N$-qubits, known as the registrar or the bath,  is in a maximally mixed state $\mathbb{I}_2$. %with $\mathbb{I}_2$ being the identity operator on  $\mathbb{C}^2$.
%% dimensional Hilbert space.
%%In this regard, the later can also be treated as a $N$ qubit bath, interacting  with a one qubit system.
% Therefore, the initial state, characterizing the system-bath properties together, can be expressed as
\begin{eqnarray}
\rho^{in}_{N+1}=\big(\frac{1}{2} \mathbb{I}_2+\alpha |0\rangle \langle 0|-\alpha |1\rangle \langle 1| \big)_1 \otimes \mathbb{I}_{2^{ N}}/2^N,
\end{eqnarray}
with $\alpha \geq 0$. As schematically depicted in figure~\ref{fig:schematic_DQC1}, the first qubit (called as ``system") is %attributed as the target bit and%
 subjected to a Hadamard gate\footnote{The Hadamard gate is defined as the unitary operator that transforms $|0\rangle\rightarrow |+\rangle$ and $|1\rangle \rightarrow |-\rangle$, where $|0\rangle$ and $|1\rangle$ are eigenvectors of $\sigma_z$ and $|+\rangle$ and $|-\rangle$ are eigenvectors of $\sigma_x$.} and is followed by  a unitary operation on the entire set-up. The collection of $N$ qubits other than the system is referred to as the ``bath". The unitary on the entire set-up  is a controlled-$U_N$, where $U_N$ is a unitary operator on the bath.  
 %, and is followed by a  controlled unitary $U_N$ on the qubits belonging to the $N$-party bath where the control operation is performed by a system of single qubit. 
 Hence, the initial state $\rho^{in}_{N+1}$ would lead to the final state given by  
\begin{eqnarray}
\label{dqc1-final}
\rho^{f}_{N+1}&=&\frac{1}{2^{N+1}}(\mathbb{I}_{ 2^{ N + 1}} +\alpha |0\rangle_1 \langle 1|_1\otimes U_N^{\dagger}\nonumber\\&+&
\alpha |1\rangle_1 \langle 0|_1\otimes U_N).
\label{eqn:final}
\end{eqnarray}
It is interesting to study the behavior of QC of the final state to understand whether the nonclassical efficiencies of the algorithm are related to QC.  
To find out the trace of $U_N$, one can now calculate the expectation values of the observables $\sigma^x$ and $\sigma^y$ of the system. These are given by
\begin{eqnarray}
 \langle\sigma^x \rangle_1 &=&\text{tr}(\sigma^x \rho^{f}_{1})= \frac{\alpha}{2^N}\text{Re [tr}(U_N)], ~ \nonumber \\ \langle\sigma^y \rangle_1&=&\text{tr}(\sigma^y \rho^{f}_{1})= \frac{\alpha}{2^N}\text{Im [tr}(U_N)],
  \end{eqnarray}
  where  $\rho^f_1=\frac{1}{2}\mathbb{I}_2+\frac{1}{2^{N+1}}\big(\alpha |0\rangle\langle 1| ~\text{tr}(U^{\dagger}_N)+\alpha |1 \rangle \langle 0|~\text{tr}(U_N)\big)$, as obtained from Eq.~(\ref{eqn:final}) by tracing out the $N$-qubit bath.
 
%The above findings seem to  have a remarkable significance, as mentioned before, since there is 
There is no known classical algorithm which can compute the trace of an arbitrary unitary operator in an efficient way~\cite{classical_unitary_trace,dqc1_ent,Dutta_thesis,Dutta-Shaji}.
%In general, estimation of trace of an arbitrary unitary operator requires resource which scales exponentially with the system size $N$, whereas using DQC1, the aforementioned task can be accomplished using at most polynomial steps. Implementation of the controlled operation is basically the only non-trivial part concealed in the DQC1 circuit. However, it is assumed that the controlled unitary operation can efficiently  be implemented using a number of one or two qubit quantum gates which lead to polynomial consumption of resource.  Hence, the DQC1 algorithm is an efficient tool to compute the normalized trace of an arbitrary unitary operators, over any classical computers.   \
 The circuit of DQC1 involves a controlled unitary gate which can be realized by several single- and two-qubit gates  with polynomial resources. 
% At the same time, the existence of classical algorithm with such resource is believed to be unlikely. Like other quantum algorithms, such as factorization~\cite{Shore_factorization}, Deutsch-Jozsa~\cite{Jozsa}, it is then natural to ask: ``what are the properties of the state that having an exponential speed up with quantum circuits gives the power of such computation?"
 At this point, it is probably natural to expect that
%   Towards that answer, the literature of quantum information strongly indicates the generation of 
   entanglement generated in the state is the key resource for success of the process\footnote{It was shown that a large amount of entanglement does not ensure increase of speed-up in an algorithm~\cite{no_entyanglement_speed_up3}.
   % Hence entanglement is necessary requirement for the outperformance of quantum algorithms, but is not sufficient
   The question however is whether entanglement or other QC are a necessary ingredient (see e.g.~\cite{entanglement_speed_up,no_ent_algorithm,no_entyanglement_speed_up2}).}. Surprisingly, it was found that the final state is separable in the system-bath bipartition for any $U_N$ and for all $\alpha >0$. This can be seen by putting $U_N=\sum_je^{i\phi_j}|e_j\rangle\langle e_j|$ in Eq.~(\ref{dqc1-final}), which gives the final state of the form
 %  $\rho_{N+1}^f$, which reduces the state of the form 
 $\rho^f_{N+1}=\frac{1}{2^{N+1}} \sum_j (|\psi_j\rangle\langle\psi_j|+|\psi_j'\rangle\langle\psi_j'|)\otimes |e_j\rangle \langle e_j|$, where $\{|e_j\rangle\}$ is an eigenbasis of the unitary $U_N$, $\phi_j$ are real,  $|\psi_j\rangle=\cos\theta~ |0\rangle+e^{i\phi_j} \sin\theta~ |1\rangle$,  and $|\psi_j'\rangle=\sin\theta~|0\rangle+e^{i\phi_j}\cos\theta~ |1\rangle$, with $\sin2\theta=\alpha$.
 
 The above example seems  to imply  that there are quantum algorithms  involving mixed states where the computational advantage  over  classical protocols  does not depend on entanglement, and hence there exists a possibility of different quantum properties of the multipartite state, independent of entanglement,
 which behave as resources. To explore such a prospect, 
%one should naturally look for  possible explanations for such exceptional speedup in the computational efficiency. In this regard,
Datta~\emph{et al.}~\cite{Dutta-Shaji} computed QD
in the splitting between the system qubit and the bath.  As we have discussed in Sec.~\ref{sec:Computability}, the main difficulty in computing QD lies in the fact that  it is not easy to find the optimal measurement basis involved. To overcome the difficulty,  a random unitary operator was generated, uniformly with respect to the Haar measure over the space of unitary operators on the Hilbert space of the system, and it was shown that  the choice of the basis plays an insignificant role in the evaluation of QD across the system-bath bipartition. Hence, the  result can be obtained by using  a measurement basis chosen from the $x$-$y$ plane. By choosing the optimal measurement basis as the eigenbasis of  $\sigma^x$, for large 
$N$,
%the Authors found,   when the normalized trace of the unitary operator $\text{tr}(U)/2^N$ is very small,
QD  can be approximated as
  \begin{eqnarray}
  \mathcal{D}_{\text{DQC1}}=2-h\left(\frac{1-\alpha}{2}\right)-\log_2\left(1+\sqrt{1-\alpha^2}\right)\nonumber\\
  -({1-\sqrt{1-\alpha^2}})\log_2 {e},
  \label{QD_DQC1}
  \end{eqnarray}
 where $h(\alpha)$ is the Shannon binary entropy, defined as
\begin{equation}
\label{binary-shannon}
h(\alpha)=-\alpha \log_{2}\alpha - (1-\alpha)\log_{2}(1-\alpha).
\end{equation} 
Note that the measurement involved in the definition is carried out on the Hilbert space of the system.  
Moreover, $\mathcal{D}_{\text{DQC1}}$ is independent of $N$, and to obtain Eq.~(\ref{QD_DQC1}), one assumes that for a typical unitary  $U_N$, real and imaginary parts of $\text{tr}(U_N)$ are small. Therefore, for any $\alpha\geq0$, QC, in the form of QD, is present in the system-bath bipartition (see figure~{\color{red}1} in Ref.~\cite{Dutta-Shaji}). The results indicate that the efficiencies in DQC1 may have a 
connection with QD, thereby providing an avenue towards establishing QD as resource. 

This work leads to several theoretical~\cite{DQC1_extra3,
DQC1_extra5,DQC1_extra8, DQC1_extra9,DQC1_extra12} (see also~\cite{other_geometric_discord1,dqc1_exp3}) and experimental studies~\cite{dqc1_exp3,dqc1_exp4,gqd-bures-abar,dqc1_exp,dqc1_exp2,dqc1_exp6} (see also~\cite{other_geometric_discord1}) that explored the possibility of identifying QD as a resource in DQC1. 

The main criticism  of the above result is that implementation of the controlled-unitary operation, which would typically  be simulated by several single-and two-qubit quantum gates~\cite{unitary_decomposition},
%controlled unitary operation, which requires implementation of several basic quantum gate, 
may generate entanglement as well as QD in the intermediate steps of the process~\cite{DQC1_extra2}. 
%All the investigation in this case consider only QC of the final state, not the states generated from the intermediate steps. On the other hand,  
%%To support such statement, it 
%it was shown by considering Grover's algorithm~\cite{grover}, that  to obtain polynomial speed-up using quantum mechanics, neither  entanglement nor QD are required~\cite{DQC1_extra2}. Hence, the resource for speed-up in quantum algorithm require more careful analysis.
Another counter-argument \cite{Vedral-against-Dutta-Shaji} was found by using a  condition on the final state to possess non-vanishing QD.
% is not the only resource. To argue that one 
 First notice that  Eq.~(\ref{eqn:final}) can alternatively be re-written as  
%  Down the avenue, some contradictory results have also been reported~\cite{Vedral-against-Dutta-Shaji,DQC1_fanchini}. Daki\'c, Vedral, and  Brukner~\cite{Vedral-against-Dutta-Shaji}  considered a class of unitaries of the form $U=e^{i\phi}A$, (where $A^2=\mathbb{I}$ is a binary operator) for which all the correlations at the output of the DQC1 circuit are indeed classical. The Authors  proposed a quantum witness using commutation of certain Hermitian operators and showed that for an alternative form of  Eq.~(\ref{eqn:final})  given below
  \begin{eqnarray}
\rho^f_{N+1}=\frac{1}{2^{N+1}}\big(\mathbb{I}_2 \otimes \mathbb{I}_{2^N} &+& \frac{1}{2} \sigma_1^x \otimes (U_N+U_N^{\dagger})\nonumber \\ &+&  \frac{1}{2i} \sigma_1^y \otimes (U_N-U_N^{\dagger})\big).
\label{DQC1_alter}
\end{eqnarray}
By using the necessary and sufficient condition of QD discussed in Sec.~\ref{sec:Computability}, it was shown that  QD vanishes across the system-bath bipartition of the state $\rho_{N+1}^f$ in Eq. (\ref{DQC1_alter}) iff $U_N^{\dagger}=k~U_N$~\cite{Vedral-against-Dutta-Shaji}. Such unitaries exist, as seen by choosing  $U_N=e^{i\phi} A$ where $A^2=\mathbb{I}_{2^N}$. Moreover, it is believed that the trace of this unitary operator cannot be simulated by a classical algorithm with polynomial resources. Therefore, this example  opens up a debate on the results in \cite{Dutta-Shaji}  about the identification of QD as a resource in DQC1 (cf.~\cite{fanchini,DQC1_extra1}).
%for the above class of unitaries. The procedure goes as follows. Consider Alice and Bob share a general mixed quantum state $\rho_{AB}$. Allowing the local measurements on the subparts, one can show that $\rho_{AB}$ is a zero discord state, iff  there exists a PV ($\Pi_k=|\psi_k\rangle\langle\psi_k|$), such that
%\begin{eqnarray}
%  \sum_k \Pi_k \otimes \mathbb{I}_B ~\rho_{AB}~\Pi_k\otimes I_B=\rho_{AB}.
%  \label{operational_discord}
%\end{eqnarray}
%The above property provides an operational meaning of QD and  can further be used to define a witness for QD, in the following way.

%Another study~\cite{DQC1_fanchini} considers a purification  of $\rho_{N+1}^f$ which is written as $|\psi\rangle= \frac{1}{\sqrt{2^{N+1}}} \sum_j (|0\rangle+e^{i\phi_j}|1\rangle) |e_j v_j\rangle$, where $\{|e_j\rangle\}$ and $\{|v_j\rangle\}$ are respectively eigenbases of unitary  and purifying system. By using a relation of Koashi-Winter~\cite{koashi_winter}
% which will be discussed in Sec.~\ref{discord monogamy-EOF}, it was argued that after the controlled  unitary operations, increase of QD occurs and bipartite as well as multipartite entanglement in different bipartitions also have been generated. It offers another perspective that both the bipartite QD and multipartite entanglement are possibly responsible for quantum speed-up. {\color {red}See also~\cite{dqc1_dekhre}.}
 
 Recently, it was also shown that the trace of any unitary operator and the complexity of the DQC1 circuit are connected to the notion of ``entangling power"~\cite{DQC1_extra7}.
 
% In this way, a witness of QD can be defined. If one considers the output state of the DQC1 circuit  in Eq.~(\ref{DQC1_alter}), one can see that the operators $\sigma^i$ do not commute with each other. Hence, in order to have zero QD, the operators $(U_N+U_N^{\dagger})$ and $(U_N-U_N^{\dagger})$ need to be linearly independent. Then,
%operator such that $A^2=\mathbb{I}_2$. For  all such unitaries acting on the  $N$ qubit bath, the correlation at the output of the DQC1 circuit turns out to be classical. Moreover,  even it appears as  quantum computation in absence of entanglement, it has been shown that  there exists a certain amount of multipartite entanglement between the system and the bath  which is responsible for the nonzero QD~\cite{DQC1_fanchini}. This suggests that QD is not merely  responsible for speedup in the efficiency of DQC1 circuit.
%
%
%
% 

\subsection{Remote state preparation}
\label{RSP}
Remote state preparation (RSP)~\cite{pati,rsp_bennet,Lo} (see also~\cite{rsp_new_new1,rsp_new_new2,rsp_new_new3}, and see Refs.~\cite{rsp_non_ent,rsp_exp1,rsp_exp2,rsp_exp3,rsp_exp4,rsp_exp5,rsp_exp6,rsp_exp7,rsp_exp8,rsp_exp9,rsp_discord} for experimental developments) is a quantum communication scheme, related to the quantum teleportation protocol~\cite{Bennett_teleport}, for sending a partially unknown qubit. Like in teleportation, the RSP scheme also requires a shared state between a sender, Alice, and a receiver, Bob. %, to send a known qubit to Bob.
Suppose Alice wants to send $|\Phi\rangle=\frac{1}{\sqrt{2}}(|0\rangle+ e^{i\phi}|1\rangle)$, $0\leq \phi < 2 \pi$ to Bob,  and they share a singlet state. The set $\{|0\rangle+e^{i\phi}|1\rangle: 0 \leq \phi < 2\pi\}$ is referred to as the equatorial qubit. %$|0\rangle$ and $|1\rangle$ are eigenstates of $\sigma_z$. 
Bob knows that the sent qubit is equatorial. 
Alice knows further: She knows  even the value of the $\phi$ of the equatorial qubit that she intends to send to Bob.  
Notwithstanding her knowledge of $\phi$, since $\phi$ is a real number in $[0,2 \pi)$, to send  it via a classical communication channel, say phone call, Alice will need an infinite amount of communication, if the shared entangled state is not used. In case  the entangled state is used, a measurement by Alice in the $\{|\Phi\rangle, |\Phi^{\perp}\rangle\}$ basis\footnote{Here, $|\Phi^{\perp}\rangle=\frac{1}{\sqrt{2}}(|0\rangle-e^{i\phi}|1\rangle)$.} - this is why Alice needs to know $\phi$ - and one bit of classical communication from Alice  to Bob can help Bob to get his part of the singlet state in the input state $|\Phi\rangle$.
%In that case, it can be shown that by performing measurement in the $\{|\Phi\rangl, e,|\Phi^{\perp}\rangle\}$ basis by Alice and one bit of classical communication from Alice to Bob, Bob can recover the state after performing suitable unitary operation in his part. It was later shown~\cite{rsp_bennet} if Alice and Bob share a priori large amount of entangled states, the asymptotic classical communication required for successfully completing the task is one bit per qubit. \

One can generalize this protocol to the case when Alice and Bob  share an arbitrary state $\rho_{AB}$, and Alice's aim is to send a  qubit, with Bloch vector $\vec{s}$, which is perpendicular to a given unit vector $\vec{\beta}$. Alice knows  $\vec{s}$, but Bob just has the knowledge of $\vec{\beta}$. After performing a local generalized measurement at her side, Alice sends  a bit of classical communication to Bob. The task of Bob is to perform a suitable quantum operation such that the fidelity between the output and input states, averaged over the unit circle on the Bloch sphere made by all the vectors in the plane perpendicular to $\vec{\beta}$ and passing through the center of the sphere  is maximized.

There have been claims that there may be separable states whose RSP fidelity  is higher than that of certain entangled states, leading to the possibility of identifying QD as a resource in RSP~\cite{Vedral-against-Dutta-Shaji} (cf.~\cite{Giorgi2,shor_rsp,RSP_others1,RSP_others2,RSP_GHZ}). The  claims however have been countered~\cite{rsp_horodechki} (cf.~\cite{Giorgi2,geomet_discord_problem2}). In particular, it has been shown that the fidelity of RSP, when carefully defined, gives a higher value for any entangled state than all unentangled ones.

\subsection{Connection with local  broadcasting}
Quantum mechanical postulates prohibit copying of 
%Another important aspect of the quantification of QD of a quantum state is its application in no local broadcasting theorem. It is an well known fact that there is no physical operation in the realm of quantum mechanics by which
an unknown pure quantum state, and even  two nonorthogonal pure states, by a single machine, a result known as the no-cloning theorem~\cite{No_cloning_Dieks, No_cloning_Wooters}.
%But in case of mixed state there are many possibilities to consider the cloning procedure, for example $i)$ it can be think as the product output states, $\rho \otimes |B\rangle \langle B| \rightarrow \rho \otimes \rho$ for some 
%blank state  $|B\rangle$, and $ii)$ $\rho_S \otimes |B\rangle\langle B|_A \rightarrow \tilde{\rho}_{SA}$ with $\text{tr}_S(\rho_{SA}) = \rho$ and $\text{tr}_A(\rho_{SA}) = \rho$ i.e., the marginals provide appropriate states \cite{Barnum_braodcast}.
%The notion of broadcasting has been extended from the idea of cloning where 
In a general setting, an ensemble of mixed quantum states $\{ p_i, \rho^i\}$ is provided and the question is whether it is possible to find a  quantum operation $\Lambda$  such that $\text{tr}_S\Lambda(\rho_S^i\otimes \sigma_E) =\text{tr}_E \Lambda(\rho^i_S \otimes \sigma_E)=\rho^i$ where  $\Lambda$ is independent of $i$.
Such an operation, called broadcasting of states, exists if and only if the states $\{\rho_i\}$ in the ensemble  $\{ p_i, \rho_i\}$
%with some probability $p_i$ need to be cloned by either any one of the above procedures $\forall ~ i$ and the allowable quantum operations can be any linear operations irrespective of unitary evolution.
%It was shown that the broadcasting of unknown quantum state is not possible even in terms of the marginals unless the states $\{\rho_i\}$
are mutually commuting ~\cite{Barnum_braodcast}.
%
%\begin{equation}
%(\rho_i)_S \otimes |B\rangle\langle B|_A \xrightarrow{{\cal F}} \tilde{\rho}_{SA},
%\end{equation}
%where , i.e., without being the output $\rho \otimes \rho$, we seek such operation which . It was also shown that the broadcasting
The above procedure of broadcasting  has been generalized by Piani \emph{et al.}~\cite{Piani_nolocalbroad} for shared bipartite states, and named as local broadcasting.
A bipartite state  $\rho_{AB}$ is locally broadcastable (LB), if there exists local operations $\Lambda_{AA'}$ and $\Lambda_{BB'}$ such that 
\begin{eqnarray}
&& \text{tr}_{AB}\Lambda_{AA'}\otimes \Lambda_{BB'} (\rho_{AB}\otimes \sigma_{A'}\otimes \sigma'_{B'})\nonumber\\
&=& \text{tr}_{A'B'}\Lambda_{AA'}\otimes \Lambda_{BB'} (\rho_{AB}\otimes \sigma_{A'}\otimes \sigma'_{B'})\nonumber\\
&=&\rho_{AB}\nonumber.
\end{eqnarray}
%$ such that $(\Lambda_A \otimes \Lambda_B)(\rho_{AB}) = \sigma_{AA'BB'}$ is a broadcast state of $\rho_{AB}$.}
%Means $\text{tr}_{A'B'} = \rho_{AB}$ and $\text{tr}_{AB}$ is same as $\rho_{AB}$.
Note that in LB, no classical communication is allowed. Communication of quantum states is certainly not allowed. It is also worth mentioning that entangled states are not LB
%in this procedure no known quantum states is LB, as it was shown in Ref.
~\cite{Yang_noentLB,broadcasting_extra1,broadcasting_extra2,broadcasting_extra3,
broadcasting_extra4,broadcasting_extra5,broadcasting_extra6,broadcasting_extra7,
broadcasting_extra8}.
The question is whether separable states having non-zero QD are suitable for LB. In this regard, the following theorem~\cite{Piani_nolocalbroad} states that this is not the case. \\
% that even LOCC can not clone a known entangled state. Now for this case the no local LB theorem states \\
{\bf Theorem 3} \cite{Piani_nolocalbroad}: {\it A state  $\rho_{AB}$ is LB if and only if it is a classical-classical state}. \\
The above theorem gives an operational interpretation of c-c states (for other variations of the LB theorem, see Refs.~\cite{Luo_LB, Sazim_braodcusting,Luo_broad}).

The no-cloning  and no-broadcasting theorems tell us about which states cannot be cloned and broadcast by global quantum engines, and it is known that such states are useful in quantum technologies~\cite{broadcasting_extra9}. It may similarly be hoped that the results on no local-cloning and no local-broadcasting by local quantum engines will be useful  for applications in quantum process (see~\cite{broadcasting_extra10} in this regard).

\section{Quantum discord in quantum spin systems}
\label{Sec:manybody}

Quantum correlations, in both entanglement as well as discord-like avatars, have turned out to be  
important for understanding QIP schemes that are more efficient than their classical analogs \cite{Horo_RMP, Kavan-rmp}. In order to achieve such tasks, one requires to identify certain realizable physical systems - substrates - in which they can be implemented in the laboratory. Interacting spin models \cite{Auerbachbook,Schollwöck}, which can naturally be found in solid-state systems~\cite{AshcroftMermin,
KittelBook}, are  one of the potential candidates for such realizations.
With currently available technology, it is  also possible to realize such systems in optical lattices~\cite{RMP_Bloch,AdP_Ujjwal, ROPP_Lewenstein, ROPP_Windpassinger, ROPP_LiLiu, lab32}, trapped ions \cite{ lab11, lab12, Blatt_review}, superconducting qubits \cite{RMP_Nori, Devoret04, Clarke08}, NMR~\cite{NMR_review}, etc.  Therefore, along with its
fundamental importance,  investigating QC in these spin models is also important from the perspective of applications.

Over the years, it was shown that a  change in  the behavior of  correlations can indicate occurrence of certain co-operative phenomena. In particular, at zero temperature, certain variations in the correlation
functions or their derivatives can  infer  rich phenomena like quantum
phase transitions (QPT)~\cite{Sachdev}. In a pair of seminal papers, Osterloh {\it et al.}~\cite{Osterloh}, and
Osborne and Nielsen~\cite{Osborne_Ni} showed that the first  derivative,
with respect to the control parameter, of nearest-neighbor bipartite entanglement, as quantified by concurrence (see Appendix~\ref{append: concurrence}, for a definition), of
 the zero-temperature state in the one-dimensional (1D) quantum 
transverse Ising model  can capture the signature of QPT present in this
model. Furthermore, it was demonstrated that there are quantum many-body
systems in which ``localizable entanglement length" diverges while the classical
correlation length remains finite~\cite{Verstraete04,Verstraete04a}. QPTs
are traditionally uncovered by the divergence of ``length" of classical
correlation functions~\cite{WenBook}. It has also been realized that it
is useful to understand the role of entanglement in classical simulation
of quantum many-body systems~\cite{Verstraete_PEPS,
Vidal_classisimulation, SWVC08}. These have led to a significant amount
of effort being given to the analysis of the behavior of entanglement,
mainly bipartite,  of zero-temperature states
in isotropic Heisenberg rings~\cite{Arnesen01,Gunlycke01,OConnor01} and in
various other quantum many-body systems~\cite{RMP_manybody, AdP_Ujjwal,
Kavan-rmp} (see also~\cite{CKW, Nielsen_thesis, Preskill_many, Wang01,
Dennison01, Wang_Heisen, Meyer_glabal, Wang_aniso, Wooters_paratrans,
Wooters_entchain, Zanardi_fermion, Zanardi_fermion2, Bose_entjump}) which
include  both ordered as well as disordered quantum spin-$\frac{1}{2}$
models with nearest-neighbor, next-nearest-neighbor as well as long range
interactions~\cite{Gu, Glaser,Roscilde1,Roscilde2,VidalLMG04,
Yang3-body05,  Guj1j204}.   Similar studies have also been carried out  in
higher-dimensional many-body systems~\cite{RMP_manybody, AdP_Ujjwal}.
In 2003, Vidal {\it et al.}~\cite{Kitaev} went beyond bipartite
entanglement and investigated  the behavior of entanglement entropy
between two disjoint blocks of a  spin chain (block entanglement), which
was later extended to several other systems~\cite{RMP_manybody}. It was
found that block
 entanglement tends to follow an ``area law'' in gapped systems~\cite{Eisert10}.

 Apart from the zero-temperature states, bipartite as well as multipartite
entanglement have also been used to investigate thermal equilibrium
states in different spin models either by analytical or numerical
techniques.
 It has in particular been found that entanglement can  sometimes be
nonmonotonic   with the increase of temperature~\cite{Arnesen01,
Brennen04, aditi_dynamics, PrabhuErgo3, Chanda, Maziero, Fumania16},
which is in contrast to the fragile nature of entanglement in the
presence of environment.

 In all the above works, tools from quantum information theory have been
employed to understand,  mainly, equilibrium co-operative phenomena
present in quantum many-body systems. However, realizing QIP tasks in
these systems normally demand investigation of trends of entanglement
with time, both bipartite as well as multipartite, in the time-evolved
state. Such studies have led to an important area in quantum computation,
called one-way quantum computer \cite{one_way,one_way_nat} which has been
extensively  investigated, both theoretically and experimentally.
Moreover,  statistical mechanical properties like the question of
ergodicity of bipartite entanglement was investigated in  $XY$ and  $XYZ$
spin systems~\cite{AditiErgo04, aditi_dynamics, PrabhuErgo1, PrabhuErgo2,
PrabhuErgo3, Sadiek12, PAM06, PAM07, SAA10}.
The dynamics of entanglement has further been explored in Refs.~\cite{Cincio07, Sengupta09, Pollmann10, Polkovnikov11, LK08} for different
types of quenches. See also~\cite{Amico04,Huang05}.
The dynamical  behavior of  block entropy after a sudden quench was
considered by Calabrese and Cardy~\cite{Calabrese1}   in the transverse
Ising chain and then later investigated in various other models~\cite{Eisert10}.

As argued in this review, QC are not limited to entanglement.
It is interesting to check whether the behavior of  QC beyond entanglement
 can also identify the key phenomena of  these systems. In this section,
we outline the works in this direction.

\subsection{Models}
Let us first briefly review the quantum spin systems in which QD has been  investigated in recent years. 
%It  includes
%quantum  spin models, %, strongly interacting ferminonic and bosonic models, 
%cold gas systems, etc. % Briefly after introducing the models, we will discuss about the patterns of QD in those systems. %Till dates most of the studies are mainly limited to 1D systems. 
Interacting systems of localized spins provide a paradigm where QD is capable of detecting natural phenomena like QPT, ground state factorization, ergodicity, etc. %Like entanglement, QD also plays an important role in critical systems. 
So far, most of the studies of QD in critical systems are limited to 1D spin  systems. Similar to entanglement, non-analyticity of the derivative of QD near the critical point can indicate the QPT in the system. 

The Hamiltonian of a 1D nearest-neighbor ``$XYZ$ spin model" with transverse magnetic field  can be written as 
\begin{align}
H = J \sum_{\langle i,j \rangle} [(1+\gamma)S^x_i S^x_j + (1-\gamma)S^y_i S^y_j + \Delta S^z_i S^z_j ]& \nonumber \\  + h_z \sum_{i} S^z_i,&
\label{eqn: Hamil}
\end{align}
where ${\langle\cdot,\cdot\rangle}$ denotes the sum over nearest-neighbor spins and $S^{\alpha}_i$ $(\alpha = x,y,z)$ are  spin operators of appropriate dimension at the $i^\text{th}$ site of the system. $J$ and $\gamma$ are respectively  the exchange coupling and the anisotropy parameter in the $x$-$y$ plane. $\Delta$ and $h_z$ are the  coupling constant  and the strength of external magnetic field in the $z$-direction, respectively. 
Here, $J$ and $h_z$ have the unit of energy while $\Delta$ and $\gamma$ are dimensionless. When $J>0$ the model is antiferromagnetic, while $J<0$ corresponds to a ferromagnetic system. We assume $1\le i,j \le N$. On top of that, periodic boundary condition requires $S_{N+1} = S_1$. %For periodic boundary condition $(N+1)^\text{th}$ site and $N^\text{th}$ site are the same. 
The thermodynamic limit can be obtained by taking the $N\rightarrow\infty$ limit. 

For spin-$\frac 12$ systems, $S^{\alpha}_i$ $(\alpha = x,y,z)$ are proportional to the Pauli spin matrices. Though the Hamiltonian, even for the spin-$\frac 12$ case, is not solvable in general,  it can  be diagonalized exactly in certain special cases in 1D. %Note that, $\gamma=\Delta =0$ case  solvable via successive Jordan-Wigner, Fourier and Bogoliubov transformation ~\cite{LSM, barouch1, barouch2}. 
When $\Delta=0$ and $\gamma\ne 0$, the model reduces to the ``transverse $XY$ spin model", whose eigenenergies and eigenvectors can be obtained exactly by using  successive Jordan-Wigner, Fourier and Bogoliubov transformations~\cite{LSM, barouch1, barouch2,barouch3}. Similar method can also be employed to find the entire spectrum of the above Hamiltonian with $\gamma=\Delta=0$, known as the ``$XX$ model".
 The case when $h_z=0$ and $\Delta \ne 0$, known as the ``$XXZ$ model" without magnetic field can also be diagonalized analytically by using  thermodynamic Bethe ansatz~\cite{Takasaki}.
%The $\Delta \ne 0$ and $h_z=0$ case is  known as the $XYZ$ model in zero field which is also analytically tractable by the thermodynamic Bethe ansatz~\cite{Takasaki}. Now we will discuss the equilibrium behavior of QD in some of special cases of the Hamiltonian in Eq. (\ref{eqn: Hamil}).

%

\subsection{Statics}

If the Hamiltonian does not have any explicit time dependence, and we are interested with static states of the system such as the ground or thermal states, we refer to the analysis as the ``static" case.
% the system does not evolve with time which we refer here as the static case. In this situation, depending on the temperature, the system can either be in its ground state or in the canonical equilibrium state.
  If the system is at zero temperature,  its properties, including any change of phase, are completely driven by quantum fluctuations. % and hence any phenomena like change of phase observed are entirely due to quantum {\color{red} mechanics}. 
  However, at any finite temperature, this is not the case. In particular, a thermal state is a  mixture of the ground state as well as all the excited states, with appropriate probabilities which are fixed by  the temperature, and hence the properties of a thermal equilibrium state is driven by the % While at the zero temperature, the system is subject to only quantum fluctuations, as the temperatures increases there is 
interplay between  quantum and thermal fluctuations. When the system reaches high enough temperature, only the thermal fluctuations dominate.
% and hence it is expected that quantum mechanical properties subsequently vanish. 
Below we briefly discuss the properties of ground and thermal states by using QD in some of the well studied spin models. 

\subsubsection{Spin-$\frac 12$ systems}

Let us first concentrate on the Hamiltonian in Eq.~(\ref{eqn: Hamil}) for the spin-$\frac 12$ case.  To study QD between the $i^\text{th}$ and $(i+r)^\text{th}$ sites of a state, we first need to calculate the two-site reduced
 density matrix, $\rho_{i,i+r}$, by tracing out all the sites except $i,i+r$ from the ground or thermal state of the system. Let us also assume the periodic boundary condition which implies that all reduced density matrices are the same. As discussed in Sec.\ \ref{sec:Computability}, systems having $\mathbb{Z}_2$-symmetry, enjoy some simplifications and the two-site reduced density matrix of such Hamiltonians can then be written as 
\begin{align}\label{Eq:two_partyx}
\rho_{i,i+r} = \frac{1}{4}\Big[\mathbb{I}_4 +& \langle \sigma^z\rangle(\sigma^z_i \otimes \mathbb{I}_2 +\mathbb{I}_2 \otimes \sigma^z_{i+r}) \nonumber \\
&+ \sum_{\alpha=x,y,z}\langle \sigma^{\alpha}_{i}\sigma^{\alpha}_{i+r }\rangle  \sigma^{\alpha}_i\sigma^{\alpha}_{i+r}\Big],
\end{align} 
%which is equivalent to the symmetric `$X$'state as given in Eq.\ (\ref{eq:X}) with $b=c$. Here, $\langle \sigma^{\mu}\rangle=\langle \psi_g \sigma^{\mu}\psi_g \rangle$ is the magnetization along the direction $\mu$ and $\langle \sigma_i^{\mu}\sigma_j^{\nu}\rangle=\langle \psi_g |\sigma_i^{\mu}\sigma_j^{\nu}|\psi_g\rangle$ are the elements of the correlation tensor where $|\psi_g\rangle$ is the ground state and $\mu,\nu=\{x, y, z\}$. 
where  $\langle \sigma^{\mu}\rangle = \text{tr}(\sigma^{\mu}\rho_i),$ with $\rho_i$ being the single site reduced density matrix,  denotes the magnetization of the system and $\langle \sigma_i^{\mu}\sigma_{i+r}^{\nu}\rangle=\text{tr}(\sigma^{\mu}\sigma^{\nu}\rho_{i, i+r})$ are the elements of the correlation tensor, as defined after Eq. (\ref{eq:gen1}), with $\mu,\nu=\{x, y, z\}$.
%The ground state obtained from canonical equilibrium state is known as thermal ground state. This represents the 
The above state is the ``symmetric $X$" state (i.e., the state in Eq.~(\ref{eq:X}) with $b=c$). If the ground state is degenerate,  one may consider the  equal mixture of all the degenerate ground states which can be called the symmetry-unbroken ground state or the zero-temperature thermal state, and is given by 
% and can be obtained from  the thermal state taking the $\beta \rightarrow \infty$ limit: 
\begin{align}
\rho_\text{eq} = \lim_{\beta\rightarrow \infty} \frac{e^{-\beta H}}{Z},
\end{align}
where $Z=\text{tr}\big[ \exp{ (-\beta H) } \big]$ is the partition function and $\beta=1/{k_B T}$ with  $k_B $ being Boltzmann constant and $T$ being the absolute temperature. 
Such mixtures retain the form of the reduced two party state as in Eq. (\ref{Eq:two_partyx}). 
In particular, the magnetizations in the $x$- and $y$-directions still vanish. 
The same properties can be retained in certain symmetric pure superpositions of the degenerate ground states.
% This is the symmetry unbroken ground state.
On the other hand, when one considers the symmetry-broken state, the magnetizations in the $x$- and $y$-direction remain non-zero. %which is not the case for unbroken ones.  
%However, for the symmetry broken state, many of the correlators become nonzero. Since our Hamiltonian is real, the correlators $\langle \sigma^y\rangle$, $\langle \sigma^x \sigma^y\rangle$,  $\langle \sigma^y \sigma^x\rangle$ and $\langle \sigma^y\sigma^z\rangle$,  $\langle \sigma^z \sigma^y\rangle$  vanish.  The correlations for the $XXZ$ model can be obtained from the thermodynamic Bethe ansatz \cite{Takasaki} while for the $XY$ model the correlators are obtained via successive Jordan-Wigner, Fourier and Bogoliubov transformation \cite{barouch1, barouch2}.% To obtain the symmetry broken 

{\bf $\boldsymbol{XXZ}$ chain ($\boldsymbol{\gamma=0}$ and $\boldsymbol{h_z=0}$):} 
It is known that the $XXZ$ model undergoes QPTs at $\Delta = \pm 1$~\cite{Takasaki}. 
For \(J>0\), when the system crosses  $\Delta =-1$ from  $\Delta<-1$,
% with positive $J$, 
a ferromagnetic-to-$XY$ (spin flopping) transition occurs, while at $\Delta = 1$, the system undergoes an $XY$-to-antiferromagnetic transition. It is known that the former is an infinite-order QPT, whereas the latter is a first-order one. The \(\Delta = -1\) transition is detected by the discontinuity in the derivative of QD, while it is the discontinuity of QD that itself detects the transition at \(\Delta =1\) \cite{Dillenschneider, Sarandy, Kundu2013}.  
%Here,  $\gamma=0$ and $h_z=0$ case is considered as the $XXZ$ model where the coupling strengths along $x$ and $y$ direction are uniform while $z$-coupling is different.  In Ref.~\cite{Dillenschneider, Sarandy}, the Authors  found that the QD of the nearest-neighbor is discontinuous at $\Delta = \pm 1$. 
Recently, Huang \cite{HuangScalingQD} provided an analytical expression of QD between two distant neighbors of the system. 
In case of entanglement, the transitions at \(\Delta = -1\) and \(+1\) are detected respectively by the change of entanglement from vanishing to non-vanishing and by its being maximal~\cite{GLL03, Syljuåsen03}.
%An interesting point to note here is that for $\Delta<-1$, the state is separable and hence entanglement in this model cannot infer the transition at $\Delta=-1$. 
%Werlang {\emph et al.}~\cite{Werlang}  showed that unlike entanglement, 
QD, however, can detect the QPT at $\Delta = -1$ until  some finite temperature with $k_BT \leq 3$ \cite{Werlang}.
% While at any finite temperature, the entanglement of the nearest-neighbor thermal state fails to identify the critical point, QD of the same state can locate the QPT even at moderately high temperature ($k_BT\le3$). 
Such study has also been carried out  
% The Authorsin Ref.~\cite{Werlang2} extended the result 
  for the $XXZ$ chain with an external magnetic field and  has been shown that  QD is a faithful  critical point detector also for this system at zero as well as finite temperatures \cite{Werlang2}.
  %when the system is in equilibrium with a thermal reserver at temperature $T$.

\begin{figure}[t]
\includegraphics[angle=0,width=7.5cm]{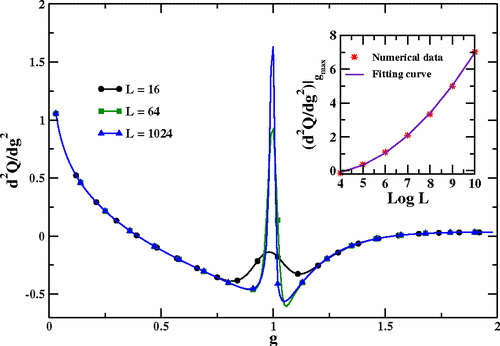}
\caption{Quantum discord in the transverse Ising model. Behavior of the second derivative of QD is plotted against $g~(= {h_z}/{J}$ in our notation) in the transverse Ising model. In the figure, 
quantum discord is denoted by \(Q\), while it is ${\cal D}$ in our notation. Also the system size is $L$ here while it is $N$ in the text. The vertical axes are in bits while the horizontal ones are dimensionless. [Reprinted from Ref.~\cite{Sarandy} with permission. Copyright 2009 American Physical Society.] }
\label{fig: Ising}
\end{figure}

{\bf Transverse field Ising and $\boldsymbol{XY}$ chains:} 
In Eq. (\ref{eqn: Hamil}), if we consider $\Delta = 0$, $\gamma \neq 0$, the Hamiltonian then describes a system, which is known as the $XY$ model. By setting $\gamma = 1$, it further reduces to the transverse Ising model. 
%In Eq.~(\ref{eqn: Hamil}), if the $zz$ interaction is dropped  by taking $\Delta=0$, the Hamiltonian is known as the transverse $XY$ Hamiltonian. This model is  exactly solvable\cite{LSM, barouch1,  barouch2}. The 1D transverse Ising Hamiltonian can be obtained from Eq. (\ref{eqn: Hamil}) by setting $\gamma=1$ and $\Delta=0$, which is a special case of the transverse XY model.    
Dillenschneider~\cite{Dillenschneider} was among the first to study QD in the transverse Ising model to identify the QPT present in this model\footnote{The Ising transition point at \(h_z/J =1\) separates the antiferromagnetic (AFM, \(J>0\)) or ferromagnetic (FM, \(J<0\)) phase from the paramagnetic (PM) one, while the anisotropy transition separates the AFM or FM order along the $x$-direction from the same along the $y$-direction.} at \(h_z/J=1\). 
It was shown that the next-nearest-neighbor QD (but not the nearest-neighbor QD) has its maximum value near the critical point, where the monogamy bound for concurrence squared is conjectured to be saturated~\cite{Osborne_Ni}. % to maximize the entanglement content between the subparts.
% Unlike entanglement where next-nearest-neighbor (and not nearest-neighbor) entanglement reaches a maximum at the critical point $h_z/J = 1$, both
However, unlike quantum entanglement, both the nearest-neighbor and next-nearest-neighbor QD are maximal in the region close to the QPT, but not exactly at the quantum critical point. 
It was shown that although nearest and next-nearest-neighbor QD are continuous, the first derivative of the QD of nearest-neighbor spins shows an inflexion, while, interestingly, the first derivative of the latter has divergence at $h_z/J = 1$~\cite{Sarandy}. %  implying the {\color{orange} second order} quantum phase transition \cite{Sarandy}.
It was also  pointed out in the same work that the second derivative of QD of the nearest-neighbor sites has quadratic logarithmic divergence, and the corresponding scaling analysis has also been performed (see figure~\ref{fig: Ising}).
%Note that, instead of the nearest-neighbor QD, the derivative of the next nearest-neighbor QD shows a divergence at the critical point. Later, Sarandy ~\cite{Sarandy} pointed out that the presence of the QPT is reflected by a inflection point of first derivative of the nearest-neighbor QD. So, the second derivative of the nearest-neighbor QD shows a kink at the quantum critical point (See Fig. \ref{fig: Ising}).  In general, the derivatives of QD display the characteristic (logarithmic) divergence of the critical Ising model. 

The Ising QPT at $h_z/J=1$ in the anisotropic $XY$ model is   marked by a divergence in the derivative of the nearest-neighbor QD with respect to the external field \cite{Maziero}. 
For the symmetry-broken state, this divergence is present in the entire Ising universality class $(0<\gamma\le 1)$, while for the thermal ground state, it holds for all $\gamma$ except $\gamma=1$~\cite{Tomacello}. At $\gamma=1$, as discussed above, instead of the first derivative, the second derivative of the nearest-neighbor QD diverges~\cite{Sarandy}. It is worthwhile to mention here that in the case of entanglement, the Ising QPT is always characterized by a divergence of the first derivative of bipartite entanglement for the entire Ising universality class ($\gamma \in (0, 1]$) including 
%and there exists no exception for 
the Ising model $(\gamma=1)$. This is expected because the Ising transition
% at $h_z/J=1$ 
is seen for the entire Ising universality class which includes the Ising model. %and $\gamma=1$, Ising model is included within it. 
The reason behind the different behavior of QD seen for the Ising QPT in the Ising model is still not clear.

Apart from the quantum criticalities, viz. the Ising and the anisotropy transitions, it has been revealed that in the $XY$ model, there exists a point in which entanglement, both bipartite as well as multipartite, vanishes, with the corresponding point being known as the factorization point, given by $h_z = h_z^f = J\sqrt{1 - \gamma^2}$ \cite{Kurmann}.  
% $XY$ model gets special attention due to the presence of a factorization point~\cite{Kurmann} in addition to the Ising and anisotropy transition. 
 The factorization point and its neighborhood refer to a region where entanglement is low, which is an important information for possible implementation of QIP in this system. %for implementation of QIP.

Up to now, we were discussing about the role of bipartite QD in  physical phenomena of quantum many-body systems. We will now discuss whether discord length, i.e. the behavior of QD between two sites, with increasing lattice distance between the sites, has significance in physical phenomena of these systems.
Note that entanglement vanishes for pairs which are farther than next-nearest-neighbors  in the transverse field $XY$ model, and this is independent of whether the system is at the factorization point. Interestingly this is not the case for QD. Specifically, 
% while it is included in Ising universality class, is still not clear. 
  in Ref.~\cite{Maziero}, the Authors showed that  just like nearest-neighbor QD,  QD between farther neighbors can still characterize QPTs. In Refs.~\cite{HuangScalingQD, Maziero2, Campbell}, five regions  in the parameter space of $XY$ model have been identified   
  %showed that while entanglement between next to next neighbor and beyond vanishes..  the Authors identified five regions 
  where scaling of two-site QD with the distance between the sites are different. 
%   Analytical expressions are provided for the regions identified by 
  They are $ h_z/J>1,h_z/J=1, \sqrt{1-\gamma^2}<h_z/J<1, h_z/J = \sqrt{1-\gamma^2}$ and $0<h_z/J<\sqrt{1-\gamma^2}$. 
  
  At the factorization point, $h_z^f = J \sqrt{1-\gamma^2}$, the ground state is doubly degenerate~\cite{Kurmann}. If we take %symmetric ground state i.e. 
  the thermal state in the $T\rightarrow0$ limit, the two-site QD becomes scale-invariant, i.e. the QD between the $i^\text{th}$ and $(i+r)^\text{th}$ spins, denoted by ${\cal D}_r$, remains constant for any $r$,   
  % ~(r\le N/2)$ 
   which leads to violation of monogamy for QD~\cite{Ciliberti,Sadhukhan}. We will discuss the issue of monogamy of quantum correlations in Sec.~\ref{sec:monogamy}.    
However, the situation changes if one takes the symmetry-broken ground state. Tomasello \emph{et al.}~\cite{Tomacello} showed that if one considers a symmetry-broken state, as obtained by a negligible perturbation of longitudinal field ($h_x$), all the two-site QDs, ${\cal D}_r$, vanish at the factorization point for all system sizes. Moreover, it was numerically found that in the symmetry-broken phase, close to the factorization point, the two-site QD between the $i^\text{th}$ and $(i+r)^\text{th}$ sites scales as
\begin{align}
{\cal D}_r \sim (h_z - h_z^f)^2 \times \Big( \frac{1-\gamma}{1+\gamma}\Big)^r.
\end{align}
\begin{figure}[t]
\includegraphics[angle=0,width=8cm]{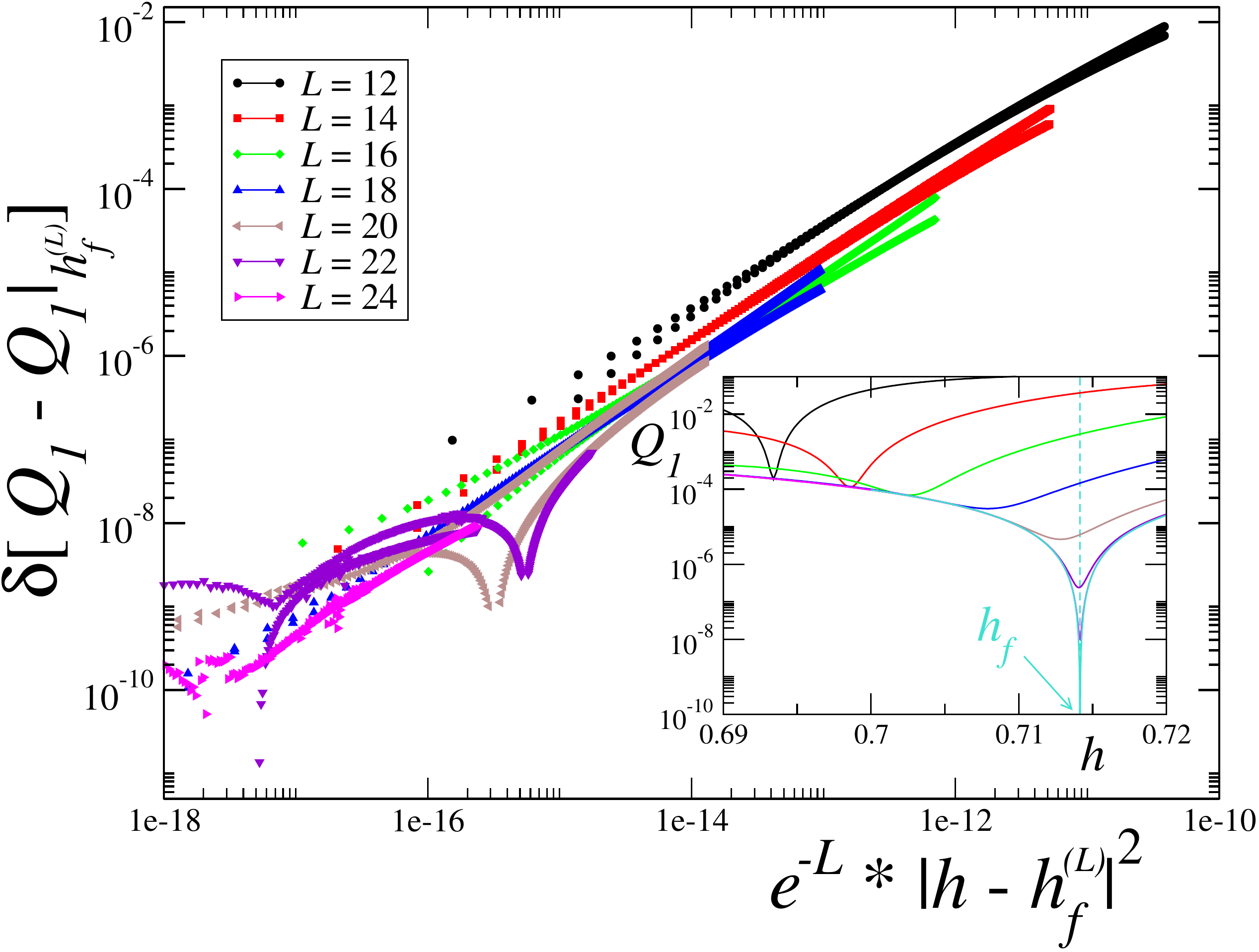}
\caption{Scaling of nearest-neighbor QD (denoted by $Q_1$ in the figure) is analyzed close to the factorization point for $\gamma = 0.7$. % quantum discord close to a factorization point. Scaling of ${\cal D}_r$ (denoted by $Q_r$ in the figure) close to the factorizing field $h_f$ for $\gamma = 0.7$. 
 The system size is $L$ here while it is $N$ in the text. It is observed that $Q_1^{(L)}$, which is the value of $Q_1$ for a system of size $L$, converges in the thermodynamic limit as % An exponential convergence to the thermodynamic limit, with a universal behavior according to  
 $e^{-\alpha L} (h-h_f^{(L)} ),~ \alpha \approx 1$. Here $h$ replaces $h_z/J$, and $h_f^{(L)}$ is the value of $h$ at the factorization point for a system of size $L$. The quantity plotted in the vertical axis is $\Big(Q_1^{(L)}-Q_1^{(L)}\Big|_{h=h_f^{(L)}}\Big)-\Big(Q_1^{(L\rightarrow \infty)}-Q_1^{(L\rightarrow \infty)}\Big|_{h=h_f^{(L\rightarrow \infty)}}\Big)$.  
% In the figure, the subscript of the transverse field is dropped and $h_f$ instead of $h_z^f$ is used. Here,   $h_f^{(L)}$ denotes the effective factorizing field in transverse direction at size $L$ and  $\delta({\cal D} ) \equiv {\cal D}^{(L)} - {\cal D}^{(L\rightarrow \infty)} \equiv Q^{(L)}_1 - Q^{(L\rightarrow \infty)}_1 $. 
Due to the extremely fast convergence to the asymptotic value, differences with the thermodynamic limit are comparable with density matrix renormalization group~\cite{white92, white93} accuracy, already at $L \sim 20$. Inset: Raw data of $Q_1$ as a function of $h$. The cyan line is for $L = 30$ for which, up to numerical precision, the system behaves as being at the thermodynamic limit. The vertical axes are in bits, while the horizontal ones are dimensionless. [Reprinted from Ref.~\cite{Tomacello} with permission. Copyright 2011 IOP Publishing.] }
\label{fig: facXY}
\end{figure}

\noindent The scaling of the symmetry-broken QD near the factorizing point  is plotted in figure~\ref{fig: facXY}, which is consistent with the results obtained in Ref.~\cite{Baroni07}. 

The temperature-dependence of nearest-neighbor QD of the $XY$ model have been 
studied in Refs. \cite{Werlang2, Maziero, Sadhukhan2, Mishra16}. 
Like entanglement~\cite{Arnesen01}, non-monotonicity of QD with the increase of temperature have also been reported~\cite{Maziero}. 
% In general, QD is more robust than entanglement against thermal noise \cite{Werlang2, aro khujte hbe}. 
  It was shown that the nearest-neighbor QD is a better indicator of the Ising transition ($h_z/J=1$) than  the two-site entanglement at  finite temperature. % $T$. 
On the other hand, at low temperatures   the anisotropy transition ($\gamma=0$) can be correctly  detected  both by entanglement and QD. With the increase of temperature, QD turns out to be a better physical quantity to identify the Ising transition point %, both Ising and the anisotropy transitions 
than pairwise entanglement~\cite{Werlang2}.  

Dhar \emph{et al.}~\cite{Dhar} studied long range QD between the non-interacting end spins of an open quantum $XY$ spin chain, with the end spins weakly coupled to the bulk of the chain. It was  shown that when the end couplings are adiabatically varied below a certain threshold, QD between the end spins remains frozen. The interval of the freezing can detect the anisotropy transition in  the chain. 

{\bf Other models:} The behavior of QD has recently been investigated in an 
%Chanda \emph{et al.}~\cite{Chanda} investigated  QD in an 
alternating field $XY$ model~\cite{Chanda} where the external magnetic field is not uniform for all the sites but has an alternating nature. 
Interestingly, with the introduction of variations of  local transverse fields, 
it was found that apart from the AFM/FM to PM transition, the system undergoes a dimer to AFM/FM transition.
Both the transitions are shown to be detected by the divergence of first derivatives of the nearest-neighbor QD. 
%The phase boundaries for the AFM-PM  and the Dimer-PM transitions in the system can be characterized from the divergence of the first derivative of nearest-neighbor QD. 
The effect of thermal fluctuation on QD  has also been studied in this system and nonmonotonic behavior of QD with respect to temperature is found in the AFM/FM and PM phases
while the dimer phase does not show any non-monotonic behavior.

Patterns of QD in several other 1D quantum spin systems have been carried out. The systems include 
%  where QD is shown to be useful to identify the QPT, especially for the long range spins. Similar studies include 
$XY$ spin chain~\cite{FZ2012} with  three-spin interaction~\cite{Li, Ben-Qiong11} and with Dzyaloshinskii-Moriya (DM)  interaction~\cite{Liu, Ma, Chen, Zhu}, $XYZ$ model with inhomogeneous interaction~\cite{Yang2012},  symmetric spin trimer and a tetramer~\cite{Pal}, Dicke model, and the Lipkin-Meshkov-Glick model~\cite{WangZhang12}.
% For DM interactions, QD even for the third-nearest-neighbor can indicate the QPT there. 
%   Surprisingly, the dimer phase lacks this kind of nonmonotonic behavior.

QD has also been studied in the 
%In Ref.\ \cite{Sadhukhan2}, the Authors took 
the quenched disordered quantum $XY$ model~\cite{Sadhukhan2} where the coupling constant  is chosen randomly from a Gaussian distribution. % with mean $J$ and  standard deviation $\sigma$. 
The disorder is assumed to be quenched which imply that the time scale of the dynamics of the system is much shorter than the equilibration time of the disorder. 
Although disorder may intuitively seem to  suppress physical quantities of the system, it turns out that QD can be enhanced with the introduction of disorder - an instance of the ``order-from-disorder" phenomena~\cite{Sadhukhan2, Sadhukhan, Niederberger08, Niederberger10, Villain80, Aharony,Minchau85, Feldman98, Abanin07,  Volovik06, Wehr06, Pradhan}. 
The disorder-induced enhancement is observed both at   
%Order from disorder phenomenon of QD is observed for both the
 zero  and  finite temperatures. Moreover, it was shown that in some parameter regimes, thermal fluctuation interfere constructively to generate a more pronounced order-from-disorder in QD.  
It was also shown that  
 the long-range behavior of QD can be improved by introducing disorder in the $XY$ spin chain~\cite{Sadhukhan}. However, the scale invariant behavior of QD at the factorization point of the ordered system is absent in its quenched disordered counterpart.

\subsubsection{Spin-1 systems}

\begin{figure}[h]
\includegraphics[angle=0,width=8cm]{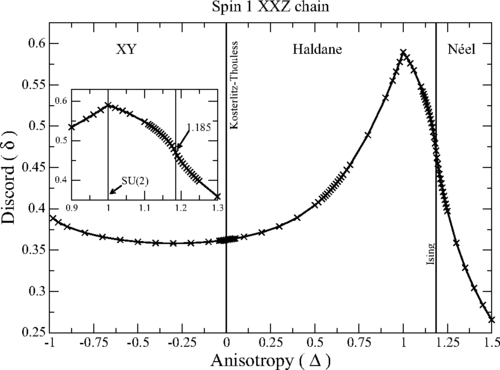} 
\caption{Quantum discord detects a Haldane - N\'{e}el transition. The figure shows QD versus the $zz$-interaction strength   in the 1D spin-1 $XXZ$ model. In the figure, QD and $zz$-interaction strength are respectively denoted by $\delta$ and $\Delta$, while these are denoted by ${\cal D}$ and $\Delta/J$ in the text, respectively. Different phases, transition points, and the $SU(2)$ symmetry point are shown. The vertical axes are in bits, while the horizontal ones are dimensionless. [Reprinted from Ref.~\cite{Malvezzi16} with permission. Copyright 2016 American Physical Society.] }
\label{fig: xxzSpin1}
\end{figure}

As already discussed in Sec.~\ref{sec:Computability}, computation of QD of higher-dimensional states, is difficult and hence most of the studies on the behavior of QD in spin models are limited to systems consisting of spin-$\frac 12$ particles. However, there are some methods (see Sec.~\ref{subsec:Higherdim}) which can be employed to deal with two-qutrit states, originating say, from spin-1 systems.  
QD in the ground state of the 
%In Ref. \cite{Malvezzi16}, the Authors calculated QD 
 spin-1 $XXZ$ chain and a spin-1 bilinear quadratic chain has been studied~\cite{Malvezzi16}.  
% In contrast to the previous case, here QD is computed by 
For optimizing over the projective measurements, generation of  random unitary matrices is employed as an initial step~\cite{Zyczkowski94, Pozniak98}. The Hamiltonian of the spin-1 $XXZ$ model can be obtained from Eq.~(\ref{eqn: Hamil}) by setting $\gamma=h_z=0$ and by taking the $S_i$'s as spin-1 operators. The model is known to have several quantum critical points with respect to the strength of the $zz$-interaction $\Delta$ as we walk from low to high $\Delta$: (i) $ \Delta = \Delta_{c_1} \equiv -1:$ FM $\rightarrow$ XY phase as in the spin-$\frac 12$ $XXZ$ chain; 
(ii) $ \Delta = \Delta_{c_2}\approx 0:$ $XY$ $\rightarrow$ Haldane; 
(iii) $\Delta = \Delta_{c_3}\approx 1.185:$ Haldane $\rightarrow$ N\'eel phase~\cite{Alcaraz92, Kitazawa96, Sakai90}. The first and third transitions are respectively 1st and 2nd order while the second one is 
%Here, the QPT at $\Delta_{c1}= -1$ is a first order transition while QPT at $\Delta_{c3}$ is a second order phase transition within the 2D Ising universality class. The transition at $\Delta_{c2}$ is 
a Kosterlitz-Thouless (KT) transition of infinite order. % {\color{orange} The Authors have found the ground state of this model by the standard DMRG infinite system method \cite{white92, white93}. }
The behavior of QD against $\Delta$ within $-1<\Delta<1.5$ is shown in figure~\ref{fig: xxzSpin1}~\cite{Malvezzi16}. While QD seemingly fails to capture the infinite order KT transition (and the situation is the same with  other QC measures), it can accurately detect the second order Haldane-N\'eel phase transition. The QD indeed shows an inflection point at $\Delta_{c_3}\approx 1.185$ which results in a kink in the derivative of QD. Moreover, the model has a  SU(2) symmetry point at $\Delta=1$ which can also be observed from the sudden change of QD, happening due to the change of the optimal measurement basis.

Another model considered in  Ref.~\cite{Malvezzi16} is a 1D spin-1 bilinear biquadratic model. The  Hamiltonian in this case is given by 
\begin{align}
H_\text{BB} = \sum_{\langle i,j \rangle} \Big[  \cos{\theta} ({\bf S}_i \cdot {\bf S}_{j}) + \sin{\theta} ({\bf S_i} \cdot {\bf S}_{j})^2  \Big],
\end{align}
where $\theta \in [0,2\pi)$ is an angle that modulates the coupling strength of the nearest-neighbor spins. By tuning $\theta$, the system undergoes four different kinds of QPTs - a KT transition at $\theta_{c1}=0.25 \pi$ from Haldane to trimerized phase, a first order transition at $\theta_{c2}=0.5 \pi$ from trimerized to ferromagnetic, and  another first order transition at $\theta_{c3}=1.25 \pi$ from ferromagnetic to dimerized phase, and finally  a second order transition at $\theta_{c4}=1.75 \pi$ from trimerized to Haldane phase. There is difference in opinions about other transitions which have been suggested for this model. % are still unresolved. 
Like the $XXZ$ spin-1 chain, this model also has a special SU(3) symmetry point at $\theta = 0.25 \pi$. 
In this case, QD has been computed for  system of up to 12 spins with open boundary conditions. Despite the small system size, QD can actually detect the critical points at $\theta_{c2}$ and $\theta_{c3}$. However, it is failed to identify the KT transition and the second order transition from dimerized to Haldane phase.
%{\color{orange} It was noticed that although the spin system of moderate  size can not detect second order transition, behavior of correlations in the two-sites in the bulk which is obtained via density matrix renormalization group infinite system method can detect such transition.}
  Moreover, like the $XXZ$  model, a sudden change in QD occurs, due to a drastic change of measurement basis, at the SU(3) symmetry points. 
For further attempts to calculate QD and related measures for two-qutrit spin systems with different magnetic fields, see Refs.~\cite{Higher-dim,Higher-dim1,Higher-dim2}. 

Another spin-1 Hamiltonian where QD has been studied is given by~\cite{Power15} 
% However, these works are limited to just two parties which does not reveal the so called `many-body' feature. In this regard, the first attempt to calculate QD in moderate size spin-$1$ systems has been made in Ref. \cite{Power15}. {\color{red} \{ Ref. \cite{GQD_many_body4} calculate QD in a XYZ spin-1 chain, but only in the vicinity of the factorizing field\}} ~With the help of the symmetry present in the system, the Authors  evaluated the bipartite version of global QD $({\cal D}_{global}(\rho_{A_1 A_2}))$ (See Sec.\ \ref{globalQD}) for the 1D spin-1 Heisenberg chain with a  uniaxial field(UF) upto 256 spins.  The corresponding Hamiltonian is given by  
\begin{align}
{H_\text{UF}} =  \sum_{\langle i,j \rangle}S^x_i S^x_j + S^y_i S^y_j + + S^z_i S^z_j + U  \sum_{i} (S^z_i)^2,
\label{eqn:spin1U}
\end{align}
where $U$ is the strength of the uniaxial field which is the same for  all lattice sites.  %Global QD for general case $({\cal D}_{global}(\rho_{A_1, \cdots ,A_N}))$  was also computed for this system but for a relatively smaller system size (upto 8 spins). 
The ground state of the Hamiltonian is known to have three phases, namely, a N\'{e}el AFM phase ($U<-0.315$), a Haldane phase ($-0.315 \le U \le 0.968$) and a {``large-D''} phase  ($U > 0.968$)~\cite{Botet83, Glaus84, Schulz86}. Note that $U$ is a dimensionless quantity. These three phases have been studied  by the block entropy and entanglement spectrum~\cite{Degli03, Pollmann10, Sanpera11, Hu11, Lepori13}. In Ref.~\cite{Power15}, QD of  the reduced density matrix of the two central spins of the zero-temperature state of the Hamiltonian in Eq.~(\ref{eqn:spin1U}) with open boundary condition has been analyzed by using density matrix renormalization group. The quantities calculated here are symmetric QD (see Sec.~\ref{bipartite global qd} and global QD (see Sec.~\ref{Sec:Global_QD}). Taking into account the symmetry of the Hamiltonian, it was argued that the optimization over measurements, required for estimating the global QD, 
can be made efficient by reducing the number of parameters over which the optimization is performed~\cite{Campbell13,GQD_many_body4}. See the discussion in Sec.~\ref{subsec:Higherdim} in this regard. See also~\cite{sun-global,Lio-globalQD}.   

%As discussed in Sec.\ \ref{subsec:Higherdim}, the parametrization of local orthogonal measurements for spin-$1$ systems requires six coefficients. Since the ground and thermal states of the above Hamiltonian can be chosen to be real, one can always  consider the optimal basis to be real \cite{Campbell13}. This results,  reduction of some of the optimizing parameters: $\gamma = \psi = \phi = \phi_0 = 0$. However, this constraint is only applied to the optimization of global QD. For the bipartite version of global QD a full minimization with the full set of angles is numerically possible.

\begin{figure}[h]
\includegraphics[angle=0,width=4.25cm]{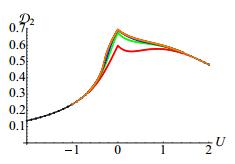} 
\includegraphics[angle=0,width=4.25cm]{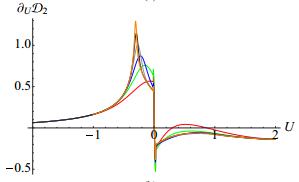} 
\caption{(Left panel) N\'{e}el AFM - Haldane transition by using symmetric QD between nearest-neighbor sites, and (right panel) its derivative for the reduced state of the two central spins of an open-ended chain, described by the Hamiltonian in Eq.~(\ref{eqn:spin1U}) of lengths  8 (red), 16 (green), 32 (blue), 64 (gray), 128 (black), and 256 (orange) going from bottom to top. The symmetric QD in the vertical axis is denoted by ${\cal D}_2$ in the figure while it is  ${\cal D}_{sym}$ in the text.  The vertical axes are in bits while the horizontal ones are dimensionless. [Reprinted from Ref.~\cite{Power15}  with permission. Copyright 2015 American Physical Society.]%Notice that the curves for L = 32, 64, 128, and 256 are almost indistinguishable except near $U ∼ −0.3$. 
}
\label{fig: Spin1U}
\end{figure}
Figure~\ref{fig: Spin1U} depicts the variation of the symmetric QD, ${\cal D}_{sym})$ with $U$. 
%The bipartite version of global quantum discord, ${\cal D}_2$, for the above-mentioned reduced density matrix is plotted in Fig.\ \ref{fig: Spin1U}
% The cusp at $U=0$ corresponds to the change in the optimizing angle for evaluating ${\cal D}_2$. For $U<0$, ${\cal D}_2$ is optimized when both spins are measured using the angles $\theta=\alpha=\beta=0$ which corresponds to a projection onto the eigenbasis of $S_z$. Similarly,  $U > 0$ the optimizing angle is $\theta=\pi/2, \alpha=\beta=0$ which corresponds to a projection onto the eigenbasis of $S_x$. However, $U=0$, both yields same result. This sudden change at $U=0$ signifies the switching from easy axis ($U<0$) to easy plane ($U>0$) and hence the point $U=0$ is not a critical point. Just like the spin-$\frac 12$ systems, here also
 The N\'{e}el AFM-Haldane phase transition, a second order QPT, is signalled by a discontinuity of   the first derivative of symmetric QD of the zero-temperature state and scaling analysis predicts the transition point to be  at $U=-0.3156$, which is consistent with result obtained in  Ref. \cite{Sanpera11} from the analysis of entanglement in this system.  However, the Haldane - large-D transition at $U=0.968$ \cite{Hu11} is known to be Gaussian, a third-order transition, and is hard to detect. % by any physical quantities. %, the first derivative of the bipartite global QD shows  a point of inflection. Finding the minimum of
 By performing  the second derivative of the symmetric QD,  this transition is predicted to be at $U=0.994$, and the critical exponent is found to be $1.6$.  Both the results are in good agreement with previous calculations~\cite{Hu11} with $20000$ spins. 
 %using finite size extrapolation which is not exactly the transition point The Authors cited that locating the QPT accurately will require 
It was argued that the results obtained by using symmetric QD are with at most $256$ spins, and hence can be improved substantially by considering larger system sizes.% beyond their computational capacities. However, the Authors found that for system size greater than 32, the second order derivative of the bipartite global QD for different system sizes crosses each other at the same point $U=0.9667$.  A scaling analysis of the same reveals that the value of the critical exponent associated with the Haldane - large-D transition matches closely with the value of the critical exponent obtained in Ref. \cite{Hu11}.

\subsection{Dynamics}
Let us now move on to discuss the behavior of QD in the time-evolved state of different systems. 
The considerations are often for time-dependent Hamiltonians, and the initial state that is usually considered is the canonical equilibrium state at the initial time\footnote{For the behavior of entanglement in the time-evolved state in the system described by the time-independent quantum transverse $XY$ model, with the evolution starting off from a 
non-thermal state, see Refs.~\cite{Amico04, Subrahmanyam04,PAOF04, CFMMP04,VPA04,Cao2005, CZ06}.}, denoted by  
 %when a system evolves under the influence of a time dependent Hamiltonian. %The ground state of a physical state evolves and eventually may incur a transition, known as the dynamical phase transition \cite{Heyl,Canovi,Macieszczak, aditi_dynamics}. 
%Any suitable observable like two point correlations, entanglement, etc.\ are  capable of detecting such transitions which occurs during the time evolution of the ground state of quantum many-body systems.  The time evolution of the ground state is given by 
%\begin{equation} \label{eqn:dynamics}
%\rho(t) = \exp{\big[ -i\int_0^t\hat{H}(\tau)d\tau\big]} \rho_\text{eq}(0)\exp{\big[i\int_0^t\hat{H}(\tau)d\tau \big]}
%\end{equation}
%where $\int_0^t\hat{H}(\tau)d\tau$ is the time ordered integral of the Hamiltonians at different times. $\hat{H}(\tau)$ denotes the Hamiltonian at time $\tau$. Here, 
%In order to study the time evolved density matrix, the canonical equilibrium state  
$\rho_\text{eq}(t=0)$. For example, the 
 dynamics of entanglement has been studied after a sudden quench in the  transverse field of a 1D quantum $XY$ model~\cite{aditi_dynamics, AditiErgo04, Huang05}. The transverse field is given by 
% Need to include Ref. Amico04
\begin{eqnarray}
\label{eqn:dyna}
 h_z(t) &=& a~ \text{(constant)~ for} ~t \le 0, \nonumber\\
  &=& b ~(\neq a)  ~\text{for} ~t > 0. 
\end{eqnarray} 
In this case, evaluation of the time-evolved state does not require a time-ordered integral, since the Hamiltonian after $t = 0$ becomes time-independent. % and hence the time-ordered integral is not needed. In Ref. \cite{aditi_dynamics} a special case, $b=0$, was considered which leads to even more simplifications. Depending on the initial value of the transverse field, $a$, the nearest-neighbour entanglement, as measured by logarithmic negativity~\cite{Werner_LN}, ${\cal L}$, collapses and revives with time \cite{aditi_dynamics}.  
Putting $b=0$~\cite{aditi_dynamics}  and the initial temperature $\rightarrow 0$, the behavior of entanglement,
as quantified by logarithmic negativity~\cite{Werner_LN}, ${\cal L}$ (for definition, see Appendix~\ref{LN}), has been investigated with respect to the initial field and time. For fixed (relatively) short times,
 entanglement shows several collapses and revivals with the increase of the initial transverse field -  dynamical phase transitions. Such revivals cannot be seen for  larger times. 
%( See Ref.  \cite{Amico04} for another kind of entanglement propagation in the $XY$ model). 
  It was found that the behavior of nearest-neighbor QD can predict such collapse and revival of entanglement~\cite{Himadri} (see also Ref. \cite{Dhar_WD}). % For a fix value of the system parameter $a$, at the particular snap when $\cal L$ encounters a sudden death, i
Specifically, it was observed that at the vicinity of collapse, if  QD is increasing i.e.\ if the slope of QD with respect to $a$ is positive, then entanglement will  revive for a certain larger value of the initial field. Mathematically, 
% Now the QD of the nearest neighbour spin pairs basically predicts the possibility of the revival of $\cal L$ with the system parameter $a$, for a particular snap of time and it is given by
\begin{equation}
\tilde{a} \frac{\partial {\cal D}(\rho_{i,i+1}(t))}{\partial \tilde{a}} \Big|_{\tilde{a}_c} > 0 \Rightarrow {\cal L} > 0 ~\text{for some} ~|\tilde{a}| > |\tilde{a}_c|,
\end{equation}
where $\tilde{a} = a/J$ and $\tilde{a}_c$ is the value of the transverse field where $\cal L$ vanishes for any fixed time $\tilde{t} = t/J$.
%Thus by calculating the derivative of QD one can anticipate the possibility of the revival of quantum entanglement of nearest neighbour spin pair. 
At $t > 0$, instead of switching off the magnetic field, one can also quench the system by fixing the magnetic field to some other constant value,  $ b$, where $b\ne a$. The effect of such quench in QD has   recently been investigated \cite{Mishra16}. %In the alternating field $XY$ model where the transverse field are of an alternating nature, QD also known 
In the $XY$ Hamiltonian, given in Eq. (\ref{eqn: Hamil}) with $\Delta = 0$, if the transverse field at the 
$i^\text{th}$ site is replaced by $h_1(t) + (-1)^i h_2(t)$ where $h_1(t)$ and $h_2(t)$ respectively possess a
 non-zero value  and then both are switched off at a latter time, it was shown that QD also undergoes several revivals and collapses   % with evolution time if both the alternating fields are turned off at $t=0$
with the system parameters $h_1(t)/J$ and $h_2(t)/J$ for relatively short times \cite{Chanda}. However, the nature and count of the revivals and collapses depend on the initial values of the  alternating fields. In the $XXZ$ models, the dynamics of QD has also been investigated after a sudden quench in the $zz$-interaction strength \cite{Ren12}. Similar to the  Ising and $XY$ models, QD is found to be oscillating initially and finally saturating to a constant value.

Instead of taking a step-function-like quench, given in Eq.\ (\ref{eqn:dyna}), the quench can also be taken as a linear ramp \cite{Polkovnikov05, ZDZ05, Dziarmaga05, Damski05, Cherng06, Mukherjee07} driven across the quantum critical point. The ramp-like quench through  the  Ising transition point at a finite and steady rate is considered previously in Refs.~\cite{Polkovnikov05, Dziarmaga05, Cherng06} and can be written as % given by 
\begin{align}
h_z(t) = t/\tau,
\end{align}
where $\tau $ is related to the characteristic time-scale for the rate of quenching and $t$ is varied from $-\infty$ to $\infty$. In the 1D transverse  XY model with a linear  quench in the magnetic field, QD vanishes for both the limits $\tau \rightarrow -\infty$ and $\infty$, but has a peak at an intermediate value of the inverse quenching rate $\tau$ \cite{Tanay11, Tanay13}. It was found that both 
entanglement as well as QD behave similarly and exhibit a power-law scaling with the slow variation of $\tau$. %the rate of quenching in the slow quenching limit. However, the Authors could not obtain a closed form of the power law scaling. In Ref. \cite{Tanay13}, the Authors extended t
QD has also been studied with quenching of the field in the transverse Ising model with  three-spin interactions \cite{Tanay13}.

\begin{figure}[h]
\includegraphics[angle=0,width=8.5cm]{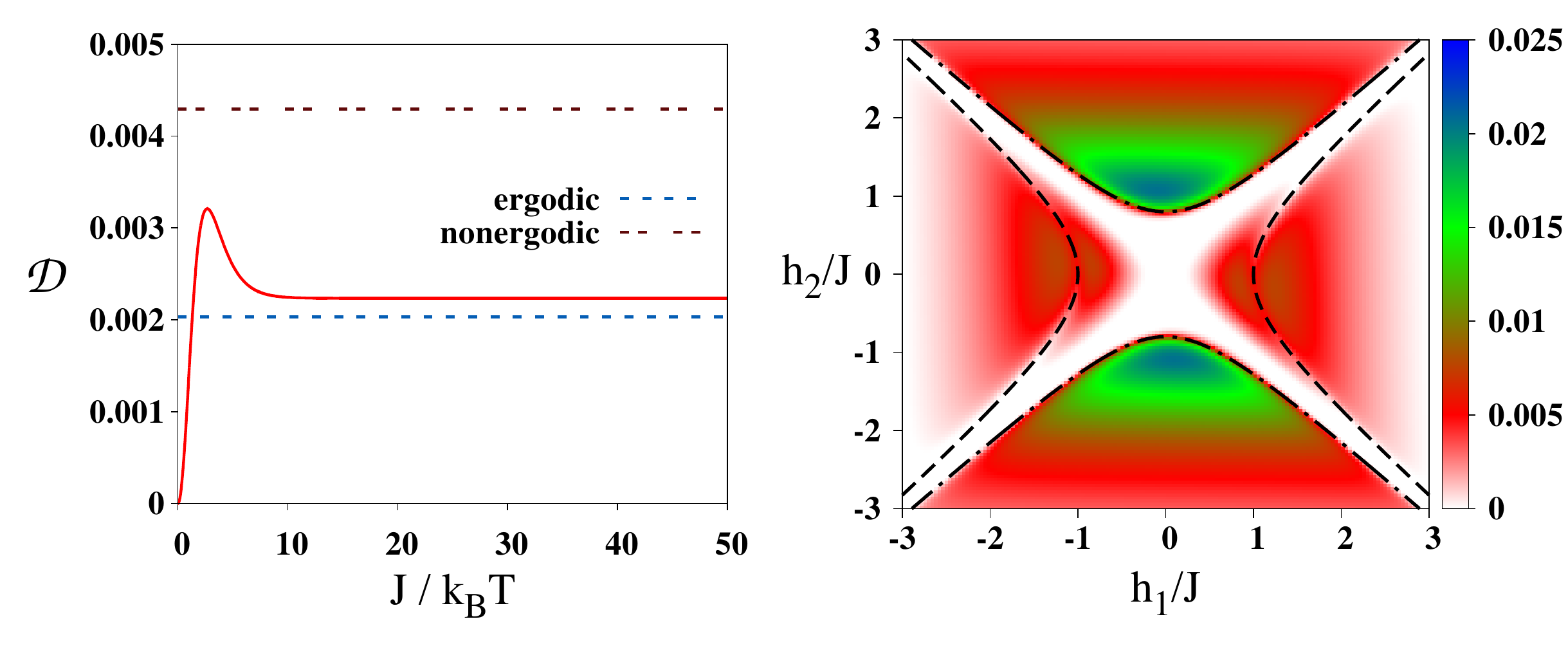}
\caption{Ergodicity of quantum discord of 1D alternating transverse field
$XY$ model. The Hamiltonian in this case is the usual 1D transverse-field \(XY\) model, with the transverse field being of the form \(h_1(t) +(-1)^ih_2(t)\) at site \(i\). Left panel: The red solid line represents the trends of  QD
for  the canonical equilibrium state of the Hamiltonian at large time
against $J/k_B T$. The blue dashed  and black double-dashed lines
correspond to  QD of  the time-evolved states of the same Hamiltonian at large time,
where the initial states are chosen to be the thermal equilibrium states
of the initial-time Hamiltonian  
with  ${h_1(t =0)}/{J}=0.0,~ {h_2(t =0)}/{J}=0.15$ and
${h_1(t = 0)}/{J}=2.5,~ {h_2(t = 0)}/{J}=1.0$ respectively. Here $\gamma
=0.8$, the temperature of the initial state is  given by $J / k_B T = 100$, and the
transverse fields $h_1(t)$ and $h_2(t)$ are switched off for  $t>0$.  
% and
%the driving field is given in Eq. (\ref{eqn:dyna}).
QD of the evolved state matches with that of the thermal state for
some temperature in the case of the blue dashed line, implying
an ergodic nature of QD, while in the other case (black double-dashed line), a nonergodicity of
QD is obtained. The vertical axis is in bits while the horizontal axis is
dimensionless.
%The ergodic case corresponds to a stiuation where the QD value for the
%time-evolved state at $t\rightarrow\infty$ (the blue-dashed line)
%coincides with QD value of the canonical equilibrium state at the initial
%time at least for one temperature (here, $T =  1.3137k_B / J $ K). The
%red-solid line represents the QD value of the initial state, given by the
%canonical equilibrium state of the alternating transverse field $XY$
%Hamiltonian with $\frac{h_1}{J}(t =0)=0.00, \frac{h_2}{J}(t = 0)=0.15$ at
%different temperature. The nonergodic situation (black-doubledashed line)
%corresponds to a case where we find no matching QD value of the canonical
%equilibrium state of the alternating transverse field $XY$ Hamiltonian
%with $\frac{h_1}{J}(t = 0)=2.50, \frac{h_2}{J}(t = 0)=1.00$ for all
%temperature.
Right panel: Map of  ergodic regions on the
$(\frac{h_1}{J},\frac{h_2}{J})$ plane of the same Hamiltonian. The white
regions are ergodic, as quantified by the ergodicity score, given by
$\eta^{\cal D}(\frac{h_1}{J},\frac{h_2}{J}) = \max \left\{0, {\cal
D}_{\infty}\big(T,\frac{h_1(0)}{J},\frac{h_2(0)}{J}\big) -\max_{T'} {\cal
D}_{eq}\big(T',\frac{h_1(\infty)}{J},\frac{h_2(\infty)}{J}\big) \right\}$, where $\mathcal{D}_{\infty}$ and $\mathcal{D}_{eq}$ denote the quantum discords of the time-evolved  and the canonical equilibrium states respectively at large time. A nonzero ergodicity score implies nonergodicity 
of QD. 
The temperature of the canonical state, from which the evolution starts off, is given by 
\(J/k_BT=100\). The anisotropy \(\gamma\) and the nature of the
driving field remain the same as in the left panel.
%For all the cases the initial temperature at $t = 0$ is taken to be $J /k_B T = 100$ at $\gamma = 0.8$.
%After $t = 0$, the initial transverse field is switched off.
%
Both the axes in the right panel are dimensionless.
 [Adapted from Ref.~\cite{Chanda} with permission. Copyright 2016 American
Physical Society.] }
\label{fig: ergo_nonergo}
\end{figure}
%In the course of the evolution of a quantum system, if the evolution time is large enough we can talk about the ergodic behaviour of an observable.
 
We now consider the situation when the evolution time is large enough, so that statistical-mechanical questions like ergodicity can be asked. A physical quantity is said to be ergodic if the time-average of the quantity is equal to its
ensemble-average. 
More precisely, a physical quantity $\mathbb{A}$ is said to be ergodic if the following two 
values match. One of these is the value of $\mathbb{A}$ in the time-evolved state at large time, where the evolution starts off from the canonical equilibrium state\footnote{There is of course a quenching in some physical parameter, e.g. in the magnetic field,
at the initial or some intermediate time before the ``large'' time.} at the initial time with a temperature $T$. 
%the value of $\mathbb{A}$ in the time-evolved state at a large time\footnote{There is of course a quenching in some physical parameter, e.g. in the magnetic field,
%at the initial or some intermediate time before the ``large'' time.}, matches 
The other is the value of $\mathbb{A}$ in the equilibrium state of the large-time Hamiltonian at some temperature $T'$~\cite{barouch1,barouch2,barouch3, PrabhuErgo1, PrabhuErgo2, PrabhuErgo3, Mishra16, Chanda, Sadiek12, HK05, Mazur,suzuki-paper,PAM06,AditiErgo04}. 
The difference of these two quantities has been denoted by $\eta^\mathbb{A}$ and called the ergodicity score,
where a maximization over $T'$ has already been carried out~\cite{PrabhuErgo2}. 
Often, $T'$ is constrained to be within an order of magnitude of $T$~\cite{PrabhuErgo1, PrabhuErgo3}. Sometimes the states being compared have been required to lie on the same energy surface~\cite{AditiErgo04}.  
In Ref.~\cite{barouch1}, it was shown that the transverse magnetization of the transverse $XY$ model is nonergodic. The case of quantum correlations was taken up in Refs.~\cite{AditiErgo04,PrabhuErgo1,PrabhuErgo2,PrabhuErgo3, Mishra16, Chanda}.
In particular, for the transverse \(XY\) model with the transverse field being given by
Eq.~(\ref{eqn:dyna}), it was shown that while bipartite entanglement is always ergodic (within the numerical accuracy used), QD can be ergodic as well as nonergodic~\cite{PrabhuErgo2}.
It was further found~\cite{Mishra16} that for the same  model, 
if $a$ and $b$ are chosen in such a way that the system is quenched from the antiferromagnetic to deep inside the paramagnetic phase, QD is enhanced, while it gets faded out during the paramagnetic to antiferromagnetic quench. Moreover, a quench within  same phase was found to cause enhancement of QD.
In
Ref.~\cite{PrabhuErgo3}, the Authors extended the results of the $XY$
model to the $XYZ$ model with the time-dependent magnetic field in the
$z$-direction as given in Eq.\ (\ref{eqn:dyna}), in 1D, ladder
(quasi-two), and 2D lattices. It was shown that tuning the interaction
strength can initiate a nonergodic to ergodic transition of QD.
For the 1D alternating field $XY$ model, QD can also
% also be both ergodic or nonergodic, indicating
 exhibit ergodic-nonergodic transitions with the variations of the system
parameters (see figure \ref{fig: ergo_nonergo})~\cite{Chanda}. 

\subsection{Geometric quantum discord in many-body systems}
\label{subsec:GQD_manybody}
After the success of QD in describing cooperative quantum phenomena, in following years 
it was seen that  geometric formulations of QD can also characterize  the  properties of various interacting systems~\cite{GQD_many_body1,GQD_many_body2,GQD_many_body3,GQD_many_body4,
GQD_many_body5,GQD_many_body6,GQD_many_body7,GQD_many_body8}.  In particular, in Ref.~\cite{GQD_many_body1}, the  $XXZ$ model with an external field and the  $XXX$ model with DM interaction were considered, and the dependencies of GQD on the system parameters were studied. 
Similar work in this 
direction has also been reported by  Cai \emph{et al.}~\cite{GQD_many_body2} where   
the DM interaction has been included along with an $XXZ$ interaction. In addition to this, 
 the Authors of Ref.~\cite{GQD_many_body3}  employed the quantum renormalization group method
 in the $XXZ$ and the anisotropic $XY$ models, and showed with several iterations of the renormalization that QD as well as GQD can faithfully detect the 
phase transition points present in the systems. Ground state properties of  a 1D Heisenberg system with next-nearest-neighbor  interaction has  been characterized by using GQD in Ref.~\cite{GQD_many_body6}, and was shown, for  4-site and 6-site systems, that there exists a one-to-one connection between the energy spectrum and GQD. 
%Additionally, in the finite-sized system, GQD is capable of capturing  the first order as well as the infinite order QPT \cite{GQD_many_body6}. 
In the cyclic $XX$ chain with uniform transverse magnetic field~\cite{GQD_many_body7}, it has been shown that at $T=0$, GQD possesses a non-zero value  for all pair-separations, $r=|i-j|$, if the
external field, $h_z$, lies below a certain critical value, $h^c_z$, and decaying only as $r^{-1}$ for large $r$. 
On the other hand, it remains non-zero for all temperatures, decaying as $T^{-2r}$ for sufficiently high $T$. The topological quantum phase transition observed in the ground state of Kitaev's 1D p-wave spinless quantum wire model has also been detected by using GQD~\cite{GQD_many_body9}. In particular, it has been reported that the derivative of
GQD is nonanalytic at the critical point, in both zero and finite temperature cases.
GQD has also been studied in the atom-cavity system modeled by the Jaynes-Cummings (JC) model \cite{JC_model}, and found that it persists in the atom versus cavity partition while entanglement vanishes~\cite{GQD_JC1,GQD_JC2} (cf.~\cite{JC_ent1,JC_ent2}). See also~\cite{JC_ent3,JC_ent4,JC_ent5} in this regard.  
Several other studies have been conducted along these lines, which have shown that the  study of GQD can provide important insight into cooperative physical phenomena~\cite{GQD_JC1,GQD_JC2,QKD_discord3,GQD_other_QC}.
%In particular, GQD is studied in a system interacting via the celebrated Jaynes-Cumming (JC) model~\cite{JC_model}, a system  consisting of Janes-Cumming atom, an isolated atom, and a cavity. It was shown to be nonvanishing where entanglement vanishes~\cite{GQD_JC1,GQD_JC2} and hence GQD helps to characterize the entire system in a better fashion. %In other word, in course of evolution, though  quantum entanglement fails to quantify the amount of QC shared between the subparts, GQD provides a better understanding of the QC present in the system.

\section{Quantum discord and open quantum systems}
\label{sec:open_QD}

Until now,  we have either considered the ideal scenario of an isolated system, i.e. a  system that is not affected by the environment and the 
properties of the system only depend on its own parameters, or a system at which one looks only at a given instant of time without considering how it arrived to that instant. The general situation is however far richer, and naturally leads one to consider the dynamics of open quantum systems \cite{frz2,Davis76,Alicki87,frz1,frz3}.
%In general, it is not the case. Studying the system,
Literature on QD in open quantum systems include
% considering the interactions of environment with it leads to a new formalism of open systems~\cite{frz2,Davis76,Alicki87}. 
 %\textcolor{gray}{The experimental as well as theoretical results
Refs.~\cite{yu1, yu2, death1, death2, death3, death4, death5, death6, death7, death8, death9, open_extra_new6, death10,open_extra_new7,open_extra_new8,open_extra_new9,open_extra_new29,open_extra_new32,open_extra_new33,open_extra_new31,open_extra_new30, open_extra_new28,open_extra_new27,open_extra_new22,open_extra_new23,open_extra_new24, open_extra_new25,open_extra_new26,open_extra_new20,open_extra_new19,open_extra_new18,open_extra_new17, open_extra_new16,Altintas-QD}.
 %strongly suggest that QC is extremely fragile even when certain system parameters are changed. It is therefore important to investigate the dynamics of QC when the system-environment duo is active.} % Eta bad deoya hoyeche...

%A complete isolation of a system is a mere idealization. Practically, a quantum system is always surrounded by some environments and thus its interaction with the surrounding environments are unavoidable \cite{frz2, Davis76, Alicki87}. Such systems are commonly referred as open quantum systems.
%{\color {cyan} Since QC are known to be fragile and get easily destroyed by the unavoidable interactions with the environment, e.g. decoherence, the characterization of the QC in a composite open quantum systems is important for the QIP tasks. The dynamical behaviour of the QC is essential to design a QIP protocol since they are the resource of such protocols. Apart from the decoherence which is primarily a quantum effect, environmental effects can also cause dissipation which can be considered as both classical as well as quantum effect. However, in this section, we are interested in typical quantum features in describing dissipation which shares a close connection with the quantum thermodynamics. }

An open system consists of a system $S$, and an environment $E$. The Hilbert space of the composite system, $S+E$, is the tensor product space ${\mathbb C}^m\otimes{\mathbb C}^n$, where $m=\text{dim}(S)$ and $n=\text{dim}(E)$. 
Typically, the environment is considered to be very large  compared to the system, and it is not always possible to have access to the entire environment. %and one can only measure on the system.
%, it is important to describe the dynamics of the observable of the system by staying within the level of the system's Hilbert space  only.
The physical state of the composite system is denoted by the density matrix $\rho_{SE}$, whereas the system state can be obtained by tracing out the environment i.e.  $\rho_S = \text{tr}_E \rho_{SE}$.

In general, the composite system $S+E$ can be described by the Hamiltonian
\begin{align}
H_{SE} = H_S \otimes \mathbb{I}_E + \mathbb{I}_S\otimes H_E  + H_{int},
\end{align}
where $H_{S(E)}$ is the Hamiltonian of the system (environment) and $H_{int}$ denotes the Hamiltonian describing the interaction between the system and the environment. Below we briefly discuss the formalism to study the evolution of $\rho_S $ in  presence of the environment $E$.

\subsection{Dynamical maps}
The system-environment state, which can together form an isolated system, is considered to be in a joint initial state $\rho_{SE}(0)$. Since the dynamics of an isolated quantum system is predicted by the Schr\"odinger equation, the time evolved state reads 
\begin{eqnarray}
\rho_{SE}(t)=U_{SE}(t)~\rho_{SE}(0)~U_{SE}^{\dagger}(t),
\end{eqnarray}
where $U_{SE}(t)$ is the unitary operator acting both on the system and the environment. %satisfying the time-independent Schr\"odinger equation
%\begin{align}
%\frac{d}{dt} U_{SE} &= -\frac{i}{\hbar}[H_{SE},U_{SE}],
%\intertext{resulting}
%U_{SE} (t) &= \exp(-iH_{SE}t/\hbar).
%\end{align}
% dynamics of a closed quantum system is predicted by the Schr\"odinger equation. So, the composite system, $S+E$, together form an isolated system, described by the joint initial state $\rho_{SE}(0)$.
% We are interested in the reduced dynamics of the subsystem $S$ which is in contact with the environment $E$. The state $\rho_S$ at time $t$ is governed by the following quantum dynamical process (QDP)
%\begin{align}
%\label{eqn:evo}
%\rho_S(t) = \text{tr}_E[\rho_{SE}(t)] =  \text{tr}_E [U_{SE}(t) \rho_{SE}(0) U_{SE}(t)^{\dagger} ],
%\end{align}

%However, as mentioned earlier, we are mainly interested in the dynamics of the system $S$ which is in contact with the environment $E$. Therefore, 
The evolved state $\rho_S(t)$, at time $t$, can now be expressed as
\begin{align}
\label{eqn:evo}
\rho_S(t) = \text{tr}_E[\rho_{SE}(t)] =  \text{tr}_E [U_{SE}(t) \rho_{SE}(0) U_{SE}(t)^{\dagger} ],
\end{align}
which is obtained by tracing out the environment part from the joint state $\rho_{SE}$.  If the initial density matrix is of the form $\rho_{SE}(0)=\rho_S\otimes |0\rangle \langle0|_E$, 
% where  $|0\rangle\langle 0|_E$ is the fixed state of environment, 
 then the final state can be expressed as $\rho_S(t)=\sum_i \langle i|U_{SE}|0\rangle~ \rho_S(0)~ \langle 0|U_{SE}|i\rangle_E$ with $\{|i\rangle\}$ being an orthonormal basis of the environment. It leads to the introduction 
 of a  linear map $\Phi_t$, 
% which transforms linear operators to a  linear operator, called a superoperator~\cite{Preskil}, 
 given by~\cite{Preskil, Sudarshan_dyna, Choi_map, Choi_map2, Kraus, Kraus83, Jamiolkowski}
\begin{eqnarray}
\rho_S(t)=\Phi^S_t(\rho_S)=\sum_i \mathcal{K}_i~ \rho_S(0)~\mathcal{K}_i^{\dagger},
\label{kraus_eqn}
\end{eqnarray}
where the Kraus operators $\mathcal{K}_i$ satisfy $\sum_i\mathcal{K}^{\dagger}_i\mathcal{K}_i=\mathbb{I}^S_m$ with $\mathbb{I}^S_m$ being the identity operator of the Hilbert space $\mathbb{C}^m$.
%and it is said that the time evolved density matrix $\rho_S(t)$, has the form of the Kraus representation,~\cite{Kraus83,frz2}, providing a formalism  to deal the effect of decoherence on the system. 
The dynamical map providing the evolution of the system due to the interaction with environment satisfies certain properties  like linearity, hermiticity, positivity, and trace preservation  and on top of that they are completely
 positive\footnote{A map $\Phi^S$ acting on density matrices on $\mathbb{C}^m$ is said to be completely positive if any possible extension, $\Phi^S_t\otimes \mathbb{I}^E_n$, of $\Phi^S$ to a bigger Hilbert space $\mathbb{C}^m\otimes\mathbb{C}^n$ is also positive.} (CP).
  The complete positivity of the system is guaranteed by the assumption of an uncorrelated product initial state of the system and environment. The presence of classical correlation as well as QC may lead to non-complete 
  positivity of the system~\cite{Pechukas94,Alicki95,open_extra_new1,
open_extra_new2,open_extra_new3,Stelmachovic01}.
%Assuming a non-evolving environmental state $\rho_E(0)$, we can define a linear map, $\Phi_t$,  on system's state space $S(\mathbb{C}^m)$ to describe Eq.\ (\ref{eqn:evo})
%\begin{align}
%\Phi_t ~:~ S(\mathbb{C}^m) \rightarrow S(\mathbb{C}^m)
%\end{align}
%such that it maps any initial state $\rho_S(0)$ of the system to the  state  $\rho_S(t)$, at time $t$ of corresponding composite system:
%\begin{align}
%\rho_S(0) \rightarrow \rho_S(0) = \Phi_t\rho_S(0).
%\end{align}
% However,
%the characterization of the evolution of an open quantum system is a long standing issue. In literature, attempts have been made  to give a general map description for an open quantum system which is not limited to initially uncorrelated system and environment. %, e with initial shared correlations between system and bath.% which applicable to many fields of physics.
%%The s can describe the dynamics of
%It was well known that an open quantum system, in contact with an environment, can be described by completely positive (CP) maps when they possess no  initial correlations and therefore the map admits a Kraus representation \citep{Kraus83, frz2}.  This issue of CP, QC the system-environment initial state, the nature of evolution are all inter-related and probably require further study.
Specifically, further investigations in this direction  reveal that  initially entangled system-bath states can lead to non-CP maps~\cite{Pechukas94, Pechukas95, Jordan04, Carteret08}. 
%Hence, one may argue that the presence of entanglement between the system and the environment  is responsible for the non-CP dynamical evolutions. 
In subsequent years, it was found that there exists nonclassical correlations other than entanglement which, when existing between system and environment, can also result in non-CP dynamical maps~\cite{Rodríguez10}.

In recent works~\cite{complete-positive,Breuer_RMP,Shabani09,Brodutch13,Buscemi14,Dominy16,Rodríguez10}, efforts have been made to describe properties of the initial system-environment duo that can assure CP-ness 
of the reduced dynamics. In particular, it was proven that a dynamical map is CP if the initial system-environment state is a  $c$-$c$ state~\cite{complete-positive}, which, of course, have vanishing QD. 
However, this does not necessarily imply that a non-zero QD in the initial system-environment state will lead to non-CP dynamical map of the system.
Indeed,  Brodutch \emph{et al.}~\cite{Brodutch13} constructed a separable state with non-vanishing QD, that when considered as a initial state of the system-environment pair, can be written in the Kraus representation from (Eq. (\ref{kraus_eqn})), so that the dynamical map of the system is CP. 
% Subsequently, it was claimed that under certain additional
% constraints, QD is also necessary for complete positivity~\cite{Shabani09}. However, recent studies show that the  necessary part shown in Ref.~\cite{Shabani09} has counter examples~\cite{Brodutch13,Buscemi14,Dominy16}. In this regard,  Bradutch \emph{et al.}~\cite{Brodutch13}  further considered an initial state  $\rho_{SE}(0)=\frac{p}{3}\big( |0\rangle \langle 0|\otimes \rho_E^0+|1\rangle\langle 1|\otimes \rho^1_E+|+\rangle \langle +| \otimes \rho_E^+\big)+\sum_{i=2}^n p_i |i\rangle \langle i| \otimes \rho^i_E$ with $p=1-\sum_{i=2}^n p_i$, which has non-vanishing QD and showed that $\rho_S(t)$ can be written by using Kraus operators, given in Eq.~(\ref{kraus_eqn}). 
 Buscemi~\cite{Buscemi14} followed this up by constructing an example of a class of maps which are CP, and for which it is possible for the system-environment states to be entangled.
\begin{center}
\begin{table}%[hbt]
\begin{tabular}{|c|c|}
\hline
& $\textrm{Kraus operators}$                                         \\ \hline \hline
& \\
BF   & $E_0 = \sqrt{1-p/2}\, I , E_1 = \sqrt{p/2} \,\sigma_1$                        \\ \hline
& \\
PF   & $E_0 = \sqrt{1-p/2}\, I , E_1 = \sqrt{p/2}\, \sigma_3$                        \\ \hline
& \\
BPF & $E_0 = \sqrt{1-p/2}\, I , E_1 = \sqrt{p/2} \,\sigma_2$                        \\ \hline
& \\
GAD   &
$E_0=\sqrt{p}\left(
\begin{array}{cc}
1 & 0 \\
0 & \sqrt{1-\gamma} \\
\end{array} \right) ,
E_2=\sqrt{1-p}\left(
\begin{array}{cc}
\sqrt{1-\gamma} & 0 \\
0 & 1 \\
\end{array} \right)$  \\
& \\
& $E_1=\sqrt{p}\left(
\begin{array}{cc}
0 & \sqrt{\gamma} \\
0 & 0 \\
\end{array} \right) ,
E_3=\sqrt{1-p}\left(
\begin{array}{cc}
0 & 0 \\
\sqrt{\gamma} & 0 \\
\end{array} \right)$  \\ \hline
\end{tabular}
\caption[table1]{Kraus operators for some well-known quantum channels: bit flip (BF), phase flip (PF), bit-phase flip (BPF), and generalized amplitude damping (GAD), where $p$ and $\gamma$ are decoherence parameters, with $0 \leq p, \gamma \leq 1$. }
\label{t1}
\end{table}
\end{center}
\subsection{Prototypical open systems}
Studying the patterns of QD under environmental effects is the main objective  in this part of the review. An open quantum system can be modeled in different ways which may represent situations such as decoherence under dissipative environment, repeated quantum interactions, spin-boson models, etc.

We start the discussion with the dynamics of QD between subparts of a system under  Markovian as well as non-Markovian noisy channels.  When the system passes through a channel, channel acts as an environment. In this
 scenario, it may be natural to assume that the system  and the given channel are in a product state, and hence Kraus representation is valid here. In Table~\ref{t1}, we tabulate the Kraus operators for some well-known channels 
 which will be relevant in this review.
 % for some different  quantum depolarizing channels, namely bit flip (BF), phase flip (PF), bit-phase flip (BPF), and generalized amplitude damping (GAD), with $p$  and $\gamma$ being decoherence probabilities.

\subsubsection{Correlation dynamics under decoherence}
%In this section, we will discuss about the freezing phenomenon of QD. It is well known that
QC in a system, in general, decreases while interacting with the environment.
%In the open quantum system,
The fragile  nature of QC with time is one of the main obstacles  in the implementation of  quantum information tasks. For example,  entanglement disappears completely after a finite time, for many dynamical maps, a phenomenon referred to as sudden death of entanglement~\cite{yu1, yu2, death1, death2, death3, death4, death5, death6, death7, death8, death9, death10}. On the contrary, QD, typically, asymptotically decays with time
~\cite{Auccaise_witnessexp,GQD_other_QC,Geo_QD_decoherence5, werlang1,ferraro,Fanchini2,decay1,decay2,decay3,decay4, decay5,decay6,decay7,decay9,decay10,decay11, decay12,decay13,decay15,decay16,decay17,death-abar1,haikka-frz,sarandy-dec,mazzola1,freezing-exp1,frz-dekhechi,decay18,open_extra_new13,open_extra_new12,open_extra_new11}.
 Moreover, there exists some special cases, when QD of the evolved state  remains constant over a finite interval of  time - a phenomenon known as {\it freezing} of QD. 
% Here, we briefly discuss these two important phenomena of QD under various depolarizing channels.
The dynamics of QD can also be such that it shows a kink in its profile which causes a finite discontinuity in its derivative, a phenomenon known as  \emph{sudden change} of QD. We briefly discuss below both these phenomena, and the conditions   on the states  and channels leading to these events.
\begin{figure}[t]
% \centering
\includegraphics[width=1.0\columnwidth,keepaspectratio]{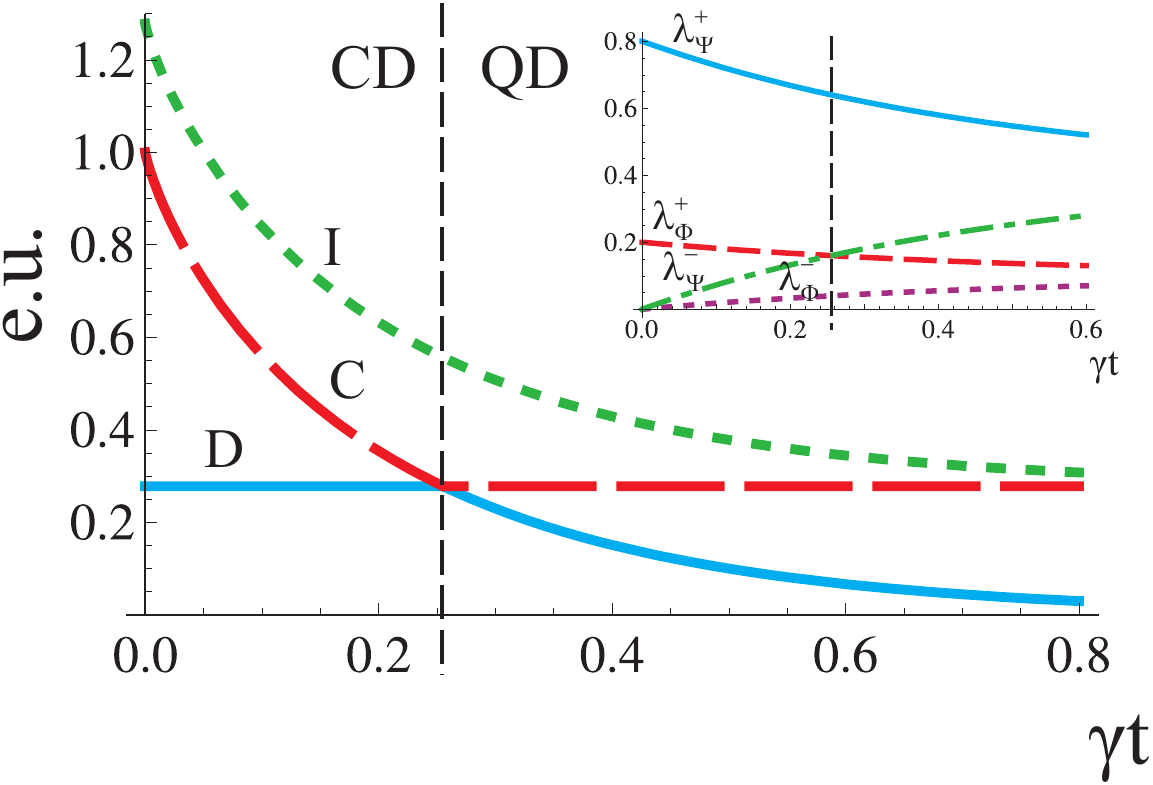}
\caption{ Freezing of quantum discord. Dynamics of mutual information (green dotted line), classical
correlation (red dashed line), and QD (blue solid line) as functions of
$\gamma t$ under the conditions $T_{11}(0)=1$, $T_{22}(0)=-T_{33}$ and $T_{33}=0.6$. In the inset,  the eigenvalues $\lambda_{\Psi}^{+}$ (blue solid line), $\lambda_{\Psi}^{-}$ (green dash-dotted line), $\lambda_{\Phi}^{+}$ (red dashed line), and $\lambda_{\Phi}^{-}$ (violet dotted line) are plotted as  functions of $\gamma t$ for the same parameter values.  In the figure, ``CD'' and ``QD'' represent regimes of ``classical decoherence'' and  ``quantum decoherence'' respectively. ``e.u.'' stands for ``entropic units''. The horizontal axes are dimensionless. The vertical axis of the inset is also dimensionless.
[Reprinted from Ref.~\cite{mazzola1}   with permission. Copyright 2010 American Physical Society.]}
\label{freezing1}
\end{figure}

\subsubsection*{a. Freezing of quantum discord}

In 2010, Mazzola \emph{et al.} \cite{mazzola1}  discovered that for certain Bell-diagonal states\footnote{For Bell-diagonal states, see Eq.~(\ref{bell-diagonal}).},
%class of initial states of the form given in Eq.~(\ref{eq:gen1}) with $x_i=y_i=0$ and $T_{ij}=\delta_{ij} c_{ii}$ $\forall$ $i, j \in \{1,2,3\}$,
%\begin{equation}
%\rho_{AB}=\frac{1}{4} (\mathbb{1}_{AB}+\sum_{i=1}^3 c_i \sigma_i^A \otimes \sigma_i^B),
%\end{equation}
QD does not decay for a finite time interval  in presence of a noisy environment.
More precisely, when 
%Phase damping dynamics is considered here which characterizes the type of quantum noise process that induces
 the input state is subjected to local PF channels, given in Table~\ref{t1}, the time evolved state is given by
% uantum coherence but without energy exchange.
% For a two-level system described by $\rho(0)$, the dynamical evolution of the state under PDC is described as
%\begin{equation}
%\rho (t)=\left(
%\begin{array}{ll}
%\rho _{11}(0) & d(t)\rho _{12}(0) \\
%d(t)\rho _{21}(0) & \rho _{22}(0)
%\end{array}
%\right) ,  \label{dephasing1}
%\end{equation}
%where $d(t)$ represents the degradation of the coherence and it often takes an exponential form for a Markovian dephasing process.
%Hereafter, we shall introduce a notation $q\equiv1-d(t)$ and treat it is as a parametrized time.
%
%%Here $c_i$ is the real no. s. t. $0 \leq |c_i| \leq 1$ for each $i$. Note that this class of states includes the Werner states when $|c_1|=|c_2|=|c_3|=c$ and the Bell states when $|c_1|=|c_2|=|c_3|=1$.
%
%Now if the initial state is in the  Bell-diagonal form (see Eq.~(\ref{bell-diagonal})),  and subjected to PDC, one can  express the time evolved state as
\begin{eqnarray}
\rho_{AB}(t)=\lambda_{\psi^{+}} (t) |\psi^+ \rangle \langle \psi^+|+\lambda_{\phi^{+}} (t) |\phi^+ \rangle \langle \phi^+|+ \nonumber\\
\lambda_{\phi^{-}} (t) |\phi^- \rangle \langle \phi^-|+
\lambda_{\psi^{-}} (t) |\psi^- \rangle \langle \psi^-|,
\end{eqnarray}
which is of the form $\frac{1}{4} (\mathbb{I}_4+ \sum_{i} T_{ii} \sigma_i \otimes \sigma_i)$. Here,
\begin{eqnarray}
\lambda_{\psi^{\pm}} (t)=\frac{1}{4} [1 \pm T_{11}(t) \mp T_{22}(t)+ T_{33}(t)], \nonumber\\
\lambda_{\phi^{\pm}} (t)=\frac{1}{4} [1 \pm T_{11}(t) \pm T_{22}(t)- T_{33}(t)],
\end{eqnarray}
where $T_{11}(t)=T_{11}(0) (1-p)^2$, $T_{22}(t)=T_{22}(0) (1-p)^2$,
$T_{33} (t)=T_{33}(0) \equiv T_{33}$. The channel parameter, $p$, of the PF channel (see Table~\ref{t1}) is related to the elapsed time by the relation $p=1-\exp(-\gamma t)$, where $\gamma$ is referred to as the phase damping rate.
 %are the time dependent co-efficients 
%and $p$ is a channel parameter called the phase damping rate.
%$|\psi^\pm \rangle=(|00\rangle \pm |11\rangle)/\sqrt{2}$, $|\phi^\pm=(|01\rangle \pm |10\rangle)/\sqrt{2}$ are the four Bell states. \\
%\begin{figure}[ht!]
%\includegraphics[width=6.0cm]{frz2.eps}
%\caption{Description of the trajectory for the freezing phenomenon
%and sudden transition dynamics of quantum discord in the parametric space.
%The three blue-dashed lines represents the Bell-diagonal states
%with zero discord and they share a common node with the two leaf-shaped curves
%determined by Eqs. (\ref{process1}) and (\ref{process2}).
%The trajectory of the solid line proceeding along
%$I\rightarrow T\rightarrow F$ illustrates the evolution of the Markovian
%phase damping process and it moves forward along the dotted line $F\rightarrow T'$ for the non-Markovian dephasing process and oscillates damply between the two specified leafs which implies multiple transitions of the discord dynamics.}
%\label{frz2}
%\end{figure}
Now under the conditions $T_{11}(0)=\pm 1$, $T_{22}(0)=\mp T_{33}$, $|T_{33}|<1$, 
% with $t<t'=-\frac{1}{2 \gamma} \ln (|T_{33}|)$, 
the mixture of four Bell states is a mixture of two Bell states. Using that as the initial state sent through the local phase damping channel, the QD for $t<t' =-\frac{1}{2 \gamma} \ln (|T_{33}|)$ is given by
%reduces to a mixture of two Bell states, and the QD is given by
\begin{equation}
\mathcal{D}(\rho_{AB} (t))=\sum_{j=1}^2 \frac{1+(-1)^j T_{33}}{2} \log_2 [1+(-1)^j T_{33}].
\end{equation}
%which  remains  constant for $t<t'$.
This is independent of time and we remember that it is valid only for $t<t'$.
 This is known as freezing of QD (see figure~\ref{freezing1}). Figure~\ref{freezing1} shows another interesting feature - QD is constant and classical correlation $J_{A|B}$, decays for $t<t'$ while QD decays and classical correlation does not change with time for $t>t'$.
%From the above equation, one can clearly see that for that interval of time, QD is constant of time. In Fig.~\ref{freezing1}, the plot of time evolution of the QD, the classical correlations, and the mutual information have been plotted which implies that at $t=t'$, there is a sharp transition from the classical to quantum decoherence regime.
Moreover, this behavior of QD  has been observed experimentally using photonic~\cite{freezing-exp1} and NMR two-qubit states~\cite{Auccaise_witnessexp, frz-dekhechi}.

%Furthermore, You \emph{et al.}~\cite{mark} obtained the
A necessary and sufficient condition for obtaining the freezing phenomenon of QD with the Bell-diagonal state as the input to local PF channels is provided in Ref.~\cite{mark}, and given in the following theorem.\\
{\bf Theorem 4}~\cite{mark}: {\it The Bell-diagonal states
%\begin{equation}
%\rho_{AB}=\sum_{i=1}^4 \lambda_i |\psi_i \rangle %\langle \psi_i|,
%\end{equation}
% where $\lambda_i \geq 0$ and
%$|\psi \rangle=\{|\phi^\pm \rangle, |\psi^\pm \rangle \}$
given in Eq.~(\ref{bell-diagonal}) can exhibit freezing of QD under local PF channels if and only if $\lambda_i$'s either satisfy
\begin{equation}
\label{process1}
\lambda_1 \lambda_4=\lambda_2 \lambda_3, (\lambda_1-\lambda_4) (\lambda_2-\lambda_3) >0
\end{equation}
or,
\begin{equation}
\label{process2}
\lambda_1 \lambda_2=\lambda_3 \lambda_4, (\lambda_1-\lambda_2) (\lambda_4-\lambda_3)>0.
\end{equation}
}

\begin{figure}
\begin{center}
\includegraphics[width=0.45\textwidth]{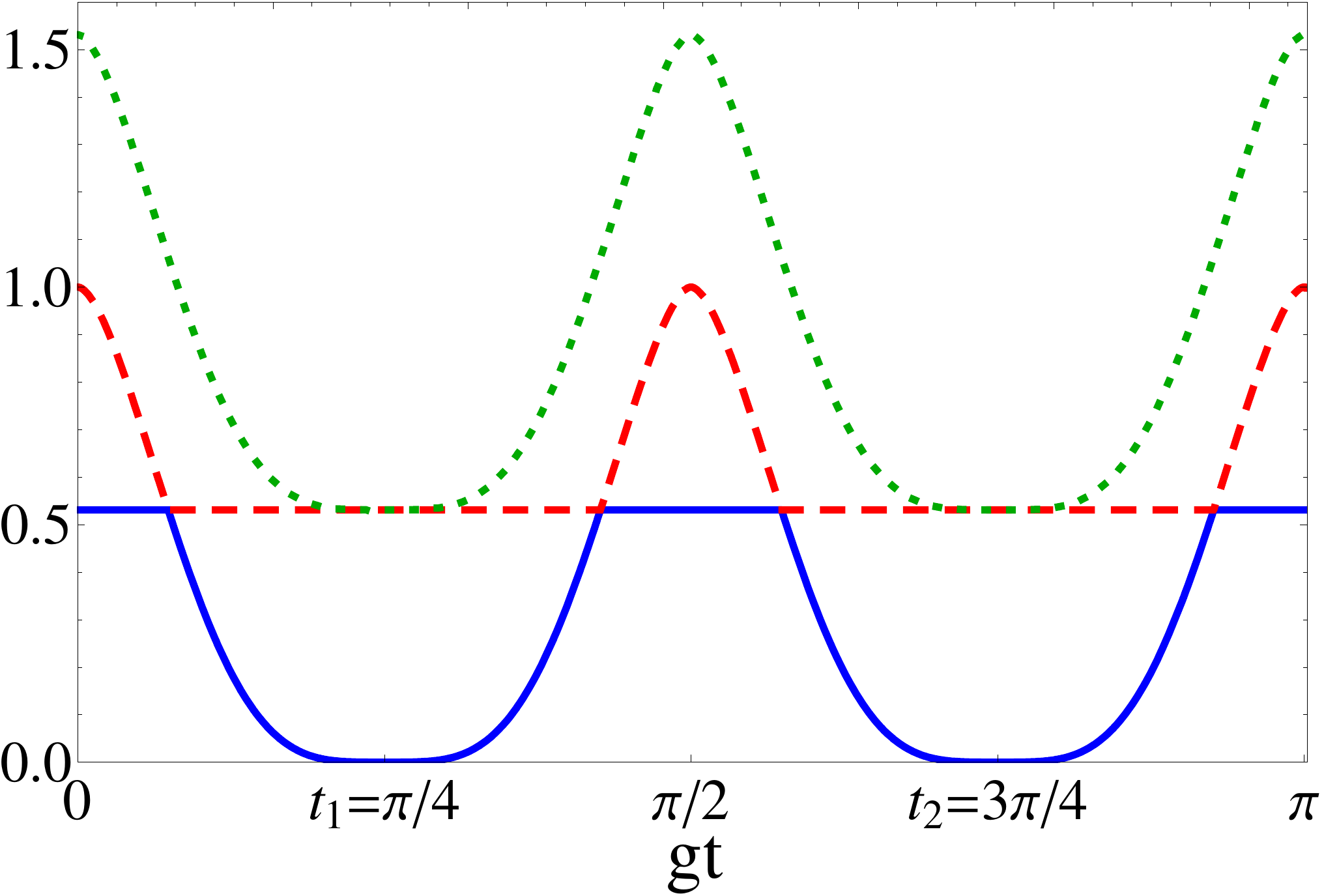}
\caption{ Dynamics of quantum discord $\mathcal{D}$ (blue solid line), classical correlation $J$ (red dashed line) and total correlation $I$ (green dotted line) for an initial Bell-diagonal state of the form 
$\rho_{AB}(0)=\lambda_1(0) |\psi^+ \rangle \langle \psi^+|+\lambda_2(0) |\psi^- \rangle \langle \psi^-| 
+
\lambda_3 (0) |\phi^+ \rangle \langle \phi^+|+
\lambda_4 (0) |\phi^- \rangle \langle \phi^-|$ with 
$\lambda_1(0)=0.9$, $\lambda_2(0)=0.1$ and $\lambda_3 (0)=\lambda_4 (0)=0$,
passing through local channels, with each channel being modeled by an interaction between the corresponding qubit and a classical field. The interaction strength is given by 
\(g\hbar\).
The vertical axis is in bits while the horizontal axis is dimensionless.
[Reprinted from Ref.~\cite{franco-frz}   with permission. Copyright 2012 American Physical Society.]}
\label{frz4}
\end{center}
\end{figure}

%Starting with Bell-diagonal states, it was shown that 
%Reverting back to the case of Bell-diagonal states as initial states of a noisy evolution, let us mention that 
It was also shown that there exists a form of non-Markovian dynamics under which QD remains invariant for all time~\cite{haikka-frz}.
 Importantly,  multiple intervals of recurring frozen QD~\cite{mark1,franco-frz,mannone-frz} can also be observed when the dynamics is considered in non-Markovian regime. 
  In another work~\cite{franco-frz}, the initial Bell-diagonal state passes through local channels, where each channel modeled by an interaction of the corresponding qubit with a classical field.
  %, with the interaction strength $g \hbar$.
   It was found that both entanglement and QD show collapse and revival, and that QD exhibits multiple freezing intervals (see figure ~\ref{frz4}). 
%  is affected by the local interactions of single classical field with random phase and it was observed that QD  as well as entanglement revive after collapse, as  shown in Figure~\ref{frz4}.
    For an experimental demonstration, see Ref.~\cite{xu-exp}.
It was noticed that when QD is frozen, classical correlation is  oscillating and vice versa (cf.~\cite{yan-dis}) and 
was also argued that revival  of QC is related to the non-Markovian nature of the evolution. 
%\textcolor{red}{ TWO TIMES OCCURENCE Dynamical decoupling protocol, that protects QD from environment-induced errors~\cite{DD1}, also exhibits freezing behavior of QD~\cite{DD2}  which has further been  demonstrated experimentally in Ref.~\cite{DD3}.}

For certain channels like BF, PF, BPF,  a  necessary and sufficient condition for freezing of QD was provided~\cite{Titas_freeze} for bipartite as well as multipartite states under local noisy channels. In this regard, 
``canonical initial (CI) states" of the  form\footnote{An arbitrary two-qubit state, given 
in Eq.~(\ref{eq:gen1}), reduces to 
 \begin{eqnarray}
 \rho_{AB}&=&\frac{1}{4} \Bigg[\mathbb{I}_2\otimes \mathbb{I}_2
 +\sum_{i=1}^{3} T_{ii}\sigma_{A}^{i}\otimes\sigma_{B}^{i}\nonumber \\
 && +\sum_{i=1}^{3}\left(x_i \sigma_{A}^{i}\otimes \mathbb{I}_2
 +\sum_{i=1}^{3}y_i \mathbb{I}_2\otimes\sigma_{B}^{i}\right) \Bigg], \nonumber
 \end{eqnarray}
up to local unitary transformations.  
Since the magnetizations other than \(x_1\) and \(y_1\) decay under the bit-flip channel,
it is expected that they do not contribute to the freezing phenomenon involving the same channel, and hence they are set to zero in Eq.~(\ref{twoqubitstate}).}
\begin{eqnarray}
 \rho_{AB}&=&\frac{1}{4} \Big[\mathbb{I}_2\otimes \mathbb{I}_2
 +\sum_{i=1}^{3} T_{ii}\sigma_{A}^{i}\otimes\sigma_{B}^{i}\nonumber \\
 && +\left(x_1 \sigma_{A}^{1}\otimes \mathbb{I}_2
 +y_1 \mathbb{I}_2\otimes\sigma_{B}^{1}\right)\Big]
 \label{twoqubitstate}
\end{eqnarray}
%T_{10}=x_1, T_{01}=y_1
have considered to investigate the freezing phenomenon of QD. %under the BF channel. 
As discussed in Sec.~\ref{sec:Computability}, for most two-qubit states among CI states, optimization of QD occurs in the  eigenbases of $\sigma^1$, $\sigma^2$, or $\sigma^3$, and such states are called  special CI (SCI) states. 
%A closed analytic form of QD can be obtained for SCI states, \textcolor{red}{ and the optimization in Eq.~(\ref{cond_entropy}) occurs at certain ``regular" values of $\theta$ and $\phi$, viz. $\{\theta=0,\pi \}$, $\{\theta=\pi/2,\phi=\pi/2,3 \pi/2\}$, and $\{\theta=\pi/2, \phi=0,\pi\}$.}  
 %For a large section of special CI (SCI) states, a closed analytical form of QD can be obtained. Since for these states, the optimization in Eq.~(\ref{cond_entropy}) occurs at certain ``regular" values of $\theta$ and $\phi$,  $\{\theta=0,\pi \}$, $\{\theta=\pi/2,\phi=\pi/2,3 \pi/2\}$, and $\{\theta=\pi/2, \phi=0,\pi\}$.  
%Which is indeed the optimization in the eigenbases of $\sigma_x$, $\sigma_y$ and $\sigma_z$, discussed earlier in the contrained QD in Sec. \ref{sec:Computability}.
A necessary and sufficient criteria for freezing of QD for the two-qubit SCI states as inputs to local BF channels is given below.\\
\noindent\textbf{Theorem 5}~\cite{Titas_freeze}: \textit{ A necessary and sufficient condition for freezing of QD of the output of a local BF channel where the input is a two-qubit SCI state, over a finite interval of time is given by any of the following sets of equations:}
\begin{eqnarray}
   \begin{cases}
     (i)  & (T_{22}/T_{33}) = -(x_1/y_1) = -T_{11},\\
     (ii) & T_{33}^2 + y_1^2  \leq  1,\\
     (iii)& F\left( \sqrt{T_{33}^2 + y_1^2}\right) <  F(T_{11}) + F (y_1) - F(x_1);
   \end{cases}\nonumber\\ 
\label{necsuf1}
\end{eqnarray}
\begin{eqnarray}
   \begin{cases}
     (i)  & (T_{33}/T_{22}) = -(x_1/y_1) = -T_{11},\\
     (ii) & T_{22}^2 + y_1^2  \leq  1,\\
     (iii)& F\left( \sqrt{T_{22}^2 + y_1^2}\right) <  F(T_{11}) + F (y_1) - F(x_1).
   \end{cases}\nonumber \\
\label{necsuf2}
\end{eqnarray}
Here $F(y)=2\left(h(\frac{1+y}{2})-1\right)$ and  $p=\gamma$ for the BF channel.
\begin{figure}
\includegraphics[scale=0.34]{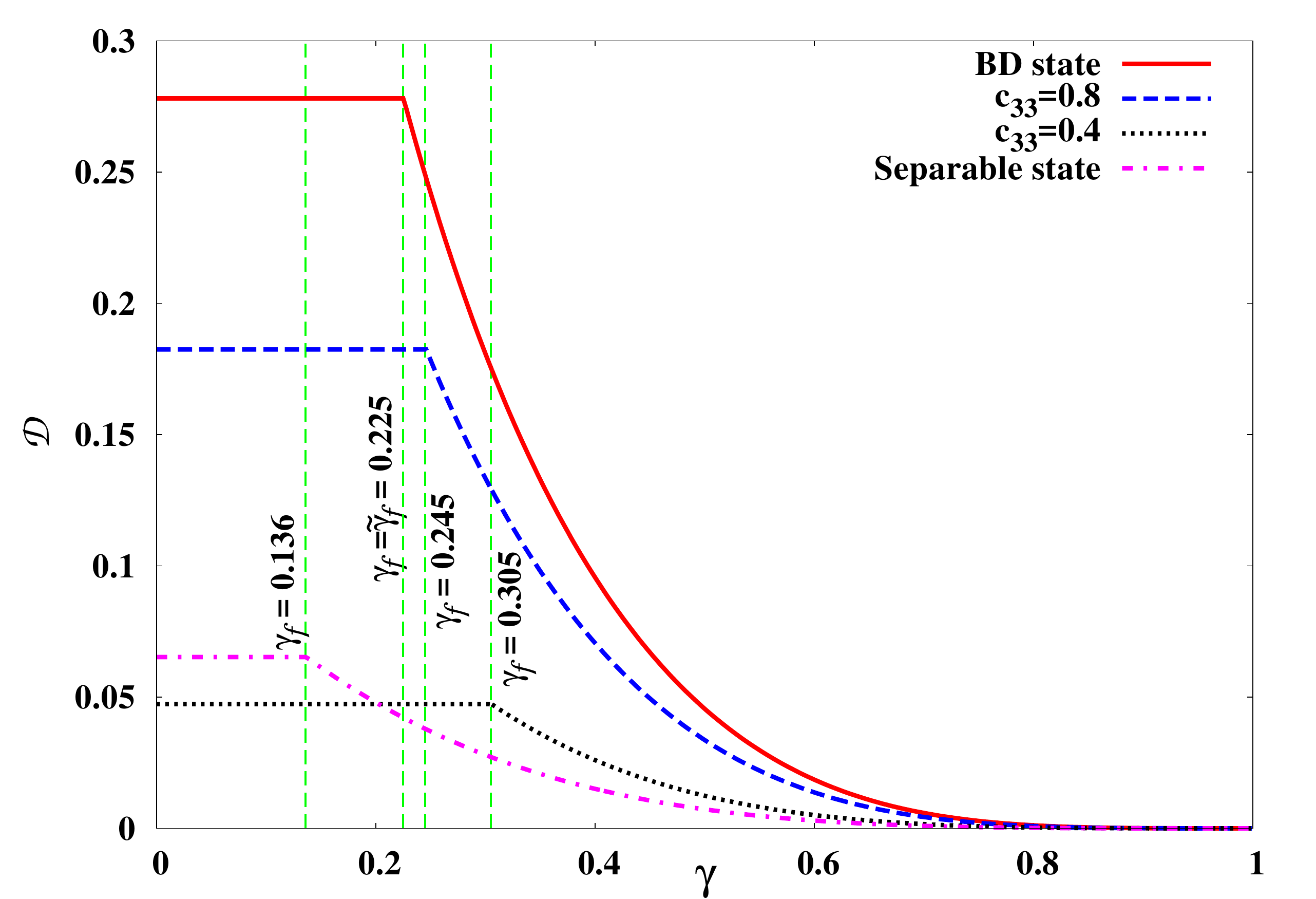}
\caption{Dynamics of quantum discord for two-qubit SCI  states under local BF channels for $|T_{11}|=0.6$ and $T_{33}^2 + y_1^2 = 1$. Note that a CI state reduces to a BD state under the assumption, $y_1=x_1=0$. The red solid line represents the  BD state with $|T_{22}|=0.6$ and $|T_{33}|=1$.  
The vertical axis is in bits, while the 
horizontal one is dimensionless. 
[Adapted from Ref.~\cite{Titas_freeze} with permission. Copyright 2015 American Physical Society.]
}
\label{twoqubitphase}
\end{figure}
Figure~\ref{twoqubitphase} exhibits the dynamics of QD for certain SCI states including BD states. 
%as well as the Bell-diagonal states and separable states. 
%Note that for Bell-diagonal state $|c_{11}|=|c_{22}|=0.6$, $|c_{33}|=1$, $c_{01}=c_{10}=0$. 
The ``freezing terminal" $(p_f)$, representing the time at which the freezing behavior of quantum correlation in the decohering state vanishes, can be much larger for some CI states compared to the Bell-diagonal states. Moreover, a complementarity relation between the frozen value of the QD and the freezing terminal has been  proposed.
% Also the set of states exhibiting freezing of QD forms a nonconvex set.
Importantly, it is possible to define a freezing index, quantifying the goodness of freezing for states having very slow decay rate of QC,  that can also capture the QPT in the quantum $XY$ spin model~\cite{Titas_freeze}. 

\begin{figure}[ht!]
\includegraphics[scale=0.83]{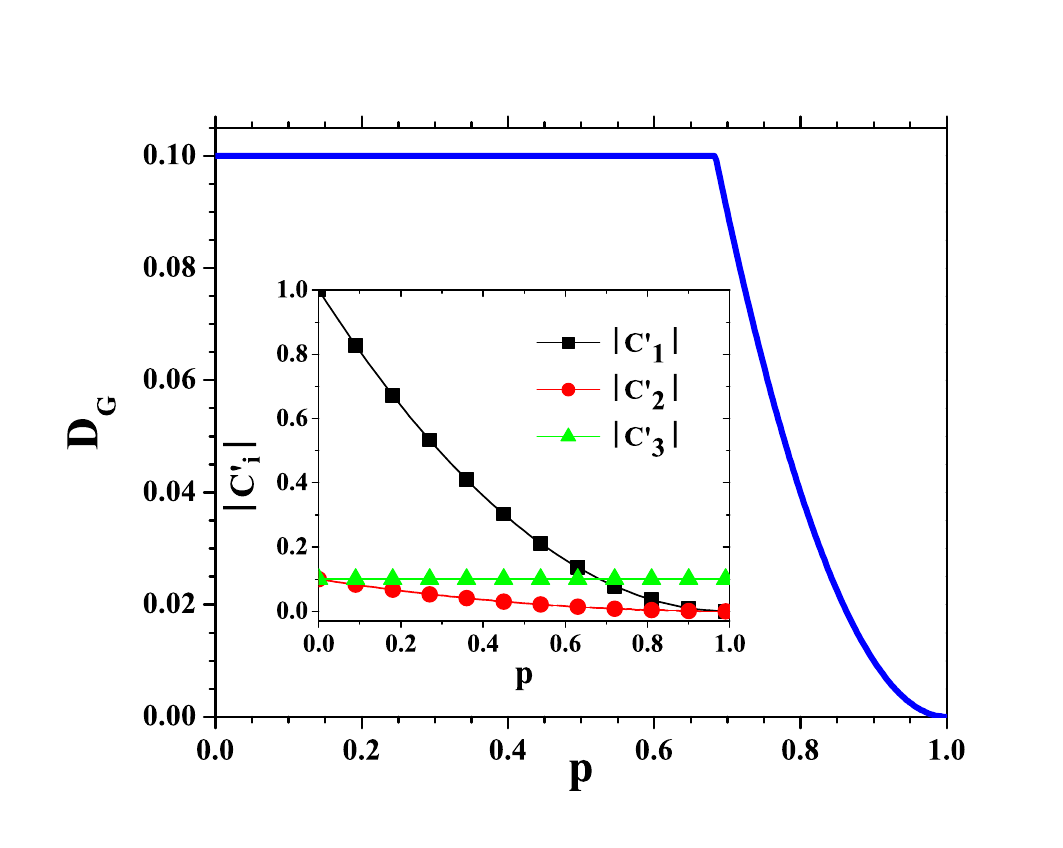}
\caption{Freezing behavior of trace norm GQD with the BD state given in Eq.~(\ref{bell-diagonal}) as the initial state with $T_{11}=1$, $T_{22}=-0.1$ and $T_{33}=0.1$ under the local PF channel. Here $D_G$ 
%and $p$ correspond to the 
represents the trace norm GQD.
% and decoherence probability. 
Inset: Absolute values of the correlators as functions of $p$. Here $|C'_1|=-|T_{11}| (1-p)^2$,  $|C'_2|=-|T_{22}| (1-p)^2$, and $|C'_3|=|T_{33}|$.  %where $C_1 = T_{11}$, $C_2 = T_{22}$, and $C_3=T_{33}$. 
[Reprinted from  Ref.~\cite{sarandy-dec}   with permission. Copyright 2015 American Physical Society.] }
\label{freezing3}
\end{figure}

The trace norm and Hilbert-Schmidt-norm GQD under the effect of Markovian channels also exhibit freezing.
See figure~\ref{freezing3}. 
The conditions on correlators $T_{ij}$, leading to the freezing of QD for the BF, PF, and BPF channels have been provided for   BD states as initial states \cite{sarandy-dec}. 
% However, for the Hilbert-Schmidt-norm GQD, two-qubit Bell-diagonal state not 
% showing any freezing phenomenon is also reported.

A necessary and sufficient condition for freezing of GQD has been provided  for $X$-states as input under local dephasing noise~\cite{song-frz}.
%\cite{karmakar-local}
%The  local filtering may remove the freezing  of QD and one-norm GQD for the Bell-diagonal states under the different Markovian channels~\cite{karmakar-local}.
Local filtering can remove the system's ability to have a frozen QD in evolution  \cite{karmakar-local,freezing_new_new1}.
%Also the set of states exhibiting freezing of QD forms a nonconvex set.

The freezing behavior of global QD of a multiparty version of the  BD state, given by
\begin{equation}
\rho=\frac{1}{2^N} \left(\mathbb{I}^{\otimes N}+\sum_{i=1}^3 T_{ii} (\sigma^i)^{\otimes N} \right),
\end{equation}
%T_{11} (\sigma^1)^{\otimes N}+T_{22} (\sigma^2)^{\otimes N}+T_{33} (\sigma^3)^{\otimes N}\right),
has been discussed in Ref.~\cite{Jianwei-Xu1} under local PF channels.
It has been found that a variety of discord-like QC measures, under certain conditions, and with Bell-diagonal state as initial state, exhibit freezing~\cite{Aaronson-frz}.
In recent years, Cianciaruso \emph{et al.}~\cite{cianciaruso-frz} demonstrated the freezing phenomenon of Bures distance-based GQD for a specific class of BD states, each party of which is 
independently interacting with a non-dissipative decohering environment.
%Under other environmental conditions like  dynamical coupling protocols,
 The conditions for choosing a set of initial states to freeze QD, have also been investigated for other environmental conditions \cite{DD1,DD2,DD3}.

\subsubsection*{b. Sudden change phenomenon}
We have  seen that for certain classes of quantum states, despite the environmental effects, QD can remain fixed over a finite interval of time. Interestingly, the decay profile of QD may experience a sudden change, so that the derivative of QD has a finite discontinuity~\cite{decay1}.
%  - another interesting phenomenon. The concept of sudden change phenomenon was first reported in Ref.~\cite{decay1} where the Authors considered the
Such behavior of QD can be seen when an initial state is chosen to be a Bell-diagonal state, which is sent through a local PF channel.
%evolution of QD of a two-qubit system which is  initially in a Bell-diagonal state.
% The dissipative phase damping channels that were used to model the environmental interactions are listed in Table \ref{t1}. Here, the Authors identify t
The conditions on the correlators of the initially prepared Bell-diagonal state for it to experience sudden change has also been provided~\cite{decay1}. This phenomenon has been realized using the  polarization degrees of freedom of two photons ~\cite{freezing-exp1} and also in NMR  experiments~\cite{Auccaise_witnessexp, SMMdBSC11}.
%  have later been proposed. It has  also been observed that the sudden change in decay rate of QD can happen even multiple numbers of times~\cite{Fanchini2, RWD14}.
%
Considering a pure state  as the initial state, for certain interactions with the bath, QD undergoes several sudden changes\footnote{It is not yet clear whether  
pure inputs will be able to provide a frozen QD after passing through a noisy channel.}~\cite{Fanchini2, RWD14}. 
It was argued that abrupt change of QD occurs due to the change of optimal basis, required to compute classical correlation~\cite{decay5}. 
For a different variant of QD measure, namely, trace norm GQD, it has also been reported that when a Bell-diagonal state is sent through a local 
 PF and or a local GAD channel, trace norm GQD changes twice  while  the associated classical correlation encounter only one sudden change in the decay process, as also observed in NMR experiments~\cite{frz-dekhechi, sarandy-dec}.

\subsubsection{ System coupled with a spin-chain environment}

The system-reservoir dynamics that we have considered up to now  only deals with the map  which describes the final state of the system after interaction. We will now consider a scenario where the bath consists of a collection of quantum spin-$1/2$
particles interacting according to some Hamiltonian, which represents a quantum spin model having quantum critical points. One may now study 
%In this study, our aim is to see 
how  QC  between parts of the system gets affected by certain  phenomena like QPT in the spin model environment, when we put on a system-environment coupling.
% Let us describe it briefly.
  Suppose that the  initial state of the system-environment duo is unentangled i.e., $\rho_{SE}(0)=\rho_S(0)\otimes \rho_E(0)$. 
%  where $\rho_S$ is a two-level system while  
  The Hamiltonian  of the reservoir can, for example, be  the $XY$ spin chain, given in Eq.~(\ref{eqn: Hamil}) with $\Delta=0$. The total Hamiltonian in this
case is  given by  $H = H_I + H_E$, where $H_I = \frac 12 J{(S_A^z + S_B^z)}\sum_i \sigma_i^z$ is the Hamiltonian  for the interaction, and $H_E$ is the quantum $XY$ spin model. A and B are the parts of the system, $S_A^z$ and  $ S_B^z$ are spin-operators of the system and $J$ is proportional to the system-environment coupling strength. 
Suppose the system is  initially prepared in a Werner state given in Eq.~(\ref{eqn:werner}) and the total state $\rho_{SE}$  evolves  via $H$.  The behavior of QD over finite time has been investigated in this case and  it was observed that near the critical point of the  spin model, QD gets minimized~\cite{LSZ10}.
Interestingly, when the initial state  does not possess any entanglement i.e. when $p \in [0, 1/3]$, entanglement, being zero, fails to detect any QPT of the environment, while QD can characterize it.
In Ref.~\cite{SLS10}, for some special choice of initial state with vanishing QD and the transverse Ising model as environment, it has been reported  that at the Ising transition point, QD rapidly increases from zero to a finite steady value, while  it oscillates in the  paramagnetic region,  and in the ferromagnetic case, it saturates to a very
 low value. 
% All the results confirm that  the properties of spin  model influences the physical quantity of the system.  
 There are several  other spin models  like three-site interaction in spin chain, DM interaction in $XY$
 chain, isotropic Heisenberg chain,   long-range  Lipkin-Meshkov-Glick model, etc., that have been considered  as 
 environments,  and their effects on QD of a pure as well as mixed state as initial states have been investigated~\cite{TZQ12threesite,YTQ11DM,AZCM15DM2,Hu10DM3,Guo16threesite,Hao10,YQY11LMG}.  

 External control, sometimes, is useful to extend the coherence of the system~\cite{VL98,YL08, Rossini07}. 
 The role of a bang-bang pulse (a train of instantaneous pulses) on the quantum correlation of two non-interacting qubits coupled to independent reservoirs is investigated in Ref.~\cite{XuXu11}. In another scenario, the effect of quantum zeno and antizeno effects on two non-interacting qubits coupled to a common bosonic reservoir was considered in Ref.~\cite{Francica10}.  

 Later, Luo {\it et al.\ }investigated a slightly different system-bath scenario, called the two spin-star model, where  two central qubits, initially prepared in an $X$-state, are coupled to their own spin baths that are of $XY$ type~\cite{LLXY11}. There is no interaction between the two central qubits. Both spin baths are modeled by the ferromagnetic 1D transverse $XY$ spin chain. The Hamiltonian for the entire setup is given by $H = H_S + H_E + H_I$ where 
\begin{align}
H_S &= \tau^z_A + \tau^z_B, \nonumber \\
H_E &= - \frac J2 \sum_{k = A', B'} \sum_{l=1}^N \Big( (1+\gamma) \sigma^x_{l,k}\sigma^x_{l+1,k} \nonumber \\ 
& \hspace{1cm}+ (1-\gamma)\sigma^y_{l,k}\sigma^y_{l+1,k} + 2\lambda \sigma^z_{l,k} \Big), \nonumber \\
H_I &= J \delta \sum_{l=1}^N (\tau^z_A\otimes\sigma_{l,A'}^z + \tau^z_B\otimes\sigma_{l,B'}^z),
\end{align}
where  $\tau^z = |e\rangle\langle e| + |g\rangle\langle g|$ with $|e\rangle$ and $|g\rangle$ being the excited and the ground states of each qubit of the central two-qubit system. Here, $A$ and $B$ denote the qubits of the
 system while $A'$ and $B'$ represent the spin-baths which are taken to be periodic $XY$ models coupled with respective qubits of the system. At the initial time $t=0$, the state of the central two-qubits is taken to be a Bell-diagonal state, described in Eq.~(\ref{bell-diagonal}) with $T_{ii} \in [-1,1]$.
 % {\color{red}(why $c_i \in [0,1]$ in original Eq. )}.
Since the $XY$ model can be solved analytically, an exact expression for the reduced density matrix of the central two-qubit system can be obtained at any finite time $t$, and then QD of the same can be obtained with a 
measurement on qubit $B$. It was shown that a freezing phenomenon followed by a sudden transition can be observed for an appropriate choice of the initial state parameters. 
 In particular, lowering the value of  $T_{33}$, results in increasing the freezing time, but with a pay-off in the value of QD, which gets decreased when $T_{33}$ is lowered. It was also argued that the sudden transition is closely related to the QPT of the $XY$ model. Next, the Authors investigated the effect of a bang-bang pulse~\cite{VL98} (see also~\cite{YL08, Rossini07, XuXu11}) applied on the system to suppress the decoherence. As expected, the
bang-bang pulse is shown to be useful to enhance the freezing time (related to what has been termed as ``freezing terminal"~\cite{Titas_freeze}) and thereby delays the sudden change in  time. 

%It was found that when the spin-bath approaches $\gamma =0$ i.e. the isotropic $XX$ model, QD of the central 2-qubit system remains constant over time for some special initial states. 

%Apart from the Markovian dynamics described above, the behavior of QD has also been investigated in 
%non-Markovian dynamics in
% Further work on QD in open systems include Ref. \cite{ }.
% which we will not discuss here. 
Quantum discord has also been calculated in various other prototypical models including spin-bosonic systems~\cite{Ge10, Man11, open_extra_new5, Wall17},  detuned  harmonic oscillators in a common heat bath~\cite{Manzano13}, dissipative cascaded systems~\cite{Lorenzo15}, qubits in a dissipative cavity \cite{Zhang11}, impurity qubits in BEC reservoir~\cite{McEndoo13}, continuous variable systems~\cite{Isar11, Isar12, Isar13, Isar13_1, Suciu15, Vasile10}, etc.   

% \subsubsection{Spin boson systems}

% An spin boson system is essentially comprise of an central spin system which is interacting with with boson reservoir. In Ref. \cite{Ge10}, the Authors studied a central two-spin system which are interacting independently with their own bosonic reservoirs and there is no interaction between the two system qubits as well as the two boson reservoirs. The Hamiltonian of the composite system is given by 

% \begin{align} 
% H^{SB} = \sum_{i=A,B} \Big( \frac {\omega}{2}\sigma_i^z &+ \sum_{k=1}^N \omega_k b_{i,k}^{\dagger}b_{i,k}  \nonumber \\
% &+ g_{i,k} \big( \sigma_i^- b_{i,k}^{\dagger} + \sigma_i^+ b_{i,k}\big) \Big) 
% \end{align}
% where $\sigma^z_i$ is the Pauli Z operator of the $i$th central spin while $\sigma^{+(-)}$ are the raising (lowering) operator. Here, $\omega$ is the Zeeman splitting of the central spins A and B and $b_{i,k}$ is the annihilation operator of the $k$th mode (/model) in the $i$th reservoir with frequency $\omega_k$. $g_{i,k}$ represents the coupling strength between the $k$th model of the $i$th reservoir and the corresponding qubit of the system.   The initial state is taken to be $| \psi(0)\rangle = (\alpha |00\rangle + \beta |11\rangle )_{S_1,S_2} \otimes |00\rangle_{R_1,R_2}$ where the subscript $S_i$ and $R_i$ are respectively denotes the $i$th system and the reservoir. For the reservoir, the state $|0\rangle_{R_i} = \Pi_{j=1}^N|0_j\rangle_{R_i}$, represents the state with no excitation  Here, $|\alpha|^2 + |\beta|^2 = 1$. 

Experimentally, in an explicit open system scenario, QD has been investigated in various substrates e.g. photons~\cite{freezing-exp1, Xu10PRA, Cialdi14, TangGessner15, Benedetti13}, ions~\cite{Gessner13NP}, NMR systems~\cite{Soares10NMR}, open solid systems~\cite{Rong13}, etc. %Apart from QD, 

Just like the usual quantum discord, the behavior of Gaussian QD has also been explored for various system-bath models like 
resonant harmonic oscillators coupled to a common environment~\cite{Vasile10, Freitas12}, non-resonant harmonic oscillators under weak and strong dissipation~\cite{Correa12}, two-mode Gaussian systems in a thermal environment~\cite{Isar11, Isar13}, two-mode squeezed thermal state in contact with local thermal reservoirs~\cite{Marian15}, bipartite Gaussian states in independent noisy channels~\cite{Cazzaniga13}, double-cavity opto-mechanical system~\cite{Qars15}, etc. Experimentally,  the behavior of Gaussian QD has been investigated in Refs.~\cite{Valente15, Hosseini14, Vogl13, Blandino12, Buono12, Laura16}.  

\subsection{Geometric quantum discord in open systems: Further issues}
Investigations similar to those for QD in open quantum systems, 
as discussed above, have also been carried out using one-norm and two-norm GQDs.  It was discovered that QD and GQD do not necessarily imply the same ordering for arbitrary two-qubit $X$-states~\cite{GQD_QD}. That is, for a pair of such states, say $\rho_{AB}$  and $\rho_{AB}'$, $\mathcal{D}(\rho_{AB})\leq  ~\mathcal{D}(\rho'_{AB})$ does not  guarantee $\mathcal{D}_G(\rho_{AB})\leq \mathcal{D}_G(\rho'_{AB})$.
Such examples have been seen to be present in situations where 
\(\rho_{AB}\) and \(\rho'_{AB}\) are respectively the initial and final states of a 
system-bath duo, with the bath being either  Markovian or non-Markovian
% bath, such difference between QD and GQD are observed
\cite{Geo_QD_decoherence3,Geo_QD_decoherence4,Geo_QD_decoherence5}.

As is the case for QD, it has been shown that there are instances for  which
GQD  provides better understanding of the dynamics of the system than that
by entanglement, when the system is  subjected to 
environmental perturbation~\cite{Geo_QD_decoherence6,Geo_QD_decoherence7,Geo_QD_decoherence8,
Geo_QD_decoherence9,Geo_QD_decoherence10,Geo_QD_decoherence11}. Starting with a pure state,  ways of protecting GQD, as measured by the Hellinger distance or the Bures distance, of the evolved state under non-Markovian structured bosonic reservoir have also been found~\cite{Geo_QD_decoherence11}.

 \section{Monogamy of quantum correlations}
\label{sec:monogamy}
 When a quantum state is shared between many parties, the amount of classical correlation between all pairs of parties can be maximal.
Consider for example, a system composed of $N$ spin-$\frac{1}{2}$ particles, in a state which is the equal mixture of all spin-up and all spin-down, in the $z$-direction. 
All two-particle states are then $\frac{1}{2}\left(\left|\uparrow_z \uparrow_z\right\rangle\left\langle \uparrow_z \uparrow_z\right| +  \left|\downarrow_z \downarrow_z\right\rangle \left\langle \downarrow_z\downarrow_z\right| \right)$, which is certainly maximally classically correlated, independent of the value of $N~(>2)$. 
In particular, for three-party system shared between Alice (A), Bob (B) and Charu (C), Alice can simultaneously be maximally classically correlated with Bob and Charu\footnote{Charu can be the name of a woman or man in parts of South Asia.}.
  However, in a similar scenario, QC    
%It is an well known fact that the sharability of CC among many parties is unrestricted, whereas the quantum one 
cannot be freely shared.
%Such restrictions on sharability leads to a quantification of QC in a multipartite domain which we  will discuss in this section by considering QD as QC measures.
%Before such discussion, let us introduce monogamy concept for arbitrary QC measure. 

Let us again consider a tripartite scenario, where three parties, $A$, $B$, and $C$ share a quantum state  $\rho_{ ABC}$, 
it can happen that Alice and Bob share a singlet and Alice and Charu share another singlet, so that Alice-Bob as well as the Alice-Charu pair share a maximally quantum correlated state.  
We are assuming here that the measure of quantum correlation being used is maximal in $\mathbb{C}^2 \otimes \mathbb{C}^2$ for the singlet state. 
This is true, for example for entanglement of formation~\cite{Bennet_EoF, woot-eof, Hill_EOF,Hayden_eof,Wooters}, quantum discord~\cite{Henderson1, Oliver1}, and quantum work deficit \cite{Oppenheim1,Sir-mam2}.
%If any two parties have maximal classical correlation, the other pair can, in principle, be simultaneously maximally classically correlated. %with their two parties.
% are maximally correlated two each other, then the individual parties may in principal be in a maximal correlation with a  other one only if the correlation is classical. 
Moreover, the system shared by Alice, Bob, and Charu is assumed to be in $\mathbb{C}^4 \otimes \mathbb{C}^2 \otimes \mathbb{C}^2$.
However, if A, B and C share a system in $\mathbb{C}^m \otimes \mathbb{C}^m \otimes \mathbb{C}^m$, a maximally quantum correlated state between A and $B$ will imply, for all quantum correlated measures (satisfying a certain set of intuitively reasonable axioms), that A and B share a pure state.
This in turn implies that the Alice-Bob pair must be as a product with the state of Charu, so that Alice cannot have any correlation, classical or quantum, with Charu~\cite{Ekert91, Bennett_monogamy, CKW, Osborne, koashi_winter, QICgroup_monogamyreview, fanchini, Bai_discordmono,Giorgi}.
This property of bipartite  quantum correlation in the multiparty scenario has been termed as the monogamy of quantum correlation.
As we see, it is a ``qualitative" version of the monogamy. 
This qualitative version can and has been quantified in a seminal paper by Coffman, Kundu, and Wootters~\cite{CKW}.
Unless stated otherwise, we will henceforth deal with monogamy only for states in  $\mathbb{C}^m \otimes \mathbb{C}^m \otimes \mathbb{C}^m$ for some specific or arbitrary $m$.

There are several ways to quantify monogamy, and we will follow the one in Ref.~\cite{CKW}.
For further discussions on this matter, see~\cite{Barry_monoreview,QICgroup_monogamyreview}. 
Following Ref.~\cite{CKW} (see also~\cite{Aditi_monogamydef, Prabhu_state_discrimi, Prabhu_lightcone,manab_mono}),
  for a given bipartite QC measure $\cal Q$, we call an arbitrary $N$-party quantum state $\rho_{1 2 \ldots N}$, as  ``monogamous" if it satisfies the inequality
\begin{equation}
\label{Eq:monogamy_relation}
{\cal Q}_{1:\text{rest}} \geq \sum_{j = 2}^N {\cal Q}_{1:j},
\end{equation}
where ${\cal Q}_{1:\text{rest}} \equiv {\cal Q}(\rho_{1:2\ldots N})$ in the 1:rest bipartition and  ${\cal Q}_{1:j} \equiv {\cal Q}(\rho_{1j})$ denotes the QC between the parties $1$ and $j$. 
 Here ``rest" comprises of all the other parties except the first one.
If $\cal Q$ satisfies the above relation for all states,  then $\cal Q$ is called a monogamous QC measure. Relation~(\ref{Eq:monogamy_relation}) is known as the monogamy inequality for a bipartite QC measure $\cal Q$.
It is clear that in the relation~(\ref{Eq:monogamy_relation}), the party ``$1$" has been given a special status since it reveals the sharability constraints of QC of party ``$1$" with other constituent parties of the multipartite state. We call the party ``$1$" as the ``nodal" observer. 
%In a similar fashion, it is reasonable to write relation~(\ref{Eq:monogamy_relation}) for other parties as nodal observers. 
In this review, we discuss  all the results on monogamy using  the party ``$1$" as the nodal observer, unless stated otherwise.

It is now useful to define a 
%computable quantity  which can characterize QC in multiparty states, both pure and mixed, based on bipartite QC measures. The 
quantity, known as monogamy score~\cite{Prabhu_state_discrimi, CKW, Prabhu_lightcone, manab_mono} for any bipartite measure $\cal Q$ and any multiparty state,  given by 
\begin{equation}
\label{Eq:monogamy_def}
\delta_{\cal Q} = {\cal Q}_{1:\text{rest}} - \sum_{j = 2}^N {\cal Q}_{1:j}.
\end{equation}
Since there exists  certain bipartite QC measures \cite{Horo_RMP,Kavan-rmp} which are computable, at least numerically, it is possible to compute $\delta_{\cal Q}$ for those measures, leading to computable multiparty QC measures for multipartite mixed states which is otherwise rare. 
Non-negativity of $\delta_{\cal Q}$ implies that the state is monogamous and vice-versa and $\cal Q$ is said to be monogamous iff $\delta_{\cal Q} \geq 0$ for all states for a fixed dimension.

There are several bipartite QC measures which satisfy the monogamy inequality, while there are plenty of measures that do not. Squared concurrence~\cite{Wooters,Hill_EOF}, squared negativity~\cite{Werner_LN,Horodecki}, squared entanglement of formation~\cite{Bennet_EoF,woot-eof, Hill_EOF,Hayden_eof}, squashed entanglement~\cite{Winter_squashed} and one-way distillable entanglement~\cite{Bennet_EoF, onew_disent} satisfy relation~(\ref{Eq:monogamy_relation}) for three-qubit states~\cite{CKW, Osborne, Ou_Fan,Oliveira}. 
Concurrence and entanglement of formation violate the monogamy relation even for pure three-qubit states \cite{Oliveira,Bai_monogamy, Giorgi,nonmono_EOF}. See also Refs.~\cite{Gerardo_contangle,Gaussian_tangle,Addeso_CVent,Addeso_CVent2,asu_monogamy,
Yuan_Tsalli,Song_Renyi,KimJ_monogamy} in this regard.
It is interesting to ask whether QC measures beyond entanglement satisfy or violate the monogamy relation. 
%Information-theoretic measures like measurement-based QD and WD for pure states reduce to the von Neumann entropy of local density matrices and hence in an extreme situation, an N-party state no two bipartite  reduced states can not possess maximal QD or WD. Specifically, 
It was found that although squared QD, $({\cal D}^{\leftarrow})^2$, satisfy monogamy for three-qubit pure states~\cite{Bai_discordmono}, there exists a class of three-qubit pure states for which QD violates monogamy relations, i.e. for those states\footnote{States with negative monogamy score for QD exist, irrespective of the party in which the measurement is carried out.}, $\delta_{\cal D} < 0$~\cite{Giorgi, Prabhu_state_discrimi}. 
The behavior of QD monogamy score can be useful in different quantum information protocols which we will discuss in Sec.~\ref{sec:app_discord monogamy score}. 
%Let us now discuss the results which are independent of any QC measure. As we see, 
Since there are some measures which satisfy monogamy while there are some which violate the same, it is interesting to find  properties related to monogamy that are true for all QC measures.
%the QC measures possess which lead to such divisions in monogamy relation. 

In this respect,  
% Similarly, 
%%behavior also observed for the 
%information theoretic measures, the QD \cite{Giorgi} and the WD \cite{asu_monogamy} also violates the monogamy relation in the three qubit system.
%Depending on these observation, 
Streltsov \emph{et al.}~\cite{Streltsov} raised the following  question:  Does there exist any measure of QC  which is non-zero for separable states, but still satisfy the monogamy relation for all states?
The  answer was found to be negative. Specifically, the following result was obtained. \\
{\bf Theorem 6}~\cite{Streltsov}: {\it Suppose a bipartite QC measure, $\cal Q$, possesses the following property: $\cal Q$ is $(i)$ non-negative, $(ii)$ local unitarily invariant and $(iii)$ non-increasing under addition of a pure ancillary system. For it to satisfy monogamy, $\cal Q$ must be zero for all separable states.} \\
%Suppose, $\cal Q$ is a bipartite QC measure which is monogamous for all states and s tisfies the conditions $i) - iii)$. 
%Let us now sketch the proof of the above statement~\cite{Streltsov}.
{\it Proof}: Let $\rho_{AB} = \sum_{i} p_i |\psi_i\rangle\langle\psi_i|_A \otimes |\phi_i\rangle\langle\phi_i|_B $ be an arbitrary separable state which can always be written as a convex combination of rank-1 projectors~\cite{Werner}.
A special extension of $\rho_{AB}$ in a tripartite state is given by
\begin{equation}\label{Eq:rho_def_addeso}
\rho_{ABC} = \sum_{i} p_i |\psi_i\rangle\langle\psi_i|_A \otimes |\phi_i\rangle\langle\phi_i|_B \otimes |i\rangle\langle i |_C,
\end{equation}
 where $\langle i|j\rangle_C = \delta_{ij}$. It is local unitarily equivalent in the $BC$ part with another state $\sigma_{ABC}$, given by
 \begin{equation}\label{Eq:sigma_def_addeso}
 \sigma_{ABC} = \sum_{i} p_i |\psi_i\rangle\langle\psi_i|_A \otimes |0\rangle\langle 0|_B \otimes |i\rangle\langle i |_C.
 \end{equation}
 Now by using conditions $(ii)$ and $(iii)$, we have ${\cal Q}(\sigma_{AC}) \geq {\cal Q}(\sigma_{A:BC}) = {\cal Q}(\rho_{A:BC}) $. Since $\cal Q$ satisfies monogamy relation,  we have
 \begin{equation}
 {\cal Q}(\sigma_{AC}) \geq {\cal Q}(\rho_{AB}) + {\cal Q}(\rho_{AC}).
\end{equation}   
From Eqs. (\ref{Eq:rho_def_addeso}) and (\ref{Eq:sigma_def_addeso}), one can find that $\sigma_{AC} = \rho_{AC}$, which implies ${\cal Q}(\rho_{AB}) = 0$ (by using condition $(i)$). Since   $\rho_{AB}$ is an arbitrary separable state,  ${\cal Q} $ vanishes and hence the proof. \hfill $\blacksquare$ % for all separable states which  indicates $\cal Q$ as an entanglement measure \cite{Streltsov}.

 %have proved that quantum correlation measures which does not vanishes for separable states, i.e., measures other than entanglement, do not satisfy the monogamy relation .
%They raise a possibility over the existence of some measure similar to discord like measure which satisfies monogamy relation.
%In this context they
It was also shown that the QC measures, which are non-monogamous for a certain state, can be made monogamous  for that state 
by a proper choice of a monotonically increasing function of that measure \cite{salini}. 
More precisely, we have the following theorem. \\
  %or by considering the monogamy inequality of that measure, $\cal Q$, with higher number of parties  \cite{asu_monogamy}. % as well as by increasing the number of parties. Hence one has the following theorem, \\
%Let us discuss the former result in the following theorem: \\
{\bf Theorem 7}~\cite{salini}: {\it 
 If a bipartite QC measure $\cal Q$ is non-monogamous, for  an N-partite quantum state $\rho_{12\ldots N}$ in arbitrary finite dimensions,  i.e., ${\cal Q}_{1:\text{rest}} < \sum_{i = 2}^N {\cal Q}_{1i}$, then 
 there always exists  a non-decreasing function $f : \mathbb{R} \rightarrow \mathbb{R}$  such that }
 \begin{equation}
f({\cal Q})_{1:\text{rest}} > \sum_{i = 2}^N f({\cal Q})_{1i},
\end{equation}  
{\it provided that $\cal Q$ is monotonically non-increasing under discarding systems and under tracing out of subsystems, invariance happens only for states satisfying monogamy.} \\
%  {\it where $\mathbb{R}$ is the set of real number.} \\
{\it Proof}:
%Let us assume that, the bipartite QC measure 
Since $\cal Q$ is non-increasing under discarding of subsystems and is non-monogamous,  $\underbrace{{\cal Q}_{1:\text{rest}}}_{\tilde{x}} >  \underbrace{{\cal Q}_{1i}}_{\tilde{y}_i} \ge 0$ $\forall i$ and ${\cal Q}_{1:\text{rest}} < \sum_i {\cal Q}_{1i}$. It implies that %, so one must have
\begin{equation}
\lim_{m \rightarrow \infty}  \left(\frac{\tilde{y}_i}{\tilde{x}}\right)^m = 0 ~ \forall i.
\end{equation}
Thus for every $\epsilon_i > 0$, however small, one must have positive integers $n_i(\epsilon_i)$, $i=2,\ldots, N$,  such that 
\begin{equation}
 \left(\frac{\tilde{y}_i}{\tilde{x}}\right)^m  < \epsilon_i ~ \forall m \geq n_i(\epsilon_i).
\end{equation}
Choose $\epsilon_i < \frac{1}{N-1} ~\forall i$ and suppose $n = \max\{n(\epsilon_i)\}$, then for any integer $m \geq n$, one gets 
\begin{eqnarray}
\sum_{i=2}^N \left(\frac{\tilde{y}_i}{\tilde{x}}\right)^m < \sum_{i=2}^N \epsilon_i < 1 \Rightarrow \tilde{x}^m \geq \sum_{i=2}^N (\tilde{y}_i)^m.
\end{eqnarray}
Hence the proof. \hfill $\blacksquare$ 
\begin{figure}[t]
% \centering
\includegraphics[width=1.03\columnwidth, height = 0.65\columnwidth]{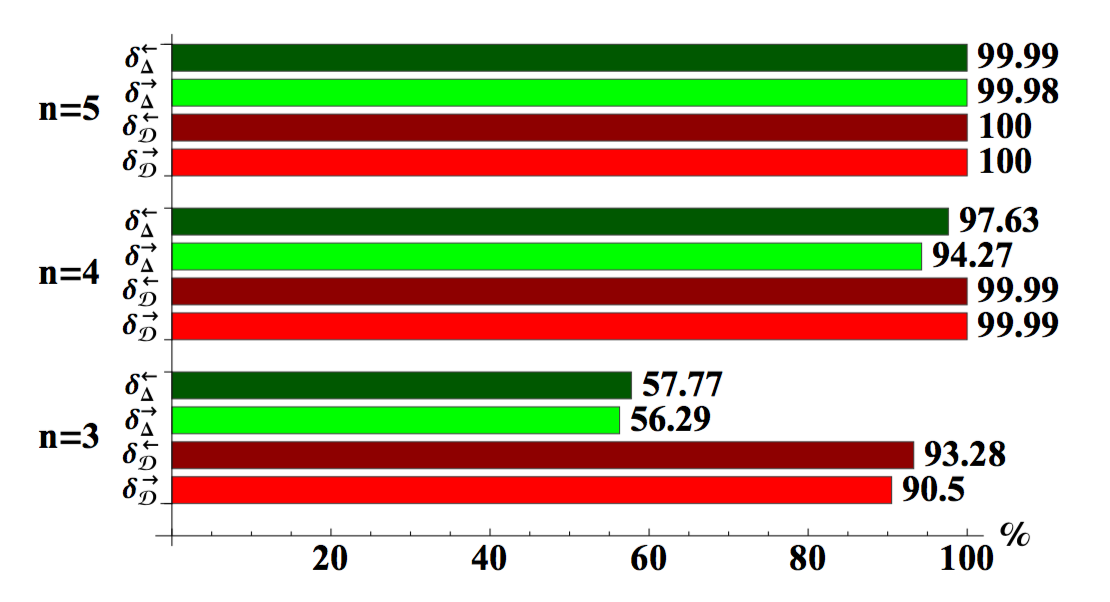}
\caption{Percentages of Haar uniformly  generated states satisfying the monogamy relations for  quantum discord and work deficit. The number of parties is denoted as ``n" in the diagram. The
 monogamy scores for ${\cal D}^\leftarrow, {\cal D}^\rightarrow, {\cal WD}^\leftarrow$, and ${\cal WD}^\rightarrow$ are respectively denoted in the diagram by $\delta_{\cal D}^\leftarrow, \delta_{\cal D}^\rightarrow,  \delta_\Delta^\leftarrow$, and $\delta_\Delta^\rightarrow$. [Reprinted from Ref.~\cite{asu_monogamy} with permission. Copyright 2015 American Physical Society.]}
\label{fig:asu_monogamy}
\end{figure}

%\noindent {\it Remark 1:} 
%%So a monotonically increasing function of $\cal Q $ ${\cal Q}^m$, has been found which can make the nonmonogamous relation to  a monogamous one.
%%It was observed that 
%\textcolor{orange}{All the known QC measures including entanglement and information-theoretic measures, satisfy the assumptions which were required for $\cal Q$ in the above proof.}\\
%This phenomenon implies the fact that although some measures does not follow the monogamy relations, they can not be freely shared among the subsystems which are different from  CC.
Note that if a QC measure is monotonically non-increasing under LOCC,  its positive powers are also non-increasing under LOCC.
%(${\cal Q}^\alpha$) with $\alpha$ being any positive number, also decreases with LOCC. 

The monogamy property of  QC measures 
 also changes from non-monogamous to monogamous  when the number of parties are increased~\cite{asu_monogamy}.
Figure~\ref{fig:asu_monogamy} depicts the percentages of states which satisfy the monogamy relations of QD and WD for a fixed number of parties up to five. The states are generated Haar uniformly. %the relative occurrence of randomly generated Haar uniform, multi qubit pure quantum monogamous states have been plotted with the total number of parties,  for both the $\delta_{\cal D}$ and $\delta_{\cal WD}$. 
The figure clearly indicates the increase in the percentage of monogamous states as one moves from three-qubit to five-qubit quantum states \cite{asu_monogamy}. 
%For example, all five-qubit randomly generated states satisfy monogamy relations for QD.  %, and for $\delta_{\cal D}$, all the five-qubit states are monogamous \

The monogamy property of  GQD has also been explored in the following years.  Streltsov \emph{et al.}~\cite{Streltsov}  has shown that for a general tripartite pure quantum state, $|\psi_{ABC}\rangle$,  GQD is  monogamous, i.e. $\mathcal{D}_G(|\psi_{A:BC}\rangle)\geq \mathcal{D}_G(\rho_{AB})+\mathcal{D}_G(\rho_{AC})$, where $\rho_{AB}$ and $\rho_{AC}$ are the reduced density matrices of $|\psi_{ABC}\rangle$. A possible extension of the monogamy relation for GQD in case of mixed quantum states has recently been  reported by Daoud \emph{et al.}~\cite{GQD_monogamy1}, where the Authors considered two  families of generalized three-qubit $X$-states.  %and studied the monogamy relation of GQD. 
Furthermore,  Cheng \emph{et al.}~\cite{GQD_monogamy2} have proven  a monogamy relation of GQD for a tripartite mixed quantum state $\rho_{ABC}$ which reads as
\begin{eqnarray}
\mathcal{D}_G(\rho_{AB})+\mathcal{D}_G(\rho_{AC})\leq \frac{1}{2}.
\end{eqnarray}
%Although several investigation has been carried out in search of making non-monogamus measures to a suitable monogamous one and among 
%states,  while the  is monogamous for the entire set of three-qubit pure states . 
%now we will briefly discuss about some of the 
%like
% the concurrence square monogamy known as N-tangle, the discord and work deficit monogamy.
%For an arbitrary three-qubit pure
% quantum state the three-tangle , where
%For example the three-tangle ($\tau$), the concurrence square monogamy score for a  $|\psi_{ABC}\rangle$ defined as
%\begin{equation}
%\tau(|\psi_{ABC}\rangle) = {\cal C}^2_{A:BC} - {\cal C}^2_{AB} - {\cal C}^2_{AC},
%\end{equation}
% is always monogamous. On the other hand the 
%Where the discord and other quantum correlation monogamy score like negativity monogamy score, can be both monogamous as well as non-monogamous. 
%Among all the non-monogamous measure of QC, discord and and work deficit involves measurement in any one of the subsystem, thus it is possible that some  monogamous state transforms into a 
%non-monogamous one under changing of measurement subsystem or vice a versa. It was shown that \cite{Giorgi} 
%$\delta_{\cal D}^{\leftarrow} < 0$, for the entire W-class of states whereas some of them became monogamous for $\delta_{\cal D}^{\rightarrow} $ \cite{TamoghnaDC}.

\section{Connecting entanglement with quantum discord-like measures}
\label{Sec:Connection}
In this section, we will discuss about relations of QD and QD-like measures with entanglement  measures.
% discussed in Sec.~\ref{Sec:quantum_correlations_def}. 
%More specifically, for multiparty quantum state we  consider the relation of the distribution of QD with the distribution of other quantum measure and then move to the relation of the 
It turns out that such relations can be used to establish connections between discord monogamy score and bipartite as well as multipartite entanglement measures for multipartite states. 
 %The relation of the discord monogamy score with one of the application of the multiparty entangled state, the multiparty dense coding capacity will be discussed later.
%The mesures of entanglement can be classified in two broad categories. The bipartite and multipartite measures. In 
%the bipartite entanglement measures, one can define entanglement if 
%Now we will briefly discuss about the EOF

\subsection{Links to entanglement of formation}
\label{discord monogamy-EOF}
%We now concentrate here mainly on the relation of 
%In this section, we will discuss the distribution of 
%QD among the subsystems of a  tripartite quantum state  with the distribution of EOF among the same. 
As we have already noted, if any two parties,  
 $A$ and $B$, of a tripartite system in $\mathbb{C}^m \otimes \mathbb{C}^m \otimes \mathbb{C}^m$ in a pure state, possess  maximal QC, then the state is nearly always of the form $|\psi_{ABC}\rangle = |\psi'_{AB}\rangle\otimes |\psi''_{C}\rangle$, thereby implying  no  correlation of party $C$ with $A$ as well as $B$. %this rules out even the existence CC of the individual parties with a third one. 
%This feature we try to discuss
In particular, there is no classical correlation, e.g. quantified by  $J_{A|C}$ between $A$ and $C$. One may now ask the extent to which $J_{A|C}$ can increase for a non-maximal  QC between $A$ and $B$. Koashi and Winter \cite{koashi_winter} derive the following useful result.\\
 %
 %The question arises when the QC between any two parties is not the maximum. 
 %
%Koashi and Winter~\cite{koashi_winter} has shown that  for an arbitrary tripartite state $\rho_{ABC}$ with arbitrary dimension, the entanglement between the two parties, restricts the amount of classical correlation  quantified in Eq.~(\ref{Eq:quant2}), that the individual party can share with the third one.   %QC between the two parties restricts the amount of CC 
%that the individual party can share with 
% the third one. It means that for a shared $\rho_{ABC}$, We have the following theorem:\\
{\bf Theorem 8}~\cite{koashi_winter}: {\it For an arbitrary tripartite state $\rho_{ABC}$,} 
 \begin{equation}\label{Eq:Koashi_relation}
  {\cal E}_{AB}  +  J_{A|C}  \leq S_A,
 \end{equation}
 {\it where ${\cal E}_{AB}$  is the EOF of the reduced state $\rho_{AB}$ while $J_{A|C}$ denotes the classical correlation of $\rho_{AC}$ (introduced in Sec.~\ref{Entanglement of formation}) with measurement being performed in $C$. $S_A$ is the von Neumann entropy of the reduced density matrix $\rho_A$.
  Here, the equality 
 %For a tripartite state, it was shown in  the distribution of quantum correlation between any two party limits the classical correlation with either one of them with the third party. 
% Here EOF was taken as a measure of quantum correlation or entanglement, and $J^{\leftarrow}$ is the classical correlation, quantifies the maximal information gain about the first party by measuring the second party, and  is given by 
% \begin{equation}\label{Eq:CC_defined}
% J^{\leftarrow}(\rho_{AB}) = S(\rho_A) - \min_{\{\Pi_j\}}\sum_j q_j S(\rho^j_A)
% \end{equation} 
%%The measure of classical correlation is quantified by  
% where $\rho_A^j = \text{tr}_B(\mathbb{I}\otimes \Pi_j \rho_{AB}\mathbb{I}\otimes \Pi_j)/\text{tr}(\mathbb{I}\otimes \Pi_j \rho_{AB}\mathbb{I}\otimes \Pi_j)$ is  the quantum state in the A's subsystem after 
%  measuring the B part and $p_j = \text{tr}(\mathbb{I}\otimes \Pi_j \rho_{AB}\mathbb{I}\otimes \Pi_j)$ is the probability of getting the $j$th outcome.
% Here the minimization is taken over all measurements applied in the subsystem B.
%Then for a pure $|\psi_{ABC}\rangle$, the distribution of two types of correlations among the parties is given by
% in Eq. (\ref{Eq:Koashi_relation}) 
 holds only when the shared state is  pure.} \\
{\it Proof:} Let us first consider an arbitrary pure state $|\psi_{ABC}\rangle$ such that  tr$_C(|\psi_{ABC}\rangle\langle\psi_{ABC}|)=\rho_{AB}$ and similarly for $\rho_{AC}$. Let us also assume that $\rho_{AB}=  \sum_i p_i  |\psi^i_{AB}\rangle\langle \psi^i_{AB}|$ where $|\psi^i_{AB}\rangle$'s form the minimum pure state decomposition required for EOF of $\rho_{AB}$. 
Let us denote the measurement at $C$ that realizes this optimal ensemble by acting n the state $|\psi_{ABC}\rangle$ as $\{M_i\}$.
%Similarly, it is obtained when the measurement $\{M_i\}$ is performed on C and the outcome $i$ clicks with probability $p_i$. 
Tracing out $B$, the same measurement on $C$ leads to the ensemble on the $A$'s part as \{$ p_i, \text{tr}_B( |\psi^i_{AB}\rangle\langle \psi^i_{AB}|)$\} \cite{HJW_theorem}, and hence from Eq.~(\ref{Eq:quant2}), we obtain 
\begin{eqnarray}\label{Eq:KS1}
 J_{A|C} &\geq & S(\rho_A) - \sum_i p_i S\big(\text{tr}_{B}(|\psi^i_{AB}\rangle\langle \psi^i_{AB}|)\big) \nonumber \\
 &=&  S(\rho_A) - {\cal E} (\rho_{AB}).
 \end{eqnarray}

% the above statement we first took a pure $\rho_{ABC}$. And the states $\rho_{AB}$ and $\rho_{AC}$ are complement to each other as both of them are derived from the same pure state $\rho_{ABC}$.
%Now recall the definition of  EOF in Eq. (\ref{Eq:EOF_def}), and suppose
% the optimum pure state decomposition $\rho_{AB}=  \sum_i p_i  |\psi_i\rangle\langle \psi_i|_{AB}$, has been achieved by doing rank one PV measurement $\{M_i\}$ in the $C$ part of the total state. 
% Now this thing can also be thing as a PV measurement $\{M_i\}$, in the $C$ part of the state $\rho_{AC}$, which results the ensemble $\{ p_i, \text{tr}_{B}(|\psi_i\rangle\langle \psi_i|_{AB})$ in the $A$ part. Hence, from Eq. (\ref{Eq:quant2}), 
 
 On the other hand, suppose that the optimum measurement performed on $C$ and required for obtaining 
 $J_{A|C}$ is $\{\tilde{M}_i\}$. It results in the output ensemble at $A$ as $\{ \tilde{p}_i, \tilde{\rho}_{A|i} = \text{tr}_{BC}(|\tilde{\psi}^i_{ABC}\rangle\langle \tilde{\psi}^i_{ABC}|)\}$. Thus 
 \begin{eqnarray} \label{Eq:KS2}
 J_{A|C} &=& S(\rho_A) - \sum_i \tilde{p}_i S(\tilde{\rho}_{A|i}) \nonumber \\
  &\leq & S(\rho_A) - {\cal E}(\rho_{AB}),
 \end{eqnarray}
 where the inequality arises from the fact that second term of  $J_{A|C}$ is higher than or equal to the EOF of $\rho_{AB}$ for all measurements\footnote{For rank-1 $\{\tilde{M}_i \}$, it is clear that the output state in the $A$ part is pure and hence, $\sum_i \tilde{p}_i S(\tilde{\rho}_{A|i}) = \sum_i \tilde{p}_i S\big(\text{tr}_{B}(|\tilde{\psi}^i_{AB}\rangle\langle \tilde{\psi}^i_{AB}|)\big) \geq {\cal E}(\rho_{AB})$.
If the measurement is not of rank-1, $\tilde{M}_i  = \sum_j \tilde{M}_{ij} $, for some rank-1  $\{\tilde{M}_{ij}\}$ with $\tilde{p}_{ij} = \text{tr}\big((\mathbb{I}^A \otimes \tilde{M}^C_{ij}) \rho_{AC}\big)$  and 
$\tilde{\rho}_{A|ij} = \text{tr}_C(\mathbb{I}^A \otimes \tilde{M}^C_{ij} \rho_{AC} \mathbb{I}^A \otimes \tilde{M}^{C}_{ij})/p_{ij}$. 
Now one can also show that $\tilde{p}_i = \sum_j  \tilde{p}_{ij}$ and $\tilde{p}_i \rho_{A|i} = \sum_{j} \tilde{p}_{ij}  \rho_{A|ij}$.  Thus from the concavity of von Neumann entropy, $\sum_i \tilde{p}_i S(\rho_{A|i}) \geq \sum_{ij} p_{ij}  S(\rho_{A|ij}) \geq {\cal E}(\rho_{AB})$.
 }. From~(\ref{Eq:KS1}) and~(\ref{Eq:KS2}), we have   $J_{A|C} + {\cal E}(\rho_{AB}) =  S(\rho_A) $ for pure tripartite  states.
 %has been obtained~ by using the Koashi-Winter relation given in Eq.~(\ref{Eq:Koashi_relation}).
  Now an arbitrary state, $\rho_{ABC}$, can be purified to form a pure four-party state  $|\psi_{ABCD}\rangle$, such that  $\rho_{ABC} = \text{tr}_D(|\psi_{ABCD}\rangle\langle\psi_{ABCD}|)$. Using the above relation for pure states and by taking $CD$ as a single party, one gets $J_{A|CD} + {\cal E}(\rho_{AB}) =  S(\rho_A)$. Note now that $J_{A|CD}$ is non-increasing under discarding the subsystem\footnote{See Appendix \ref{Sec:CC_notincreasing} for the proof.}, i.e. $J_{A|CD} \geq J_{A|C}$. Combining the above results, we obtain  Eq. (\ref{Eq:Koashi_relation}) for arbitrary tripartite states. \hfill $\blacksquare$
  
 For a tripartite state, \(\rho_{ABC}\),  a relation between QD of the reduced state \(\rho_{AB}\) and the classical correlation of \(\rho_{BC}\) can be obtained by using  Eq. (\ref{Eq:Koashi_relation}) and   is given by~\cite{Wang-monogamydekh}
  %where the following relation holds
%have showed that the QD of any two parties and  the classical correlation of the measured party party with respect to the third one is  should be sum up to von-Neumann entropy of the 
\begin{equation}
{\cal D}(\rho_{AB}) + J_{C|B} \leq S(\rho_B),
\end{equation}
where the equality holds for pure states.
% where in the calculation of classical correlation $J^{\leftarrow}(\rho_{AC})$ we measure only in the second party.
% And for a mixed $\rho_{ABC}$, the above equation turns out to be an inequality. 
 %The classical correlation defied in Eq. (\ref{Eq:quant2}), we have measured in the second subsystem, we can also defined it by measuring the first system and in this case we call it $J^{\rightarrow}(\rho_{AB})$. 
It is important to note here that the definition of quantum discord used in the Koashi-Winter result in Theorem $8$ involves an optimization over POVMs, and not merely over PV measurements. This will remain true whenever the Koashi-Winter result is used.
%  which is the difference of two inequivalent definition of quantum mutual information (see Eq. (\ref{Eq:quant1})), and the quantity defined as CC in Eq. (\ref{Eq:quant2}) is one kind of quantum mutual information.
%Form now onwards in this review unless specified $\cal D$ will denote discord for a bipartite state and measurement can be done in either any one of the two subsystems. ${\cal D}^{\rightarrow}$ and ${\cal D}^{\leftarrow}$  represents the measurement in the first and the second party respectively. 
%So we have
%\begin{equation}
%{\cal D}^{\rightarrow}(\rho_{AB}) = I(\rho_{AB}) - J^{\rightarrow}(\rho_{AB})
%\end{equation}
%and vice a versa.  
%Now recall the definition of QD in Eq. (\ref{Eq:discord1}), 

For a tripartite quantum state $\rho_{ABC}$, we have \cite{fanchini}% the QD of  $\rho_{AB}$ and $\rho_{AC}$ are summed up to 
\begin{eqnarray}\label{Eq:fanchini}
{\cal D}(\rho_{AB}) &+& {\cal D}(\rho_{AC}) \nonumber \\
&=& S(\rho_A) - J_{A|B} + S(\rho_A) - J_{A|C} + \Delta \nonumber \\
&\geq& {\cal E}(\rho_{AB}) + {\cal E}(\rho_{AC}) + \Delta, 
\end{eqnarray}
 where $\Delta = S(\rho_B) + S(\rho_C) - S(\rho_{AB}) - S(\rho_{AC})$, and where the  inequality in  (\ref{Eq:Koashi_relation}) has been used. Strong subadditivity of von Neumann entropy gives  $\Delta \leq 0$ and hence  no definite relation can be established between the EOFs and the QDs in (\ref{Eq:fanchini}).  
% can be positive as well as negative, 
However,  for a  pure state $|\psi_{ABC}\rangle$, $\Delta = 0$ since $S(\rho_B) = S(\rho_{AC})$ and $S(\rho_{C}) = S(\rho_{AB})$. Thus for a  tripartite pure state, one has ``conservation law"  given by 
\begin{equation}
{\cal D}(\rho_{AB}) + {\cal D}(\rho_{AC}) = {\cal E}(\rho_{AB}) + {\cal E}(\rho_{AC}).
\end{equation}
% between the entanglement measure and the QD was established \cite{fanchini} as the theorem states: \\
%\textbf{Theorem 7:} {\it  Given an arbitrary tripartite pure system, the sum of all possible bipartite entanglement quantified as EOF shared with a particular subsystem cannot be increased without increasing  the sum of all QD shared with these same subsystems by the same amount.}
%a quantitative connection has been established between 
% the distribution of QD as well as EOF within the subsystem \cite{fanchini}, defined from a completely different perspective of entanglement separability criteria. 
% between 
% the information theoretic approach and the entanglement separability approach of quantum correlation .
% the relation of EOF and CC was extended to  .

\subsection{Relating with multipartite entanglement}
%In this section, we will discuss the 
We now establish a connection between two multiparty QC quantifiers, namely,  discord monogamy score and a  genuine multiparty entanglement measure, known as generalized geometric measure (GGM)~\cite{GGM,GGM2,GGM_anindya} (see also~\cite{GM_shimony, GM_Barnum, GM_Wei_Goldbart}, see Appendix \ref{Sec:Multiparty_QCdef} for definition). \\
%  of an arbitrary N qubit state, and observe that a light cone like behavior emerges from it \cite{Prabhu_lightcone}. The genuinely multiparty entanglement measure, we will use here, is the  generalized geometric measure 
%(GGM) \cite{GGM}, defined in Appendix \ref{Sec:Multiparty_QCdef}.
%For the arbitrary three-qubit pure state $|\psi_{ABC}\rangle$ and for generalized GHZ state (gGHZ)  let us  now state  a theorem  connecting discord monogamy score ($\delta_{\cal D}$) and the GGM ($\cal G$). \\
%betwfollowing relation of arbitrary pure and the generalized GHZ state.\\
\textbf{Theorem 9}~\cite{Prabhu_lightcone}: {\it For all  three-qubit pure states, $|\psi_{ABC}\rangle$, whose GGM are  same as that  of the  generalized GHZ} (gGHZ) state\footnote{
An $N$-qubit gGHZ state \cite{GHZ} is given by
\begin{equation}\label{Eq:gGHZ}
|gGHZ_N\rangle  = \sqrt{\alpha}|00\ldots0\rangle_N + \sqrt{1-\alpha}e^{i \phi}|11\ldots1\rangle_N,
\end{equation}
 with $\alpha \in [0,1]$, and $\phi$ being a phase factor.}, {\it  $|gGHZ_3\rangle$,  the discord monogamy score of $|\psi_{ABC}\rangle$ is bounded above by the modulus of the discord monogamy score of the gGHZ state, i.e. }
\begin{equation}
-\delta_{\cal D} (|gGHZ_3\rangle) \leq \delta_{\cal D} (|\psi_{ABC}\rangle) \le \delta_{\cal D}(|gGHZ_3\rangle),
\end{equation}
{\it provided the maximum eigenvalue in GGM of the arbitrary state is obtained from the nodal~$:$~rest bipartition.}\\ %given arbitrary state is closest to a nongenuinely multiparty state which is separable in the nodal : rest bipartition.}\\
{\it Proof}:  Without loss of generality, let us fix the party $A$ as the nodal observer.
For an arbitrary three-qubit pure state $|\psi_{ABC}\rangle$,  $\delta_{\cal D}$ is given by
\begin{equation}\label{Eq:Dis_psi}
\delta_{\cal D}(|\psi_{ABC}\rangle) = S(\rho_A) - {\cal D}(\rho_{AB}) - {\cal D}(\rho_{AC}),
\end{equation}
and the same for  $|gGHZ_3\rangle$ is given by 
\begin{equation}\label{Eq:Dis_gGHZ}
\delta_{\cal D}(|gGHZ_3\rangle) = h(\alpha),
\end{equation}
where $h(\alpha)$ is defined in Eq.~(\ref{binary-shannon}). 
 % and an three qubit gGHZ state one can get from Eq. (\ref{Eq:gGHZ}) by putting $N = 3$.
The GGM of these two states, $|\psi_{ABC}\rangle$ and $|gGHZ_3\rangle$, are respectively given by %of the following form
\begin{equation}
{\cal G}(|\psi_{ABC}\rangle) = 1 - \max\{\lambda_A,\lambda_B,\lambda_C\},
\end{equation}
\begin{equation}
 {\cal G}(|gGHZ_3\rangle) = 1 - \alpha,
\end{equation} 
where $\lambda_i$, $i = A,~B,~C$, are the largest eigenvalues of the reduced density matrices $\rho_A$, $\rho_B$, $\rho_C$ respectively. Here  % $ = \text{tr}_{\bar{i}}(|\psi\rangle\langle \psi|_{ABC})$, $\bar{i}$ are the parties complement to $i$ and
 we assume $\alpha \geq \frac{1}{2}$. Suppose now that the GGMs for these two states are equal which leads  $\alpha = \max\{\lambda_A,\lambda_B,\lambda_C\} = \lambda_A$ as per the premises of the theorem. %assumption $\lambda_A \geq \lambda_B,\lambda_C$.
 %The $\delta_{\cal D}^{\leftarrow}$, of this two states are given by, by consider the A-party as the nodal one
%Now  from the assumption $\lambda_A \geq \lambda_B,\lambda_C$. 
%So we have $\alpha = \lambda_A$ and hence from Eq.~(\ref{Eq:Dis_psi}) 
This immediately implies $\delta_{\cal D}(|\psi_{ABC}\rangle) \leq h(\lambda_A) = \delta_{\cal D}(|gGHZ_3\rangle)$, where  we use the fact that $S(\rho_A) = h(\lambda_A)$, and ${\cal D} \geq 0$.
% for all valid density matrices. 

To obtain lower bound, we note that Eq.~(\ref{Eq:fanchini}) reduces to 
%Now for a pure  $|\psi_{ABC}\rangle$ and from the Eq.~(\ref{Eq:fanchini}), we got 
${\cal D}(\rho_{AB}) + {\cal D}(\rho_{AC}) = {\cal E}(\rho_{AB}) + {\cal E}(\rho_{AC})$ when $|\psi\rangle_{ABC}$ is pure and also the EOF (${\cal E}$) of a bipartite state is bounded above by % not greater than either of the local
 von Neumann entropies of the   local density matrices\footnote{Suppose the optimal pure state decomposition of $\rho_{AB}$, is the ensemble $\{p_i, |\psi_i\rangle\}$. Thus from Eq. (\ref{Eq:EOF_def}),
${\cal E}(\rho_{AB}) = \sum_i p_i S(\rho_X^i) \leq S(\sum_i p_i \rho_X^i) = S(\rho_X)$, where we use the concavity of von Neumann entropy. Here, $X = A,B$, $\rho_X^i = \text{tr}_{\bar{X}}(|\psi_i\rangle\langle\psi_i|)$
with $\bar{X}$ being the complement to $X \in\{A,B\}$.}. Therefore
  ${\cal D}(\rho_{AB}) + {\cal D}(\rho_{AC}) \leq 2S(\rho_A)$, which implies
 $\delta_{\cal D}(|\psi_{ABC}\rangle) \geq -S(\rho_A) =   -h(\lambda_A) = -h(\alpha)=- \delta_{\cal D}(|gGHZ_3\rangle)$.  \hfill $\blacksquare$ \\

\section{Applications of discord monogamy score}
\label{sec:app_discord monogamy score}
Over the last few years, it has been found that the discord monogamy score can be efficiently used for analysis 
and applications in different 
multiparty quantum information tasks 
 including state discrimination, distinguishing between noisy channels, classical information transfer between multiple senders and receivers and identifying different phases in many-body systems. 
 
\subsection{Quantum state discrimination}
The set of three-qubit genuinely multiparty entangled pure  states can be divided into two disjoint subsets  with respect to transformation possible by using stochastic local operations and 
classical communication (SLOCC)~\cite{Dur1}. Specifically, it was shown that states from one class cannot be converted into another at the single-copy level under LOCC with any non-zero probability. These two inequivalent classes are 
the ``GHZ" and the ``W" classes, arbitrary members of which can be expanded as % given respectively by 
 % Discord monogamy score can be used to distinguish the two  inequivalent classes in three qubit genuinely multiparty entangled states   \cite{Prabhu_state_discrimi, Giorgi}. These are  \cite{Dur1} given by
\begin{equation}
|GHZ^c\rangle = \sqrt{K} (\alpha_{0} |000\rangle + \beta_{0}e^{i \phi}|\psi_1 \psi_2 \psi_3\rangle),
\end{equation}
where $|\psi_j\rangle = \alpha_{j}|0\rangle + \beta_j|1\rangle$, with $K = \big(1 + 2\beta_0 \Pi_{i = 0}^3\alpha_{i} \cos\phi\big)^{-1}$ and
\begin{equation}
|W^c\rangle = \sqrt{a}|001\rangle + \sqrt{b}|010\rangle + \sqrt{c}|100\rangle + \sqrt{d}|000\rangle,
\end{equation}
where $a,~b,~c,~d$ are real numbers with $a + b + c + d = 1$. 
The three-party gGHZ state, $|gGHZ_3\rangle$, belong to the GHZ class, while the generalized W (gW) state, given by  
\begin{equation}
|gW_3\rangle = \sqrt{a} |001\rangle + \sqrt{b} |010\rangle + \sqrt{c}|100\rangle,
\end{equation}
is a subclass of $|W^c\rangle$. % with $d = 0$.
%It was shown that the states from one class can not be converted into the other one in a single copy level \cite{Dur1}. 
It can be easily shown  that the discord monogamy score is negative for the entire  class  of gW states while it is non-negative for the  gGHZ states~\cite{Prabhu_state_discrimi}. Furthermore, it was shown that for states of the W class, $\delta_{\cal D} < 0$ \cite{Prabhu_state_discrimi, Giorgi}, although for states of the GHZ class, $\delta_{\cal D}$ can be both positive and negative (and zero).
% Such conclusive discrimination protocol does not exist for all the states from the two inequivalent classes.
%In particular, numerically and then analytically it was shown that  
%
%, and any state from one class can not be   one
%%be distinguished into two inequivalent classes, , which can not be convertible to each other in a single copy level  
%by s. 
%This two class of states are
%Calculating $\delta_{\cal D}^{\leftarrow}$ can distinguish the two classes, as
 %$\delta_{\cal D}^{\leftarrow} < 0$, only for the entire W-class of states.
%on the other hand it was numerically found that 
%the same for GHZ-class can be 
%Prabhu \emph{et al} \cite{Prabhu_state_discrimi}, numerically showed that
% $\delta_{\cal D} < 0$, for all the states from the   W class~\cite{Prabhu_state_discrimi, Giorgi}, while  numerical simulation shows that only few states from the GHZ class violate the monogamy relation~\cite{Prabhu_state_discrimi}.
To prove the result for the W class, first notice that  the 
%   The results for the W class was later proved analytically by Giorgi \cite{Giorgi}. 
%It was shown in \cite{Prabhu_state_discrimi}, that numerical simulation shows that the  
%To prove the above relation, the results of Coffman-Kundu-Wooters \cite{CKW}, the ${\cal C}^2$ 
monogamy score of squared concurrence
 vanishes, i.e. ${\cal C}^2_{AB} + {\cal C}^2_{AC} = {\cal C}^2_{A:BC}$, for all W-class states~\cite{CKW}. Since $\cal E$  (see  Eq.~(\ref{Eq:EOF_def})) is a concave function of ${\cal C}^2$~\cite{Wooters} and  $ {\cal E, C} \in [0,1]$, ${\cal E}_{A:BC} < {\cal E}_{AB} + {\cal E}_{AC}$ for states of the W class\footnote{
For a concave function $f(x)$, with $f(0) = 0$, we can show that if $x = \sum_j y_j$, then $f(x) \leq \sum_j f(y_j)$, where equality holds when $y_j = 0, ~\forall j$ except some $y_i$. To show this, note here that for some $t \in [0,1]$, as $f$ is concave, we have $f(tz) = f(tz + (1-t)0) \geq tf(z)$. So, 
\begin{eqnarray}\label{Eq:asutoshproof}
\sum_j f(y_j) &=& \sum_j f\bigg(\Big(\sum_i y_i\Big) \frac{y_j}{\sum_i y_i}\bigg) \nonumber \\
&=& \sum_j f\big(x t_j\big) \geq \sum_j t_j f\big(x \big) = f(x),
\end{eqnarray}
where $0\leq t_j \leq 1$, and $\sum_j {t_j} = \sum_j y_j /(\sum_i y_i)  = 1$.
Equality holds only when $t_j = 0 ~\forall j$ except one. 
%implies all other $y_i$ are zero except some $i = j$.
}. % where equality holds for the bi-separable states. 
%The above inequality can be understood in the following way 
%
%as $\cal E$, is concave function of ${\cal C}^2$, one has for any $t \in [0,1]$,
%\begin{eqnarray}
%{\cal E}(t {\cal C}^2) &=& {\cal E}(t  {\cal C}^2 + (1-t)\times 0) \nonumber \\
%&\geq & t {\cal E}( {\cal C}^2) + (1-t){\cal E}(0) = t {\cal E}( {\cal C}^2).
%\end{eqnarray} 
%Thus 
Using the relation of Koashi-Winter  given in Eqs.~(\ref{Eq:Koashi_relation}), and~(\ref{Eq:fanchini}), for states of the W-class, one finds %that for a pure multiparty entangled state $|W_{ABC}^c\rangle$,
\begin{equation}
 {\cal D}^{\leftarrow}_{AB} + {\cal D}^{\leftarrow}_{AC} = {\cal E}_{AB} + {\cal E}_{AC} >  {\cal E}_{A:BC} = {\cal D}^{\leftarrow}_{A:BC}.
\end{equation}
% thereby the states from the W class always violating discord monogamy relation. 
The inequality comes from the concavity of entanglement of formation with respect to concurrence squared.
The inequality is strict, as $(i)$ ${\cal C}_{AB} = 0$ or ${\cal C}_{AC} = 0$ along with ${\cal C}^2_{AB} + {\cal C}^2_{AC} = {\cal C}^2_{A:BC}$ implies that three-qubit pure state is not genuinely multiparty entangled, and $(ii)$ the relation (\ref{Eq:asutoshproof}) is strict unless $t_j = 0 ~\forall j$ except one.
 The last equality comes from the fact that both EOF and QD reduce to the von Neumann entropy of the local density matrices~\cite{Oliver1,Bennett_monogamy} for pure states.
%Note that, since  the states from the GHZ class satisfy  the monogamy relation for ${\cal C}^2$,  the above theorem does  not hold in this case.
%Therefore, it was found that  for arbitrary three-qubit pure states,  the positivity of $\delta_{\cal D}$ ensures that the state belongs to the GHZ class while its negative value can not conclusively idenitify its set. \\
The discord monogamy score, therefore, is to the GHZ-class states as the entanglement witnesses \cite{Horodecki, Maciej_entwitness,Bruss_entwitness} are to entangled states: $\delta_{\cal D} \geq 0$ implies that the state is from the GHZ class, while $\delta_{\cal D} < 0$ is inconclusive, provided the input is promised to be a three-qubit pure genuinely multipartite entangled state \cite{Prabhu_state_discrimi, Giorgi}.

\subsection{ Quantum channel discrimination } \label{Sec:Channel_discri}
Another important aspect of the discord monogamy score is that it can distinguish between noisy channels \cite{Asu_global_local}. 
Consider a game in which we are provided with  a black box that is a quantum channel taking an arbitrary three-qubit state as an input, and which is promised to be a global noisy, or  a local amplitude damping (ADC), or a local phase damping (PDC), or a local depolarizing channel (DPC) \cite{Nielsen,Preskil}.
The game is to find out what the channel is. 
% through which a multiparty quantum state can undergo. In Ref.~,
%it was shown that quantum states subjected to the 
% global noise as well as local noisy channels such as  
%amplitude damping channel (ADC), phase damping channel (PDC) and depolarising channel (DPC)~, can be detected conclusively by studying the behavior of monogamy of QC measures. 
The input states that have been used are the three-qubit gGHZ as well as the gW states, and the monogamy scores of  QD  ($\delta_{\cal D}^\rightarrow$) and negativity ($\delta_{\cal N}$) are considered as the distinguishing ``order parameter". (See Appendix~\ref{LN} for a definition of negativity.) % see Eq. (\ref{Eq:Neg_def}), have been considered. define in , 
% the negativity  monogamy score.
%can be helpful for conclusive detection of the type of noises acts on a quantum state including 
%
%of a quantum state after passing through a channel
%relation as an 
%observable,
% global as well as  .
\begin{figure}[t]
% \centering
\includegraphics[width=1.0\columnwidth,keepaspectratio]{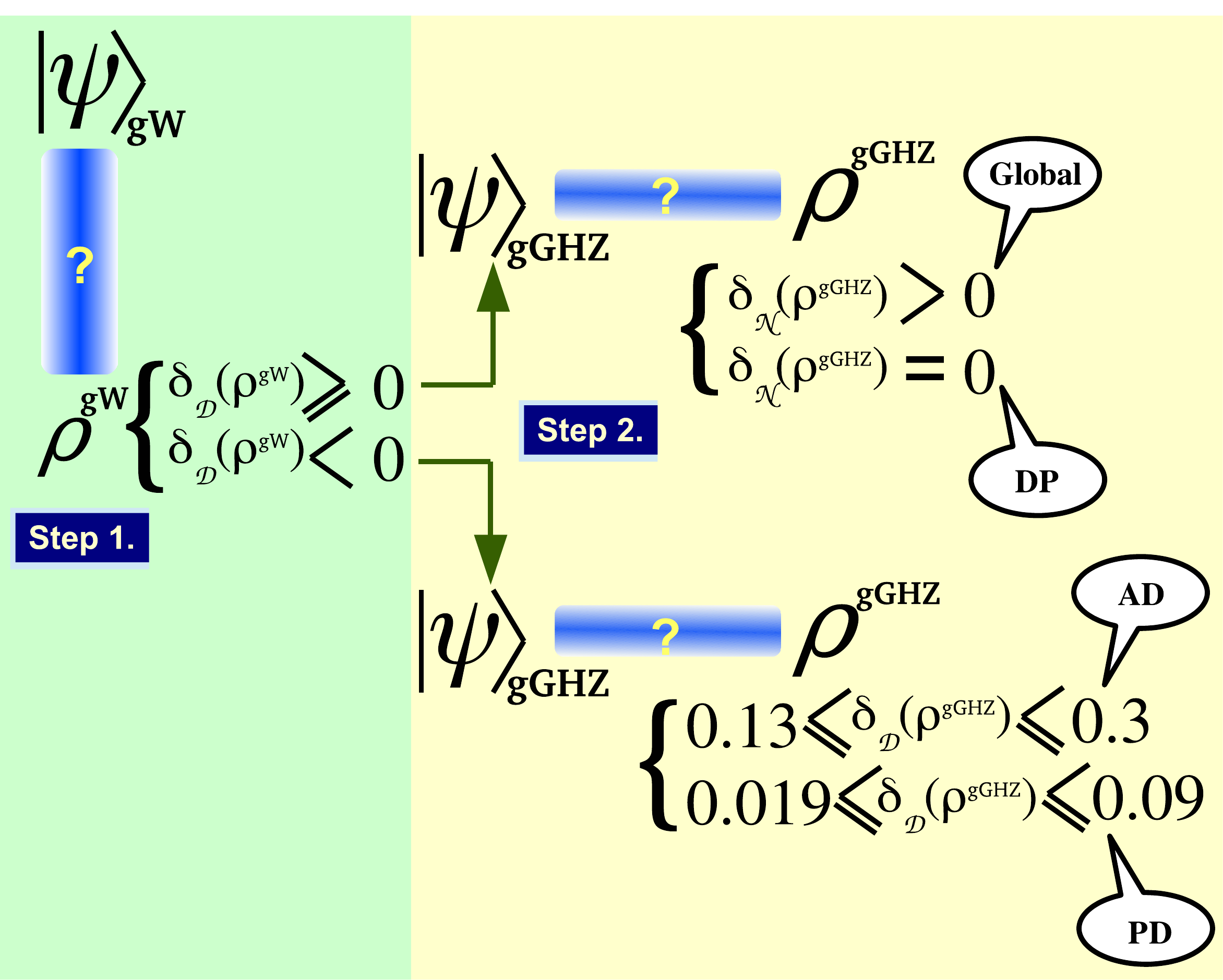}
\caption{Discrimination of quantum channels by using monogamy scores. The protocol is discussed in Sec.~\ref{Sec:Channel_discri} by calculating $\delta_{\cal D}^\rightarrow$ (denoted in the figure as $\delta_{\cal D}$) and $\delta_{\cal N}$. The states $|gGHZ_3\rangle$ and $|gW_3\rangle$ are represented in the figure as $|\psi\rangle_{gGHZ}$ and $|\psi\rangle_{gW}$ respectively. 
The corresponding outputs are respectively denoted in the figure as $\rho^{gGHZ}$ and $\rho^{gW}$. [Reprinted from Ref.~\cite{Asu_global_local} with permission. Copyright 2016 Elsevier.]
 }
\label{fig:channel_schematic}
\end{figure}
The Authors of Ref.~\cite{Asu_global_local} proposed a two-step protocol, for discriminating the global and local noises by using  the gGHZ  and gW states  (see figure~\ref{fig:channel_schematic}), where the choice of QC measure in the second step depends on the outcome of  the first step.
It works in the following way: Step 1: the gW state 
%at first gW state
 is taken as an input and after its passage through the unknown channel, $\delta_{\cal D}^\rightarrow$ is computed for the output. 
%In this protocol, the and under passing it through the unknown channel the 
%of the output state is calculated. 
Step 2: According to the value of $\delta_{\cal D}^\rightarrow$ in Step 1, $\delta_{\cal N}$ or $\delta_{\cal D}^\rightarrow$ is calculated for the output state, when 
 a gGHZ state (see Eq.~(\ref{Eq:gGHZ})) with $0.65 \leq \sqrt{\alpha} \leq \frac{1}{\sqrt{2}}$ is sent through the same channel.
  %Depending on the outcome of $\delta_{\cal D}$, in this step $\delta_{\cal N}$ or $\delta_{\cal D}$ has further been calculated. 
If $\delta_{\cal D}^\rightarrow \geq 0$ in the first step, 
$\delta_{\cal N}$ is calculated in the next step, while for negative $\delta_{\cal D}^\rightarrow$, measurement of $\delta_{\cal D}^\rightarrow$ is performed again in the second step. If $\delta_{\cal D}^\rightarrow \geq 0$ in the first step together with $\delta_{\cal N} > 0$ in the second one, the channel is global, whereas $\delta_{\cal N} = 0$ implies that it is the  DPC. 
% and a strictly positive result ensures the   channel as the
%global one. Similarly  if one gets $\delta_{\cal N} = 0$,  the channel is attributed to be the DPC.
 On the other hand, if $\delta_{\cal D}^\rightarrow< 0$ in the first step, and 
 %requires  another evaluation of $\delta_{\cal D}$, in the next step.
if the value of $\delta_{\cal D}^\rightarrow$ lies within  $[0.13,0.3]$ in the second step, the channel can be identified as ADC, while if $\delta_{\cal D}^\rightarrow \in [0.019,0.09]$ in the second step, the channel is the PDC. 
The accomplishment of the above protocol depends on two assumptions, namely $(i)$ the strength of the noises should be ``moderate" and $(ii)$ the channels can be used twice. 

It was also observed that when the three-party states are sent through these noisy channels, $\delta_{\cal D}^\rightarrow$ is always monotonically decreasing with the increase of noise when gGHZ states are used as the input, while $\delta_{\cal D}^\rightarrow$ of the resulting states behave non-monotonically with noise parameters when the input states are gW states~\cite{Asu_global_local}. 

% The Authors also discuss about the detection of this two kind of input states, the gW and gGHZ states, which belong to two inequivalent classes of three-qubit pure states \cite{Dur1}. Additionally,  gW and gGHZ states can be detected by computing the dynamics of $\delta_{\cal D}$ with respect to the noise parameters. 
% It has been reported that \cite{Asu_global_local}, dynamics of $\delta_{\cal D}$ is non-monotonic for gW, while it is  monotonic for gGHZ.
% 
% occurs, when the three-qubit  are taken as an input states. 
%violation of Bell or  was extended to the multiparty domain by means of monogamy relation, a  has been defined in 
% This phenomena leads some author to established a relation between discord and the violation of Bell inequality.  
\subsection{ Connection with dense coding } 
In the bipartite domain, efficiency of quantum communication protocols, both classical information transfer via quantum states and quantum state transmission, are related to the QC content of the shared 
quantum state. 
%\textcolor{gray}{Specifically, a bipartite pure entangled state is always better than an unentangled pure state~\cite{Hiroshima_DC,qcom, Buzek_DC,Horodecki_DCadv}
%for quantum dense coding~\cite{Bennett2} and teleportation~\cite{Bennett_teleport}. However such correspondence is missing in a multipartite domain. }
It was found that the pattern of $\delta_{\cal D}$
%In this section, we will discuss how the quantification of  $\delta_{\cal D}$ 
can be used for understanding the capacity of dense coding (DC) involving multiple senders and multiple receivers. 
We will discuss three different DC protocols, namely, Case 1: multiple senders and a single receiver~\cite{distri_DC,Bruss_mDC}, 
%\textcolor{orange}{Case 2:
%The DC protocol is the classical information transfer \cite{Bennett2} between distant observers by using a shared multipartite state. In the single sender to single receiver scenario, for all bipartite pure states a bijective mapping exists between QD and the DC capacity \cite{Bose_DC,Hiroshima_DC,qcom}, which has no extension for mixed bipartite and multipartite pure or mixed states. In the multiparty domain, the capacity of DC investigated for three different scenarios, case 1: multiple senders to single receiver \cite{Bruss_mDC,Liu_DC}, case 2: 
%multiple senders and two receivers~\cite{distri_DC},} 
and Case 2: a single sender and many receivers~\cite{Nepal}.  

The multiparty DC capacity $(C_{multi})$ \cite{distri_DC,Bruss_mDC}, of an $N$-party state  $\rho_{1 2\ldots N}$, shared between $N - 1$  senders and a single receiver, is given by
\begin{eqnarray}\label{Eq:DC_def}
C_{multi}(\rho_{1 2\ldots N})&=&\frac{1}{\log_2 d_{1 2 \ldots N}}\max \{\log_2 d_1 d_2 \ldots d_{N - 1}, \nonumber\\
&& \hspace{-0.58in} \log_2 d_{1} d_2 \ldots d_{N - 1}+ S(\rho_{N}) -  S(\rho_{12\ldots N}) \},
\end{eqnarray}
where $d_{1}, d_2, \ldots, d_{N - 1}$ are dimensions of the systems in possession of the  $N-1$ senders, and where the last party is taken to be the receiver, in possession of a system of dimension $d_N$.
We set $d_{12\ldots N} = d_{1} d_2 \ldots d_{N }$.
Here, one may note that the 
amount of information that can be sent by using a ``classical" protocol  (i.e., without using a shared quantum state) is $\log_2 d_{1}\ldots d_{N - 1}$, and hence the positivity of  the ``coherent information" \cite{Nielsen}, $S(\rho_{N}) - S(\rho_{12\ldots N})$, guarantees the advantage of using the shared quantum state in classical information transmission, and  is known as the DC advantage~\cite{Horodecki_DCadv}. %A multiparty shared quantum state will be useful for  DC, if its DC advantage is positive.
A connection between $C_{multi}$ and 
  $\delta_{\cal D}$ has been made for arbitrary pure states~\cite{TamoghnaDC}, by considering the receiver as a nodal observer, as given in the theorem below. \\
 {\bf Theorem 10}~\cite{TamoghnaDC}:  
{\it Among all multiparty pure states having equal amount of $\delta_{\cal D}$, $C_{multi}$ is bounded below by that of the  gGHZ state}.
%  It 
%was shown that  . 
\\
{\it Proof:}
For an arbitrary pure state $|\psi\rangle_{12\ldots N}$, the discord monogamy score is given by
\begin{eqnarray}
\delta_{\cal D} &=& {\cal D}_{12 \ldots N-1:N} - \sum_i {\cal D}_{i:N} \nonumber \\
&\leq & {\cal D}_{12 \ldots N-1:N} = S(\rho_N).
\end{eqnarray}
Equating $\delta_{\cal D}$ for $|\psi\rangle_{12\ldots N}$  with $\delta_{\cal D}(|gGHZ_N\rangle)$, given in Eq. (\ref{Eq:Dis_gGHZ}), one has 
\begin{eqnarray}
S(\rho_N)  &\geq& h(\alpha), \nonumber \\
 \text{which implies~} C_{multi}(|\psi\rangle_{12\ldots N}) &\geq &  C_{multi}(|gGHZ_N\rangle).\nonumber \\
\end{eqnarray}
Hence the proof. \hfill $\blacksquare$ \\
\begin{figure}[!ht]
 \begin{center}
 \includegraphics[width=1.05\columnwidth,keepaspectratio]{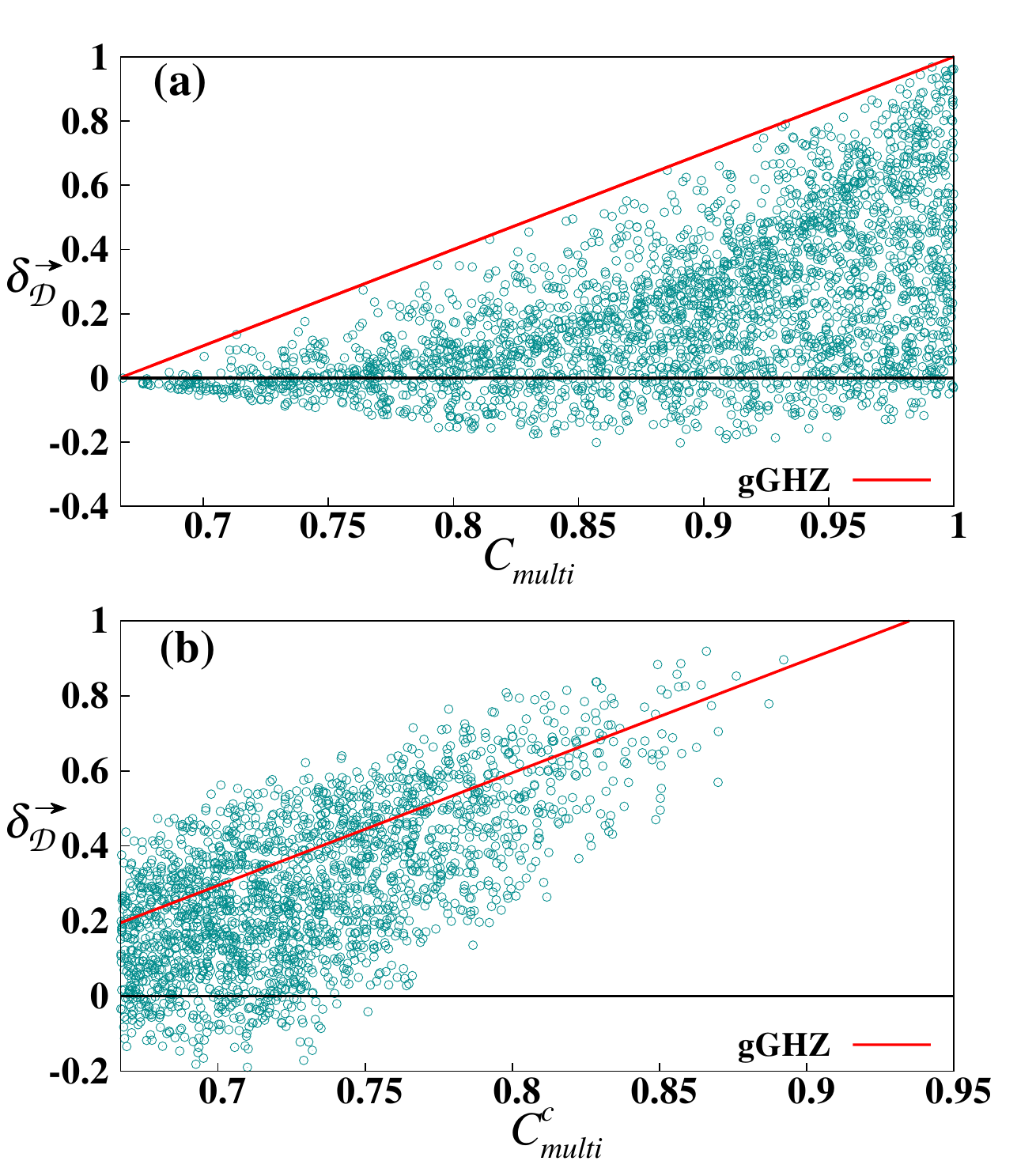} 
 \end{center}
\caption[]{ Capacity of DC vs. the monogamy score of QD ($\delta_{\cal D}$). Blue dots represent Haar uniformly generated three-qubit pure states while the solid line is for the gGHZ state.
The  ordinates and the abscissae are respectively $\delta_{\cal D}$ and $C_{\text{multi}}$.
% is plotted along the ordinate while the $C$ is plotted along the abscissa. 
In panel (a), the noiseless DC capacity of Case 1 is considered, and it is observed that all the points are bounded above by the gGHZ line, as shown in Theorem 10. On the other hand, panel (b) depicts noisy channels\footnote{The covariant noise used in the depiction of the figure is the fully correlated Pauli noise \cite{Pauli_channel}, acting on the senders subsystem in the following way
\begin{equation}
\rho_{ABC} \rightarrow \Lambda^c(\rho_{ABC}) = \sum_i p_i (\sigma_A^i \otimes \sigma_B^i \otimes \mathbb{I}_2) \rho_{ABC} (\sigma_A^i \otimes \sigma_B^i \otimes \mathbb{I}_2),
\end{equation} 
with $p_i$ being the probability or the noise parameters, $i \in \{0,x,y,z\}$ and  $\sigma^0 = \mathbb{I}_2$. The  noise parameters are taken to be $p_0 = p_3 = 0.485$ and $p_1 = p_2 = 0.015$.}  between two senders and a receiver and it is observed in this case that all the points are moving leftward, with respect to their position in panel (a), as is expected due to the interaction of noise in the channels, and at the same time they cross the gGHZ line, thereby violating the constraint set in place in the noiseless case by Theorem 10.  
In both the cases, all the QDs are calculated by performing measurement on the nodal observer, which is the receiver. Both axes in both figures are measured in bits. [Adapted from Ref.~\cite{TamoghnaDC} with permission. Copyright 2014 American Physical Society.]
}
 \label{fig:DC_QD_relation}
\end{figure} 
In other words, %It means that 
to send a fixed amount of classical information, in the scenario of Case 1, the gGHZ state requires the maximal multiparty QC, as quantified by $\delta_{\cal D}$, among all % than any arbitrary 
pure multipartite quantum states. %, as it is 
This is prominently visible  from figure~\ref{fig:DC_QD_relation}(a). 
The DC capacity, given in Eq.~(\ref{Eq:DC_def}), has been derived under the assumption that the encoded  qubits are sent through the noiseless quantum channels to the receivers. %considered to be noiseless and the $\delta_{\cal D}$ is independent of the parties being measured.
%one should note that in the monogamy score no mesurement party 
%If the noise in the DC protocol is considered, then it can act in  two different ways. 

Let us now move to a scenario where the channels between the senders and the receivers are noisy. There are at least two different ways in which the noise can act. 
Firstly, it can affect the shared quantum state at the time of sharing the multipartite state. Secondly, noise can be present in the quantum channel by which 
the senders send their encoded part~\cite{Shadman,Shadman2,Shadman3,Shadman4} to the receiver. 
The first case has already been incorporated in the capacity given in Eq.~(\ref{Eq:DC_def}). %For the second situation of noisy channels, i
The second case is not easy to handle, and a compact form  of DC capacity for an arbitrary noisy channel is not known.  However, for  the covariant noisy channel\footnote{
Covariant noise \cite{Holevo_covariant,Shadman2}, $\Lambda^c$, in a quantum channel is a completely positive trace preserving (CPTP) map which ``commutes" with any one complete set of unitary operators $\{W_i\}$, defined on the same Hilbert space of operators which contains the state, that the channel will carry in the following sense: %and its action is given by
\begin{equation*}
\Lambda^c(W_i \rho W_i^{\dagger}) =W_i  \Lambda^c(\rho) W_i^{\dagger}, ~ \forall i,
\end{equation*}
where $\rho$ is a quantum state passing through the quantum channel.}, 
 the capacity for DC can be obtained\footnote{
The DC capacity \cite{Shadman,Shadman2,Shadman3,Shadman4} of a shared quantum state $\rho_{12\ldots N}$, under the covariant noise $\Lambda^c$ between $N-1$ senders and a single receiver, where noise acts after the encoding, is given by
\begin{eqnarray*}
C_{multi}^c(\rho_{12\ldots N}) = \frac{1}{\log_2 d_{12\ldots N}}\max \{ \log_2 d_{12\ldots N-1}, \\ \log_2 d_{12\ldots N - 1} + S(\rho_{N}) - S(\tilde{\rho}_{12\ldots N})\},
\end{eqnarray*} 
where $\tilde{\rho}_{12\ldots N} = \Lambda^c\big( (U_{12\ldots N-1}^{\min}\otimes\mathbb{I}_N) \rho_{12\ldots N} (U_{12\ldots N-1}^{\min \dagger}\otimes\mathbb{I}_N) \big)$, with $U_{12\ldots N-1}^{\min}$ being a unitary operator in the sender's subsystems. The unitary operator can be global or local depending on the type of encoding, and ``min" in the superscript of $U^{\min}_{12\ldots N-1}$ indicates that the unitary operator  minimizes the von Neumann entropy of $\tilde{\rho}_{12\ldots N}$.
}, in a useful form. Moreover, the capacity can be 
connected with $\delta_{\cal D}$, which establishes a relation between the noisy DC capacity of an arbitrary state with that of the gGHZ state, as has been obtained in the noiseless case in Theorem 10 (see figure~\ref{fig:DC_QD_relation}(b))~\cite{TamoghnaDC}.
%In both the cases, the relation between noisy DC capacity with $\delta_{\cal D}$ has been established and a violation of the above connection has been reported \cite{TamoghnaDC} (see figure \ref{fig:DC_QD_relation}(b)).
%A relation has also been made between 
  %The Authors provides a necessary condition over the vanishing QD, and also obtained that
 % the loss in DC capacity is bounded below by the sum of quantum discord for tripartite quantum states.
%in the senders subsystem of the DC protocol, and with a further assumption that the noise acts when  it was mentioned that the above stated relation got altered \cite{TamoghnaDC}. keepaspectratio

%The capacity of DC for single sender to many receiver has also been investigated and it was defined as the 
Let us now consider a different classical information transmission protocol, viz. that corresponding to Case 2. Suppose that an $N$-party state $\rho_{12\ldots N}$ is shared between a single sender (``1") and $N-1$ receivers, and where the sender individually sends classical information to each receiver~\cite{prabhu_exclusion,Nepal}. In this case, the 
DC advantage (C$_{adv}$)~\cite{Nepal} reads as
\begin{equation}
C_{adv}(\rho_{12\ldots N}) = \max\left[\left\{ S(\rho_i ) - S(\rho_{1i})| i = 2, \ldots, N\right\}, 0\right].
\end{equation}
%reads as $C_{adv}(\rho_{12\ldots, N})$. 
%of a shared $\rho_{12\ldots N}$, with the first party being the sender. 
 The connection between $\delta_{\cal D}$ and C$_{adv}$ has also been analyzed. It was found that for three-qubit pure states, a complementary relation  exists between the DC advantage and $\delta_{\cal D}$~\cite{Nepal}. 
 Moreover, the equality of that relation is attained by an
 %It was shown that among all three qubit pure states with arbitrary but equal amount of $\delta_{\cal D}$, the $C_{adv}$ can not be arbitrarily large, there exists an upper bound on it.
  one-parameter family of states, given by  $|\psi_{\alpha}\rangle = \frac{1}{\sqrt{2 (1 + \alpha^2})}\big(|111\rangle + |000\rangle + \alpha(|101\rangle + |010\rangle) \big)$, with $\alpha \in [0,\frac{1}{2}]$, within the GHZ-class of states,  and which has been called the ``maximally-dense-coding-capable" states.
%   provides the upper bound.
% Thus we can conclude that quantification of  $\delta_{\cal D}$ of multiparty shared quantum state can ensure classical information transfer between multiple senders to single receiver upto a certain amount. Additionally its maximum 
%DC advantage can also be attained when single sender to multiple receiver is present.
 %
 %
%  opposite relation has been reported. from the previous one. The authors have proved that  among all arbitrary three-qubit pure states having equal amount of $\delta_{\cal D}^{\leftarrow}$, the gGHZ state having the maximum C$_{adv}$ \cite{Nepal}. 
%%it was shown that  a completely different relation exists between the  . 

\subsection{Discord monogamy score in cooperative phenomena}
We now briefly discuss the behavior of
% In this section we will briefly discuss the usefulness of the QD monogamy score in a 
 QD monogamy score in cooperative quantum phenomena. %has been investigated.
Such many-body system include one-dimensional spin models and a biological model, mimicking the photosynthesis process. 
 %At first we consider 1D spin chains and calculate its $\delta_{\cal D}$ for any three spin reduced density matrix then we move on to a biological system where we investigate the photosynthesis effect on the $\delta_{\cal D}$ of several sites.
%In the multiparty domain, to calculate the monogamy score of any QC measure $\cal Q$, for mixed state, we need to calculate it for an arbitrary dimension $d_1 \otimes d_2$. There are very few entanglement measures exists which can be calculated for such a scenario. 
%% discussed, it is very hard to compute for a mixed state. 
%For example no definition of concurrence exists  which can be used to calculate the EOF for a mixed state of arbitrary dimension. In such cases discord come to a rescue and it is possible to calculate it for arbitrary dimensional mixed state if we can optimize the measurement on a $d$-dimensional Hilbert space. In the rest of the section we will consider the 
%discord monogamy of multiparty qubit system, where measurement need to be done on a 2-dimensional Hilbert space.

\subsubsection{Many-body systems}
We have already discussed in Sec.~\ref{Sec:manybody} about the effectiveness of QD as  detector of  different phases in many-body system. 
In this subsection, we will discuss whether QD monogamy score can also detect quantum critical points.
%{\color{orange} Although two party QC, in general, can identify quantum phenomena seen in the ground and the thermal state of the spin models, there are certain spin systems for which bipartite QC fails~\cite{Yang3-body05,Qian05}. 
%The potential candidate for studying such system is the multiparty QC measures.} 
The monogamy scores are the one of the few QC measures which quantify QC in a multiparty domain, that are relatively  easy to compute. % for many cases.
%Quantification of QD in a physical system has been investigated for a long time as it is very much helpful to detect the QPT, in the many body system and can be realised in the laboratory. 
%Computation of QD in many body system reveals the amount of QC present in  and can be made in an experiment. 
%Although several investigations \cite{RMP_manybody}, have been made in the direction of quantifying the QC of the ground state as well as the thermal state of the reduced density matrix of pair sites (or spin pairs), 
%behavior of QC in multiparty domain remain unexplored.
%rather than quantifying the QC present in multiple sites (or among many spins) of a many body Hamiltonian. 
The monogamy score of squared QD has been used to analyze the QPT at $\Delta = 1$ of the $XXZ$ model (which is obtained by setting $\gamma = h =  0$ in the Hamiltonian in Eq.~(\ref{eqn: Hamil})~\cite{Liang}. 
%In this regard, one of the widely used physical system is the quantum XYZ model of spin-$\frac{1}{2}$ particles of length $N$, given by the Hamiltonian in Eq.~(\ref{eqn: Hamil}).
%where $\gamma$ is the anisotropy parameter and $J$ is the field strength. The $<ij>$, implies the nearest neighbour terms and $S^{\alpha}_i, ~ \alpha \in \{x,y,z\}$ are the spin angular momentum operators. 
%By fixing $\gamma = h =  0$, the Hamiltonian can be reduces to the well known Heisenberg XXZ model. 
%To detect 
 %the QPT of this model at $\Delta = 1$, monogamy score of squared QD, $\delta_{{\cal D}^2}$ is studied by considering a tripartite pure state obtained from the ground state of this model~\cite{Liang}. 
 %for the zero temperature state of three consequtive spins .
 %The ground state of the finite length $XXZ$ model with periodic boundary condition are obtained by using  the tools of quantum renormalizing group (QRG) \cite{Kargarian1,Kargarian2}.
  %It was also shown that the above result holds even in the finite temperature.
%Discord monogamy score has also been investigated to characterize the multipartite QC of physical systems. 
%The Hamiltonian in Eq.~(\ref{eqn: Hamil}) reduces to the transverse field Ising model for $\gamma = 1$ and $\Delta = 0$. 
In a triangular configuration, by varying $J$ from positive to negative, the transverse Ising model ($\gamma = 1$ and $\Delta = 0$ in Eq.~(\ref{eqn: Hamil})) changes from a frustrated to a  non-frustrated phase at $J = 0$.
The ground state of this model has been simulated in the laboratory in an NMR system~\cite{Koteswara}.
% and multiparty QC has been calculated in terms of $\delta_{\cal D}$.
It was reported  that the value of $\delta_{\cal D}$ is much higher in the non-frustrated regime than the frustrated one. Moreover, the transition point was accompanied with the vanishing of  
$\delta_{\cal D}$.  
%with an adiabatic approximation and along with the process of state tomography, by both theoretically and experimentally.
%  in  that the evaluation of  of the ground state of the triangular Ising model is able to detect the phase transition, and it was experimentally simulated in the laboratory by the  experiment
%an experimental setting 
%The $\delta_{\cal D}^{\leftarrow}$, has been evaluated in the laboratory for three spins frustraed system by considering the   \cite{Koteswara}. and it was shown in the experiment that $\delta_{\cal D}^{\leftarrow}$, can detect the phase transition between the frustrated to non-frustrated system. 

Another  investigation of monogamy of QD has been carried out for   strongly correlated electrons in the bond charge exended 1D Hubbard model \cite{Allegra}.
The ground state of the model possesses three different phases. Varying the system parameters, it was shown that  the discord monogamy score, which is always negative in this case, behaves differently, depending on the phases in which the system lies. 
For example, in one phase where off-diagonal long-range order is present, the ground state violates the monogamy relation maximally. 
% The $\delta_{\cal D}$ of ground state has been calculated and it was shown that for some values of the system parameters of the above models, the ground state forms a permutationaly invariant pure state of zero or doubly occupied sites. For a system size of length $N$, the zero temperature state is given by
% \begin{equation}
% |\psi(N_d,L)\rangle = \binom{N}{N_d}^{-\frac{1}{2}}\sum_P P[|N_d,N-N_d\rangle],
%\end{equation}   
%where $N_d$ is the number of doubly occupied sites, and the summation $P$ is over all  possible permutations of $N_d$ doubly occupied and $N-N_d$, number of zero occupied sites. It is very clear that the state in the above equation reduces 
%to the W state for $N = 3$ and QD  is non-momogamous  for such state. For any $N > 3$ the $\delta_{\cal D}$ of the ground state of the bond charge Hubbard model was shown  to be non-monogamous for all $N_d$ \cite{Allegra}. 
%reduce to the following form for the permutationally invariant state. And 
%it was shown  that there exists some values of $N_d$ for which 

%The ground state property of that model in the permutation invariant state is showing non-monogamous property in the multiparty domain.

\subsubsection{Quantum biological systems}
Recent developments suggest that QC can play an important role in biological processes including the light harvesting protein complexes responsible for photosynthesis, avian magnetoreception, and tunnelling through enzyme-catalysed reactions~\cite{Sarovar,QB_magrec,Lambert_QB, Sai_QDFMO, Titas, Mahdian_QB, Saberi_QDFMO, Jevtic_enzyme}. This, however, is still being debated. 
 In the photosynthesis process, as modeled by the Fenna-Matthews-Olson (FMO) light-harvesting pigment-protein complexes~\cite{Lambert_QB}, it was claimed that quantum coherence measures~\cite{Coherence_measure, Coherence_measure2}, QD, and the Leggett-Garg inequality~\cite{Legget_garg, Legget_Garg_ROPP} can help to understand the energy transfer mechanism~\cite{Engel_FMO,Sai_QDFMO, Wilde_Leggett,Coherence_FMO, Coherence_FMO3,Coherence_FMO2}. 
%{\color{red} Although there is a lot of debates regarding the role of quantum mechanics in this phenomena, 
%%  debatable question whether several organisms took advantage of quantum mechanics to gain a leap over some other competitor organisms but 
%it is quiet clear that in several biological systems quantum mechanics can explain their efficiencies more reliably than the classical models. We concentrate here mainly on photosynthesis process in which Femma-Mathew-Olson (FMO) light harvesting pigment protein complexes~\cite{Lambert_QB} are responsible for the energy transfer from the antenna of the bacteria to the reaction center~\cite{Engel_FMO}.
%It was shown that quantum coherence measures \cite{Coherence_measure, Coherence_measure2}, QD, Legett Garg inequality \cite{Legget_garg} can help to to understand energy transfer in this complex. }

%do some task more efficiently than the widely uses classical form.
% And for that physiological time scale 
%In this section we will discuss about the quantum mechanical effect on the photosynthesis of some of the biological objects \cite{Titas, Mahdian_QB, Giorgi_QB}. 
%In case of  green sulfur bacteria which can sustain even for very low light condition due to their large chlorosome antenna, 
%the QD  between  various sites of the which is responsible for the energy transfer from the antenna of the bacteria to the reaction center   has been calculated.
%Recently, the study of entanglement has been carried out for the  which are the fully connected network and the  complex. 
\begin{figure}[h]
 \begin{center}
 \includegraphics[width=1.\columnwidth,keepaspectratio]{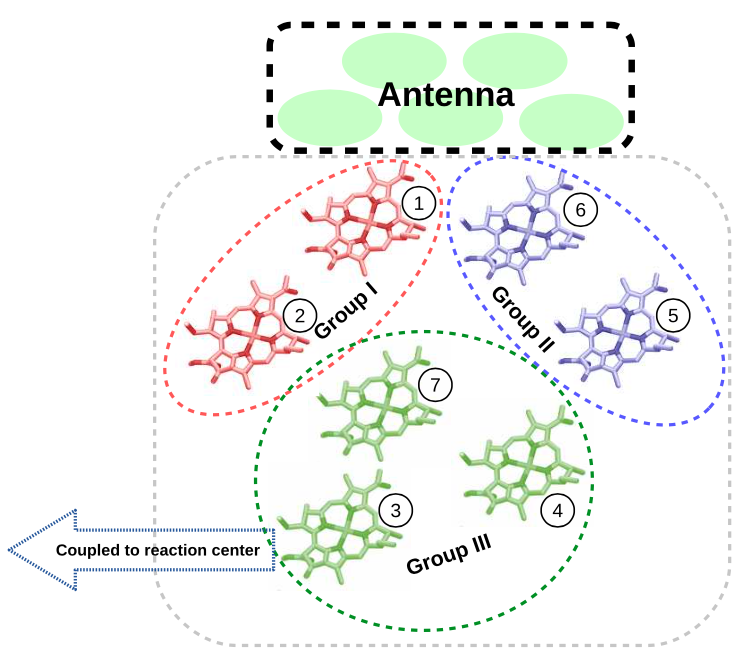} 
 \end{center}
\caption{Schematic structure of the FMO complex and the group classifications of different sites, as inferred from the dynamics of quantum correlations.
[Reprinted from Ref.~\cite{Titas} with permission.]}
\label{Fig:FMO_QB}
\end{figure}
%The FMO complex has been revealed to have seven inequivalent chromophore sites and a sink. 
%It was known  by ultrafast spectroscopy that the quantum coherence for such biological system sustained for a time scale of the order of 
%\cite{Engel_FMO,Cheng_QB}.
 A recent study shows that the time-dynamics of discord monogamy scores between different sites of FMO complexes is useful for indicating 
% and fully connecteed network\footnote{A simplified version of FMO complex.} help us to btained  
 the pathway of energy transfer from the pigment-protein antenna to the reaction center in the photosynthetic FMO complex~\cite{Titas}. 
The evolution was taken to be Markovian,  represented by the Lindblad master equation with  dissipative and dephasing effects. 
The initial state of the evolution is chosen to be an excited pure state at one of the sites closer to the antenna or an equal mixture of them. See figure~\ref{Fig:FMO_QB} for a schematic diagram of the FMO complex.
 
 %and with the initial state being taken to be a pure state of the sites closer to the antenna or equal mixture of them. 
%It was also observed that the time dynamics of entanglement and the QD  shows a complementary relation in the presence of noise for both the model, with the absolute value of QD monogamy being very small with the bipartite QD.
%The presence of QC has also been observed in Ref. \cite{•}
%Here we discuss basically the time dynamics of the discord monogamy score of 
%Moreover,  patterns of discord monogamy score with different nodal observer of the evolved state can  distribute the seven chromophore sites of the complex into three groups, as depicted in figure \ref{Fig:FMO_QB}.
%In the photosynthesis process the energy has been captured by the pigment protein antennas and 
% Quantification of $\delta_{\cal Q}$, helps us to revel the structure of network sites and we are able to divide them in three groups. 
%The maximum of the $\delta_{\cal Q}$ with time by taking several sites as the nodal observer, basically dictates the 
%energy transfer from the antenna to the reaction center for photosynthesis.  
% 
% The FMO complex was studied was for the light harvesting protein complexes of green sulphur bacteria. The dynamics contains of a coherent and a dissipative, dephasing and sink operator term. 

\subsection{Linking with Bell inequality violation} 
Monogamy of QD has also been connected to violation of Bell inequalities for multipartite pure states. 
For two-party system, all pure entangled states violate a Bell inequality~\cite{bell-original, Bell, CHSH, Gisin_Bellv, Popescu_Bellv}. This one-to-one correpondence is however missing in the case of multiparty pure states~\cite{Multiparty_BV, Multiparty_BV2} (cf. \cite{Yu_entnoBV, Liang_entnoBV}). 
For an arbitrary two-qubit state $\rho_{AB}$, which is possibly mixed, the maximal amount of violation of the Bell-CHSH~\cite{bell-original, CHSH} inequality is given by~\cite{Horo_Bell}
%$B^V_{AB}$ is defined as
\begin{equation}
B^V_{AB} = B^V(\rho_{AB}) = \max \{2\sqrt{M(\rho_{AB})} -2,0 \},
\end{equation}
where $M(\rho_{AB})$ is the sum of the two largest eigenvalues of the Hermitian matrix $T^T T$, with $T$ being the classical correlation matrix $T_{ij}$ (see Eq.~(\ref{eq:gen1})). Here the quantity $B^V_{AB}$ is shifted, so that it is vanishing for Bell-inequality-satisfying states and non-vanishing otherwise.

For an $N$-party state $\rho_{1 2 \ldots N}$, one may define a  %multipartite set, one can similarly defined 
monogamy score for violation of Bell inequality (BVM) %of an $N$-party state $\rho_{12\ldots N}$ 
as
\begin{equation}
\delta_{B^V} = B^V(\rho_{1:\text{rest}}) - \sum_{i = 2}^N B^V_{1:i}.
\end{equation}

Let us begin by noting that for any $N$-qubit state, at most one reduced two-qubit state can violate the Bell-CHSH inequality~\cite{Bell_monogamy}. Consider now a subset of $N$-qubit pure states, called ``non-distributive"
 states, for which no two-qubit reduced state violates the Bell-CHSH inequality. It was shown in Ref.~\cite{Kunal}
that among all non-distributive $N$-qubit pure states having the same discord monogamy score, the BVM of a gGHZ state is the least. Restricting to the three-qubit case, but for all pure states, whether distributive or not, it
was numerically found~\cite{Kunal} that the lower bound was provided by the gGHZ state or the ``special GHZ" state, depending on whether $\mathcal{D}^\leftarrow$ or $\mathcal{D}^\rightarrow$ is used to calculate the discord monogamy score. Here, the special GHZ state is given by $|sGHZ_N\rangle = \frac{1}{\sqrt{2}}\big(|00\ldots 0\rangle_N + |11\rangle (\sqrt{\beta}|00\ldots0\rangle + \sqrt{1 - \beta}|11\ldots1\rangle)_{N-2}\big)$. 
A numerically obtained complementarity relation between monogamy of Bell inequality violation and discord monogamy score was also reported in Ref.~\cite{Avijit_Bell} (cf.~\cite{GQD_monogamy2}).

A connection between GQD and a maximum violation of  CHSH inequality has  also been established \cite{GQD_CHSH1, GQD_CHSH2}.
For example, it was shown that in case of Bell-diagonal states for a given GQD, the violation of CHSH inequality \cite{CHSH} is bounded between $4\sqrt{\mathcal{D}_G}$ and $2\sqrt{1 + 2\mathcal{D}_G}$.
 
% by  Batle \emph{et al.}~\cite{GQD_CHSH1} and also by Yao \emph{et al.} \cite{GQD_CHSH2}. It has been reported that there exists a clear tendency of nonlocality to increase, as one considers two-qubits states exhibiting increasing values of GQD.
%Apart from this, study of  geometric measure of QD for a class of quantum states with bound entanglement~\cite{Bound_ent_GQD} has also been carried out. It has been shown that  there exists a non dynamic sudden change in QD accompanying transition from bound entanglement to free entanglement.

 \section{Multiparty measures}
 \label{Disccord_other_multiparty}
% In the previous section, we have discussed discord monogamy score which is a measure of multiparty QC. Additionally, it also quantifies the distribution of QC among its subsystems. 
%Introduction of discord-like measures for quantifying QCs in a bipartite system turns out to extremely important to understand quantumness in a system which is independent from entanglement, thereby giving a different perspective in quantum information theory.
It is natural to extend the notion of QC beyond entanglement to the %such notion in a 
multipartite regime, and this is the main aim in this section. Discord monogamy score, discussed in the preceeding section, is one approach to capture QC in multipartite states. Several other investigations have been carried out in search of multiparty QC beyond entanglement, including Refs.~\cite{relative1,Dissension, rulli-sarandy1,modi-vedral1,CHI-GLOBAL, xu-ggqd-dekho,Jianwei-Xu1,coto-discord,qiang-discord}.
% In recent times,  several investigations have been  carried out in search of the extension of QD and discord-like measures~\cite{relative1,Dissension, rulli-sarandy1,modi-vedral1}. 
% In most of the cases,  the existing correlation measures has been extended to the multiparty domain and some  new quantity have been coined from it.  For example, 
%Prominant examples include relative entropy-based QD and global quantum discord for multiparty quantum states which we will discuss in this section.
%GQD  which are already discussed  in the previous section, can also be extended in the multiparty domain. Now here we will discuss global QD and Dissonance for multiparty quantum state.
 %definition of total and classical correlation measure of classical and nonclassical correlations are revisited and extended for the case of multipartite state. 

\subsection{Global quantum discord}\label{Sec:Global_QD}
%Recently, the generalizations of quantum discord to multiparty systems have been acknowledged in different scenerios \cite{modi-vedral1}. In this direction, 
Rulli and Sarandy \cite{rulli-sarandy1} proposed a multipartite measure for QC, called global QD, by extending symmetric QD for bipartite systems (introduced in Sec.~\ref{bipartite global qd}) to multipartite states.

Let us consider an $N$-party quantum state $\rho_{1 2 \ldots N}$ on which a set of local measurements $\{\Pi^{1}_{j_1} \otimes \ldots \otimes \Pi^{N}_{j_N} \}$ has been performed. The global QD  for $\rho_{1 2 \ldots N}$ is then defined as
\begin{eqnarray}
\label{global1}
\mathcal{D}_{global}(\rho_{1 2\ldots N})=\min_{\{\Pi_k\}}\{S(\rho_{1 2\ldots N}||\phi(\rho_{1 2\ldots N})) \nonumber\\
-\sum_{i=1}^N S(\rho_{i}||\phi_i(\rho_{i}))\}. \nonumber\\
\end{eqnarray}
Here $\phi_i(\rho_{i})=\sum_{j_i} {\Pi^{i}_{j_i} \rho_{i} \Pi^{i}_{j_i}}$ and
$\phi(\rho_{1 2\ldots N})=\sum_k \Pi_k \rho_{1 2\ldots N} \Pi_k$ with $\Pi_k=\Pi^{1}_{j_1} \otimes \ldots \otimes \Pi^{N}_{j_N}$, $k$ being the indices 
$j_1 \ldots j_N$. By definition, the measure is symmetric with respect to exchange of subsystems, and it was shown that it is non-negative for an arbitrary multipartite state. 
%For example, let us consider a tripartite state, the GHZ state with white noise:

\begin{figure}[t]
\includegraphics[width=1.0\columnwidth,keepaspectratio]{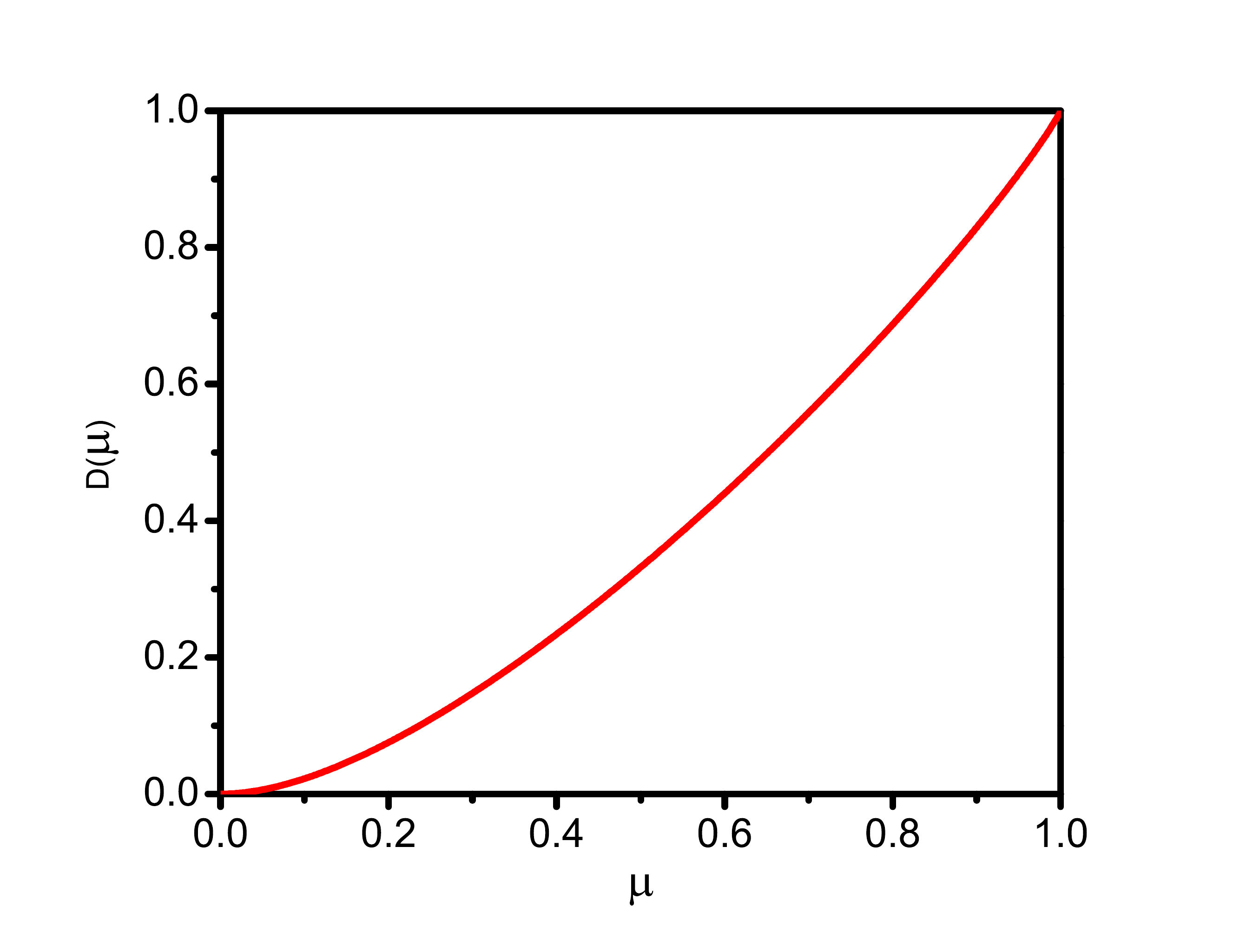}
\caption{Global QD for noisy GHZ states. Tripartite global QD for  the GHZ state, admixed with white noise, is plotted as a function 
of the mixing parameter $\mu$. In this plot, we have considered $\mu = p$ in the text. Note that global QD 
%Here GQD is nonvanishing for $\mu\ne 0$. Moreover, 
 is a monotonic function of $\mu$. Global QD is denoted as $D(\mu)$ in the figure while it is 
 \(\mathcal{D}_{global}\) in the text. All quantitites plotted are dimensionless. [Reprinted from Ref.~\cite{rulli-sarandy1} with permission. Copyright 2011 American Physical Society.]}
\label{f2}
\end{figure}

The optimization in the definition can be performed analytically for the tripartite mixed state
%The global QD has been calculated for some tripartite states and used to detect QPT in the spin chain. Let us study the behavior of global QD for tripartite mixed state, 
given by
\begin{equation}
\label{GHZ-noise}
\rho_{ABC}=\frac{1-p}{8} \mathbb{I}_8+p |GHZ_3\rangle \langle GHZ_3|,
\end{equation}
where $0\leq p \leq 1$ and $|GHZ_3\rangle=\frac{1}{\sqrt{2}} (|000 \rangle+|111 \rangle)_{ABC}$. 
%Here $|0\rangle$ and $|1\rangle$  denote the computational basis. 
%It was shown that for the above state, optimization over measurements can be performed analytically and hence 
The expression of global QD for this state takes the form
\begin{eqnarray} 
\label{state-global}
\mathcal{D}_{global} (\rho_{ABC})=-\frac{1}{4} (1+3 p) \log_2(1+3 p)+\nonumber\\
\frac{1}{8} (1-p) \log_2(1-p)
+\frac{1}{8} (1+7 p) \log_2(1+7 p).
\end{eqnarray}
Note that $\mathcal{D}_{global}=0$ for the maximally mixed state (with $p=0$), while it is maximum for the GHZ state $(p=1)$ (see figure~\ref{f2}).
The results can be generalized to the case of $N$-qubit GHZ states admixed with white noise.
Comparing figures~\ref{werner_plot} and~\ref{f2}, we notice that the trends of QD for the Werner state are similar to that of the global QD for the GHZ state admixed with white noise.

%From Eq.~(\ref{state-global}) it follows, for $\mu=0$, $\rho_{ABC}$ becomes a completely mixed state, resulting $\mathcal{D}_{global} (\rho_{ABC})=0$. Additionally, for $\mu=1$, one can have $\mathcal{D}_{global} (\rho)=1$. Therefore, $\rho$ turns out the $GHZ$ state.  \\
%Jianwei Xu \cite{Jianwei-Xu1} in 2012 proposed an equivalent expression for global QD which is given by
%The GQD of a state $\rho_{A_1 A_2 \dots A_N}$ can be expressed as

Another class of $N$-qubit states, for which it is possible to analytically compute global QD, is given by
%Furthermore, the Authors \cite{Jianwei-Xu1} showed that for a particular $N$-qubit state
 \begin{equation}
\rho_{1 2 \ldots N} =\frac{1}{2^{N}}\left(\mathbb{I}_2^{\otimes N}+\sum_{i=1}^3 c_{i}(\sigma^{i})^{\otimes N}\right),
 \end{equation}
% c_{1}\sigma _{x}^{\otimes N}+c_{2}\sigma _{y}^{\otimes N}+c_{3}\sigma _{z}^{\otimes N}),
%the global QD becomesis evaluated and is given by 
and the corresponding global QD is
 \begin{equation}
\mathcal{D}_{global}(\rho_{1 2 \ldots N} )=f(\rho_{1 2 \ldots N} )-g(\rho_{1 2 \ldots N} ).
 \end{equation}
Here $f(\rho_{1 2 \ldots N} )=-\frac{1+c}{2}\log _{2}\frac{1+c}{2}-\frac{1-c}{2}\log _{2}\frac{1-c}{2}$ with $c=\max\{|c_{1}|,|c_{2}|,|c_{3}|\}$. And $g(\rho_{1 2 \ldots N} )=-\frac{1+d}{2}\log _{2}\frac{1+d}{2}-\frac{1-d}{2}\log _{2}\frac{1-d}{2}$ with $d=\sqrt{c_{1}^{2}+c_{2}^{2}+c_{3}^{2}}$ for odd values of $N$, while for even $N$, $g(\rho_{1 2 \ldots N} )=-1-\sum_{i=1}^{4}\lambda_{i}\log _{2}\lambda _{j}$, where
\begin{eqnarray}
\lambda _{1}=[1+c_{3}+c_{1}+(-1)^{N/2}c_{2}]/4,\nonumber\\
\lambda _{2}=[1+c_{3}-c_{1}-(-1)^{N/2}c_{2}]/4,\nonumber\\
\lambda _{3}=[1-c_{3}+c_{1}-(-1)^{N/2}c_{2}]/4,\nonumber\\
\lambda _{4}=[1-c_{3}-c_{1}+(-1)^{N/2}c_{2}]/4. \nonumber\\
\end{eqnarray}
Here  $c_i$'s, $i=1, 2, 3$
are real numbers constrained by $0 \le \sum_{i=1}^3 c_i^2 \leq 1$, when $N$ is odd, or $0 \leq \lambda_i \leq 1$, $i=1, 2, 3, 4$, when $N$ is even.

Symmetric QD can be written in terms of mutual information as given in Eq.~(\ref{sym-qd3}). Similarly, global QD
% for multipartite state given in Eq.~(\ref{global1}) 
can also equivalently be written as~\cite{Jianwei-Xu1,okrasa1}
\begin{eqnarray}
\label{gqd11}
\mathcal{D}_{global}(\rho_{1 2 \ldots N})=\min_{\phi} [I(\rho_{1 2 \ldots N})-\nonumber\\
I(\phi(\rho_{1 2 \ldots N}))],
\end{eqnarray}
where the $N$-party mutual information is given by $I(\rho_{1 2 \ldots N})=\sum_{i=1}^N S(\rho_{i})-S(\rho_{1 2 \ldots N})$.
% Here $\phi_{1 2 \ldots N}$ denotes a set of local projective measurement performed over all the subsystem $1 2 \dots N$. From 
Like symmetric QD, Eq.~(\ref{gqd11}) can be used to interpret as the minimal loss of mutual information  due to local measurements. 
%An analytical expression for global QD has been obtained for a mixture of $N$-qubit GHZ state with white noise, generalization of Eqs.~(\ref{GHZ-noise}) and~(\ref{state-global}). Moreover, the global QD for
%\textcolor{red}{The authors in \cite{okrasa1} investigated the eq. (\ref{gqd11}).} 
%Additionally, in Ref.~\cite{saguia1}, Saguia \emph{et. al.} proposed a multipartite witness operatorbased on local measurements, leading to a definition of classcial states which turns out to be a set of states having vanishing global QD.
 % based on their disturbance under local measurements which provides a sufficient condition for non classicality that is in agreement with a non vanishing global QD. 

\subsection{Quantum dissonance}
\label{dissonance}
In Secs.~\ref{bipartite global qd} and~\ref{relative_QD}, we have seen that the relative entropy distance can be used to conceptualize measures of QC beyond entanglement in the bipartite case.  Similar definitions are possible in the multipartite case. The options here are far more than in the bipartite case, partially due to the multitude of sets of states that can be identified as sets of ``classically correlated" states.  

An $N$-party quantum state will be called a product state if it is of the form 
\begin{equation}
\Pi_{12\ldots N} = \rho_1 \otimes \rho_2 \otimes \ldots \otimes \rho_N.
\end{equation}
 Clearly, the product state  does not have any kind of correlation (classical or quantum). 
%  while a classical state $\chi_{12\ldots N}$ and separable state $\sigma_{12\ldots N}$ already defined in Eqs. (\ref{Eq:classical_state}) and  (\ref{Eq:seperable-defn}) respectively.
%  while  a 
%classical state $\chi_{1,2,\ldots,N} = \sum_{i_1, i_2, \ldots, i_N} p_{i_1, i_2, \ldots, i_N} |i_1, i_2, \ldots, i_N\rangle\langle i_1, i_2, \ldots, i_N|$, with $\langle i'_j | i_j\rangle = \delta_{i,i'}$, which can be prepared by local operation and classical communication. 
%On the other hand a separable state is  $\sigma_{1,2,\ldots,N} = \sum_{i_1, i_2, \ldots, i_N} p_{i_1, i_2, \ldots, i_N} \sigma_1^{i_1} \otimes \sigma_2^{i_2} \otimes \ldots \otimes \sigma_N^{i_N}$ and the $\sigma_j$ is the quantum state in the jth subsystem. 
One may note that the set of product states are a subset of $N$-party c-c states $\chi_{12\ldots N} = \sum_{i_1 i_2 \ldots i_N} p_{i_1i_2\ldots i_N}|i_1i_2\ldots i_N\rangle \langle i_1i_2\ldots i_N|$ with $\langle i_j|i'_j\rangle = \delta_{ii'}$ for $ j = 1,2, \ldots, N $. 
Compare with Eq.~(\ref{Eq:classical_state}). % for bipartite state.
The  set of separable states 
$\sigma_{12 \ldots N} = \sum_{i_1 i_2 \ldots i_N} p_{i_1i_2\ldots i_N} \rho_{i_1} \otimes \rho_{i_2} \otimes \ldots \otimes \rho_{i_N}$ (see Eq.~(\ref{Eq:seperable-defn}) for  the bipartite case) is a convex set and the 
sets of product as well as $N$-party c-c states are subsets of it.  
 However,  if a  state  cannot be written in the separable form, then 
it can be called a multiparty entangled state. 
%Here, a product state is one which  a classical introduced  to define multipartite version of quantum discord. 
% In 2010 
% the definition of classical correlation and non-classical correlation has been extended for multipartite state  by 
The minimum relative entropy distance of an $N$-party state, $\rho_{1 2 \ldots N}$, from the set of separable states, and from the set of $N$-party c-c states lead to two definitions of QC, and they are respectively a measure of relative entropy of entanglement and a measure of relative entropy-based discord. 
 
 Modi \emph{et al.} \cite{relative1}  came up with another definition of  
  nonclassical correlation, called ``dissonance",
  in the following way.
  Suppose that for an arbitrary state $\rho_{12\ldots N}$, the relative entropy of entanglement, defined above,
  is attained for the separable state $\sigma_{\rho_{12\ldots N}}$. 
  We now find the relative entropy-based quantum discord, as defined above, for $\sigma_{\rho_{12\ldots N}}$, 
  and suppose that this minimum is attained at $\chi_{\sigma_{\rho_{12\ldots N}}}$.
  The last quantity is referred to as the dissonance of $\rho_{12\ldots N}$. 
  %Mathematically, t
  Therefore the dissonance of  $\rho_{12\ldots N}$ is given by
\begin{equation}
Q=\min_{\chi} S(\sigma_{{\rho}_{12\ldots N}}||\chi_{1 2\ldots N}),
\end{equation}
where the minimization is over all \(N\)-party c-c states. 
It was shown in Ref. \cite{relative1} that $Q$ can be rewritten as % for an arbitrary state 
%$\rho_{12\ldots N}$, the closest classical state is $\chi_{\rho}=\sum_{k'}|k'\rangle\langle k'| \rho |k'\rangle\langle k'|$, where $\{|k'\rangle=|k_1 k_2 \ldots k_n\rangle\}$ are the eigen vectors of $\chi_{{\rho}_{1 2\ldots N}}$. With this result, the dissonance reduces to
\begin{equation}
Q=\min_{|k'\rangle} S\left(\sum_{k'}|k'\rangle\langle k'| \sigma_{{\rho}_{12\ldots N}} |k'\rangle\langle k'|\right) - S(\sigma_{{\rho}_{12\ldots N}}),
\end{equation}
where $\{|k'\rangle=|k_1 k_2 \ldots k_n\rangle\}$.
On the other hand, the relative entropy-based QD of $\rho_{12\ldots N}$ is given by
\begin{equation}
{\cal D}_{rel}= \min_{|k'\rangle} S\left(\sum_{k'}|k'\rangle\langle k'| \rho_{12\ldots N} |k''\rangle\langle k'|\right) - S(\rho_{12\ldots N}).
\end{equation} 
%where $\chi_{\rho}$ denotes the closest classical state of $\rho$ instead of $\sigma$ as in dissonance. 
%\textcolor{gray}{In a similar spirit, the total mutual information, $T_{\sigma}$, was introduced which is a minimum relative entropy distance of $\sigma$ from a product state $\Pi_{\sigma}$. Moreover, the relation between total mutual information for $\sigma$, dissonance and classical correlation which is a distance of $\chi_{\sigma}$ from the closest product state $\Pi_{\sigma}$ were introduced.}

Dissonance has been evaluated for certain classes of multipartite pure states \cite{relative1}. 
%$|k\rangle$ and $|k'\rangle$  are the eigenvalues of $\rho$ and $\sigma$ with $|k\rangle = |k_{1}, \ldots, k_{N}\rangle$. In particular, from the above definition, it is prominent that the dissonance plays the role of QD for the separable states. Thus the introduction of a new quantity in the multiparty domain helps us to understand how the total correlation of a quantum state is divided into entanglement, CC and dissonance. The total correlation in \cite{relative1} was taken to be the minimum relative entropy distance from a product state and it can be easily proved that for a bipartite state $\rho_{AB}$, the minimum distance is indeed the mutual information (see Eq. (\ref{Eq:quant1})), where the product state achieves the minimum is $\rho_A \otimes \rho_B$. % with $\rho_{A/B}$ are the reduced density matrix of $\rho_{AB}$ in A and B part.
For  example, if one considers  
$|W_3\rangle=\frac{1}{\sqrt{3}} (|100\rangle +
|010\rangle+|001\rangle)$, the closest separable state %which provides the 
for obtaining the relative entropy of entanglement is  
$\sigma_3 = \frac{8}{27} |000\rangle\langle 000| + \frac{12}{27}|W_3\rangle\langle W_3| + \frac{6}{27}|\bar{W}_3\rangle\langle \bar{W}_3| + \frac{1}{27}|111\rangle\langle 111|$, with $|\bar{W}_3\rangle=\frac{1}{\sqrt{3}} (|011\rangle +|101\rangle+|110\rangle)$ \cite{Wei_rel}. Now, $\chi_{\sigma_3}$
% does not have product eigenbasis and hence it is not a classical state, 
 is obtained by dephasing $\sigma$ in the $x$-basis, %resulting its relative entropy based QD and
 resulting in the dissonance of $|W_3\rangle$ to be  approximately % ${\cal D}_{rel} = 1.58$ and 
 $Q = 0.94$. On the other hand, for the cluster states of 4 qubits~\cite{Briegel_cluster, Nielsen_cluster}, given by
$|C_4\rangle = |0+0+\rangle + |1+1+\rangle + |0-1-\rangle + |1-0-\rangle$, the closest separable state is $\sigma_4 = \frac{1}{4}\big( |0+0+\rangle\langle 0+0+| + |1+1+\rangle\langle 1+1+| + |0-1-\rangle\langle 0-1-| + |1-0-\rangle\langle1-0-|\big)$, which is a four-party c-c state, leading to vanishing dissonance for $|C_4\rangle$.
The possibility of using dissonance as resource in unambiguous quantum state discrimination was considered in Ref. \cite{Dissonance_Roa}. 

\section{Miscellaneous}
\label{sec:miss-QD}
%This section contains the results of QD, which we have not discussed in previous sections. In particular we will briefly discuss the relation established between QD with Benford's law \cite{Benford} and also its role in the uncertainty relation \cite{Robertson}.
A set of disparate aspects of QD are collated in this section.

\subsection{Quantum discord and Benford's law}
%The Benford distribution also calculated for quantum discord
%uantum discord follow Benford law.
Benford's law is  an empirical law of distribution of the first significant digits of data obtained from natural sources or models and from mathematical sequences. The first significant digits of such data may intuitively be expected to be uniformly distributed. 
%distribution of a data sets ranging from  one to nine. An arbitrary set of data is completely unexpected to follow any kind of universal rule.
% or one may think that there distribution over the first significant digits over a huge data point may be uniform.
Benford's law proposes to rule out such intuition.
By analyzing  huge collections of data sets from different origins, Simon Newcomb in 1881~\cite{Newcomb}
and later, Frank Benford in 1938~\cite{Benford}, discovered that the relative frequency distribution of the
%conclude that
% law which
%was later  This law  predicts the probability
%out the expectancy of any randomly generated set of data obtained from of arbitrary
%sources to be completely random, i.e.,  a similarity
 the first significant digits, $d$,  which can take values from $1$ to $9$, is given by
$p_b(d) = \log_{10}(1 + 1/d)$.

%Suppose there are certain data sets for which the first significant digit of the entire set is same, leading to trivial violation of Benford's law.
To bypass certain trivialities, for any data set representing a quantity $q$, one defines the quantity  $q_b = \frac{q-q_{\min}}{q_{\max} - q_{\min}}$,
where $q_{\min}$ and $q_{\max}$ respectively denote the minimum and maximum values of $q$.
%The Benford's law has to be checked with this new rescaled data set.
Data sets ranging from biological phenomena to financial models in economy and astronomical data satisfy the law. However, there exists  data sets which may violate Benford's law,
and it turns out that the violation amount  can be used to detect certain phenomena like the onset of earthquake~\cite{Benford_earthquake}, QPT~\cite{Aditi_Benford_law}, etc.  
The Benford violation parameter (BVP) can be defined as
\begin{equation}\label{Eq:BVP_def}
v_{md} = \sum_{d = 1}^9 \bigg|\frac{p_0(d) - p_b(d)}{p_b(d)}\bigg|,
\end{equation}
where $p_0(d)$ and $p_b(d)$ are respectively the observed relative frequency distribution and that predicted by Benford's law. The BVP can be seen as a distance between the two distributions.
% The distance between two distributions can be measured by different distance metrics, like mean deviation used in Eq. (\ref{Eq:BVP_def}).
Other distance metrics  such as  the Bhattacharya metric~\cite{Bhattacharya_metric} has also been considered ~\cite{ Titas_benford,Utkarsh_Benford}.

In  Ref.~\cite{Titas_benford}, the first significant digit distributions for several entanglement as well as information-theoretic QC measures have been calculated using  Haar-uniformly  generated two-qubit
states of varied ranks.
%  Comparing QC measures, independent than entanglement,
 It was observed that the distribution for QD is closer to the Benford prediction than for  quantum WD.
%with $d$ being the first significant digit of a given data set.
%This law was later rediscovered by Benford in 1938 \cite{Benford}.
%Now as the law predicts the distribution of the leading digits of any data, so it must be normalized to be within $[0:1]$.
%Thus for any data points, expectation value of any observable or for any random variable with outcome $x$, with $x \in [x_{min},x_{max}]$,
%% set which ranging from some number $x_1$ to $x_2$,
% where $x_{min}$ and $x_{max}$ both are different from zero and one, a normalization process is required
%and the new value is $x_b = \frac{x - x_{min}}{x_{max} - x_{min}}$.
%  and  one need to require a set of number ranging from zero to one for calculating its first si
% \cite{Titas_benford}. The Benford violation parameter (BVP) was calculated for all those measures and states, which quantifies the deviation of digit distribution from Benford  by using
%the three distance metrics,
%Where they found that among all measures, quantum discord follow
%Benford law much better than the other measures.
%and with the increasing rank of states the benford violation parameter quantified  almost remain same which is not the case for rest of them. 
%%Where they found that among all other measures the quantum discord follow  distribution much better than the other in a sense that the occurrence of first significant digit
%as $1$ is the highest and for the rest it gradually diminishing.
Moreover, it was also shown that for the  transverse field $XY$ model (Eq.~(\ref{eqn: Hamil}) with $\Delta = 0$), one can detect the  QPT by considering the leading digit distribution of ${\cal D}$ of the nearest-neighbor spin pairs of the zero-temperature state.
Unlike entanglement measures, the observed frequency distribution, $p_0(d)$, for QD changes its pattern  from a decreasing one (decreasing with respect to $d$) to an increasing one in the two phases, namely the antiferromagnetic and the paramagnetic phases. However, the BVP of ${\cal D}$ cannot detect the phase transition present in the $XXZ$ model (Eq.~(\ref{eqn: Hamil}) with $\gamma=0$) and remains unchanged at the critical point.

% as the probability $p_d$, profile alters completely as one move from one phase to another.
%with the increasing rank of the arbitrary two-qubit state  the distribution of all the QC measures tends to obey Benford law with the increasing rank by using different distance metric except the discord which remain almost fix with time.
%Whereas for the XXZ model (see Eq. (\ref{eqn: Hamil}) with $\gamma = 0$) the BVP of ${\cal D}$ can not detect phase transition and remain unchanged at the critical point.% but BVP of EoF changes
%% first significant digits  with $1$ is getting
%the maximum probability.

\subsection{Uncertainty relation}
Uncertainty relations form one of the basic tenets  of quantum mechanics. Entropic Uncertainty relations (EUR)~\cite{EUR_review1,EUR_review2} were
initially formulated by Deutsch~\cite{EUR1} and latter improved by Massen and  Uffink~\cite{EUR5}.
% % has been introduced which includes Shannon entropy of the probability distribution, as a quantification of the ignorance of knowledge. In this regard,
For  an arbitrary pair of observables $X$ and $Y$,  the EUR reads
%entropic constraints were initially formulated  by Deutsch~\cite{EUR1} and in the following years,  improved  by  others~\cite{EUR2, EUR3, EUR4, EUR5,EUR_review2}. The mathematical form of EUR obtained in this way can be   expressed as
\begin{eqnarray}
H(X)+H(Y)\geq -\text{log}_2 c_{X,Y}.
\label{EUR1}
\end{eqnarray}
\begin{figure}[!ht]
% \centering
\includegraphics[width=0.85\columnwidth,keepaspectratio]{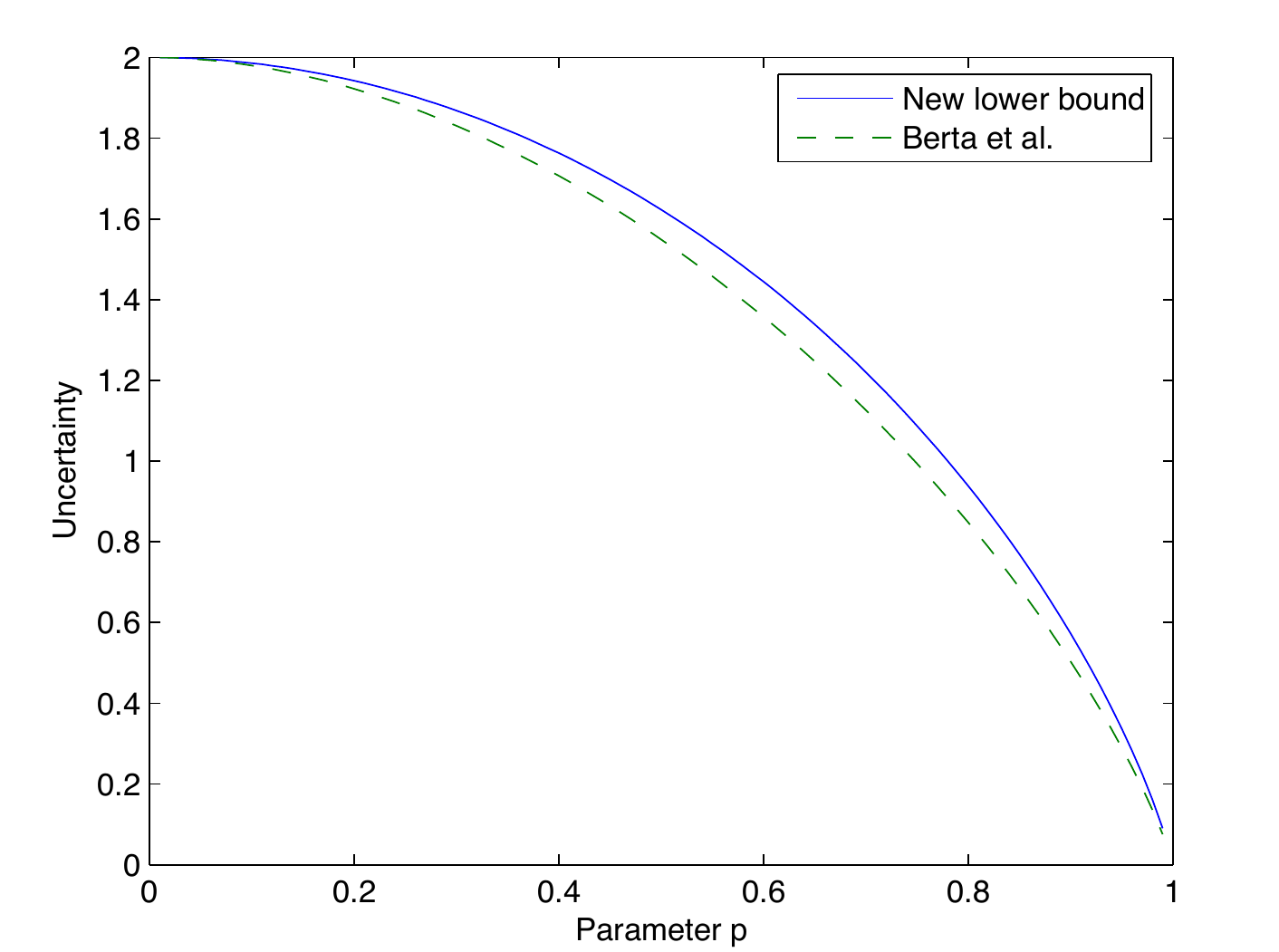}
\caption{ Plot of the right-hand side of the EUR by Berta \emph{et al.}~\cite{EUR_barta} as mentioned  in ~(\ref{Eq:EUR1a}) (dashed green) and  the right-hand side of the improved EUR derived by Pati~\emph{et al.}~\cite{QKD_pati}  as defined in~(\ref{EUR3}) (solid blue line), as functions of the  system parameter $p$, when the system-memory state is in a two-qubit Werner state $\rho_W(p)$. The horizontal axis is dimensionless, while the vertical one is in bits. [Reprinted from Ref.~\cite{QKD_pati} with permission. Copyright 2012 American Physical Society.]}
\label{fig:EUR3}
\end{figure}
%\noindent where, $H(i)$ is the Shannon entropy of the probability distribution  of the outcomes obtained by the measurements of the observable ``$i$" and for non-degenerate observables,  $X$ and $Y$,  $c_{X,Y} $ is defined as
Here $H(X)$ denotes the Shannon entropy of the probability distribution of the outcomes obtained by measuring the observable $X$ on a quantum state $\rho$. Note that we have used the same notation to denote the Shannon entropy of a probability distribution corresponding to a classical random variable $X$ in Sec.~\ref{meas}. $H(Y)$   represents the same for the observable $Y$ on the same quantum state $\rho$. And
$c_{X,Y} =\text{max}_{i,j}~|\langle x_i|y_j\rangle|^2$ with $\{|x_i\rangle\},~ \{|y_i\rangle\}$ being the eigenbases of  $X$ and $Y$  respectively.
The right hand side of (\ref{EUR1}) gives a non-trivial lower bound when $X$ and $Y$ do not share any common eigenstate. This formulation of the uncertainty relation does not
%The above formulation of  EUR is useful in the area of quantum cryptography~\cite{EUR_QKD_first}; although the above relation does not
incorporate
%However, the above relation does  not  incorporate
the  possibility of the system being measured  having
a quantum memory. To overcome such disadvantage,
%Hence, the direct implication of the above EUR relation on the  security issue of a cryptographic scheme was still not prominent. In this regard,
Berta \emph{et al.}~\cite{EUR_barta} provided a reformulation that incorporates
% had reformulated Eq.~(\ref{EUR1}) by  incorporating
a quantum memory.
Consider a scenario in which the system $A$ which performs the measurements and the memory $B$  share a quantum state $\rho_{AB}$. The EUR in this case was proven to be of the form
\begin{equation}
\label{Eq:EUR1a}
S_{X|B} + S_{Y|B} \geq -\log_2 c_{X,Y} + \tilde{S}_{A|B}.
\end{equation}
Here $\tilde{S}_{A|B}=S(\rho_{AB})-S(\rho_B)$. $S_{X|B}$ and $S_{Y|B}$ are defined as follows. After measurement in the basis $\{|x_i\rangle\}$, post-measurement state is given by
\begin{eqnarray}
\rho_{XB}=\sum_i p_i |x_i\rangle \langle x_i|\otimes \rho^i_B,
\end{eqnarray}
where $\rho_B^i=\frac{\text{tr}_A(\langle x_i|\rho_{AB}|x_i\rangle)}{p_i}$, with
$p_i=\text{tr}_{AB} \langle x_i|\rho_{AB}|x_i\rangle.$ Then $S_{X|B}$  is given by $S_{X|B}=S(\rho_{XB})-S(\rho_B)$. $S_{Y|B}$ is similarly defined\footnote{The derivation of ~(\ref{Eq:EUR1a}) also considers POVM measurement carried out by $A$.}.
%Note here that even when the quantum memory is not present, Eq.~(\ref{Eq:EUR1a}) reduces to  $H(X)+H(Y)\geq -2~\text{log}_2 c_{X,Y} + H(A)$ which is stronger than Eq. (\ref{EUR1}).
%Moreover, a non-trivial lower bound is obtained from Eq.~(\ref{Eq:EUR1a}) whenever the system is entangled with the memory except when they are maximally entangled.
%%attached to Bob's system (receiver) which is entangled with Alice (sender of the secret key). This extension allows  Bob to beat the lower bound obtained above, to some extent.
%%In that case,  refinement of the Eq.~(\ref{EUR1}) leads to
%%\begin{eqnarray}
%%H(X)+H(Y)\geq -2~\text{log}_2 c_{X,Y} +S_{A|B},
%%\label{EUR2}
%%\end{eqnarray}
%%where $S_{A|B}$ is the conditional von Neumann entropy defined in Eq.~(\ref{cond_entropy}).
%The  generality  of the above equation  opens  up  a wide  variety of applications of EUR, ranging from
%entanglement witness schemes~\cite{EUR_Ewitness1,EUR_Ewitness2}, to the security analysis of  QKD~\cite{EUR_QKD1,EUR_QKD2,EUR_QKD3,EUR_QKD4}. \
%
%
%

It was shown recently~\cite{QKD_pati} that the lower bound obtained in~(\ref{Eq:EUR1a})
%Soon it was observed that  presence of quantum entanglement between the Alice's subsystem  and Bob's quantum memory, may not exhaust all possibilities which tightends the uncertainty bound. Pati \emph{et al.}~\cite{QKD_pati} showed that the lower bound obtained above
can be improved further.
Precisely, it was shown that
% by considering QD and classical correlation. The Authors proved that  along with the entanglement, discord and the classical correlation between the system and the memory play and important role as
%the non-negative contribution obtained  from the  difference between QD and the classical correlation $J_{A|B}$, yields  following modification of the right-hand side of Eq.~(\ref{EUR2}) which is  given by
\begin{eqnarray}
S_{X|B}+S_{Y|B}&\geq & -\text{log}_2 c_{X,Y}+\tilde{S}_{A|B}
\nonumber \\  && \hspace{-3.5em} + ~\max\{0,\mathcal{D}^{\rightarrow}(\rho_{AB})-J_{B|A}\}.
\label{EUR3}
\end{eqnarray}
%where $J_{A|B}$ is the classical mutual information  defined in Eq.~(\ref{Eq:quant2}) and  all other quantities have the usual meaning.
In particular, it has found that the  sum of the LHS of~(\ref{EUR1}), for the two-qubit Werner state (see Eq.~(\ref{eqn:werner})) coincides  with  the lower bound  in~(\ref{EUR3}), for  $X = \sigma_x$ and $Y = \sigma_z$, clearly showing the improvement achieved in~(\ref{EUR3}) over~(\ref{Eq:EUR1a}), as also depicted
in figure~\ref{fig:EUR3}.
% Interestingly, the above bound in Eq.~(\ref{Eq:EUR1a}) is equally tight, even in the case of higher-dimensional Werner and isotropic states.
Comparative studies of trends of the above two EURs ((\ref{Eq:EUR1a}) and (\ref{EUR3})) for two-qubit states under different local decoherence models have been carried out~\cite{EUR2_EUR3_compare}.
%The Author consider a two-qubit composite system interacting with two independent environments,
% In particular both in Markovian and non-Markovian regimes,
% it has been shown that  the lower bound in Eq.~(\ref{EUR3}) gives better lower bound  on the uncertainty since in most of the cases, $\Delta^{\rightarrow}_{\rho_{AB}} > J_{A|B}$.
%  when discord is bigger than classical correlation, which is often the case in practice.
% \textcolor{orange}{ The Authors also argued that the relation between QC and the uncertainties is subtle, since a certain reduction in the uncertainty may also happen in presence of small quantum correlations, even  without the presence of quantum entanglement.
%Quite interestingly, in Ref.~\cite{EUR_discord_upper_bound} Hu \emph{et al.} consider completely opposite scenario to that has been addressed by Pati \emph{et al.} }
The EUR in~(\ref{EUR3}) turns out to be useful to obtain an upper bound of QD.
%presented above can  constrains the magnitude of QD. It has been shown taht from the uncertainty principle represented in Eq.~(\ref{EUR3}), one can eventually derive certain improved upper bounds for QD. In  general, it is known from the literature that
For a two-qubit state  $\rho_{AB}$, QD is bounded above by the von Neumann entropy of the measured subsystem (see Theorem 3 of Ref.~\cite{Dutta_thesis}) i.e., % a tight upper bound for QD is proven to be
\begin{eqnarray}
\mathcal{D}^{\rightarrow}(\rho_{AB})\leq S(\rho_{A}).
\end{eqnarray}
%where equality holds for the state which allows following decomposition of the Hilbert space of the  local part $B$, given by  $\mathbb{C}^{B}=\mathbb{C}^{B^L}\otimes \mathbb{C}^{B^R}$ such that $\rho_{AB}=(|\psi\rangle \langle \psi|)_{AB^L}\otimes \rho_{B^R}$.
However,  applying ~(\ref{EUR3}), one obtains a
% show  the applications of the EUR eventually leads to  a slightly
stronger upper bound of QD~\cite{EUR_discord_upper_bound}, as given by
% {\color{orange}
%\begin{eqnarray}
%\mathcal{D}^{\rightarrow}(\rho_{AB})\leq \text{min}\{S(\rho_A), S(\rho_A)-S_{A|B}\},
%\label{Dis_EUR_bound1}
%\end{eqnarray}
%The above relation holds for several natural bipartite states for example pseudo pure state in $\mathbb{C}^d\otimes\mathbb{C}^d$~\cite{Psudeo_pure} given by
%\begin{eqnarray}
%\rho_s(r)=\frac{1-r}{d^2-1}\mathbb{I}_d+\frac{r~d^2-1}{d^2-1} |\psi\rangle \langle \psi|,
%\end{eqnarray}
%where $|\psi\rangle=\sum_i u_i |ii\rangle$ and $\sum_i u_i^2=1$.
%Further improvement yields modification of the above equation in the following way}
\begin{eqnarray}
\mathcal{D}^{\rightarrow}(\rho_{AB})\leq \min \{S(\rho_A),I_{AB},\Lambda_T \},
\label{Dis_EUR_bound2}
\end{eqnarray}
where $I_{AB}$ is the total correlation defined in Eq.~(\ref{Eq:quant1}) and $\Lambda_T$ is given by $\Lambda_T=\frac{1}{2}(I_{AB}+S_{X|B}+S_{Y|B}+\text{log}_2 c_{X,Y}-S_{A|B})$.
%  where  $S(i|B)$'s are the  conditional entropies after measurement on subsystem $A$ and definition of the terms $S_{A|B}$ and $c_{X,Y} $ remain same as before. However, one should note that  the bound of QD obtained above is weaker than that obtained using  Eq.~(\ref{Dis_EUR_bound1}), but the later  may be favored
%for their ease of experimental accessibility~\cite{EUR_Ewitness1,EUR_Ewitness2}. Moreover,
%recently, in Ref.~\cite{tighther_pati_EUR}, Ma \emph{et al.} argued, as the term $c_{X,Y}$ defined above,  quantifies the compatibility of the two observables,  thus accounts for specific information concerning the measurements carried out.
Moreover, an observable-independent lower bound of the memory-assisted EUR,  has recently been proposed~\cite{tighther_pati_EUR}, and is given by
\begin{eqnarray}
S_{X|B}+S_{Y|B}\geq 2~S_{A|B}+2~\mathcal{D}^{\rightarrow}(\rho_{AB}).
\label{EUR4}
\end{eqnarray}
It turns out to be less
%Additionally, the Authors claim that the EUR obtained in this way is often leads to a
tight  than that  obtained in~(\ref{EUR3}), as can be illustrated by considering the Werner state and higher-dimensional isotropic states. %As an example, for two-qubit Werner state  defined in Eq.~(\ref{eqn:werner}), one can show that the left hand side of the above EUR, iss

\subsection{Complementarity between quantum discord and purity}
\label{complement-QD}
A complementarity relation between purity and QC measures for multipartite states has recently been obtained~\cite{bera-comp}. The purity of a part of the system is shown to have connection with a quantum characteristic of that part with the remainder of the system. It was found to have potential connection with quantum cryptography~\cite{Gisin-rmp-crypto}. 
%holds both for  multipartite pure as well as for mixed states and also irrespective of
%the dimensions of the subsystems and the total number of parties. The possible scenario can be a QKD protocol involving three parties, namely, Alice$(A)$, Bob$(B)$, and Charu$(C)$ sharing a quantum state $\rho_{ABC}$ where 
%$A$ and $B$ represent the legitimate users of the protocol whose aim is to share identical keys by using entangled state~\cite{Ekert91}, while $C$ represents the potential eavesdropper whose motive is to share identical keys like Alice and Bob. 
%Let us consider an arbitrary tripartite quantum state $\rho_{ABC}$. Then for 
Let us concentrate on a bipartite QC measure, $\cal Q'$ such that ${\cal Q'}(\rho_{AB:C}) \leq S(\rho_{AB})$, for a three-party quantum state $\rho_{ABC}$. The complementarity relation then reads  
\begin{equation}
\label{comple1}
\mathcal{P}(\rho_{AB}) +\mathcal{Q}(\rho_{AB:C}) \leq 1,~\text{when} \log_2 d_1 d_2 \leq \log_2 d_3.
\end{equation}
For  $\log_2 d_1 d_2 > \log_2 d_3$, if we additionally assume $0 \leq {\cal Q'}(\rho_{AB:C}) \leq \log_2 d_3$, we get 
\begin{eqnarray}
\label{comple2}
\mathcal{P}(\rho_{AB}) +\mathcal{Q}(\rho_{AB:C}) \leq
2- \frac{\log_2 d_3}{\log_2 d_1 d_2}.
\end{eqnarray}
Here ${\cal P}(\rho_{AB})=\frac{\log_2 d_1 d_2-S(\rho_{AB})}{\log_2 d_1 d_2}$ quantifies the normalized purity of the system in the $AB$ part and ${\cal Q}(\rho_{AB:C})=\frac{{\cal Q'}(\rho_{AB:C})}{\min\{\log_2 d_1 d_2, \log_2 d_3\}}$  represents the normalized QC measures of the system in the $AB\):\(C$ bipartition. 
\begin{figure}[!ht]
\centering
\includegraphics[width=0.95\columnwidth,keepaspectratio]{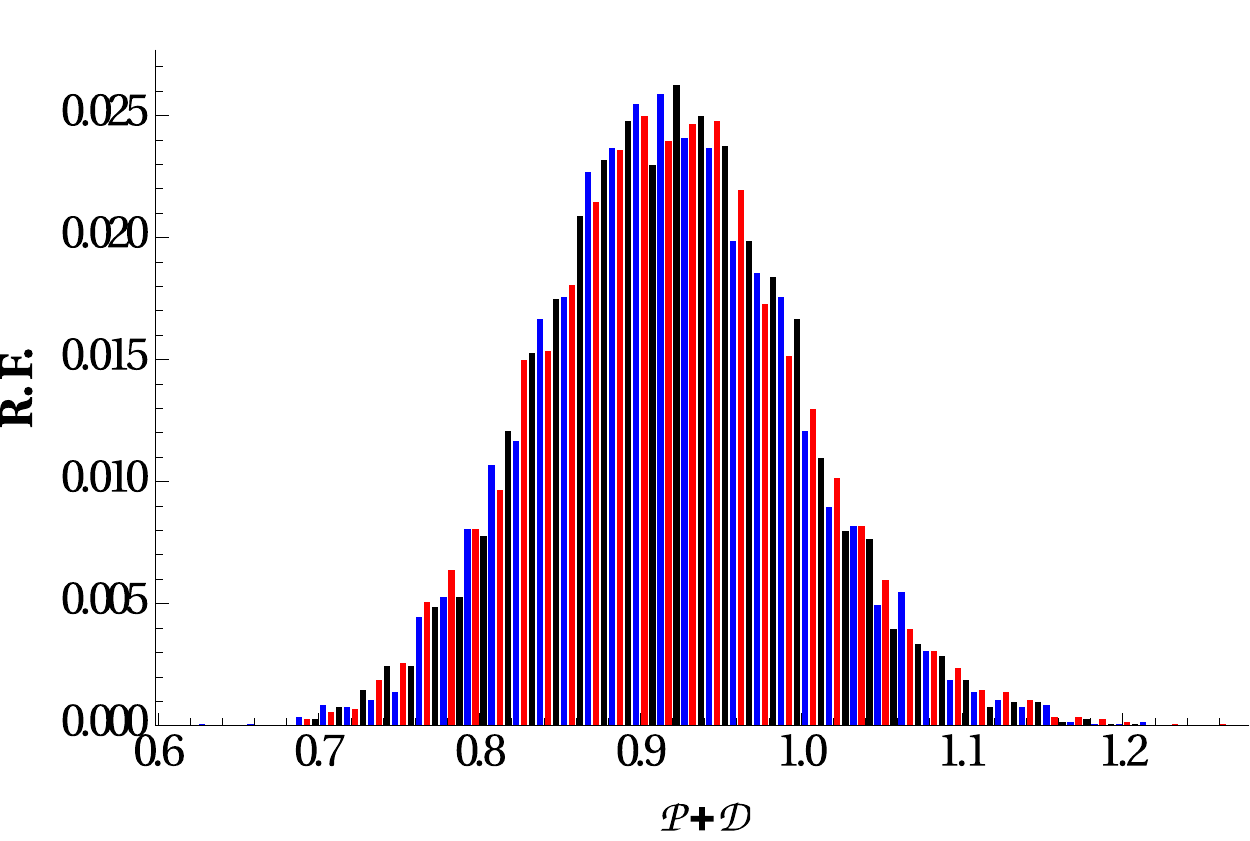}
\caption{The complementarity relation for three-qubit rank-2 states. The histogram exhibits the sum of the normalized purity $\mathcal{P}$ and  normalized QD. See text for the definitions.
 The vertical axis represents the relative frequency (R.F.) of occurrence of a Haar uniformly  generated rank-2   three-qubit state in the
corresponding range of the sum of the two quantities on the horizontal axis. All quantities are dimensionless.
 [Adapted from Ref.~\cite{bera-comp} with permission. Copyright 2016 American Physical Society.]}
\label{comple_plot} 
\end{figure}
Here $d_{1}d_2$ and $d_3$ are the dimensions of the Hilbert spaces of $AB$ and $C$ respectively.
% The above relation holds for entanglement
%measures like negativity, logarithmic negativity as well as  information-theoretic measures like QD and quantum work deficit. It is important to mention here that the relation~(\ref{comple1}) holds for QD only when the measurement is performed on the joint subsystem $AB$. Moreover, 
Calculating QD for $\rho_{ABC}$ in the $AB:C$ bipartition, and by measuring in the $AB$ part, it follows that the QD is bounded by $S(\rho_{AB})$, and consequently the above relations are true for this variety of 	QD.
%For $\rho_{ABC}$ on $\mathbb{C}^{d} \otimes \mathbb{C}^{d} \otimes \mathbb{C}^{d}$, 
When $d_1=d_2=d_3$, a dimension-independent complementarity bound can be obtained:
\begin{equation}
\label{indep-comple}
\mathcal{P}(\rho_{AB}) +\mathcal{Q}(\rho_{AB:C}) \leq \frac{3}{2}.
\end{equation}
%In addition 
%to this, it was shown numerically that several other measures also obey the relation~(\ref{indep-comple}) for three-qubit states generated haar uniformly with ranks ranging from 1 to 4.
The complementary relation has also been numerically checked for measures not satisfying  the entropy bound~\cite{bera-comp}. Figure~\ref{comple_plot} shows a histogram of the relative frequency distribution of the sum of the purity and QD for rank-2 three-qubit states.

\section{Conclusion}
\label{end}
Quantum discord, and measures resembling it, were first conceptualized about a decade and a half earlier. In the  ensuing years, the concepts have been seen from a variety of approaches. 
The notions have also been criticized from several angles. One such is based on the fact that almost all two-party quantum states have a non-zero QD~\cite{Vedral-against-Dutta-Shaji}, the criticism being that if some quantity is present in almost all states, it cannot be useful for any task.  One may however note that almost all pure states are coherent superpositions of a chosen basis of pure states. Such superpositions  
%One of the arguments against this criticism is as follows. How do we know that if some quantity is present in almost all states, it cannot be useful for any task --
%indeed, given a fixed basis of orthonormal states for a given quantum system, almost all pure states are coherent superpositions of the basis elements, and 
%such superpositions 
are known to be useful, for example, for security in quantum cryptography~\cite{BB84, Ekert91, B92, Gisin-rmp-crypto}. 

Among the diverse topics that have been considered within the realm of QD and related measures, there are quite a few which have not been possible to cover within the limited span of this review.
%, include~\cite{demon-braga,zurek-comple,cakmak-discord,Fedorov-discord,Jara-power,dan-discord,xu-
%lazy,kanno-discord,Wang_DC}.
%It was recently shown  that  the loss of DC capacity under decoherence is bounded below by the sum of QD~\cite{Wang_DC}.

\acknowledgements

A.B. acknowledges support of the Department of Science and Technology
(DST), Government of India, through the award of an INSPIRE fellowship. We thank all past and present
members of the quantum information and computation group at the Harish-Chandra Research Institute, and all our collaborators and teachers.

\section{Appendix: Entanglement measures}
%\appendix
%\addcontentsline{toc}{part}{\appendixname}
%\begin{appendices}
%\chapter{Measures of entanglement}\label{Sec:QC_mesures}

%\subsection{}
\label{Sec:QC_mesures}

In this Appendix, we define certain bipartite and multipartite entanglement measures which we have used in different parts of this review. Bipartite  QC measures can be classified into two broad categories. 
One contains those which are  based 
on the entanglement-separability paradigm \cite{Horo_RMP} and the other consists of those which are defined from an information-theoretic perspective. The latter was the main focus of this review. 

%
%\section{}\label{Sec:QC_mesures}

\subsection{Bipartite entanglement measures}\label{Sec:quantum_correlations_def}
%In the entanglement separability regime the entanglement has been quantified from the separability of of any bipartite quantum state. The measures in this category are the  entanglement of formation (EOF),  distillable entanglement (DE), relative entropy of entanglement (RE) and logarithmic negativity (LN). Where the information theoretic measures basically quantifies the difference of two equivalent definition of some measures in classical information theory to the quantum domain. The measures fall in this category are the quantum discord and quantum work deficit. Now we will briefly discuss about them. 
Among bipartite entanglement measures, entanglement of formation (EOF)~\cite{Bennet_EoF},  concurrence~\cite{Wooters}, logarithmic negativity (LN)~\cite{Werner_LN}, and relative entropy of entanglement (RE)~\cite{knight1} are defined below. 
\subsubsection{Entanglement of formation}
\label{Entanglement of formation} 
Entanglement of formation~\cite{Bennet_EoF, Hill_EOF, Wooters,woot-eof} of an arbitrary bipartite quantum state $\rho_{AB}$ is defined as the minimum number of singlet 
states % $\frac{1}{\sqrt{2}}(|01\rangle - |10\rangle )$, 
required to prepare $\rho_{AB}$ by LOCC. For a pure bipartite state $|\psi_{AB}\rangle$, EOF is defined as % EOF is given by
\begin{equation}
{\cal E}(|\psi_{AB}\rangle) = S(\rho_A) ~\text{or}~ S(\rho_B),
\end{equation}
which is the minimal asymptotic rate at which singlets are required to create $|\psi_{AB}\rangle$ by LOCC~\cite{Bennett_monogamy}.
%where $S(\rho) = -\text{tr}(\rho\log_2 \rho)$, is the von-Neumann entropy and $\rho_i$ is the reduced density matrix in the A or B part, defined as $\rho_i = \text{tr}_j (|\psi\rangle\langle\psi|_{AB})$, where $i\neq j \in \{A,B\}$. 
For a mixed  state $\rho_{AB}$, the EOF is defined by using the EOF of pure states and a convex roof extension, so that %it reads as
\begin{equation}\label{Eq:EOF_def}
 {\cal E}(\rho_{AB}) = \min_{\{p_i, |\psi_i\rangle\}} \sum_i p_i S(\text{tr}_B |\psi_i\rangle\langle\psi_i|),
 \end{equation}
 where the minimization is taken over all possible pure state decompositions  of $\rho_{AB} = \sum_i p_i |\psi_i\rangle\langle\psi_i|$. \\
 \subsubsection{ Concurrence} 
 \label{append: concurrence}
 The  EOF for an arbitrary mixed state, discussed in Eq.~(\ref{Eq:EOF_def}),  is not easy to compute due to the minimization involved in the definition. For two-qubit systems, the minimization has been carried out~\cite{Wooters,Hill_EOF,woot-eof}, and is represented by 
\begin{equation}\label{Eq:EoF}
{\cal E}(\rho_{AB}) = h\bigg(\frac{1 + \sqrt{1 - {\cal C}^2}}{2}\bigg),
\end{equation}
where $h(x)$ is given in Eq.~(\ref{binary-shannon}). ${\cal C}$ is the ``concurrence" defined as
\begin{equation}
{\cal C}(\rho_{AB}) = \max\{0, \sqrt{\lambda_1} - \sqrt{\lambda_2} - \sqrt{\lambda_3} - \sqrt{\lambda_4}\},
\end{equation}
where $\{\lambda_i: i = 1, \ldots, 4\}$ are the eigenvalues of the non-Hermitian matrix $\rho_{AB}\tilde{\rho}_{AB}$ in descending order, with $\tilde{\rho}_{AB} = \sigma^y\otimes \sigma^y \rho_{AB}^*\sigma^y\otimes \sigma^y$ being the spin-flipped state. The complex conjugation of $\rho_{AB}$ is in the computational basis. %with a fix basis.  
In case of a pure state  $|\psi_{AB}\rangle$, we have 
 %above takes the form %of concurrence can be extended to $2 \otimes d$ dimensional system and it 
%takes {\color{orange}the form} 
${\cal C} = 2\sqrt{\text{det}\rho_A}$. 
 \\
\subsubsection{Negativity and logarithmic negativity} 
\label{LN}
 The negativity, ${\cal N}$,~\cite{peres1996,Horodecki,Werner_LN,sanpera11,LN_Lee,LN_Plenio} of a bipartite quantum state $\rho_{AB} $  is based on the partial transposition criterion \cite{peres1996,Horodecki}. The partial transposition of $\rho_{AB} = \sum_{i,j,\mu,\nu} p_{ij}^{\mu \nu}| ij \rangle\langle \mu \nu |$ with respect to subsystem \(A\), denoted by $\rho^{T_A}_{AB}$, is defined as $\rho^{T_A}_{AB} = \sum_{i,j,\mu,\nu} p^{\mu \nu}_{ij}|  \mu j\rangle\langle i \nu |$, and similarly with respect to B. 
 % on any one of the subsystem.  
 A partial transposed state of a 
 %A bipartite quantum state will be called 
 separable state % if it can be written as
 $\rho_{AB} = \sum_i p_i \rho^i_A\otimes \rho^i_B$ is always %. Thus for such  separable states $\rho_{AB}^{T_A}$ is always 
 positive semidefinite. The negativity of $\rho_{AB}$ is then defined as
 \begin{equation}\label{Eq:Neg_def}
 {\cal N}(\rho_{AB}) = \frac{ ||\rho_{AB}^{T_A}||_1 - 1}{2},
 \end{equation}
where $||\rho||_1$ is the trace norm, defined as $||\rho||_1 = \text{tr}(\sqrt{\rho^{\dagger}\rho})$. Therefore, the negativity is obtained by adding the moduli of all negative eigenvalues of the partial transposed state. 
%If the state is entangled, then $\rho_{AB}^{T_A}$ must have at least one negative eigenvalue, thereby leading to non-zero % not positive semi definite, and ${\cal N}$. 
On ${\mathbb{C}^2\otimes \mathbb{C}^2}$, a non-zero negativity is a necessary and sufficient condition for entanglement. %  gives non zero value for that.

Logarithmic negativity (LN) is then defined as  
\begin{equation}
{\cal LN}(\rho_{AB}) = \log_2\big(1 + 2{{\cal N}(\rho_{AB})}\big) = \log_2 ||\rho_{AB}||_1.
\end{equation}
%\textcolor{gray}{It is an entanglement monotone, and it is one of the entanglement measures which does not reduce to von Neumann entropy of local density matrices in case of pure states~\cite{Werner_LN}.}
It is interesting to note that $\cal LN$ is additive on tensor products of bipartite states, i.e.\ ${\cal LN} (\rho_{AB} \otimes \sigma_{AB}) = {\cal LN} (\rho_{AB}) + {\cal LN} (\sigma_{AB})$, while $\cal N$ is not.  \\
\subsubsection{Relative entropy of entanglement}\label{Sec:ReoE}
Relative entropy of entanglement~\cite{knight1, VedralRMP02,vedral-plenio1} of an arbitrary bipartite quantum state $\rho_{AB}$ is the minimum relative entropy distance of $\rho_{AB}$ from the set of separable state $\cal S$, and is given by
\begin{equation}
{\cal E}_R(\rho_{AB}) = \min_{\sigma_{AB} \in {\cal S}} S(\rho_{AB} || \sigma_{AB}),
\end{equation} 
where $\sigma_{AB}$ is a bipartite separable state.
%. with $S(\rho || \sigma) = \text{tr}(\rho\log_2\frac{\rho}{\sigma})$ is the von-Neumann relative entropy, a distance measure. 
%A bipartite mixed state $\sigma_{AB}$, will be  called separable, if it can be produced by local operation and classical communication (LOCC) \cite{Horo_RMP}.
It  satisfies many of  the properties required of entanglement measures, and reduces to local von Neumann entropy for pure bipartite states. The asymptotic relative entropy of entanglement is bounded below and above, respectively, by distillable entanglement and entanglement cost. See Ref.~\cite{HHH00} in this regard.
The definition of relative entropy of entanglement can 
be extended to the multiparty domain by considering the minimum distance from a suitable set of multipartite separable states~\cite{Plenio01}. 

\subsection{Multiparty entanglement measures}\label{Sec:Multiparty_QCdef}
%In this section we will talk about some of the definition of multiparty entanglement measure. 
Let us now move on to the multipartite scenario. We have already mentioned that multiparty entangled measures can be defined using the relative entropy distance. 

Here we  use another distance measure, and restrict to only pure states. 
Moreover, we try to identify a quantity to measure genuine multiparty entanglement. 
An $N$-party pure state $|\psi_N\rangle$ is said to be genuinely multiparty 
entangled if it is not a product across any bipartition of the \(N\) parties.
%A pure multipartite state is gunuinely multiparty entangled if it is not product accross any bipartition. We will now define a genuine multiparty entanglement measure, namely generalized geometric measure 
%(GGM)~\cite{GGM, GGM2}. %In the multiparty domain also the entanglement of any multiparty state can be defined by taking the distance from a multiparty separable state and a genuinely multiparty entanglement measure, generalized geometric measure (GGM) have been defined \cite{GGM,GGM2}. 
%in other ways \\
%Where a minimum distance of a multiparty quantum state is taken from a state which is not 
%genuinely multiparty entangled. 
%For an arbitrary $N$-party pure state $|\psi_N\rangle$, t
The generalized geometric measure  (GGM) of $|\psi_N\rangle$ is given by \cite{GGM, GGM2,GGM_anindya} (see also \cite{GM_shimony,GM_Barnum,GM_Wei_Goldbart})
%defined as the minimum distance from a pure state which is not genuinely multiparty entangled. Mathematically, it reads as %, leads to
\begin{equation} \label{Eq:GGM}
{\cal G}(|\psi_N\rangle) = 1 - \max_{|\chi\rangle} |\langle \chi|\psi_N\rangle|^2,
\end{equation} 
where the maximization is over all $N$-party pure states, $|\chi\rangle $, that are not genuinely multiparty entangled. It is a measure of genuine multiparty entanglement.
 %The minimum is taken over the  set NGE which are separable across atleast  one bipartition. 
The distance  measure used here is known as the Fubini-Study metric~\cite{Anandan_fubini}. Eq.~(\ref{Eq:GGM})
reduces to a simplified form, given by
\begin{equation}
{\cal G}(|\psi_N\rangle) = 1 - \max\{ \lambda_{A:B}| A\cup B = \{1,2,\ldots, N\}, A\cap B = \emptyset \},
\end{equation}
where $\lambda_{A:B}$ is the largest eigenvalue of the marginal density matrix $\rho_{A}$ or $\rho_B$ of $|\psi_N\rangle$. This makes it computable in any dimension and for an arbitrary number of parties.  It can also be shown to be non-increasing under LOCC~\cite{GGM}. 

Multiparty entanglement measures can also originate from the concept of monogamy of bipartite QC measures. Examples include the tangle or the monogamy score of squared concurrence,  $\delta_{{\cal C}^2}$ \cite{CKW}, and the squared negativity monogamy score, $\delta_{{\cal N}^2}$~\cite{Ou_Fan}.

\section{Appendix: Classical correlation does not increase under discarding}\label{Sec:CC_notincreasing}
We prove here 
%Here we will give the proof 
that the quantity $J$ is not increasing under discarding  a subsystem. Precisely, for a tripartite state, $\rho_{ABC}$, we wish to show that $J$ follows the relation given by
\begin{equation}
J_{A|BC} \geq J_{A|B}.
\end{equation} 
From the definition of $J$, given in Eq.~(\ref{Eq:quant2}), one has $J_{A|BC} = S(\rho_A) - S_{A|BC}$, with the conditional entropy $S_{A|BC} =  \min_{\{\Pi^{BC}_i\} } \sum_i p_i S(\rho_{A|i}),$ where \(\rho_{A|i} = \mbox{tr}_{BC} (\mathbb{I}^A \otimes \Pi^{BC}_i \rho_{ABC} \mathbb{I}^A \otimes \Pi^{BC}_i)/p_i\), and \(p_i = \mbox{tr} (\mathbb{I}^A \otimes \Pi^{BC}_i \rho_{ABC} \mathbb{I}^A \otimes \Pi^{BC}_i)\). %It was also shown in Ref. \cite{Huang2} that t
The conditional entropy can also be written as~\cite{Huang2}
\begin{equation}\label{Eq:condE_ABC} 
S_{A|BC} = \min_{\{\Pi^{BC}_i\} }\big[S(\rho'_{ABC}) - S(\rho'_{BC})\big],
\end{equation}
where $\rho'_{ABC} = \sum_i \mathbb{I}^A \otimes \Pi^{BC}_i \rho_{ABC} \mathbb{I}^A \otimes \Pi^{BC}_i.$ From the strong subadditivity of von Neumann entropy~\cite{Lieb2}, one gets
\begin{equation}\label{Eq:subsdditivity}
S(\rho'_{ABC}) - S(\rho'_{BC}) \leq S(\rho'_{AB}) - S(\rho'_{B}),
\end{equation}
where $\rho'_{AB} = \text{tr}_C (\rho'_{ABC}) = \sum_k \mathbb{I}^A \otimes \Pi'^{B}_k \rho_{AB} \mathbb{I}^A \otimes \Pi'^{B\dagger}_k$, for some measurements $\{\Pi'^{B}_k\}$  derived from $\{\Pi^{BC}_i \}$. 
%To obtain the $S_{A|BC}$ and $S_{A|B}$, the optimization over the respective measurement operators is required. 
Suppose the optimization 
%over $\Pi'^{B}$, 
in $S_{A|B}$ is achieved
%, and it is
for $\{\tilde{\Pi}^{B}_k\}$. As one can always have an extension of it in the higher-dimensional Hilbert space of $BC$,
% in terms of $`\tilde{\Pi}^{BC}$, 
so from (\ref{Eq:subsdditivity}), one has
\begin{equation}\label{Eq:B4}
S_{A|B} = S(\rho''_{AB}) - S(\rho''_{B}) \geq S(\rho''_{ABC}) - S(\rho''_{BC}) \geq S_{A|BC},
\end{equation} 
where $\rho''_{AB} = \sum_k \mathbb{I}^A \otimes \tilde{\Pi}^{B}_k \rho_{AB} \mathbb{I}^A \otimes \tilde{\Pi}^{B\dagger}_k$ and $\rho''_{ABC} = \sum_i \mathbb{I}^A \otimes \tilde{\Pi}^{BC}_i \rho_{ABC} \mathbb{I}^A \otimes \tilde{\Pi}^{BC\dagger}_i$, and where the equality and the last inequality in (\ref{Eq:B4}) were obtained from Eq.~(\ref{Eq:condE_ABC}). Hence the result.

\onecolumngrid

\vspace{0.5cm}
\PRLsep

%\newpage
\section*{Acronyms}
\begin{leftitemize}
\item[ADC] Amplitude damping channel

\item[AFM] Antiferromagnetic 

\item[BB84] Bennett and Brassard quantum cryptography protocol  in 1984

\item[BD] Bell-diagonal

\item[BF] Bit flip

\item[BPF] Bit-phase flip

\item[BV] Bell inequality violation

\item[BVP] Benford violation parameter

\item[BVM] Bell inequality violation monogamy score

\item[B92] Bennett quantum cryptography scheme in 1992

\item[CC]  Classical correlation

\item[CHSH] Clauser-Horne-Shimony-Holt inequality

\item[CI] Canonical initial

\item[CLOCC] Closed local operations and classical communication

\item[CO] Closed operation

\item[CP] Completely positive

\item[CPTP] Completely positive trace-preserving

\item[CV] Continuous variable

\item[DC] Dense coding

\item[DE] Distillable entanglement 

\item[DM] Dzyaloshinskii-Moriya 

\item[DPC] Depolarizing channel

\item[DQC1] Deterministic quantum computation with single qubit

\item[EOF] Entanglement of formation

\item[EPR] Einstein-Podolsky-Rosen

\item[EUR] Entanglement uncertainty relation

\item[EW] Entanglement witness

\item[E91] Ekert quantum cryptography protocol in 1991

\item[FM] Ferromagnetic

\item[FMO] Fenna-Matthews-Olson

\item[GAD] Generalized amplitude damping

\item[GGM] Generalized geometric measure

\item[gGHZ] Generalized Greenberger-Horne-Zeilinger state

\item[GHZ] Greenberger-Horne-Zeilinger state

\item[GQD] Geometric quantum discord

\item[gW] Generalized W state

\item[JC] Jaynes-Cummings model

\item[LB] Locally broadcastable

\item[LOCC] Local operations and classical communication

\item[LN] Logarithmic negativity

\item[LU] Local unitary

\item[MIN] Measurement-induced nonlocality

\item[NMR] Nuclear magnetic resonance

\item[NPPT] Non-positive partial transpose 

\item[PDC] Phase damping channel

\item[PF] Phase flip 

\item[PM] Paramagnetic

\item[POVM] Positive operator valued measurements 

\item[PPT] Positive partial transpose

\item[PV] von Neumann projective measurement 

\item[QC] Quantum correlation 

\item[QD] Quantum discord

\item[QDP] Quantum dynamical process

\item[QIP] Quantum information processing

\item[QKD] Quantum key distribution

\item[QPT] Quantum phase transition

\item[RE] Relative entropy of entanglement

\item[RSP] Remote state preparation

\item[SCI] Special canonical initial

\item[SLOCC] Stochastic local operation and classical communication 

\item[SPPT] Strong positive partial transpose

\item[SVD] Singular value decomposition

\item[UF] Uniaxial field

\item[WD] Quantum work deficit

\item[1D] One-dimensional
\end{leftitemize}

\twocolumngrid


\begin{thebibliography}{10}
\bibitem{ErwinSchroedinger} Schr{\" o}dinger E 1935 Discussion of Probability Relations between Separated Systems 
\emph{Mathematical Proceedings of the Cambridge Philosophical Society} \textbf{31} 555

\bibitem{Horo_RMP} 
Horodecki R, Horodecki P, Horodecki M and  Horodecki K 2009 Quantum entanglement \emph{Rev. Mod. Phys.} {\bf  81} 865 

\bibitem{Nielsen}
Nielsen M A and Chuang I L 2000 \emph{Quantum Computation and Quantum Information} (Cambridge University Press, Cambridge)

\bibitem{Gisin-rmp-crypto} 
Gisin N, Ribordy G, Tittel W and Zbinden H 2002 Quantum cryptography \emph {Rev. Mod. Phys.} {\bf 74} 145

\bibitem{qcom} Sen(De) A and Sen U 2010 Quantum advantage in communication networks physics \emph{Physics News} {\bf 40} 17

\bibitem{MWildebook}  
Wilde M M 2013 \emph{Quantum Information Theory} (Cambridge University Press, Cambridge)

\bibitem{lab4} Makhlin Y, Sch{\"o}n G and Shnirman A 2001 Quantum-state engineering with Josephson-junction devices \emph{Rev. Mod. Phys.} {\bf 73} 357

\bibitem{lab11}
Leibfried D, Blatt R, Monroe C and Wineland D 2003 Quantum dynamics of single trapped ions \emph{Rev. Mod. Phys.} {\bf 75} 281 

\bibitem{lab31} Bloch I 2005 Exploring quantum matter with ultracold atoms in
optical lattices \emph{J. Phys. B: At. Mol. Opt. Phys.} {\bf 38} S629

\bibitem{lab32} Treutlein P, Steinmetz T, Colombe Y, Lev B, Hommelhoff P, Reichel J, Greiner M, Mandel O, Widera A, Rom T, Bloch I and H{\"a}nsch T W 2006 Quantum information processing in optical lattices and magnetic microtraps \emph{Fortschr. Phys.} {\bf 54} 702

\bibitem{lab12} 
H{\"a}ffnera H, Roosa C F and Blatt R 2008 Quantum computing with trapped ions \emph{Phys. Rep.} {\bf 469} 155

\bibitem{lab2}
Northup T E and Blatt R 2014 Quantum information transfer using photons \emph{Nat. Photonics} {\bf 8} 356

\bibitem{Knill1}
Knill E  and Laflamme R 1998 Power of one bit of quantum information \emph{Phys. Rev. Lett.} { \bf 81} 5672 

\bibitem{gordon-enrico}
Gordon J P 1964 Noise at optical frequencies; information theory in \emph{Proc. Int. School 
Phys. ``Enrico Fermi", Course XXXI}   156 (Academic Press, London)

\bibitem{levitin-inf}
Levitin L B 1969 On the quantum measure of the amount of information in \emph{Proc. VI National Conf. Inf. Theory} Tashkent p. 111

\bibitem{Holevo} Holevo A S 1973 Bounds for the Quantity of Information Transmitted by a Quantum Communication Channel  \emph{Probl. Pereda. Inf.} {\bf  9} 3  [\emph{Probl. Inf. Transm.} {\bf 9} 110]

\bibitem{helstrom-book}
Helstrom C W 1976 \emph{Quantum Detection and Estimation Theory} (Academic Press, New York)

\bibitem{Schumacher-channel}
Schumacher B,  Westmoreland M and Wootters W K 1996
Limitation on the Amount of Accessible Information in a Quantum Channel \emph{Phys. Rev. Lett.} {\bf 76} 3452

\bibitem{yuen-measurement}
Yuen H P 1997 in \emph{Quantum Commun. Comput.
Measurement} 
(Plenum, New York)


\bibitem{Bennett3}
Bennett C H, DiVincenzo D P, Fuchs C A, Mor T, Rains E, Shor P W, Smolin J A and Wootters W K 1999 Quantum nonlocality without entanglement \emph{Phys. Rev. A} {\bf 59} 1070

\bibitem{UPB1} Bennett C H, DiVincenzo D P, Mor T, Shor P W, Smolin J A and Terhal B M 1999 Unextendible Product Bases and Bound Entanglement \emph {Phys. Rev. Lett.} {\bf 82} 5385

\bibitem{UPB2} DiVincenzo D P, Mor T, Shor P W, Smolin J A and Terhal B M 2003 Unextendible Product Bases, Uncompletable Product Bases and Bound Entanglement \emph {Comm. Math. Phys.} {\bf 238} 379

\bibitem{WHSV} Walgate J, Short A J, Hardy L and Vedral V 2000 Local distinguishability of multipartite orthogonal quantum states \emph{Phys. Rev. Lett.} {\bf 85} 4972

\bibitem{Virmani} Virmani S, Sacchi M F, Plenio M B and Markham D 2001 Optimal local discrimination of two multipartite pure states \emph {Phys. Lett. A} {\bf 288} 62 

\bibitem{Chen1} Chen Y-X and Yang D 2001 Optimal conclusive discrimination of two nonorthogonal pure product multipartite states through local operations \emph {Phys. Rev. A} {\bf 64} 064303

\bibitem{Chen2} Chen Y-X and Yang D 2002 Optimally conclusive discrimination of nonorthogonal entangled states by local operations and classical communications \emph {Phys. Rev. A} {\bf 65} 022320



 


\bibitem{Horodecki2003a}
Horodecki M, Sen(De) A,  Sen U and Horodecki K 2003 Local Indistinguishability: More Nonlocality with Less Entanglement \emph{Phys. Rev. Lett.} {\bf 90} 047902






 \bibitem{Sir_maam_new_new}
 Sen(De) A and Sen U  2003   Can there be quantum correlations in a mixture of two separable states? \emph{J. Mod. Opt.}  {\bf 50} 981 
 
 \bibitem{horo_new_new}
Horodecki M, Sen (De) A and  Sen U 2007 Quantification of quantum correlation of ensembles of states
\emph{Phys. Rev. A} {\bf 75} 062329

\bibitem{fuchs_new}
 Fuchs C A and  Peres A 1996 Quantum-state disturbance versus information gain: Uncertainty relations for quantum information  \emph{Phys. Rev. A} {\bf 53} 2038
 
 \bibitem{englart_new}
Englert B G 1996  Fringe Visibility and Which-Way Information: An Inequality
  \emph{Phys. Rev. Lett.} {\bf 77} 2154 

\bibitem{peres_new_new}
Peres A and Wootters W K 1991 Optimal detection of quantum information
 \emph{Phys. Rev. Lett.} {\bf 66} 1119 
 
 \bibitem{popescu_new_new}
Massar S and  Popescu S 1995 Optimal Extraction of Information from Finite Quantum Ensembles
  \emph{Phys. Rev. Lett.} {\bf 74} 1259

\bibitem{rsp_new_new1}
Gisin N and  Popescu S 1999 Spin Flips and Quantum Information for Antiparallel Spins
 \emph{Phys. Rev. Lett.} {\bf 83} 432

\bibitem{rsp_new_new2}
 Bu$\check{z}$ek V, Hillery M and Werner R F 1999  Optimal manipulations with qubits: Universal-NOT gate
 \emph{Phys. Rev. A} {\bf 60} 2626(R)
 
 
 \bibitem{rsp_new_new3}
Ghosh S,  Roy A  and  Sen U 2000 Antiparallel spin does not always contain more information
  \emph{Phys. Rev. A} {\bf 63} 014301 

\bibitem{mani_new}
Mani A, Karimipour V and  Memarzadeh L 2015 Comparison of parallel and antiparallel two-qubit mixed states
\emph{Phys. Rev. A} {\bf 91} 012304

\bibitem{huelga_new}
 Huelga S F, Macchiavello C, Pellizzari T,  Ekert A K, Plenio M B and  Cirac J I 1997 Improvement of Frequency Standards with Quantum Entanglement
 \emph{Phys. Rev. Lett.} {\bf 79} 3865



\bibitem{Acin-Durt}
Ac{\' i}n A, Durt T, Gisin N and Latorre J I 2002 Quantum nonlocality in two three-level systems \emph {Phys. Rev. A} {\bf 65} 052325


  
\bibitem{Horo_asym_ent} Horodecki M, Sen(De) A and Sen U 2003 Rates of asymptotic entanglement transformations for bipartite mixed states: Maximally entangled states are not special \emph {Phys. Rev. A} {\bf 67} 062314  
  

\bibitem{review-qd2}
C{\'e}leri L, Maziero J and Serra R M 2011 
Theoretical and experimental aspects of quantum discord and related measures \emph{Int. J. Quantum Inf.}  {\bf 09} 1837 


\bibitem{Kavan-rmp} Modi K, Brodutch A, Cable H, Paterek T and Vedral V 2012 The classical-quantum boundary for correlations: Discord and related measures \emph {Rev. Mod. Phys.} {\bf 84} 1655

\bibitem{review-qd1}
Zhang J-S and Chen A-X 2012 Review of quantum discord in bipartite and multipartite systems \emph{Quantum Phys. Lett} {\bf 1} 69
 


\bibitem{review-qd3}
Aldoshin S M, Fel'dman E B and Yurishchev M A 2014
Quantum entanglement and quantum discord in magnetoactive materials \emph{Fiz. Nizk. Temp.} {\bf 40} 5
 
\bibitem{review-qd4}
Modi K 2014 A pedagogical overview of quantum discord \emph{Open Syst. Inf. Dyn.} {\bf 21} 1440006


\bibitem{zurek-conditional}
Zurek W H 2000 Einselection and Decoherence from an Information Theory Perspective \emph{Ann. Phys. Lpz.} {\bf 9} 855

\bibitem{Oliver1}
Ollivier H and  Zurek W H 2001 Quantum Discord: A Measure of the Quantumness of Correlations \emph{Phys. Rev. Lett.} {\bf 88} 017901

\bibitem{Henderson1}
Henderson L and Vedral V 2001 Classical, quantum and total correlations \emph{J. Phys. A: Math. Gen.} {\bf 34} 6899


\bibitem{nou-do-igyara} 
Cover T M, Thomas J A 2006 \emph {Elements of Information Theory}  (John Wiley \& Sons, New Jersey)


 


\bibitem{Oppenheim1}
Oppenheim J, Horodecki M, Horodecki P, and Horodecki R 2002 Thermodynamical Approach to Quantifying Quantum Correlations \emph{Phys. Rev. Lett.} {\bf 89} 180402

\bibitem{Sir-mam2} 
Horodecki M, Horodecki K, Horodecki P,  Horodecki R, Oppenheim J, Sen(De) A and Sen U 2003 Local Information as a Resource in Distributed Quantum Systems \emph{Phys. Rev. Lett.} {\bf 90} 100402

 \bibitem{Sir-mam3}
Horodecki M,  Horodecki P,  Horodecki R,  Oppenheim J, Sen(De) A, Sen U and  Synak-Radtke B 2005 Local versus nonlocal information in quantum-information theory: Formalism and phenomena \emph{Phys. Rev. A} {\bf 71} 062307






\bibitem{Devetak1}
Devetak I 2005 Distillation of local purity from quantum states \emph{Phys. Rev. A} {\bf 71} 062303


\bibitem{QDefi1} 
 Rajagopal A K and Rendell R W 2002 Separability and correlations in composite states based on entropy methods \emph{Phys. Rev. A} {\bf 66} 022104 
 
\bibitem{QDefi2} Luo S 2008 Using measurement-induced disturbance to characterize correlations as classical or quantum \emph{Phys. Rev. A} {\bf 77} 022301

\bibitem{MIN} 
 Luo S and Fu S 2011 Measurement-Induced Nonlocality \emph {Phys. Rev. Lett.} {\bf 106} 120401

\bibitem{Bennet_EoF} Bennett C H,  DiVincenzo D P, Smolin J and  Wootters W K 1996 Mixed-state entanglement and quantum error correction \emph{Phys. Rev. A} {\bf 54} 3824

\bibitem{rains_new}
Rains E M 1999 Rigorous treatment of distillable entanglement
\emph{Phys. Rev. A} {\bf  60} 173


\bibitem{shared-anindyada}
Biswas A, Sen(De) A and Sen U 2014 Shared purity of multipartite quantum states
 \emph{Phys. Rev. A} {\bf 89} 032331



\bibitem{bound-ent}
Horodecki M, Horodecki P and Horodecki R 1998 Mixed-State Entanglement and Distillation: Is there a ``Bound" Entanglement in Nature? {\it Phys. Rev. Lett.}  {\bf 80} 5239

\bibitem{concave1} 
Wehrl A 1978 General properties of entropy \emph{Rev. Mod. Phys.} {\bf 50} 221

\bibitem{cerf-negative} Cerf  N J and Adami C 1997 Negative Entropy and Information in Quantum Mechanics \emph{Phys. Rev. Lett.} {\bf  79} 5194



\bibitem{HOW05} Horodecki M, Oppenheim J and Winter A 2005  Partial quantum information {\it Nature} {\bf 436} 673

\bibitem{HOW07} Horodecki M, Oppenheim J and Winter A 2007  Quantum State Merging and Negative Information  {\it Comm. Math. Phys.} {\bf 269} 107 

\bibitem{Li07} Li N, Luo S and Zhang Z 2007 Quantumness of bipartite states in terms of conditional entropies {\it J. Phys. A: Math. Theor.} \textbf{40} 11361  
 
\bibitem{Vedral-against-Dutta-Shaji} 
Daki{\' c} B, Vedral V and Brukner \v{C} 2010 Necessary and sufficient condition for nonzero quantum discord \emph {Phys. Rev. Lett.} {\bf 105} 190502

\bibitem{goRar-katha} 
Groisman B,  Popescu S and  Winter A 2005 Quantum, classical, and total amount of correlations in a quantum state  \emph{Phys. Rev. A} {\bf 72} 032317 

 
 
\bibitem{koashi_winter}  
Koashi M and  Winter A 2004 Monogamy of quantum entanglement and other correlations \emph{Phys. Rev. A} {\bf 69} 022309

\bibitem{szilard}
Szilard L 1929 On the decrease of entropy in a thermodynamic system by the intervention of intelligent beings \emph{Z. Phys.} {\bf 53} 840

\bibitem{landauer1}
Landauer R 1961 Irreversibility and heat generation in the computing process \emph{IBM J. Res. Dev.} {\bf 3} 183

\bibitem{benn-demon1}
Bennett C H 1982 The thermodynamics of computation: A review \emph{Int. J. Theor. Phys.} {\bf 21} 905

\bibitem{benn-demon2}
Bennett C H 1987 Demons, engines and the lecond Law \emph{Sci. Am.} {\bf 255} 108

\bibitem{zurek2003}
Zurek W H 2003 Quantum discord and Maxwell's demons \emph{Phys. Rev. A} {\bf 67} 012320


   
\bibitem{Galve2} 
Galve F, Giorgi G L and Zambrini R 2011 Orthogonal measurements are almost sufficient for quantum discord of two qubits {\it Europhys. Lett.} {\bf 96} 40005 
 
\bibitem{povm-qd1}
Synak-Radtke B and Horodecki M  2004
Classical information deficit and monotonicity on local operations {\it J. Phys. A: Math. Gen.} {\bf 37} 11465  
 
\bibitem{Chen3} 
Chen Q, Zhang C, Yu S, Yi X X and Oh C H 2011 Quantum discord of two-qubit X states  \emph{Phys. Rev. A} {\bf 84} 042313 
 
 \bibitem{lang-caves-povm}
Lang M D, Caves C M and Shaji A 2011 Entropic measures of non-classical correlations \emph{Int. J. Quantum Inf.} {\bf 09} 1553 
 
\bibitem{povm-qd2}
Shi M, Sun C, Jiang F, Yan X and Du J 2012
Optimal measurement for quantum discord of two-qubit states \emph{Phys. Rev. A} {\bf 85} 064104


 
\bibitem{DAriano} 
D'Ariano G M, Presti P L and Perinotti P 2005 Classical randomness in quantum measurements {\it J. Phys. A: Math. Gen.} {\bf 38} 5979 
 
\bibitem{Diff_discord} 
 Liu F, Tian G-J, Qin S-J, Wen Q-Y and Gao F 2015  General bounds for quantum discord and discord distance \emph{ Quantum Inf. Proc.} {\bf  14} 1333

 \bibitem{Dutta_thesis} Datta A 2008 Studies on the role of entanglement in mixed-state
quantum computation \emph{PhD Thesis} 
arXiv:0807.4490 (University of New Mexico)

\bibitem{Wang-monogamydekh}
Xi Z, Lu X-M, Wang X and Li Y 2012
Necessary and sufficient condition for saturating the upper bound of quantum discord \emph{Phys. Rev. A} {\bf 85} 032109

\bibitem{Lieb3}
Araki H  and Lieb E H 1970 Entropy inequalities
\emph{Commun.  Math.  Phys.} {\bf 18} 160

\bibitem{Lieb1}
Lieb E H and Ruskai M B 1973 A Fundamental Property of Quantum-Mechanical Entropy
\emph{Phys. Rev. Lett.} {\bf 30} 434

\bibitem{Lieb2}
Lieb E H and Ruskai M B 1973 Proof of the strong subadditivity of quantum-mechanical entropy \emph{J. Math. Phys.} {\bf 14} 1938

\bibitem{mutual-dekh} 
Luo S, Fu S and Li N 2010 Decorrelating capabilities of operations with application to decoherence \emph{Phys. Rev. A } {\bf 82} 052122 


\bibitem{continuity-discord}
Xi Z, Lu X-M, Wang X and  Li Y  2011 The upper bound and continuity of quantum discord \emph{J. Phys. A: Math. Theor.} {\bf 44} 375301  

\bibitem{luo-li1}
Li N and Luo S 2011 Classical and quantum correlative capacities of quantum systems \emph{Phys. Rev. A } {\bf 84} 042124

\bibitem{Werner} 
Werner R F 1989 Quantum states with Einstein-Podolsky-Rosen correlations admitting a hidden-variable model \emph{Phys. Rev. A} {\bf 40} 4277

\bibitem{Luo} Luo S 2008 Quantum discord for two-qubit systems \emph{Phys. Rev. A} {\bf 77} 042303 


% %%%%%%%Demon discord reference 








 
 
 
%%%%%%%%%%%%%%%%%%%%%% 
% Gausian Quantum discord references %

\bibitem{duan-gauss}
Duan L-M, Giedke G, Cirac J I and  Zoller P  2000 Inseparability Criterion for Continuous Variable Systems \emph{Phys. Rev. Lett.} {\bf 84} 2722

\bibitem{simon-gauss}
Simon R 2000 Peres-Horodecki Separability Criterion for Continuous Variable Systems \emph{Phys. Rev. Lett.} {\bf 84} 2726

\bibitem{Giedke-thesis}
Giedke G 2001 Quantum information and continuous variable systems \emph{PhD Thesis} (Leopold-Franzens-Universit{\"a}t Innsbruck)

\bibitem{cirac1}
Cirac J I and Giedke G 2002 Characterization of Gaussian operations and distillation of Gaussian states \emph{Phys. Rev. A} {\bf 66} 032316

\bibitem{arun-book-continuous}
Braunstein S L and Pati A K 2003 \emph{Quantum Information with Continuous Variables} (Kluwer Academic, Dordrecht)


\bibitem{bowen-gauss}
Bowen W P, Schnabel R, Lam P K and Ralph T C 2004 Experimental characterization of continuous-variable entanglement \emph{Phys. Rev. A} {\bf 69} 012304

%\bibitem{Weedbrook-GAUSS}
% Weedbrook C,  Pirandola C,  Garc{\'i}a-Patr{\'o}n R,  Cerf N J, Ralph T C, Shapiro J H and Lloyd S 2012 Gaussian quantum information \emph{Rev. Mod. Phys.} {\bf 84} 621

\bibitem{Braunstein-gauss}
 Braunstein  S L and Loock P V 2005 Quantum information with continuous variables \emph{Rev. Mod. Phys.} {\bf 77} 513
 
 \bibitem{wenger-gauss}
Wenger J, Ourjoumtsev A, Tualle-Brouri R and Grangier P 2005 Time-resolved homodyne characterization of individual quadrature-entangled pulses \emph{Eur. Phys. J. D} {\bf 32} 391
 
\bibitem{mista1}
Fiur{\' a}\v{s}ek J and Mi\v{s}ta L Jr 2007 Gaussian localizable entanglement \emph{Phys. Rev. A} {\bf 75} 060302 
 
\bibitem{adesso-illu}
Adesso G and Illuminati F 2007 Entanglement in continuous-variable systems: recent advances and current perspectives \emph{J. Phys. A: Math. Theor.} {\bf 40} 7821 
 
\bibitem{pirandola1}
Weedbrook C, Pirandola S, Garc{\'i}a-Patr{\' o}n R, Cerf N J, Ralph T C, Shapiro J H and Lloyd S 2012 Gaussian quantum information \emph {Rev. Mod. Phys.} {\bf 84} 621  
 
 \bibitem{ragy-gauss}
Adesso G, Ragy S and Lee A R 2014 Continuous Variable Quantum Information: Gaussian States and Beyond \emph{Open Syst. Inf. Dyn.} {\bf 21} 1440001

%\bibitem{Thermal_squezed} Knight P L and Allen L 1983 \emph{Concepts of Quantum Optics} (Oxford: Peramon Press)
%
%\bibitem{Thermal_squezed2} Kim M S, de Oliveira F A M and Knight P L 1989 Properties of squeezed number states and squeezed thermal states
%\emph{Phys. Rev. A} {\bf 40} 2494 
%







\bibitem{paris1}
Giorda P and Paris M G A 2010 Gaussian Quantum Discord \emph{Phys. Rev. Lett.} {\bf 105} 020503

\bibitem{adesso1}
Adesso G and Datta A 2010 Quantum versus Classical Correlations in Gaussian States \emph{Phys. Rev. Lett.} {\bf  105} 030501

\bibitem{filip-gauss}
Mi{\v s}ta Jr L, Filip R and 
 Fiur{\' a}{\v s}ek  J 2002 Continuous-variable Werner state: Separability, nonlocality, squeezing, and teleportation \emph{Phys. Rev. A} {\bf 65} 062315

\bibitem{tatham-gauss}
Tatham R, Mi{\v s}ta Jr L, Adesso G and  Korolkova N 2012 Nonclassical correlations in continuous-variable non-Gaussian Werner states \emph{Phys. Rev. A} {\bf 85}  022326


\bibitem{giorda-gaussdekho}
 Giorda P, Allegra M and Paris M G A 2012
Quantum discord for Gaussian states with non-Gaussian measurements \emph{Phys. Rev. A} {\bf 86} 052328

\bibitem{Olivares-gauss}
Olivares S and Paris M G A 2013 The balance of quantum correlations for a class of feasible tripartite continuous variable states \emph{Int. J. Mod. Phys. B} {\bf 27} 1345024

\bibitem{pirandola-gaussdekho}
Pirandola S, Spedalieri G, Braunstein S L, Cerf N J
and  Lloyd S 2014 Optimality of Gaussian Discord \emph{Phys. Rev. Lett.} {\bf 113} 140405

\bibitem{guha-gauss}
Giovannetti V, Guha S, Lloyd S, Maccone L and
Shapiro J H 2004 Minimum output entropy of bosonic channels: A conjecture \emph{Phys. Rev. A} {\bf 70}
 032315

\bibitem{CERF-GAUSS}
Giovannetti V, Garc{\' i}a-Patr{\' o}n R, Cerf N J and Holevo A S 2014 Ultimate classical communication rates of quantum optical channels \emph{Nat. Photonics}
 {\bf 8} 796
 
 \bibitem{mari-gauss}
Mari A, Giovannetti V and Holevo A S 2014  Quantum state majorization at the output of bosonic Gaussian channels \emph{Nat. Commun.} {\bf 5} 3826

\bibitem{madsen-gauss}
 Madsen L S, Berni A, Lassen M and Andersen U L 2012 Experimental Investigation of the Evolution of Gaussian Quantum Discord in an Open System \emph{Phys. Rev. Lett.} {\bf 109} 030402
 
 \bibitem{meda-gauss}
Meda A, Olivares S, Degiovanni I P, Brida G, Genovese M and Paris M G A 2013 Revealing interference by continuous variable discordant states \emph{Opt. Lett.} {\bf 38} 3099
 
 \bibitem{Chille-gauss} 
Chille V, Quinn N, Peuntinger C, Croal C, Mi{\v s}ta Jr L, Marquardt C, Leuchs G and  Korolkova N 2015 Quantum nature of Gaussian discord: Experimental evidence and role of system-environment correlations \emph{Phys. Rev. A} {\bf 91} 050301(R)

\bibitem{qars-gauss}
Qars J E, Daoud M and Laamara R A 2016 Nonclassical correlations in a two-mode optomechanical system \emph{Int. J. Mod. Phys B} {\bf 30} 1650134




%%%%%%%%%%%%%%%%%%%
% Symmetric quantum discord %

\bibitem{rulli-sarandy1}
Rulli C C and Sarandy M S 2011 Global quantum discord in multipartite systems \emph{Phys. Rev. A} {\bf 84} 042109

\bibitem{symmqd3}
Girolami D, Paternostro M and Adesso G 2011 Faithful nonclassicality indicators and extremal quantum correlations in two-qubit states  \emph{J. Phys. A: Math. Theor.} {\bf 44} 352002

\bibitem{Geom_discord1}
Luo S and  Fu S 2010  Geometric measure of quantum discord \emph{Phys. Rev. A} {\bf 82} 034302

\bibitem{okrasa1}
Okrasa M and Walczak Z 2011 Quantum discord and multipartite correlations \emph{Europhys. Lett.} {\bf 96} 60003

\bibitem{Piani_nolocalbroad} 
Piani M, Horodecki P and Horodecki R 2008 No-Local-Broadcasting Theorem for Multipartite Quantum Correlations \emph{Phys. Rev. Lett.} {\bf 100} 090502

\bibitem{symmqd4}
Wu S, Poulsen U V and M{\o}lmer K 2009 Correlations in local measurements on a quantum state, and complementarity as an explanation of nonclassicality
\emph{Phys. Rev. A} {\bf 80} 032319




\bibitem{symmqd5}
Brodutch A and Modi K 2012 Criteria for measures of quantum correlations \emph{Quantum Inf.  Comput.} {\bf 12} 0721

\bibitem{symmqd1}
Maziero J, C{\'e}leri L C and Serra R M 2010 Symmetry aspects of quantum discord arXiv:1004.2082
 
\bibitem{symmqd2}
Auccaise R, Maziero J, C\'eleri L C, Soares-Pinto D O,  deAzevedo E R, Bonagamba T J, Sarthour R S, Oliveira I S and Serra R M 2011 Experimentally Witnessing the Quantumness of Correlations \emph{Phys. Rev. Lett.} {\bf 107} 070501 
 
 \bibitem{symmqd6}
 Mi\v{s}ta Jr L, Tatham R, Girolami D, Korolkova N and  Adesso G 2011 Measurement-induced disturbances and nonclassical correlations of Gaussian states \emph{Phys. Rev. A} {\bf 83} 042325

\bibitem{symmqd8}
Luo S and Zhang Q 2009 Observable Correlations in Two-Qubit States \emph{J. Stat. Phys.} {\bf 136} 165

\bibitem{symmqd7}
Zhong-Xiao  W and Bo-Bo W 2014 Symmetric quantum discord for a two-qubit state \emph{Chin. Phys. B} {\bf 23} 070305




%%%%%%%%%%%%%%%%
% Distance based quantum discord %
\bibitem{distance_based}
Bengtsson I and  Zyczkowski K 2006 Geometry Of Quantum States (Cambridge University Press, Cambridge)



\bibitem{knight1}
Vedral V, Plenio M B, Rippin M A and Knight P L 1997 Quantifying Entanglement \emph{Phys. Rev. Lett.} {\bf 78} 2275

\bibitem{vedral-plenio1}
Vedral V and Plenio M B 1998 Entanglement measures and purification procedures \emph{Phys. Rev. A} {\bf 57} 1619

\bibitem{bravyi1}
 Bravyi S 2003 Entanglement entropy of multipartite pure states \emph{Phys. Rev. A} {\bf 67} 012313

\bibitem{relative1}
Modi K, Paterek T, Son W, Vedral V and  Williamson M 2010 Unified View of Quantum and Classical Correlations \emph{Phys. Rev. Lett.} {\bf 104} 080501

\bibitem{rel-ref2}
Daoud M, Laamara R A and Kaydi W 2014 Unified scheme for correlations using linear relative entropy \emph{Phys. Lett. A} {\bf 378} 3501







%%%%%%%%%%%%%%%%%%%%%%%%%%%%%%%%Geometric Quantum Discord%%%%%%%%%%%%%%%%%%%%%%%%%%%%%%%%%%%%%%%%%%%

\bibitem{GQD_extra5}
Nakano T,  Piani M  and  Adesso G 2013 Negativity of quantumness and its interpretations \emph{Phys. Rev. A} {\bf 88} 012117 

 \bibitem{MIN_GQD}
 Lugiewicz P,  Frydryszak A and  Jakobczyk L 2016 Measurement-induced qudit geometric discord
 arXiv:1607.08753  
 
\bibitem{GQD_new_extra1}
Roga W, Spehner D and Illuminati F 2016 Geometric measures of quantum correlations: characterization, quantification, and comparison by distances and operations
{J. Phys. A: Math. Theor.} {\bf  49}  235301

\bibitem{GQD_extra4}
Jak{\'o}bczyk L, Frydryszak A and  {\L}ugiewicz P 2016 Qutrit geometric discord \emph{Phys. Lett. A} {\bf 380}  1535











\bibitem{Geo_addesso_analytic}
Abramowitz M and Stegun I A 1964 \emph{Handbook of Mathematical  Functions  with  Formulas,  Graphs,  and Mathematical Tables} (Dover, New York)

\bibitem{Geo_QD_decoherence}
Lu X M,  Xi Z J, Sun Z and Wang X 2010 Geometric measure of quantum discord under decoherence \emph{Quantum Inf. Comp.} {\bf 10} 11
 

\bibitem{Shi2}
Shi M, Jiang F and  Du J 2011 Symmetric geometric measure and dynamics of quantum discord arXiv:1107.2958

\bibitem{Geom_discord_lower_bound_adesso1}
Girolami D and Adesso G  2011 Quantum discord for general two-qubit states: Analytical progress \emph{Phys. Rev. A} {\bf 83} 052108

 
\bibitem{Geom_discord_lower_bound_adesso2}
Girolami D and Adesso G 2011  Interplay between computable measures of entanglement and other quantum correlations  \emph{Phys. Rev. A} {\bf 84}  052110

\bibitem{GQD_new_extra3}
Daoud M and Laamara R A 2012 Geometric measure of pairwise quantum discord for superpositions of multipartite generalized coherent states \emph{ Phys.  Lett. A}  {\bf 376}  2361

\bibitem{GQD_bound_new1}
Hassan A S M and Joag P S 2012 Geometric measure of quantum discord and total quantum correlations in an N-partite quantum state \emph{ J. Phys. A: Math. Theor.} {\bf 45} 34

\bibitem{Hassan} Hassan A S M,  Lari B and  Joag P S 2012 Tight lower bound to the geometric measure of quantum discord \emph {Phys. Rev. A} {\bf 85} 024302


\bibitem{Sai} Vinjanampathy S and  Rau A R P 2012  Quantum discord for qubit-qudit systems
\emph{J. Phys. A: Math. Theor.} {\bf 45} 9

\bibitem{luo_2_n}
Luo S and  Fu S 2012 Evaluating the geometric measure of quantum discord
\emph{Theor. Math. Phys.} {\bf 171} 870 


\bibitem{GQD_measurement}
Miranowicz M,  Horodecki P,  Chhajlany R W, Tuziemski J and  Sperling J 2012 Analytical progress on symmetric geometric discord: Measurement-based upper bounds \emph{Phys. Rev. A} {\bf 86} 042123  

\bibitem{Rana1}Rana S and Parashar P 2012 Tight lower bound on geometric discord of bipartite states \emph{Phys. Rev. A} {\bf 85} 024102

\bibitem{GQD_two_sided}
Xu J 2012 Geometric measure of quantum discord over two-sided projective measurements \emph{Phys. Lett.  A} {\bf 376}  320

\bibitem{Geom_discord_lower_bound2}
Girolami D and Adesso G  2012 Observable Measure of Bipartite Quantum Correlations \emph{Phys. Rev. Lett.} {\bf 108} 150403

\bibitem{GQD_tomography}
Jin J S,  Zhang F Y,  Yu  C S and  Song H S 2012  Direct scheme for measuring the geometric quantum discord \emph{J. Phys. A: Math. Theor.} {\bf 45} 11 

\bibitem{GQD_square_norm}
Bellomo B,  Giorgi G L,  Galve F, Franco R L,  Compagno G and  Zambrini R 2012  Unified view of correlations using the square-norm distance  \emph{Phys. Rev. A} {\bf  85}  032104 

\bibitem{wei1}
Wei H R, Ren B C and Deng F G 2013 Geometric measure of quantum discord for a two-parameter class of states in a qubit-qutrit system under various dissipative channels  \emph{ Quantum Inf. Proc.} {\bf 12} 1109

\bibitem{GQD_new_extra2}
Xie C M,  Liu Y M,  Xing H   and  Zhang Z J 2015 Analytic Expression of Geometric Discord in Arbitrary Mixture of any Two Bi-qubit Product Pure States  \emph{Commun. Theor. Phys.} {\bf  63} 439
 
 
\bibitem{volumehoro} 
Horodecki P 1997 Separability criterion and inseparable mixed states with positive partial transposition {\it Phys. Lett. A} {\bf 232} 333

 



\bibitem{Rana2}
Rana S and Parashar P 2013  Geometric discord and Measurement-induced nonlocality for well known bound entangled states \emph{Quantum Inf. Proc.} {\bf 12} 2523



\bibitem{Bound_ent_GQD}
Yan X Q,  Liu G H and  Chee J 2013  Sudden change in quantum discord accompanying the transition from bound to free entanglement \emph{Phys. Rev. A} {\bf  87}  022340; \emph{ibid.} {\bf 88} 039901(E) (2013)




\bibitem{GQD_witness1}
Debarba T,  Maciel T O and Vianna R O 2012 Witnessed entanglement and the geometric measure of quantum discord  \emph{Phys. Rev. A} {\bf 86} 024302 



\bibitem{GQD_witness2}
Rana S and Parashar P 2013 Comment on ``Witnessed entanglement and the geometric measure of quantum discord"  \emph{Phys. Rev. A} {\bf 87} 016301

\bibitem{Rana_reply}
Debarba T,  Maciel T O and Vianna R O 2013 Reply to ``Comment on `Witnessed entanglement and the geometric measure of quantum discord'"  \emph{Phys. Rev. A} {\bf 87} 046301 

 


\bibitem{geomet_discord_problem1}
Piani M 2012 Problem with geometric discord \emph {Phys. Rev. A} {\bf 86} 034101

\bibitem{geomet_discord_problem2}
Tufarelli T, Girolami D, Vasile R, Bose S and Adesso G 2012 Quantum resources for hybrid communication via qubit-oscillator states \emph{Phys. Rev. A} {\bf 86} 052326
 
\bibitem{geomet_discord_problem3}
Paula F M, de Oliveira T R and Sarandy M S 2013 Geometric quantum discord through the Schatten 1-norm \emph{Phys. Rev. A} {\bf 87} 064101

\bibitem{geomet_discord_problem5}
 Hu X,  Fan H, Zhou D L, and Liu W M 2013 Quantum correlating power of local quantum channels
\emph{Phys. Rev. A} {\bf 87} 032340 

\bibitem{geomet_discord_problem4}
Adesso G and Girolami D 2011 Gaussian geometric discord \emph{Int. J. Quantum Inf.} {\bf 09} 1773
 
 \bibitem{dqc1_exp3}
Passante G,  Moussa O and  Laflamme R 2012 Measuring geometric quantum discord using one bit of quantum information \emph{Phys. Rev. A} {\bf 85} 032325 



 
\bibitem{other_geometric_discord1}
Spehner D and Orszag M 2013 Geometric quantum discord with Bures distance \emph{ New J. Phys.} {\bf 15} 103001
  
 \bibitem{gqd-bures-abar}
  Spehner D and Orszag M 2014 Geometric quantum discord with Bures distance: the qubit case \emph{J. Phys. A: Math. Theor.} {\bf 47} 035302  
  
\bibitem{other_geometric_discord2}
Tufarelli1 T, MacLean T,  Girolami D,  Vasile R and  Adesso G 2013  The geometric approach to quantum correlations: computability versus reliability \emph{J. Phys. A: Math. Theor.} {\bf 46} 275308 
 
\bibitem{other_geometric_discord3}
Chang L and Luo S 2013 Remedying the local ancilla problem with geometric discord \emph{Phys. Rev. A} {\bf 87} 062303


\bibitem{Bai2}
 Bai Y K,  Zhang T T,  Wang L T and  Wang Z D 2014 Correlation evolution and monogamy of two geometric quantum discords in multipartite systems \emph{Eur. Phys. J. D} {\bf  68} 274 
 
 \bibitem{Yu_Xstate}  Yu Tand and Eberly J H 2007 Evolution from Entanglement to Decoherence of Bipartite Mixed ``X" States \emph{Quantum Inf. Comp.} {\bf 7}  459

\bibitem{x-dekho}
Rau A R P 2009 Algebraic characterization of X-states in quantum information  \emph{J. Phys. A: Math. Theor.} {\bf 42} 412002

 
 \bibitem{Ciccarello} Ciccarello F,  Tufarelli T and  Giovannetti V 2014 Toward computability of trace distance discord   \emph{New J. Phys.} {\bf 16} 013038 
 
 
\bibitem{Hilbert_trace_norm_comparision}
Jakobczyk L 2014 Spontaneous emission and quantum discord: comparison of Hilbert-Schmidt and trace distance discord
 \emph{ Phys. Lett. A} {\bf  378} 3248 

\bibitem{tsallis-dis}
Jurkowski J 2012 Discord Derived from Tsallis Entropy  arXiv:1206.0241

\bibitem{renyi-dis3}
Hou X -W, Huang Z -P and Chen S 2014 Quantum discord through the generalized entropy in bipartite quantum states \emph{Eur. Phys. J. D} {\bf 68} 1

\bibitem{comment-gen1}
Bellomo G, Plastino A, Majtey A P and Plastino  A R 2014 Comment on ``Quantum discord through the generalized entropy in bipartite quantum states" \emph{Eur. Phys. J. D} {\bf 68} 337


 \bibitem{renyi-dis1}
Seshadreesan K P, Berta M and Wilde M M 2015 R{\`e}nyi squashed entanglement, discord, and
relative entropy differences \emph{J. Phys. A: Math. Theor.} {\bf 48} 395303

\bibitem{Seshadreesan-SQUASH}
Seshadreesan K P and Wilde M M 2015 Fidelity of recovery, squashed entanglement, and measurement recoverability \emph{Phys. Rev. A} {\bf 92} 042321

\bibitem{sand-renyi-dis2}
Misra A, Biswas A, Pati A K, Sen(De) A and Sen U 2015 Quantum correlation with sandwiched relative entropies: Advantageous as order parameter in quantum phase transitions \emph {Phys. Rev. E} {\bf 91} 052125

\bibitem{rel-ref3}
Mahdian M and Arjmandi M B 2016 Comparison of quantum discord and relative entropy in some bipartite quantum systems \emph{Quantum Inf. Proc.} {\bf 15} 1569


\bibitem{renyi-main1}
R{\'e}nyi A 1961 On  measures  of  information  and  entropy \emph{Proc. Symp.  Math. Stat.  Prob.}  {\bf 1} 547 (University of California Press, Berkeley)

\bibitem{tsallis-main1}
Tsallis C 1988 Possible generalization of Boltzmann-Gibbs statistics \emph{J. Stat. Phys.} {\bf 52} 479


\bibitem{renyi-main2}
Horodecki R, Horodecki P and Horodecki M 1996 Quantum $\alpha$-entropy inequalities: independent condition for local realism? \emph{Phys. Lett. A} {\bf 210} 377

\bibitem{renyi-main3}
Horodecki R and Horodecki M 1996 Information-theoretic aspects of inseparability of mixed states \emph{Phys. Rev. A} {\bf 54} 1838

\bibitem{tsallis-main2}
Tsallis S, Mendes R S and Plastino A R 1998 The role of constraints within generalized nonextensive statistics \emph{Physica A} {\bf 261} 534
 

\bibitem{san-main2}
M\"uller-Lennert M, Dupuis F, Szehr O, Fehr S and Tomamichel M 2013 On quantum R\'enyi entropies: A new generalization and some properties  \emph{J. Math. Phys.} {\bf 54} 122203

\bibitem{san-main}
Wilde M M, Winter A and Yang D 2014 Strong Converse for the Classical Capacity of Entanglement-Breaking and Hadamard Channels via a Sandwiched R\'enyi Relative Entropy \emph{Commun. Math. Phys.} {\bf 331} 593






\bibitem{herbut-QD}
Herbut F 2005 Mutual Information of Bipartite States and Quantum Discord in Terms of Coherence Information
\emph{Int. J Quantum Inf.} {\bf 3} 691 


\bibitem{saitoh-dis}
SaiToh A, Rahimi R and Nakahara M 2008 Nonclassical correlation in a multipartite quantum system: Two measures and evaluation \emph{Phys. Rev. A} {\bf 77} 052101

\bibitem{yin-other}
Yin X, Xi Z, Lu X-M, Sun Z and Wang X 2011
Geometric measure of quantum discord for superpositions of Dicke states 
\emph{J. Phys. B: At. Mol. Opt. Phys.} {\bf 44} 245502

\bibitem{xu-Generalizations}
Xu J 2011 Generalizations of quantum discord 
\emph{J. Phys. A: Math. Theor.} {\bf 44} 445310


\bibitem{local-dis}
Xi Z, Fan H and Li Y 2012 One-way unlocalizable quantum discord \emph{Phys. Rev. A} {\bf 85} 052102 

\bibitem{Tufarelli-other}
Tufarelli T, MacLean T, Girolami D,  Vasile R and Adesso G 2013 The geometric approach to quantum correlations: computability versus reliability \emph{J. Phys. A: Math. Theor.} {\bf 46} 275308

\bibitem{doukas-discord} 
Doukas J, Brown E G, Dragan A and Mann R B 2013 Entanglement and discord: Accelerated observations of local and global modes \emph{Phys. Rev. A} {\bf 87} 012306


\bibitem{li-other}
Li B, Chen L and Fan H 2014 Non-zero total correlation means non-zero quantum correlation \emph{Phys. Lett. A} {\bf 378} 1249

\bibitem{uttam-other}
 Singh U and Pati A K 2014 Quantum discord with weak measurements \emph{Ann. Phys.} {\bf 343} 141

\bibitem{roga-response}
Roga W, Giampaolo S M and Illuminati F 2014 Discord of response \emph{J. Phys. A: Math. Theor.} {\bf 47} 365301

\bibitem{oneway-dis}
Liu S-Y, Zhang Y-R, Yang W-L and Fan H 2015 Multipartite distribution property of one way discord
 beyond measurement \emph{Ann. Phys.} {\bf 354} 157

\bibitem{beggi-otherqc}
Beggi A, Buscemi F and Bordone P 2015 Analytical expression of genuine tripartite quantum discord for symmetrical X-states 
\emph{Quantum Inf. Proc} {\bf 14} 573


\bibitem{marian-other}
 Marian P and Marian T A 2015 Hellinger distance as a measure of Gaussian discord \emph{J. Phys. A: Math. Theor} {\bf 48} 115301


\bibitem{Sanders-classic}
Gheorghiu V, de Oliveira M C and Sanders B C 2015 Nonzero Classical Discord \emph{Phys. Rev. Lett.} {\bf 115} 030403


\bibitem{seth-demon}
 Lloyd S 1997 Quantum-mechanical Maxwell's demon \emph{Phys. Rev. A} {\bf 56} 3374 
 
 
 \bibitem{laws-horo}
Horodecki M, Oppenheim J and Horodecki R 2002 Are the Laws of Entanglement Theory Thermodynamical? \emph{Phys. Rev. Lett.} {\bf 89} 240403

\bibitem{PT-Oppenheim}
Oppenheim J, Horodecki M and Horodecki R 2003 Are There Phase Transitions in Information Space? \emph{Phys. Rev. Lett.} {\bf 90} 010404

\bibitem{Reversible-horodecki}
Horodecki M, Horodecki P and Oppenheim J 2003 Reversible transformations from pure to mixed states and the unique measure of information
\emph{Phys. Rev. A} {\bf 67} 062104

\bibitem{horoall-mutual} 
Oppenheim J, Horodecki K, Horodecki M, Horodecki P and Horodecki R 2003 Mutually exclusive aspects of information carried by physical systems: Complementarity between local and nonlocal information \emph{Phys. Rev. A} {\bf 68} 022307


\bibitem{PrabhuErgo1} Prabhu R, Sen(De) A and Sen U 2013 Ergodicity from Nonergodicity in Quantum Correlations of Low-dimensional Spin {\it Euro Phys. Lett.} {\bf 102} 30001

\bibitem{lian-1wdeficit}
Lian-He S, Zheng-Jun X and Yong-Ming L 2012 Remark on the One-Way Quantum Deficit for General Two-Qubit States \emph{Commun. Theor. Phys.} {\bf 59} 285

\bibitem{one-wayqd1}
Wang Y-K, Ma T, Li B and Wang Z-X 2013
One-Way Information Deficit and Geometry for a Class of Two-Qubit States \emph{Commun. Theor. Phys.} {\bf 59} 540



\bibitem{law2} Zemansky M W 1968 \emph{Heat and Thermodynamics}  (McGraw-Hill Book Company, New York) 

\bibitem{maxwell}
Maxwell J C 1902 \emph{Theory  of  Heat} (Longmans, Green, and Co., London)

\bibitem{demon-book}
Leff H S and Rex A F 1990 \emph{Maxwell's Demon: Entropy, Information, Computing} (Princeton University Press, Princeton)

\bibitem{demon-dekhre1}
Bub J 2001 Maxwell's demon and the thermodynamics of computation \emph{Stud. Hist.  Phil. Mod. Phys.} {\bf 32} 569

\bibitem{demon-dekhre2}
Norton J D 2013 All Shook Up: Fluctuations, Maxwell’s Demon and the Thermodynamics of Computation \emph{Entropy} {\bf 15} 4432

\bibitem{alicki-information}
Alicki R, Horodecki M, Horodecki P and Horodecki R 2004 Thermodynamics of Quantum Information Systems — Hamiltonian Description \emph{ Open Syst. Inf. Dyn.} {\bf 11} 205

\bibitem{peng-demon}
Peng P-Y and Duan C-K 2010 A Maxwell Demon Model Connecting Information and Thermodynamics \emph{Chin. Phys. Lett.} {\bf 33} 080501

\bibitem{demon-Brodutch}
Brodutch A and Terno D R 2010 Quantum discord, local operations, and Maxwell's demons \emph{Phys. Rev. A} {\bf 81} 062103

\bibitem{hosoya-demon1}
Hosoya A, Maruyama K and Shikano Y 2011 Maxwell's demon and data compression \emph{Phys. Rev. E} {\bf 84} 061117

\bibitem{hosoya-demon2}
Hosoya A, Maruyama K and Shikano Y 2015 Operational derivation of Boltzmann distribution with Maxwell’s demon model \emph{Sci. Rep.} {\bf 5} 17011





\bibitem{usha-raja}
Usha Devi A R and Rajagopal A K 2008 Generalized information theoretic measure to discern the quantumness of correlations \emph{Phys. Rev. Lett.} {\bf 100} 140502



\bibitem{EPR_paper} Einstein A, Podolsky B and Rosen N 1935 Can Quantum-Mechanical Description of Physical Reality Be Considered Complete? \emph{Phys. Rev.} {\bf 47} 777

\bibitem{CHSH}  Clauser J F,  Horne M A, Shimony A and  Holt R A 1969 Proposed Experiment to Test Local Hidden-Variable Theories
\emph{Phys. Rev. Lett.} {\bf 23} 880 



%%%%%%%%%%%%%%%%
% Computability


\bibitem{Bell} Bell J S 1964 \emph{Speakable and Unspeakable in Quantum Mechanics} (Cambridge University Press, Cambridge)

\bibitem{Davies78} Davies E B 1978 Information and quantum measurement {\it IEEE. Inf. Theory} {\bf IT-24} 596



\bibitem{Hamieh} Hamieh S, Kobes R and Zaraket H 2004 Positive-operator-valued measure optimization of classical correlations \emph{Phys. Rev. A} {\bf 70} 052325

\bibitem{Sasaki99} Sasaki S, Barnett S M, Jozsa R, Osaki M and Hirota O 1999 Accessible information and optimal strategies for real symmetrical quantum sources
\emph{Phys. Rev. A} {\bf 59} 3325


\bibitem{Huang} 
Huang Y 2014 Computing quantum discord is NP-complete \emph{New J. Phys.} {\bf 16} 033027 

\bibitem{Rossignoli11} Rossignoli R, Canosa N and Ciliberti L 2011 Quantum correlations and least disturbing local measurements {\it Phys. Rev. A} {\bf 84} 052329 

\bibitem{Shi}  Shi M, Yang W,  Jiang F and  Du J 2011 Quantum discord of two-qubit rank-2 states  \emph{J. Phys. A: Math. Theor.} {\bf 44} 415304 

\bibitem{Ali} Ali M,  Rau A R P and Alber G 2010 Quantum discord for two-qubit X states \emph{Phys. Rev. A} {\bf 81} 042105; \emph{ibid.} {\bf 82} 069902(E) (2010)

\bibitem{Huang2}  Huang Y 2013 Quantum discord for two-qubit X states: Analytical formula with very small worst-case error \emph{Phys. Rev. A} {\bf  88} 014302

\bibitem{Fanchini2} Fanchini F F, Werlang T, Brasil C A, Arruda L G E and Caldeira A O 2010 Non-Markovian dynamics of quantum discord {\it Phys. Rev. A} {\bf 81} 052107

\bibitem{Celeri}  Celeri L C, Landulfo A G S, Serra R M and Matsas G E A 2010 Sudden change in quantum and classical correlations and the Unruh effect {\it Phys. Rev. A} {\bf 81} 062130 

\bibitem{Auyuanet} Auyuanet A and Davidovich L 2010 Quantum correlations as precursors of entanglement {\it Phys. Rev. A} {\bf 82} 032112 

\bibitem{LuMa} Lu X, Ma J, Xi Z and Wang X 2011 Optimal measurements to access classical correlations of two-qubit states  \emph{Phys. Rev. A} {\bf 83} 012327

\bibitem{Galve} Galve F, Giorgi G L and Zambrini R 2011 Maximally discordant mixed states of two qubits {\it Phys. Rev. A} {\bf 83} 012102; \emph{ibid.} {\bf 83}  069905(E) (2011) 

\bibitem{LiWang} Li B, Wang Z X and  Fei S M 2011 Quantum discord and geometry for a class of two-qubit states {\it Phys. Rev. A} {\bf 83} 022321

\bibitem{Lang} Lang M D and Caves C M 2010 Quantum Discord and the Geometry of Bell-Diagonal States {\it Phys. Rev. Lett.}  {\bf 105} 150501

\bibitem{YuZhang} Yu S, Zhang C, Chen Q and Oh C H 2011 Tight bounds for the quantum discord arXiv:1102.1301
 
\bibitem{Namkung}  Namkung M,  Chang J,  Shin J and  Kwon Y 2015 Revisiting Quantum Discord for Two-Qubit X States: The Error Bound to an Analytical Formula \emph{Int. J. Theor. Phys.} {\bf 54} 3340 



\bibitem{Titas_freeze} Chanda T, Pal A K, Biswas A, Sen(De) A and  Sen U 2015 Freezing of quantum correlations under local decoherence \emph{Phys. Rev. A} {\bf 91} 062119

\bibitem{Computational_complexity} Chanda T,  Das T,  Sadhukhan D, Pal A K, Sen(De) A and  Sen U 2015 Reducing computational complexity of quantum correlations \emph{Phys. Rev. A} {\bf 92} 062301



\bibitem{peres1996} Peres A 1996 Separability Criterion for Density Matrices
{\it Phys. Rev. Lett.}  {\bf 77} 1413
 
 \bibitem{Horodecki}  
Horodecki M, Horodecki P and  Horodecki R  1996 Separability of mixed states: necessary and sufficient conditions \emph{Phys. Lett. A} {\bf 223} 1

\bibitem{GQD_many_body4}
Rossignoli R,  Matera J M and  Canosa N 2012   Measurements, quantum discord, and parity in spin-1 systems \emph{Phys. Rev. A} {\bf 86} 022104 

\bibitem{Horodecki_BE} Horodecki P, Horodecki M, and Horodecki R 1999 Bound Entanglement Can Be Activated \emph{Phys. Rev. Lett.}
{\bf 82} 1056 
 
 % -------------------------------------------------------
 % QD witness References
\bibitem{Maciej_entwitness} Lewenstein M,  Kraus B,  Cirac J I and Horodecki P 2000 Optimization of entanglement witnesses \emph{Phys. Rev. A} {\bf 62} 052310
 
 \bibitem{Bruss_entwitness} Bru{\ss} D, Cirac J I, Horodecki P, Hulpke F, Kraus B,  Lewenstein M and Sanpera A 2002 Reflections upon separability and distillability \emph{J. Mod. Opt.} {\bf 49} 1399

\bibitem{Tomography} Stokes G C 1852 Can a light beam be considered to be the sum of a completely polarized and a completely unpolarized beam? \emph{Trans. Cambridge Philos. Soc.} {\bf 9} 399

\bibitem{DAriano_tomo} D’Ariano G M, Vasilyev M and Kumar P 1998 Self-homodyne tomography of a twin-beam state \emph{Phys. Rev. A} {\bf 58} 636

\bibitem{White_tomo} White A G, James D F V, Eberhard P H and Kwiat P G 1999 Nonmaximally Entangled States: Production, Characterization, and Utilization \emph{Phys. Rev. Lett.} {\bf 83} 3103
 
\bibitem{Blatt_iontrap} 
Blatt R and Wineland D 2008 Entangled states of trapped atomic ions
\emph{Nature} {\bf 453} 1008

\bibitem{Simmons} Simmons G F 1963 \emph{Introduction to Topology and Modern Analysis} (McGraw-Hill,  New York)

%\bibitem{Holmes} Holmes R B 1975 \emph{Geometric Functional Analysis and its Applications} (Berlin: Springer-Verlag)

 \bibitem{sanpera11} {\.Z}yczkowski K, Horodecki P, Sanpera A and Lewenstein M 1998 Volume of the set of separable states {\it Phys. Rev. A}  {\bf 58} 883
 
\bibitem{ent_witness2} 
Bertlmann R A, Durstberger K, Hiesmayr B C and Krammer P 2005 Optimal entanglement witnesses for qubits and qutrits \emph{Phys. Rev. A} {\bf 72} 052331  
 
\bibitem{Guhne_entreview} G{\"u}hne O and  Toth G 2009 Entanglement detection  \emph{Phys. Rep.} {\bf 474} 1

\bibitem{ent_witness}  Chru{\'s}ci{\'n}ski D, Pytel J and Sarbicki G 2009 Constructing optimal entanglement witnesses \emph{Phys. Rev. A} {\bf 80} 062314 

\bibitem{Rahimi} Rahimi R and SaiToh A 2010 Single-experiment-detectable nonclassical correlation witness
\emph{Phys. Rev. A} {\bf 82} 022314

\bibitem{Saitoh} SaiToh A, Rahimi R and Nakahara M 2012 Limitation for linear maps in a class for detection and quantification of bipartite nonclassical correlation \emph{Quantum Inf. Comp.} {\bf 12} 0944

\bibitem{LUR_hoffmann} Hofmann H F and  Takeuchi S 2003 Violation of local uncertainty relations as a signature of entanglement \emph{ Phys. Rev. A} {\bf 68} 032103

\bibitem{Nonl_entwitness} G{\"u}hne O and  L{\"u}tkenhaus N 2006 Nonlinear entanglement witnesses \emph{Phys. Rev. Lett.} {\bf 96} 170502 

\bibitem{Nonlinear_entwit2} G{\"u}hne O and L{\"u}tkenhaus N  2007
Nonlinear entanglement witnesses, covariance matrices and the geometry of separable states \emph{J. Phys. C: Conf. Ser.} {\bf 67} 012004

 \bibitem{Rahimi_witness} Rahimi R, SaiToh A, Nakahara M and Kitagawa M 2007 Single-experiment-detectable multipartite entanglement witness for ensemble quantum computing
\emph{Phys. Rev. A} {\bf 75} 032317 

\bibitem{Maziero_witness} Maziero J and Serra R M 2012 Classicality witness for two-qubit states  \emph{Int. J. Quantum Inf.} {\bf 10} 1250028

\bibitem{Auccaise_witnessexp}  Auccaise R, C\'{e}leri L C, Soares-Pinto D O, de Azevedo E R,
 Maziero J, Souza A M, Bonagamba T J, Sarthour R S, Oliveira I S and Serra R M 2011 Environment-Induced Sudden Transition in Quantum Discord Dynamics \emph{Phys. Rev. Lett.} {\bf 107} 140403   
 
 \bibitem{Bylicka_diswitness} Bylicka B and Chru{\'s}ci{\'n}ski D 2010 Witnessing quantum discord in 2$\times$N systems  \emph{Phys. Rev. A} {\bf 81} 062102 

\bibitem{SPPT} Chru{\'s}ci{\'n}ski D, Jurkowski J and Kossakowski A 2008 Quantum states with strong positive partial transpose \emph{Phys. Rev. A} {\bf 77} 022113 


\bibitem{Zhang_witness} Zhang, C,  Yu S,  Chen Q and  Oh C 2011 Detecting the quantum discord of an unknown state by a single observable \emph{Phys. Rev. A} {\bf 84} 032122

\bibitem{Yu} Yu X, Zhang C, Chen Q, Oh C H 2011 Witnessing the quantum discord of all the unknown states arXiv:1102.4710 
 

\bibitem{state_estimator}
Horodecki P 2003 Measuring Quantum Entanglement without Prior State Reconstruction
 \emph{Phys. Rev. Lett.} {\bf  90}  167901
 
 \bibitem{saguia1}
Saguia A, Rulli C C, Oliveira T R D and Sarandy M S 2011 Witnessing nonclassical multipartite states \emph{Phys. Rev. A} {\bf 84} 042123

 \bibitem{Cialdi-discord}
Cialdi S, Smirne A, Paris M G A, Olivares S and Vacchini B 2014 Two-step procedure to discriminate discordant from classical correlated or factorized states \emph{Phys. Rev. A} {\bf 90} 050301(R)
 
 
 
 
 
 %%%%%%%%%%%%%%%%%%%%%%%%%
 				%%%%%%%%%%%%%%%
 							%%%%%
 
%%%%%%%%%new reference on volume%%%%%%%%%
\bibitem{real-royden}
Royden H L 1968 \emph{Real Analysis} (The Macmillan Company, New York)
%Royden H L and Fitzpatrick P M 2010 \emph{Real Analysis} (Prentice-Hall)

\bibitem{math-gupta}
 Gupta A 2016 \emph{Introduction to Mathematical Analysis} (Academic Publishers, Kolkata)
 
 \bibitem{karol-volume}
{\.Z}yczkowski K 1999 Volume of the set of separable states.\ II \emph{Phys. Rev. A} {\bf 60} 3496

\bibitem{szarekvol} Szarek S J 2005 Volume of separable states is super-doubly-exponentially small in the number of qubits {\it Phys. Rev. A}  {\bf 72} 032304
 

 \bibitem{braunstein} Braunstein S L, Caves C M, Jozsa R, Linden N, Popescu S and Schack R 1999 Separability of Very Noisy Mixed States and Implications for NMR Quantum Computing {\it Phys. Rev. Lett.}  {\bf 83} 1054

\bibitem{ferraro} Ferraro A, Aolita L, Cavalcanti D, Cucchietti F M and Ac{\' i}n A 2010 Almost all quantum states have nonclassical correlations {\it Phys. Rev. A}  {\bf 81} 052318

\bibitem{zero-QD} Huang J-H, Wang L and Zhu S-Y 2011 A new criterion for zero quantum discord \emph{New J. Phys.} {\bf 13} 063045

\bibitem{animesh-nullity}
Datta A 2011  A Condition for the Nullity of Quantum Discord arXiv:1003.5256

%\bibitem{Gessner-ZERODISCORD}
%Gessner M, Laine E M, Breuer H-P and Piilo J 2012
%Correlations in quantum states and the local creation of quantum discord \emph{Phys. Rev. A} {\bf 85} 052122
%


\bibitem{geometry-gen} 
Shi M, Jiang F, Sun C and Du J 2011 Geometric picture of quantum discord for two-qubit quantum states \emph{New J. Phys.} {\bf 13} 073016

\bibitem{geometry-cc} Oszmaniec M, Suwara P and Sawicki A 2014 Geometry and topology of CC and CQ states \emph{J. Math. Phys.} {\bf 55} 062204
  
\bibitem{GQD_CHSH1}
Batle J, Plastino A,  Plastino A R and   Casas M 2011 Properties of a geometric measure for quantum discord  arXiv:1103.0704

\bibitem{Geo_QD_decoherence6}
 Song W,  Yu L B,  Dong P, Li D C, Yang M and Cao Z L 2013  Geometric measure of quantum discord and the geometry of a class of two-qubit states \emph{Sci. Chin. Phys. Mech. Astro.} {\bf 56} 737
 
		    %%
				 %%%%
			   %%    %%
			 %%        %%
		   %%%%%%%%
 		  %%			  %%
 		 
%%%%%%%%% APPLICATION %%%%%%%%%

\bibitem{Ekert91} 
Ekert A K 1991 Quantum cryptography based on Bell's theorem \emph{Phys. Rev. Lett.} {\bf 67} 661


\bibitem{Bennett2}
 Bennett C H  and  Wiesner S J  1992  Communication via one- and two-particle operators on Einstein-Podolsky-Rosen states \emph{Phys. Rev. Lett.} {\bf 69} 2881
 
 \bibitem{Bennett_teleport}
Bennett C H, Brassard G, Cr\'epeau C, Jozsa R, Peres  A and Wootters W K 1993 Teleporting an unknown quantum state via dual classical and Einstein-Podolsky-Rosen channels 
 \emph{Phys. Rev. Lett.} {\bf 70} 1895


\bibitem{BB84} 
Bennett C H  and Brassard G  1984 Quantum cryptography: Public key distribution and coin tossing \emph{Proceedings of IEEE International Conference on Computers, Systems and Signal Processing} {\bf 175}



\bibitem{device_ind1}
Mayers D and  Yao A C-C  1998 Quantum Cryptography with Imperfect Apparatus \emph{Proceedings of the 39th Annual Symposium on Foundations of Computer Science} (IEEE Computer Society, Washington, DC) 503

\bibitem{device_ind3}
Vazirani U and Vidick T 2004  Fully Device-Independent Quantum Key Distribution
 \emph{Phys. Rev. Lett.} {\bf 113} 140501; \emph{ibid.} {\bf 116} 089901(E) (2016)

\bibitem{device_ind2}
Ac{\'i}n A,  Brunner N, Gisin N, Massar S, Pironio S and Scarani V 2007 Device-Independent Security of Quantum Cryptography against Collective Attacks
 \emph{Phys. Rev. Lett.} {\bf 98} 230501 
 
\bibitem{device_ind4}
Pironio S, Ac{\'i}n A, Brunner N,  Gisin N,  Massar S and Scarani V 2009  Device-independent quantum key distribution secure against collective attack  \emph{New J. Phys.} {\bf 11} 045021

\bibitem{device_ind5}
Lo H K,  Curty M and  Qi B 2012 Measurement-Device-Independent Quantum Key Distribution
 \emph{Phys. Rev. Lett.} {\bf 108} 130503

 \bibitem{bersten_vazirani}
 Bernstein E and Vazirani U 1993 Quantum complexity theory \emph{Proceedings of the Twenty-Fifth Annual ACM Symposium on Theory of Computing (Association for Computing Machinery, New York)}  11

\bibitem{smolin_terhal}
Terhal B M and Smolin J A 1998 Single quantum querying of a database \emph{Phys. Rev. A} {\bf 58}  1822

\bibitem{no_ent_algorithm}
Meyer D A 2000 Sophisticated Quantum Search Without Entanglement  \emph{Phys. Rev. Lett.} {\bf  85} 2014

\bibitem{unitary_decomposition}
Barenco A,  Bennett C H, Cleve R,  DiVincenzo D P, Margolus N,  Shor P, Sleator T, Smolin J A and Weinfurter H 1995 Elementary gates for quantum computation
  \emph{Phys. Rev. A} {\bf 52} 3457
  
  
%\bibitem{Horo_nature} Horodecki M, Oppenheim J and Winter  A 2005 Partial quantum information \emph{Nature} {\bf 436} 673

\bibitem{state_merging1}
Madhok V and Datta A 2011 Interpreting quantum discord through quantum state merging \emph{Phys. Rev. A}  {\bf 83} 032323

\bibitem{state_merging2}
Cavalcanti D, Aolita L, Boixo S,  Modi K, Piani M and Winter A 2011 Operational interpretations of quantum discord
\emph{ Phys. Rev. A} {\bf  83} 032324 

 
\bibitem{uni_discrimi} Gu M, Chrzanowski H M, Assad S M, Symul T, Modi K, Ralph T C, Vedral V and 
 Lam P K 2012 Observing the operational significance of discord consumption \emph{Nat. Phys.} {\bf 8} 671 

\bibitem{Channel_discrimi_QD} 
Weedbrook C,  Pirandola S, Thompson J,  Vedral V and
 Gu M 2016 How discord underlies the noise resilience of quantum illumination \emph{New J. Phys.} {\bf 18} 043027
 
\bibitem{metrology_QD}  
Girolami D,  Tufarelli T and Adesso G 2013 Characterizing Nonclassical Correlations via Local Quantum Uncertainty \emph{Phys. Rev. Lett.} {\bf 110} 240402 

\bibitem{QD_application2}
Girolami D,  Souza A M,  Giovannetti V,  Tufarelli T,   Filgueiras J G,   Sarthour R S, Pinto D O S,  Oliveira S and  Adesso G 2014 Quantum Discord Determines the Interferometric Power of Quantum States \emph{Phys. Rev. Lett.} {\bf 112} 210401

 \bibitem{QD_application4}
Li  B,  Fei S M,   Wang Z X  and  Fan H 2012  Assisted state discrimination without entanglement \emph{Phys. Rev. A} {\bf  85}  022328
  
\bibitem{QD_application6}
Perinotti P 2012 Discord and Nonclassicality in Probabilistic Theories \emph{Phys. Rev. Lett.} {\bf  108}  120502 
 
\bibitem{QD_application3}
Brodutch A 2013 Discord and quantum computational resources \emph{Phys. Rev. A} {\bf  88}  022307
 
 \bibitem{QD_application1}
Spehner D 2014  Quantum correlations and distinguishability of quantum states \emph{J. Math. Phys.} {\bf  55}  075211

\bibitem{QD_application5}
Sab{\'i}n C,  Fuentes I and  Johansson G 2015 Quantum discord in the dynamical Casimir effect \emph{Phys. Rev. A} {\bf 92} 012314


\bibitem{Sai_QDFMO} Br{\'a}dler K, Wilde M M, Vinjanampathy S and Uskov D B 2010 Identifying the quantum correlations in light-harvesting complexes
\emph{Phys. Rev. A} {\bf 82} 062310

\bibitem{Titas} Chanda T,  Mishra U, Sen(De) A and  Sen U 2014
Time dynamics of multiparty quantum correlations indicate energy transfer route in light-harvesting complexes
arXiv:1412.6519

\bibitem{Mahdian_QB} Mahdian M and Kouhestani H  2015 Thermal Quantum Correlations in Photosynthetic Light-Harvesting Complexes \emph{Int. J. Theor. Phys.} {\bf 54} 2576

\bibitem{Saberi_QDFMO} Saberi M, Harouni M B, Roknizadeh R and Latifi H 2016 Energy transfer and quantum correlation dynamics in FMO light-harvesting complex
\emph{Mol. Phys.} {\bf 114} 14

 
 
 \bibitem{Jevtic_enzyme} Jevtic S 2013 Large consequences of quantum coherence in small systems \emph{PhD Thesis} (Imperial College, UK)

%\bibitem{quant_comp_separable_state1}
%Braunstein S L,  Caves C M,  Jozsa R,  Linden N, Popescu S and Schack R 1999  Separability of Very Noisy Mixed States and Implications for NMR Quantum Computing  \emph{Phys. Rev. Lett.} {\bf 83} 1054

\bibitem{dqc1_ent}
Datta A, Flammia S T and Caves C M 2005 Entanglement and the power of one qubit  \emph{Phys. Rev.  A} {\bf 72} 042316




\bibitem{classical_unitary_trace}
Datta A and Vidal G 2007  Role of entanglement and correlations in mixed-state quantum computation
\emph{Phys. Rev. A} {\bf 75} 042310
 


\bibitem{Dutta-Shaji}
Datta A, Shaji A and Caves C M 2008 Quantum Discord and the Power of One Qubit \emph {Phys. Rev. Lett.} {\bf 100} 050502
  

 

 \bibitem{no_entyanglement_speed_up3}
Gottesman D 1998 The Heisenberg Representation of Quantum Computers \emph{Proceed. of the XXII Int. Coll. on Group Theor. Methd. in Phys. Cambridge, MA, International Press} {\bf 32} 



\bibitem{entanglement_speed_up}
Ding S C and Jin Z 2007  Review on the study of entanglement in quantum computation speedup \emph{Chin. Sci. Bull.} {\bf 52} 2161

\bibitem{no_entyanglement_speed_up2}
 Nest M V  2013 Universal Quantum Computation with Little Entanglement
 \emph{Phys. Rev. Lett.} {\bf  110}  060504
 
 
 

 
  \bibitem{DQC1_extra3}
 Zhang C, Yu S,  Chen Q and  Oh C H 2011  Detecting the quantum discord of an unknown state by a single observable   \emph{Phys. Rev. A} {\bf  84} 032122
  
  
  
\bibitem{DQC1_extra5}
Morimae T, Fujii K and  Fitzsimons J F 2014  Hardness of Classically Simulating the One-Clean-Qubit Model
  \emph{Phys. Rev. Lett.} {\bf  112} 130502 
  
   \bibitem{DQC1_extra12}
 {\'A}vila M, Sun G H and Salas-Brito A L 2014 Scales of time Where the quantum discord allows an efficient execution of the DQC1 algorithm  \emph{Adv. Math. Phys.} { \bf 2014} 367905
 
 \bibitem{DQC1_extra8}
 Cable H,  Gu M and  Modi K 2016  Power of one bit of quantum information in quantum metrology
 \emph{Phys. Rev. A} {\bf  93} 040304(R) 

 \bibitem{DQC1_extra9}
Matera J M, Egloff D, Killoran N and  Plenio M B 2016  Coherent control of quantum systems as a resource theory
 \emph{Quantum Sci. Tech.} {\bf 1} 01LT01

 \bibitem{dqc1_exp6}
 Ryan C A,  Emerson J,  Poulin D,  Negrevergne C and  Laflamme R 2005 Characterization of Complex Quantum Dynamics with a Scalable NMR Information Processor  \emph{Phys. Rev. Lett.} {\bf 95} 250502 
  
\bibitem{dqc1_exp}
Lanyon B P, Barbieri M, Almeida M P and White A G 2008 Experimental Quantum Computing without Entanglement \emph{Phys. Rev. Lett.} {\bf 101} 200501

\bibitem{dqc1_exp2}
 Passante G, Moussa O, Trottier D A and  Laflamme R 2011 Experimental detection of nonclassical correlations in mixed-state quantum computation \emph{Phys. Rev. A} {\bf 84} 044302

 \bibitem{dqc1_exp4}
 Mansell  C W  and  Bergamini S 2014 A cold-atoms based processor for deterministic quantum computation with one qubit in intractably large Hilbert spaces \emph{New J. Phys.} {\bf  16} 053045
     
 \bibitem{DQC1_extra2}
Kay A 2013 Degree of quantum correlation required to speed up a computation
  \emph{Phys. Rev. A} {\bf  92} 062329   
  
 \bibitem{DQC1_extra1}
Datta A and Gharibian S 2009 Signatures of nonclassicality in mixed-state quantum computation
 \emph{Phys. Rev. A} {\bf  79} 042325  
 
\bibitem{fanchini}  
 Fanchini F  F,   Cornelio M F, de Oliveira M C and Caldeira A O 2011 Conservation law for distributed entanglement of formation and quantum discord  \emph{Phys. Rev. A} {\bf 84} 012313 
 
 \bibitem{DQC1_extra7}
Yu C S, Yi X X, Song H S and Fan H 2013 Entangling power in deterministic quantum computation with one qubit
 \emph{Phys. Rev. A} {\bf  87} 022322
 
 \bibitem{Lo} 
Lo H-K 2000 Classical-communication cost in distributed quantum-information processing: A generalization of quantum-communication complexity \emph{Phys. Rev. A} {\bf 62} 012313 

\bibitem{pati}
Pati A K 2000 Minimum classical bit for remote preparation and measurement of a qubit
\emph{Phys. Rev. A}  \textbf{63} 014302

\bibitem{rsp_bennet} 
 Bennett C H, DiVincenzo D P, Shor P W, Smolin J A, Terhal B M and Wootters W K 2001 Remote State Preparation \emph{Phys. Rev. Lett} \textbf{87} 077902

   
   \bibitem{rsp_exp6}
Peng X,  Zhu X,  Fang X,  Feng M, Liu M and Gao K 2003 Experimental implementation of remote state preparation by nuclear magnetic resonance \emph{Phys.  Lett. A} {\bf 306} 271

\bibitem{rsp_non_ent}
 Peters N A,  Barreiro J T,  Goggin M E, Wei T-C and  Kwiat P G 2005     Remote State Preparation: Arbitrary Remote Control of Photon Polarization \emph{Phys. Rev. Lett.} {\bf  94} 150502
 
 \bibitem{rsp_exp7}
Xiang G Y, Li  J, Bo Y, Guo G C 2005  Remote preparation of mixed states via noisy entanglement \emph{Phys. Rev. A} {\bf 72} 012315

\bibitem{rsp_exp8}
Rosenfeld  W, Berner S, Volz J, Weber M and Weinfurter H 2007  Remote Preparation of an Atomic Quantum Memory \emph{ Phys. Rev. Lett.} {\bf  98}  050504

 \bibitem{rsp_exp5}
 Liu W T, Wu W, Ou B Q, Chen P X, Li C Z and Yuan J M  2007 Experimental remote preparation of arbitrary photon polarization states. \emph{Phys. Rev. A} {\bf 76}  022308 

\bibitem{rsp_exp4}
 Barreiro J T,  Wei T C and  Kwiat P G 2010 Remote preparation of single-photon ``Hybrid" entangled and vector-polarization states  \emph{Phys. Rev. Lett.} {\bf 105} 030407 
 
 \bibitem{rsp_discord}
  Daki\'c B,  Lipp Y O,  Ma  X,  Ringbauer M, Kropatschek S,  Barz  S, Paterek S, Vedral V, Zeilinger A, Brukner C and Walther P  2012  Quantum discord as resource for remote state preparation \emph{Nat. Phys.} {\bf 8} 666 
 
 \bibitem{rsp_exp1}
 Lanyon B P ,  Jurcevic C ,  Hempel C,  Gessner M,  Vedral V,  Blatt R and  Roos C F  2013  Experimental generation of quantum discord via noisy processes \emph{Phys. Rev. Lett.} {\bf 111} 100504  
 
 \bibitem{rsp_exp9}
R\r{a}dmark M,  Wie{\'s}niak M, \.{Z}ukowski M  and  Bourennane M 2013 Experimental multilocation remote state preparation \emph{Phys. Rev. A} {\bf  88} 032304

 \bibitem{rsp_exp3}
 Zenchuk A I 2014 Remote creation of a one-qubit mixed state through a short homogeneous spin-1/2 chain 
 \emph{Phys. Rev. A} {\bf  90} 052302 
 
 \bibitem{rsp_exp2}
Ra Y S,  Lim h T and Kim Y H 2016  Remote preparation of three-photon entangled states via single-photon measurement   \emph{Phys. Rev. A} {\bf 94} 042329 
 
\bibitem{RSP_GHZ}  
Shi B S and Tomita A  2002   Remote state preparation of an entangled state \emph{J. Opt. B} {\bf 4} 380 
 
 \bibitem{shor_rsp}
Leung D W and Shor P W 2003 Oblivious Remote State Preparation \emph{Phys. Rev. Lett.} {\bf 90} 127905

\bibitem{RSP_others2}
Ye M Y, Zhang Y S and Guo G C 2004 Faithful remote state preparation using finite classical bits and a nonmaximally entangled state \emph{Phys. Rev. A} {\bf 69} 022310
 
\bibitem{RSP_others1}
   Yu C S, Song H S and Wang Y H 2006 Remote preparation of a qudit using maximally entangled states of qubits \emph{Phys. Rev. A} {\bf 73} 02234
 
\bibitem{Giorgi2}
 Giorgi G L 2013 Quantum discord and remote state preparation \emph{ Phys. Rev. A} {\bf 88} 022315 
 
 \bibitem{rsp_horodechki} 
 Horodecki P,  Tuziemski J,  Mazurek P  and  Horodecki R 2014 Can communication power of separable correlations exceed that of entanglement resource?
 \emph{Phys. Rev. Lett.} {\bf 112} 140507  
  
  

\bibitem{No_cloning_Wooters}  Wootters W K and  Zurek W H 1982 A single quantum cannot be cloned \emph{Nature} {\bf 299} 802

\bibitem{No_cloning_Dieks}  Dieks D 1982 Communication by EPR devices
 \emph{Phys. Lett.} {\bf 92A}  271

\bibitem{Barnum_braodcast} 
Barnum H, Caves C M, Fuchs C A, Jozsa R and Schumacher B 1996 Noncommuting Mixed States Cannot Be Broadcast \emph{Phys. Rev. Lett.} {\bf 76} 2818 
  

\bibitem{broadcasting_extra7}
Buzek V,  Vedral V, Plenio M B,  Knight P L and Hillery M 1997 Broadcasting of entanglement via local copying \emph{Phys. Rev. A} {\bf 55} 3327


 \bibitem{broadcasting_extra1}
  Lamoureux L P, Navez P, Fiur{\'a}$\check{s}$ek J and  Cerf N J  2004  Cloning the entanglement of a pair of quantum bits \emph{Phys. Rev. A} {\bf 69} 040301(R) 
 
 \bibitem{broadcasting_extra4}
Horodecki M, Sen (De) A and  Sen U 2004  Dual entanglement measures based on no local cloning and no local deleting  \emph{Phys. Rev. A} {\bf 70} 052326


\bibitem{broadcasting_extra5}
Ghosh S,  Kar G and  Roy A 2004 Local cloning of Bell states and distillable entanglement
\emph{Phys. Rev. A} {\bf 69} 052312

\bibitem{broadcasting_extra6}
Anselmi F,  Chefles A and Plenio M B  2004 Local copying of orthogonal entangled quantum states \emph{New J. Phys.} {\bf 6} 164 
 
\bibitem{Yang_noentLB} 
Yang D, Horodecki M, Horodecki R and Synak-Radtke B 2005 Irreversibility for All Bound Entangled States \emph{Phys. Rev. Lett.} {\bf 95} 190501    
 
 \bibitem{broadcasting_extra2}
Karpov E,  Navez P and  Cerf N J 2005 Cloning quantum entanglement in arbitrary dimensions
 \emph{Phys. Rev. A} {\bf 72} 042314 
 
 \bibitem{broadcasting_extra3}
 
Novotn{\'y} J, Alber G and Jex I 2005  Optimal copying of entangled two-qubit states
 \emph{Phys. Rev. A} {\bf 71} 042332
 
 
 \bibitem{broadcasting_extra8}
Dobrza{\'n}ski  R D, Lewenstein M, Sen(De) A, Sen U and  Bru{\ss} D  2006 Usefulness of classical communication for local cloning of entangled states  \emph{Phys. Rev. A} {\bf 73} 032313 


\bibitem{Luo_LB} 
Luo S 2010 On Quantum No-Broadcasting  \emph{Lett. Math. Phys.} {\bf 92} 143

\bibitem{Luo_broad} 
Luo S and Sun W 2010 Decomposition of bipartite states with applications to quantum no-broadcasting theorems \emph{Phys. Rev. A} {\bf 82} 012338 

\bibitem{Sazim_braodcusting} 
Chatterjee S, Sazim S and Chakrabarty I
 2016 Broadcasting of quantum correlations: Possibilities and impossibilities
\emph{Phys. Rev. A} {\bf 93} 042309 


\bibitem{broadcasting_extra9}
 Scarani V,  Iblisdir S, Gisin N and  Ac{\'i}n A 2005 Quantum cloning \emph{Rev. Mod. Phys.} {\bf 77} 1225

\bibitem{broadcasting_extra10}
Dobrza{\'n}ski R D, Sen(De) A,  Sen U and Lewenstein M 2009 Entanglement enhances security in quantum communication \emph{Phys. Rev. A} {\bf 80} 012311


\bibitem{Auerbachbook}
Auerbach A 1998 \emph{Interacting Electrons and Quantum Magnetism} (Springer-Verlag, Berlin)

\bibitem{Schollwöck}
Schollwöck U, Richter J, Farnell D and  Bishop R 2004 \emph{Quantum Magnetism} (Springer-Verlag, Berlin)

\bibitem{AshcroftMermin} Ashcroft N W and Mermin N D 1976 {\it Solid State Physics} (Holt, Rinehart and Winston)

\bibitem{KittelBook} Kittel C 2004 {\it Introduction to Solid State Physics} (Wiley)

\bibitem{AdP_Ujjwal}  Lewenstein M,  Sanpera A,  Ahufinger V,  Damski B,  Sen(De) A and  Sen U 2007 Ultracold atomic gases in optical lattices: mimicking condensed matter physics and beyond {\it Adv. Phys.} {\bf 56 } 243


\bibitem{RMP_Bloch} Bloch I, Dalibard J and Zwerger W 2008 Many-body physics with ultracold gases \emph{Rev. Mod. Phys.} {\bf 80} 885

\bibitem{ROPP_Windpassinger}  Windpassinger P and Sengstock K 2013 Engineering novel optical lattices  {\it Rep. Prog. Phys.} {\bf 76} 086401


\bibitem{ROPP_Lewenstein}   Dutta O, Gajda M, Hauke P,  Lewenstein M, L\"uhmann D-S, Malomed B A, Sowiński T and  Zakrzewski J 2015 Non-standard Hubbard models in optical lattices: a review   \emph{ Rep. Prog. Phys.} {\bf 78} 066001

\bibitem{ROPP_LiLiu} Li X and Liu W V 2016 Physics of higher orbital bands in optical lattices: a review {\it Rep. Prog. Phys.} {\bf 79} 116401

\bibitem{Blatt_review} Blatt R and Roos C F 2012 Quantum simulations with trapped ions {\it Nat.  Phys.}  {\bf 8} 277

\bibitem{Devoret04} Devoret M H, Wallraff A, and Martinis J M 2004 Superconducting Qubits: A Short Review  arXiv:0411174

\bibitem{Clarke08} Clarke J and Wilhelm F K 2008 Review Article Superconducting quantum bits {\it  Nature} {\bf 453} 1031

\bibitem{RMP_Nori} Xiang Z-L, Ashhab S, You J-Q and Nori F 2013  Hybrid quantum circuits: Superconducting circuits interacting with other quantum systems {\it Rev. Mod. Phys.} {\bf 85} 623 

\bibitem{NMR_review}  Vandersypen L M K and Chuang I L 2005 NMR techniques for quantum control and computation {\it Rev. Mod. Phys.} {\bf 76} 1037


 \bibitem{Sachdev}
 Sachdev S 2011 \emph{Quantum Phase Transitions} (Cambridge University Press, Cambridge) 

\bibitem{Osterloh} 
 Osterloh A, Amico L, Falci G and  Fazio R 2002 Scaling of entanglement close to a quantum phase transition \emph{Nature} {\bf 416} 608
 
\bibitem{Osborne_Ni} 
Osborne T J and Nielsen M A 2002 Entanglement in a simple quantum phase transition \emph{Phys. Rev. A} {\bf 66} 032110

\bibitem{Verstraete04a} Verstraete F, Popp M and  Cirac J I 2004 Entanglement versus Correlations in Spin Systems {\it Phys. Rev. Lett.} {\bf 92} 027901 

\bibitem{Verstraete04} Verstraete F, Martín-Delgado M A and  Cirac J I 2004 Diverging Entanglement Length in Gapped Quantum Spin Systems {\it Phys. Rev. Lett.} {\bf 92} 087201

\bibitem{WenBook} Wen X-G 2004 {\it Quantum Field Theory of Many-body Systems} (Oxford University Press, Oxford)

\bibitem{Vidal_classisimulation} Vidal G 2003 Efficient Classical Simulation of Slightly Entangled Quantum Computations
\emph{Phys. Rev. Lett.} {\bf 91} 147902 and references thereto.


\bibitem{Verstraete_PEPS} Verstraete F, Murg V and  Cirac J I 2008 Matrix product states, projected entangled pair states, and variational renormalization group methods for quantum spin systems \emph{Adv. Phys.} {\bf 57} 143

 \bibitem{SWVC08} Schuch N, Wolf M M,  Vollbrecht K G H and  Cirac J I 2008 On entropy growth and the hardness of simulating time evolution {\it New J. Phys.} {\bf 10} 033032


  \bibitem{Arnesen01} Arnesen M, Bose S and  Vedral V 2001 Natural Thermal and Magnetic Entanglement in the 1D Heisenberg Model {\it Phys. Rev. Lett.} {\bf 87} 017901

\bibitem{Gunlycke01} Gunlycke D, Bose S, Kendon V and Vedral V 2001 Thermal concurrence mixing in a one-dimensional Ising model {\it Phys. Rev. A} {\bf 64} 042302
 
 \bibitem{OConnor01} O'Connor K and Wootters W K 2001 Entangled rings {\it Phys. Rev. A} {\bf 63} 052302
  
  
   \bibitem{RMP_manybody} Amico L,  Fazio R, Osterloh A and Vedral V 2008 Entanglement in many-body systems {\it Rev. Mod. Phys.} {\bf 80} 517 
  
  
  
%%%% Many body bunch %%% 
\bibitem{Nielsen_thesis} Nielsen M A 1998  Quantum information theory \emph{Ph.D. Thesis} arXiv:quant-ph/0011036  (University of New Mexico, USA)  

\bibitem{Preskill_many}  Preskill J 2000 Quantum information and physics: Some future directions \emph{J. Mod. Phys.} {\bf 47} 127 

\bibitem{CKW}  
Coffman V,  Kundu J and  Wootters W K 2000 Distributed entanglement \emph{Phys. Rev. A} {\bf 61} 052306

\bibitem{Wang01} Wang X 2001 Entanglement in the quantum Heisenberg XY model {\it Phys. Rev. A} {\bf 64} 012313 

\bibitem{Dennison01}  Dennison K A and  Wootters W K 2001 Entanglement sharing among qudits {\it Phys. Rev. A} {\bf 65} 010301

\bibitem{Wang_Heisen} Wang X, Fu H and Solomon A I 2001 Thermal entanglement in three-qubit Heisenberg models \emph{J. Phys. A: Math. Theor.} {\bf 34} 11307

\bibitem{Wang_aniso} Wang X 2001 Effects of anisotropy on thermal entanglement \emph{Phys. Lett. A} {\bf 281} 101 

\bibitem{Meyer_glabal} Meyer D A and Wallach N R 2002 Global entanglement in multiparticle systems \emph{J. Math. Phys.} {\bf 43} 4273 

\bibitem{Wooters_paratrans} Wootters W K 2002 Parallel transport in an entangled ring \emph{J. Math. Phys.} {\bf 43} 4307 

\bibitem{Wooters_entchain} Wootters W K 2002 Entangled chains \emph{Contemporary Mathematics} {\bf 305} 299 

\bibitem{Zanardi_fermion} Zanardi P 2002 Quantum entanglement in fermionic lattices \emph{Phys. Rev. A} {\bf 65} 042101 

\bibitem{Zanardi_fermion2} Zanardi P and Wang X 2002 Fermionic entanglement in itinerant systems \emph{J. Phys. A: Math. Theor.} {\bf 35} 7947 

\bibitem{Bose_entjump} 
Bose I and Chattopadhyay E 2002 Macroscopic entanglement jumps in model spin systems \emph{Phys. Rev. A} {\bf 66} 062320


%%%%   Many body bunch %%%%%
\bibitem{Gu}  
Gu S-J,  Lin H-Q and  Li Y-Q 2003 Entanglement, quantum phase transition, and scaling in the XXZ chain \emph{Phys. Rev. A} {\bf 68} 042330

\bibitem{Glaser}  
Glaser U,  B\"uttner H and Fehske H 2003 Entanglement and correlation in anisotropic quantum spin systems
\emph{Phys. Rev. A} {\bf 68} 032318

\bibitem{Roscilde1} 
Roscilde T, Verrucchi P, Fubini A, Haas S and Tognetti V 2004 Studying Quantum Spin Systems through Entanglement Estimators
\emph{Phys. Rev. Lett.} {\bf 93} 167203 

\bibitem{VidalLMG04} Vidal J, Mosseri R and Dukelsky J 2004 Entanglement in a first-order quantum phase transition {\it Phys. Rev. A} {\bf 69} { 054101}

\bibitem{Guj1j204} Gu S-H,  Li H,  Li Y-Q and Hai-Qing Lin 2004 Entanglement of the Heisenberg chain with the next-nearest-neighbor interaction {\it Phys. Rev. A} {\bf 70} 052302 

\bibitem{Yang3-body05} Yang M-F 2005 Reexamination of entanglement and the quantum phase transition {\it Phys. Rev. A} {\bf 71} 030302(R) 

\bibitem{Roscilde2} 
Roscilde T, Verrucchi P, Fubini A, Haas S and Tognetti V 2005 Entanglement and Factorized Ground States in Two-Dimensional Quantum Antiferromagnets
\emph{Phys. Rev. Lett.} {\bf 94} 147208 


\bibitem{Kitaev}  
Vidal G, Latorre J I,  Rico E and  Kitaev A 2003 Entanglement in Quantum Critical Phenomena
\emph{Phys. Rev. Lett.} {\bf 90} 227902 

 \bibitem{Eisert10} Eisert J, Cramer M and Plenio M B 2010 Colloquium: Area laws for the entanglement entropy {\it Rev. Mod. Phys.} {\bf 82} 277

%%%%%%%   Many body bunch 3
\bibitem{Brennen04}  Brennen G K and Bullock S S 2004 Stability of global entanglement in thermal states of spin chains {\it Phys. Rev. A} {\bf 70} 052303

\bibitem{aditi_dynamics}
Sen(De) A, Sen U and Lewenstein M 2005 Dynamical phase transitions and temperature-induced quantum correlations in an infinite spin chain \emph{Phys. Rev. A} {\bf 72} 052319

   \bibitem{Maziero} Maziero J, Guzman H C, Celeri L C, Sarandy M S and Serra R M 2010 Quantum and classical thermal correlations in the XY spin-$\frac 12$ chain  \emph{Phys. Rev. A} {\bf 82} 012106
 

\bibitem{PrabhuErgo3} Mishra U, Prabhu R, Sen(De) A and Sen U 2013 Tuning interaction strength leads to ergodic-nonergodic transition of quantum correlations in anisotropic Heisenberg spin model {\it Phys. Rev. A} {\bf 87} 052318

 \bibitem{Chanda} Chanda T, Das T, Sadhukhan D, Pal A K, Sen(De) A and Sen U 2016 Static and dynamical quantum correlations in phases of an alternating field XY model \emph{Phys. Rev. A} {\bf 94} 042310
 
\bibitem{Fumania16} Fumania F K, Nematib S, Mahdavifarc S and Daroonehb A H 2016 Magnetic entanglement in spin-1/2 XY chains {\it Physica A} {\bf 445} 256 

\bibitem{one_way} Raussendorf R and Briegel H J 2001 A One-Way Quantum Computer
\emph{Phys. Rev. Lett.} {\bf 86} 5188

\bibitem{one_way_nat} Briegel H J, Browne D E, D{\"u}r W,  Raussendorf R and Van den Nest M 2009 Measurement-based quantum computation \emph{Nat. Phys.} {\bf 5} 19


\bibitem{AditiErgo04} Sen(De) A,  Sen U and  Lewenstein M 2004  Nonergodicity of entanglement and its complementary behavior to magnetization in an infinite spin chain {\it Phys. Rev. A} {\bf 70} 060304(R)

\bibitem{PAM06} Plekhanov E, Avella A and Mancini F 2006 Nonergodic dynamics of the extended anisotropic Heisenberg chain {\it Phys. Rev. B} {\bf 74} 115120 %– Published 26 September 2006

\bibitem{PAM07}  Plekhanov E,  Avella A and Mancini F 2007 Entanglement in the 1D extended anisotropic Heisenberg model {\it Physica B: Cond. Matt.} {\bf 403} 1282

\bibitem{SAA10} Sadiek G, Alkurtass B and Aldossary O 2010 Entanglement in a time-dependent coupled $XY$ spin chain in an external magnetic field {\it Phys. Rev. A} {\bf 82} 052337 %– Published 30 November 2010

\bibitem{Sadiek12} Sadiek G,  Xu Q and Kais S 2012 Tuning entanglement and ergodicity in two-dimensional spin systems using impurities and anisotropy {\it Phys. Rev. A} {\bf 85} 042313

\bibitem{PrabhuErgo2} Prabhu R, Sen(De) A and Sen U 2012 Dual quantum-correlation paradigms exhibit opposite statistical-mechanical properties {\it Phys. Rev. A} {\bf 86} 012336



\bibitem{Cincio07} Cincio L, Dziarmaga J, Rams M M, and Zurek W H 2007 Entropy of entanglement and correlations induced by a quench: Dynamics of a quantum phase transition in the quantum Ising model {\it Phys. Rev. A} {\bf 75} 052321 

\bibitem{LK08}  L\"{a}uchli A M and Kollath C 2008 Spreading of correlations and entanglement after a quench in the one-dimensional Bose-Hubbard model {\it  J. Stat. Mech. } P05018

\bibitem{Sengupta09} Sengupta K and  Sen D 2009 Entanglement production due to quench dynamics of an anisotropic XY chain in a transverse field {\it Phys. Rev. A} {\bf 80} 032304 


\bibitem{Pollmann10} Pollmann F, Turner A M, Berg E and Oshikawa M 2010 Entanglement spectrum of a topological phase in one dimension {\it Phys. Rev. B} {\bf 81} 064439 

\bibitem{Polkovnikov11} Polkovnikov A, Sengupta K, Silva A and Vengalattore M 2011 Colloquium: Nonequilibrium dynamics of closed interacting quantum systems {\it Rev. Mod. Phys.} {\bf 83} 863

\bibitem{Amico04} Amico L, Osterloh A, Plastina F, Fazio R and  Palma G M 2004 Dynamics of entanglement in one-dimensional spin systems {\it Phys. Rev. A} {\bf 69} 022304

\bibitem{Huang05} Huang Z and Kais S 2005 Dynamics of entanglement for one-dimensional spin systems in an external time-dependent magnetic field  {\it Int. J. Quantum Inf.} {\bf 03} 483 


\bibitem{Calabrese1} 
Calabrese P and Cardy J 2005 evolution of entanglement entropy in one-dimensional Systems \emph{J. Stat. Mech.} P04010







%%%%%%%%%%%%%%%%%%%%%%%%%%%%%%%%%%%%%%%%





%     ||*	         *||       ||%)           ||
%	    ||   *   *   ||	    ||     ))		  ||
%	    ||     *      ||	    ||%)			  ||====


\bibitem{LSM} Lieb E, Schultz T and Mattis D  1961 Two soluble models of an antiferromagnetic chain
 \emph{Ann. Phys.} {\bf 16} 407
 
 \bibitem{barouch1}  Barouch E, McCoy B M and Dresden M 1970 Statistical mechanics of the XY model. I \emph{Phys. Rev. A} {\bf 2} 1075
 
 \bibitem{barouch2} Barouch E and McCoy B M 1971 Statistical mechanics of the XY model. II. Spin-correlation functions \emph{Phys. Rev. A} {\bf 3} 786
 
 \bibitem{barouch3} Barouch E and McCoy B M 1971 Statistical Mechanics of the XY model. III \emph{Phys. Rev. A} {\bf 3} 2137
 
 \bibitem{Takasaki} Takahashi M 1999 \emph{Thermodynamics of One-Dimensional Solvable Models} (Cambridge University Press, Cambridge)
 




%         ##				  /||
%      \\    *                ||
%        \\        ----  ======
%     _   \\			     -*//
%        ##				     //__


%%% Static Spin - 1/2
\bibitem{Dillenschneider} 
Dillenschneider R 2008 Quantum discord and quantum phase transition in spin chains \emph {Phys. Rev. B} {\bf 78} 224413

  \bibitem{Sarandy}  Sarandy M S 2009 Classical correlation and quantum discord in critical systems \emph{Phys. Rev. A} {\bf 80} 022108
  
  \bibitem{Kundu2013} Kundu A and  Subrahmanyam V 2013 Distribution of quantum correlations and conditional entropy in Heisenberg spin chains \emph{J. Phys. A: Math. Theor.} {\bf  46}
435304

 \bibitem{HuangScalingQD} Huang Y 2014 Scaling of quantum discord in spin models  \emph{Phys. Rev. A} {\bf 89} 054410
   
\bibitem{GLL03}  Gu S-H, Lin H-Q and Li Y-Q 2003 Entanglement, quantum phase transition, and scaling in the $XXZ$ chain {\it Phys. Rev. A} {\bf 68} 042330 

\bibitem{Syljuåsen03} Sylju{\r a}sen O F 2003 Entanglement and spontaneous symmetry breaking in quantum spin models  {\it Phys. Rev. A} {\bf 68} 060301


    \bibitem{Werlang} Werlang T, Trippe C,  Ribeiro G A P and Rigolin G 2010 Quantum Correlations in Spin Chains at Finite Temperatures and Quantum Phase Transitions \emph{Phys. Rev. Lett.} {\bf 105} 095702
 
 \bibitem{Werlang2} Werlang T, Ribeiro G A P and Rigolin G 2011 Spotlighting quantum critical points via quantum correlations at finite temperatures \emph{Phys. Rev. A} {\bf 83} 062334
 
\bibitem{Tomacello} Tomasello B, Rossini D, Hamma A and  Amico L 2011 Ground-state factorization and correlations with broken symmetry \emph{Europhys. Lett.} {\bf 96} 27002
 
 
 \bibitem{Kurmann} Kurmann J, Thomas H and M\"uller G 1982 Antiferromagnetic long-range order in the anisotropic quantum spin chain \emph{Physica A} {\bf 112} 235  

 
   \bibitem{Maziero2}  Maziero J,  Celeri L C, Serra R M and Sarandy M S 2012 Long-range quantum discord in critical spin systems \emph{ Phys. Lett. A} {\bf 376} 1540
 
  \bibitem{Campbell} Campbell S, Richens J, Gullo N L and Busch T 2013 Criticality, factorization, and long-range correlations in the anisotropic XY model \emph{Phys. Rev. A} {\bf 88} 062305
 
    \bibitem{Ciliberti} Ciliberti L, Rossignoli R and Canosa N 2010 Quantum discord in finite XY chains \emph{Phys. Rev. A} {\bf 82} 042316
  
  \bibitem{Sadhukhan} Sadhukhan D, Singha Roy S, Rakshit D, Prabhu R, Sen(De) A and Sen U 2016 Quantum discord length is enhanced while entanglement length is not by introducing disorder in a spin chain \emph{Phys. Rev. E} {\bf 93} 012131
  
  \bibitem{white92} White S R 1992 Density matrix formulation for quantum renormalization groups {\it Phys. Rev. Lett.} {\bf 69} 2863

\bibitem{white93} White S R 1993 Density-matrix algorithms for quantum renormalization groups {\it Phys. Rev. B} {\bf 48} 10345
  
  \bibitem{Baroni07} Baroni F, Fubini A,  Tognetti V and Verrucchi P 2007 Two-spin entanglement distribution near factorized states  {\it J. Phys. A: Math. Theor.} {\bf 40} 9845
 
\bibitem{Sadhukhan2} Sadhukhan D, Prabhu R, Sen(De) A and Sen U 2016 Quantum correlations in quenched disordered spin models: Enhanced order from disorder by thermal fluctuations \emph{Phys. Rev. E} {\bf 93} 032115
 
 \bibitem{Mishra16} Mishra U, Rakshit D and Prabhu R 2016 Survival of time-evolved correlations depends on whether quenching is across critical point in XY spin chain {\it Phys. Rev. A} {\bf 93} 042322 


\bibitem{Dhar} Dhar H S, Rakshit D, Sen(De) A and Sen U 2016 Adiabatic freezing of long-range quantum correlations in spin chains  \emph{Europhys. Lett.} {\bf 114} 60007

\bibitem{FZ2012} Fel'dman E B and   Zenchuk A I 2012 Quantum correlations in different
density-matrix representations of spin-1/2 open chain \emph{ Phys. Rev. A}
{\bf  86} 012303 %manybody 

\bibitem{Li} Li Y-C and Lin H-Q 2011 Thermal quantum and classical correlations and entanglement in the XY spin model with three-spin interaction {\it Phys. Rev. A} {\bf 83} 052323

 \bibitem{Ben-Qiong11} Ben-Qiong L, Bin S and Jian Z 2011 Quantum and Classical Correlations in Isotropic XY Chain with Three-Site Interaction  {\it Commun. Theor. Phys.} {\bf 56} 46
 

\bibitem{Chen} Chen Y-X and Yin Z 2010 Thermal Quantum Discord in Anisotropic Heisenberg XXZ Model with Dzyaloshinskii-Moriya Interaction  \emph{Commun. Theor. Phys.} {\bf 54} 60

\bibitem{Liu}  Liu B Q , Shao B, Li J G, Zou J and Wu L A 2011 Quantum and classical correlations in the one-dimensional XY model with Dzyaloshinskii-Moriya interaction \emph{Phys. Rev. A} {\bf 83} 052112

\bibitem{Zhu} Zhu Y and Zhang Y 2012 Quantum discord in the three-spin XXZ chain with Dzyaloshinskii-Moriya interaction 2012 \emph{Sci. Chin. Phys. Mech. Astro.} {\bf 55} 2081 

\bibitem{Ma} Ma X S and Wang A M 2015 Quantum discord in spin-1/2 Heisenberg chains with Dzyaloshinkii-Moriya interaction \emph{Int. J. Quantum Inf.} {\bf 13} 1450043
 
\bibitem{Yang2012} Yang G-H,  Y G and  Min Y R 2012 Quantum Discord Behavior for Two
Qubits Heisenberg XYZ Chain with Inhomogeneous Magnetic Field   \emph{Int.
 J.  Theor.  Phys.} {\bf  51}  2985 %manybody

  \bibitem{Pal} Pal A K and  Bose I 2011 Quantum discord in the ground and thermal states of spin clusters {\it J. Phys. B} {\bf 44} 045101

\bibitem{WangZhang12} Wang C,  Zhang Y Y and  Chen Q H 2012  Quantum correlations in collective spin systems \emph{Phys. Rev. A} {\bf  85} 052112 %manybody


% Many body bunch 5
\bibitem{Aharony} Aharony A 1978 Spin-flop multicritical points in systems with random fields and in spin glasses {\it Phys. Rev. B} {\bf 18} 3328

\bibitem{Villain80} Villain J,  Bidaux R, Carton J-P and  Conte R 1980 Order as an effect of disorder {\it  J. Phys. France} {\bf 41} 1263 % (1980).

\bibitem{Minchau85} Minchau B J and Pelcovits R A 1985 Two-dimensional XY model in a random uniaxial field {\it Phys. Rev. B} {\bf 32} 3081 

\bibitem{Feldman98} Feldman D E 1998 Exact zero-temperature critical behaviour of the ferromagnet in the uniaxial random field {\it J. Phys. A: Math. Theor.} {\bf 31} L177 %(1998)

\bibitem{Volovik06} Volovik J E 2006 Random anisotropy disorder in superfluid 3He-A in aerogel {\it JETP Lett.} {\bf 84} 455 %(2006)

\bibitem{Wehr06} Wehr J, Niederberger A, Sanchez-Palencia L and Lewenstein M 2006 Disorder versus the Mermin-Wagner-Hohenberg effect: From classical spin systems to ultracold atomic gases {\it Phys. Rev. B} {\bf 74} 224448 %(2006)

\bibitem{Abanin07}  Abanin D A, Lee P A and Levitov L S 2007  Randomness-Induced $XY$
Ordering in a Graphene Quantum Hall Ferromagnet {\it Phys. Rev. Lett.} {\bf 98} 156801 

\bibitem{Niederberger08} Niederberger A, Schulte T, Wehr J, Lewenstein M, 
Sanchez-Palencia L and  Sacha K 2008 Disorder-Induced Order in Two-Component Bose-Einstein Condensates {\it Phys. Rev. Lett.} {\bf 100} 030403

\bibitem{Niederberger10} Niederberger A, Rams M M,  Dziarmaga J, Cucchietti F M,
Wehr J and  Lewenstein M 2010 Disorder-induced order in quantum 
$XY$  chains {\it Phys. Rev. A} {\bf 82} 013630 

\bibitem{Pradhan} Prabhu R, Pradhan S, Sen(De) A and  Sen U 2011 Disorder overtakes order in information concentration over quantum networks {\it Phys. Rev. A} {\bf 84} 042334 


\bibitem{Malvezzi16} Malvezzi A L, Karpat G, \c{C}akmak B, Fanchini F F, Debarba T and Vianna R O 2016 Quantum correlations and coherence in spin-1 Heisenberg chains {\it Phys. Rev. B} {\bf 93} 184428

\bibitem{Zyczkowski94} Zyczkowski K and Kus M 1994 Random unitary matrices  {\it J. Phys. A: Math. Gen.} {\bf 27} 4235

\bibitem{Pozniak98} Pozniak M, Zyczkowski K and Kus M 1998 Composed ensembles of random unitary matrices {\it J. Phys. A: Math. Gen.} {\bf 31} 1059

\bibitem{Sakai90} Sakai T and Takahashi M 1990 Finite-Size Scaling Study of S=1 XXZ Spin Chain {\it J. Phys. Soc. Jpn.} {\bf 59}  2688

\bibitem{Alcaraz92} Alcaraz F C and Moreo A 1992 Critical behavior of anisotropic spin-S Heisenberg chains {\it Phys. Rev. B} {\bf 46} 2896

\bibitem{Kitazawa96} Kitazawa A, Nomura K and Okamoto K 1996 Phase Diagram of S=1 Bond-Alternating XXZ Chains  {\it Phys. Rev. Lett} {\bf 76} 4038




\bibitem{Higher-dim} Li X-J, Ji H-H and Hou X-W 2013 Thermal discord and negativity in a two-spin-qutrit system under different magnetic fields {\it Int. J. Quantum Inf.} {\bf 11} 1350070

\bibitem{Higher-dim1} Liu B, Hu Z and Hou X-W 2014 Comparative study of quantum discord and geometric discord for generic bipartite states {\it Int. J. Quantum Inf.} {\bf 12} 1450027

\bibitem{Higher-dim2} Yuan Y-L and  Hou X-W 2016 Thermal geometric discords in a two-qutrit system {\it Int. J. Quantum Inf.} {\bf 14} 1650016



\bibitem{Power15} Power M J M, Campbell S, Moreno-Cardoner M and Chiara G D 2015 Nonclassicality and criticality in symmetry-protected magnetic phases {\it Phys. Rev. B} {\bf 91} 214411

%%%%New ref in many body portion%%%

 \bibitem{Botet83} Botet R, Jullien R and Kolb M 1983 Finite-size-scaling study of the spin-1 Heisenberg-Ising chain with uniaxial anisotropy {\it Phys. Rev. B} {\bf 28} 3914 

\bibitem{Glaus84} Glaus U and Schneider T 1984 Critical properties of the spin-1 Heisenberg chain with uniaxial anisotropy {\it Phys. Rev. B} {\bf 30} 215 

\bibitem{Schulz86} Schulz H J 1986 Phase diagrams and correlation exponents for quantum spin chains of arbitrary spin quantum number {\it Phys. Rev. B} {\bf 34} 6372

\bibitem{Degli03} Degli C, Boschi E, Ercolessi E, Ortolani F and Roncaglia M 2003 On $c = 1$ critical phases in anisotropic spin-1 chains {\it Eur. Phys. J. B} {\bf 35} 465

\bibitem{Sanpera11}  De Chiara D Lewenstein M and Sanpera A 2011 Bilinear-biquadratic spin-1 chain undergoing quadratic Zeeman effect {\it Phys. Rev. B} {\bf 84} 054451

\bibitem{Hu11} Hu S, Normand B,  Wang X and Yu L 2011 Accurate determination of the Gaussian transition in spin-1 chains with single-ion anisotropy {\it Phys. Rev. B} {\bf 84} 220402(R)

\bibitem{Lepori13} Lepori L, De Chiara G and Sanpera A 2013 Scaling of the entanglement spectrum near quantum phase transitions {\it Phys. Rev. B} {\bf 87} 235107




\bibitem{Campbell13} Campbell S, Mazzola L, De Chiara G,  Apollaro T J G, Plastina F,  Busch T and Paternostro M 2013 Global quantum correlations in finite-size spin chains
 {\it New J. Phys.} {\bf 15} 043033



\bibitem{sun-global}
Sun Z-Y, Liao Y-E, Guo B, Huang H-L, Zhang D, Xu J,  Zhan B-F, Wu Y-Y, Cheng H-G, Wen G-Z, Fang C, Duan C-B and Wang B 2014 Global quantum discord in infinite quantum spin chains arXiv:1412.5285

\bibitem{Lio-globalQD}
Liu S-Y, Zhang Y-R, Yang W-L and Fan H 2015 Global quantum discord and quantum phase transition in XY model \emph{Ann. Phys.} {\bf 362} 805



%							     _________
%			    			 //""""""""""""||
%  					   //						||
% 		    ____//___________||_
%			 ||----  (())---------------(())----||


%%%%% REFERENCE OF DYNAMICS START HERE  %%%%%%%%

\bibitem{Subrahmanyam04}  Subrahmanyam V 2004 Entanglement dynamics and quantum-state transport in spin chains {\it Phys. Rev. A} {\bf 69} 034304

\bibitem{PAOF04} Plastina F, Amico L, Osterloh A and Fazio R 2004 Spin wave contribution to entanglement in Heisenberg models {\it New J. Phys.} {\bf 6} 124

\bibitem{CFMMP04} De Chiara G, Fazio R, Macchiavello C, Montangero S and Palma G M 2004  Quantum cloning in spin networks {\it Phys. Rev. A} {\bf 70} 062308 

\bibitem{VPA04} Vidal J, Palacios G and Aslangul C 2004 Entanglement dynamics in the Lipkin-Meshkov-Glick model {\it Phys. Rev. A} {\bf 70} 062304 


\bibitem{Cao2005}  Cao J, Wang Y and Wang X R 2005  Entanglement of a degenerate system in adiabatic process {\it Phys. Lett. A} {\bf 353} 295

\bibitem{CZ06} Ciancio E and Zanardi P 2006 Coupling bosonic modes with a qubit: entanglement dynamics at zero and finite temperatures {\it Phys. Lett. A} {\bf 360} 49



\bibitem{Werner_LN} Vidal G and Werner R F 2012 Computable measure of entanglement \emph{Phys. Rev. A } {\bf 65} 032314



\bibitem{Himadri}  Dhar H S, Ghosh  R,  Sen(De) A and Sen U 2012 Quantum discord surge heralds entanglement revival in an infinite spin chain \emph{Europhys. Lett.}  {\bf 98}  30013 

\bibitem{Dhar_WD} Dhar H S, Ghosh  R,  Sen(De) A and Sen U 2014 Cumulative quantum work-deficit versus entanglement in the dynamics of an infinite spin chain \emph{Phys. Lett. A} {\bf 378} 1258 

\bibitem{Ren12} Ren J, Wu Y-Z and Zhu S-Q 2012 Quantum Discord and Entanglement in Heisenberg XXZ Spin Chain after Quenches {\it Chin. Phys. Lett.} {\bf 29} 060305


\bibitem{Polkovnikov05} Polkovnikov A 2005 Universal adiabatic dynamics in the vicinity of a quantum critical point {\it Phys. Rev. B} {\bf 72} 161201(R)

\bibitem{ZDZ05} Zurek W H, Dorner U and Zoller P 2005 Dynamics of a Quantum Phase Transition {\it Phys. Rev. Lett.} {\bf 95} 105701 

\bibitem{Dziarmaga05} Dziarmaga J 2005 Dynamics of a Quantum Phase Transition: Exact Solution of the Quantum Ising Model {\it Phys. Rev. Lett.} {\bf 95} 245701

\bibitem{Damski05} Damski B 2005 The Simplest Quantum Model Supporting the Kibble-Zurek Mechanism of Topological Defect Production: Landau-Zener Transitions from a New Perspective {\it Phys. Rev. Lett.} {\bf 95} 035701


\bibitem{Cherng06} Cherng R W and Levitov L S 2006 Entropy and Correlation Functions of a Driven Quantum Spin Chain {\it Phys. Rev. A} {\bf 73} 043614 

\bibitem{Mukherjee07}  Mukherjee V,  Divakaran U,  Dutta A and  Sen D 2007 Quenching dynamics of a quantum XY spin-1/2 chain in a transverse field {\it Phys. Rev. B} {\bf 76} 174303 

\bibitem{Tanay11}  Nag T,  Patra A and  Dutta A 2011 Quantum Discord in a spin-1/2 transverse XY Chain Following a Quench {\it J Stat. Mech.} {\bf 08} P08026

\bibitem{Tanay13}  Nag T, Dutta A and Patra A 2013 Quenching dynamics and quantum information {\it Int. J. Mod. Phys. B} {\bf 27} 1345036 

\bibitem{Mazur} Mazur P 1969 Non-ergodicity of phase functions in certain systems {\it Physica} {\bf 43} 533

\bibitem{suzuki-paper}
Suzuki M 1971 Ergodicity, constants of motion, and bounds for susceptibilities {\it Physica} {\bf 51} 271

\bibitem{HK05} Huang  Z and  Kais S 2005 Dynamics of Entanglement for One-Dimensional Spin Systems in an External Time-Dependent Magnetic Field   {\it Int. J. Quantum Inf.} {\bf 03} 483



%%%%% GQD IN many body

\bibitem{GQD_many_body1} 
Zhanga G F, Fanb H,  Jib A L,  Jiangc Z T,  Ablizd A and Liub W M 2011  Quantum correlations in spin models  \emph{ Ann. Phys.} {\bf 326} 10

\bibitem{GQD_many_body3}
Yao Y,  Li H W,  Zhang C M,  Yin Z Q,  Chen W, Guo G C and  Han Z F 2012  Performance of various correlation measures in quantum phase transitions using the quantum renormalization-group method \emph{Phys. Rev. A} {\bf 86} 042102 



\bibitem{GQD_many_body2}
Cai J T, Abliz A and  Li S S 2013 Various Correlations in a Two-Qubit Heisenberg XXZ Spin System Both in Thermal Equilibrium and Under the Intrinsic Decoherence  \emph{ Int. J. Theor. Phys.} {\bf 52} 576 


\bibitem{GQD_many_body5}
Canosa N,  Ciliberti L and  Rossignoli R 2013  Quantum discord and related measures of quantum correlations in XY chains  \emph{Int. J. Mod. Phys. B} {\bf  27} 1345033 


\bibitem{GQD_many_body6}
Fan C H, Xiong H N,  Huang Y and   Sun Z 2013 Quantum discord and quantum phase transition in spin-1/2 frustrated Heisenberg chain  \emph{Quantum Inf. Comp.} {\bf 13} 5 

\bibitem{GQD_many_body7}
Ciliberti L,  Canosa N and  Rossignoli R 2013 Discord and information deficit in the XX chain \emph{Phys. Rev. A} {\bf  88} 012119
  
\bibitem{GQD_many_body8}
Muthuganesan R and Sankaranarayanan R 2016 Nonlocal Correlation in Heisenberg Spin Models arXiv:1604.02655
 

\bibitem{GQD_many_body9}
 Shan C J, Cheng W W, Liu J B, Cheng Y S and  Liu T K 2014 Scaling
of Geometric Quantum Discord Close to a Topological Phase Transition
\emph{Sci Rep.} {\bf  4} 4473

\bibitem{JC_model}
Jaynes E T and  Cummings F W 1963  Comparison of quantum and semiclassical radiation theories with application to the beam maser \emph {Proc. IEEE.} {\bf  51} 89 
  
  \bibitem{GQD_JC1} 
Qiang W C,  Zhang L and Zhang H P 2015 Geometric quantum discord of a Jaynes-Cummings atom and an isolated atom arXiv:1504.08253 


 
\bibitem{GQD_JC2}
Raja S H, Mohammadi H  and  Akhtarshenas S J 2015 Geometric discord of the Jaynes-Cummings model: pure dephasing regime \emph{Eur. Phys. J. D} {\bf 69} 14


\bibitem{JC_ent1}

Rendell  R  W and  Rajagopal A K 2003 Revivals and entanglement from initially entangled mixed states of a damped Jaynes-Cummings model  \emph{Phys. Rev. A} {\bf 67} 062110

\bibitem{JC_ent2}
  Li Z J,  Li j Q,  Jin Y H and  Nie Y H 2007  Time  evolution and transfer of entanglement between an isolated atom and a Jaynes-Cummings  atom \emph{J.  Phys.  B:  At.  Mol.  Opt.  Phys.} {\bf 40} 3401
  
  
  \bibitem{JC_ent3}
{\"Y}onac M,  Yu T and Eberly J H 2006 Sudden death of entanglement of two Jaynes-Cummings atoms \emph{J. Phys. B: At. Mol. Opt. Phys.} {\bf 39} 621 

  \bibitem{JC_ent4}
Sainz I and  Bj{\"o}rk G 2007 Entanglement invariant for the double Jaynes-Cummings model \emph{Phys. Rev. A} {\bf 76} 042313 

\bibitem{JC_ent5}
 Pandit M, Das S,  Singha Roy S,  Dhar H S and  Sen U 2016 Effects of cavity-cavity interaction on the entanglement dynamics of a generalized double Jaynes-Cummings model  	arXiv:1612.01165

  \bibitem{GQD_other_QC}
Xu Z,  Yang W,  Xiao X and  Feng M 2011  Comparison of different measures for quantum discord under non-Markovian noise \emph{J. Phys. A: Math. Theor.} {\bf 44} 395304

\bibitem{QKD_discord3}
Wu X and Zhou T 2015  Geometric discord: A resource for increments of quantum key distribution through twirling   \emph{Sci. Rep.} {\bf 5} 13365
 


\bibitem{Davis76} Davis E B 1976 {\it Quantum Theory of Open Systems} (Academic Press, London)

 \bibitem{Alicki87} Alicki R and Lendi K 1987 {\it Quantum Dynamical Semigroups and Applications} Lecture Notes Physics  Vol. 286 (Springer, Berlin) 
 
  \bibitem{frz2} Breuer H-P and Petruccione F 2002 \emph{The Theory of Open Quantum Systems}  (Oxford University Press, Oxford) 

\bibitem{frz1} Rivas \'{A} and Huelga S F 2011 \emph{Open Quantum Systems: An Introduction} (Springer, Heidelberg)

\bibitem{frz3} Rivas \'{A}, Huelga S F and Plenio M B 2014 Quantum non-Markovianity: characterization, quantification and detection \emph{Rep. Prog. Phys.} {\bf 77} 094001




  

%%%% Open bunch 1 contain 32 reference

\bibitem{death1} \'{Z}yczkowski K, Horodecki P, Horodecki M and Horodecki R 2001 Dynamics of quantum entanglement \emph{Phys. Rev. A} {\bf 65} 012101 
 
\bibitem{death2} Di\`osi L 2003 Progressive decoherence and total environmental disentanglement \emph{Lect. Notes Phys.} {\bf 622} 157

\bibitem{yu1} Yu T and Eberly J H 2004 Finite-Time Disentanglement Via Spontaneous Emission \emph{Phys. Rev. Lett.} {\bf 93} 140404
 
 \bibitem{death3} Dodd P J and Halliwell J J 2004 Disentanglement and decoherence by open system dynamics  \emph{Phys. Rev. A} {\bf 69} 052105

\bibitem{yu2} Yu T and Eberly J H 2006 Quantum Open System Theory: Bipartite Aspects \emph{Phys. Rev. Lett.} {\bf 97} 140403
 
   
\bibitem{death4} Almeida M P, de Melo F, Hor-Meyll M, Salles A, Walborn S P, Ribeiro P H S and Davidovich L 2007 Environment-Induced Sudden Death of Entanglement \emph{Science} {\bf 316} 579

\bibitem{death7} Bellomo B, Franco R L, Maniscalco S and Compagno G 2008 Entanglement trapping in structured environments \emph{Phys. Rev. A} {\bf 78} 060302(R)
 
\bibitem{death5} Salles A,  de Melo F, Almeida M P, Hor-Meyll M, Walborn S P, Ribeiro P H S and Davidovich L 2008 Experimental investigation of the dynamics of entanglement: Sudden death, complementarity, and continuous monitoring of the environment \emph{Phys. Rev. A} {\bf 78} 022322
 
 \bibitem{death6} 
 Yu T and Eberly J H 2009 Sudden Death of Entanglement \emph{Science} {\bf 323} 598
 
\bibitem{open_extra_new6} 
Lu X M, Wang X  and  Sun C P 2010 Quantum Fisher information flow and non-Markovian processes of open systems \emph{Phys. Rev. A} {\bf 82}  042103  
 
     \bibitem{open_extra_new30}
Streltsov A,  Kampermann H and Bru{\ss} D 2011 Behavior of Quantum Correlations under Local Noise  \emph{Phys. Rev. Lett.} {\bf 107} 170502

\bibitem{open_extra_new28}
Gessner M and  Breuer H-P 2011 Detecting Nonclassical System-Environment Correlations by Local Operations
  \emph{Phys. Rev. Lett.} {\bf 107}  180402
   
\bibitem{open_extra_new31}
Hu X, Gu Y,  Gong Q and Guo G  2011 Necessary and sufficient condition for Markovian-dissipative-dynamics-induced quantum discord \emph{Phys. Rev. A} {\bf 84} 022113

\bibitem{death8} Franco R L, D’Arrigo A, Falci G, Compagno G and Paladino E 2012 Entanglement dynamics in superconducting qubits affected by local bistable impurities \emph{Phys. Scr.} {\bf T147} 014019
 
 \bibitem{open_extra_new22}
 Xu J 2012 Creating quantum discord through local generalized amplitude
damping channel \emph{Int. J. Quantum Inf.} {\bf  10}  1250071

\bibitem{open_extra_new29}
Ciccarello F and  Giovannetti V 2012 Creating quantum correlations
through local nonunitary memoryless channels   \emph{Phys. Rev. A} {\bf 85}  010102

 \bibitem{open_extra_new33}
Hu X,  Fan H,  Zhou D L and  Liu W M 2012 Necessary and sufficient conditions for local creation of quantum correlation  \emph{Phys. Rev. A} {\bf 85} 032102 

\bibitem{open_extra_new27}
Yang Q,  Yang M,  Li D C  and  Cao Z L 2012 Quantum Discord of Two-Qubit in Dephasing Model   \emph{ Int.  J.  Theor.  Phys.} {\bf  51}  2160 
 
 \bibitem{open_extra_new23}
 Wu  T and  Ye L  2012 The dynamics of quantum discord and entanglement of two atoms coupled to two spatially separate cavities in cavity QED
\emph{Eur. Phys. J. D} {\bf  66}  261


 \bibitem{open_extra_new19}
 Coles P J 2012 Unification of different views of decoherence and discord
 \emph{ Phys. Rev. A} {\bf  85}  042103
 
\bibitem{open_extra_new20}     
 Alipour S,  Mani A   and   Rezakhani A T 2012 Quantum discord and non-Markovianity of quantum dynamics   \emph{ Phys. Rev. A} {\bf  85}  052108

 \bibitem{open_extra_new24}
 Guo J L,   Mi Y J and  Song H S 2012 Quantum discord dynamics of two-qubit system in a quantum spin environment   \emph{Eur. Phys. J. D} {\bf  66} 24

\bibitem{Altintas-QD}
Altintas F, Kurt A and Eryigit R 2012 Classical memoryless noise-induced maximally discordant mixed separable steady states \emph{Phys. Lett. A} {377} 53 
 
 \bibitem{open_extra_new25}    
 Xiao R H,  Guo Z Y,  Zhu S Q  and  Fang J X 2013 Dynamics of Quantum Discord of Two-Qubit Coupled with a Vacuum Cavity \emph{Int. J. Theor.
Phys.} {\bf  52} 1721

 \bibitem{open_extra_new26} 
 Yu  R M and  Yang G H 2013  Steady Quantum Discord Behavior for Two Qubits Heisenberg XYZ Chain with Intrinsic Decoherence \emph{Int. J. Theor.
Phys.} {\bf  52} 1621

\bibitem{open_extra_new16}
  Behzadi N and  Ahansaz A 2013 Thermal tripartite quantum correlations: quantum discord and entanglement perspectives  \emph{Eur. Phys. J. D} {\bf  67}  112

 \bibitem{open_extra_new17}
 Yang Y and   Wang A M 2013 Quantum Discord for a Qutrit-Qutrit System under Depolarizing and Dephasing Noise \emph{Chin.  Phys. Lett.} {\bf  30} 080302
 
 
 \bibitem{open_extra_new18}
 Wang C and  Chen Q H 2013 Quantum discord dynamics of two qubits in single-mode cavities \emph{Chin.  Phys. B} {\bf  22} 040304

\bibitem{death9} D'Arrigo A, Franco R L, Benenti G, Paladino E and Falci G 2014 Recovering entanglement by local operations \emph{Ann. Phys.} {\bf 350} 211
 
 \bibitem{death10} D'Arrigo A, Benenti G, Franco R L, Falci G and Paladino E 2014 Hidden entanglement, system-environment information flow and non-Markovianity \emph{Int. J. Quantum Inf.} {\bf 12} 1461005

\bibitem{open_extra_new9}
Mishra U, Sen(De) A  and   Sen U 2015  Local decoherence-resistant quantum states of large systems   \emph{Phys. Lett. A} {\bf  379}  261

\bibitem{open_extra_new7} 
Chunfang S C,  Chen Z,  Wang G,   Wu C,  Xue K  and  Kwek  L C 2016
Protection of quantum correlations against decoherence \emph{Quantum Inf. Proc.} {\bf  15}  773

\bibitem{open_extra_new8} 
Krzywda  J and Roszak K 2016 Phonon-mediated generation of quantum correlations between quantum dot qubits \emph{ Sci. Rep.} {\bf  6} 23753

\bibitem{open_extra_new32}
Carnio E G, Buchleitner A and  Gessner M 2016 Generating and protecting correlated quantum states under collective dephasing \emph{New J. Phys.} {\bf 18} 073010 

\bibitem{Sudarshan_dyna}  Jordan T F and Sudarshan E C G 1961 Dynamical Mappings of Density Operators in Quantum Mechanics \emph{J. Math. Phys.} {\bf 2} 772 

\bibitem{Kraus}   Kraus K 1971 General state changes in quantum theory \emph{Ann. Phys.} {\bf 64}  311


\bibitem{Jamiolkowski}  
Jamio{\l}kowski A 1972 Linear transformations which preserve trace and positive semidefiniteness of operators \emph{Rep. Math. Phys.} {\bf 3} 275


\bibitem{Choi_map} Choi M D 1972 Positive linear maps on C*-algebras \emph{Canad. J. Math.} {\bf 24} 520 

\bibitem{Choi_map2} Choi M D 1975 Completely positive linear maps on Complex Matrices \emph{Linear Algebra Appl.} {\bf 10} 285
   
\bibitem{Kraus83} Kraus K 1983 \emph{States, Effects, and Operations} (Springer, Berlin)

\bibitem{Preskil} 
Preskill J  \emph{Lecture notes on Quantum Information and Computation} \url{http://www.theory.caltech.edu/people/preskill/ph229/}
 
%%%% open system bunch 2
\bibitem{Pechukas94} Pechukas P 1994 Reduced dynamics need not be completely positive {\it Phys. Rev. Lett.} {\bf 73} 1060 

\bibitem{Alicki95} Alicki R 1995 Comment on ``Reduced dynamics need not be completely positive" \emph{Phys. Rev. Lett.} {\bf 75} 3020

\bibitem{open_extra_new1}
 Royer A 1996 Reduced dynamics with initial correlations, and time-dependent environment and Hamiltonians  \emph{Phys. Rev. Lett.} {\bf 77} 3272 
 
\bibitem{Stelmachovic01} {\v S}telmachovi{\v c} P and Bu{\v z}ek V 2001 Dynamics of open quantum systems initially entangled with environment: Beyond the Kraus representation {\it Phys. Rev. A} {\bf 64} 062106;  {\it ibid.} {\bf 67} 029902(E) (2003)

\bibitem{open_extra_new3}
Salgado D and  Gomez J L S 2002 Comment on ``Dynamics of open quantum systems initially entangled with environment: Beyond the Kraus representation" arXiv:quant-ph/0211164

 \bibitem{open_extra_new2}
Rodr\'{i}guez-Rosario C A,  Modi K,   Mazzola L and Guzik A A 2012 Unification of witnessing initial system-environment correlations and witnessing non-Markovianity \emph{Europhys. Lett.} {\bf  99}  20010

%%%%%   

\bibitem{Pechukas95} Pechukas P 1995 Pechukas Replies: {\it Phys. Rev. Lett.} {\bf 75} 3021 

\bibitem{Jordan04} Jordan T F, Shaji A and  Sudarshan E C G  2004 Dynamics of initially entangled open quantum systems {\it Phys. Rev. A} {\bf 70} 052110 

\bibitem{Carteret08} Carteret H A, Terno D R and \.Zyczkowski K  2008 Dynamics beyond completely positive maps: Some properties and applications {\it Phys. Rev. A} {\bf 77} 042113 

\bibitem{Rodríguez10} Rodr{\'i}guez-Rosario C A, Modi K and  Aspuru-Guzik A 2010 Linear assignment maps for correlated system-environment states {\it Phys. Rev. A} {\bf 81} 012313


\bibitem{complete-positive} Rodriguez-Rosario C A, Modi K, Kuah A M, Shaji A and Sudarshan E C G 2008 Completely positive maps and classical correlations {\it J. Phys. A: Math. Gen.} {\bf 41} 205301

\bibitem{Shabani09} Shabani A and Lidar D A 2009 Vanishing quantum discord is necessary and sufficient for completely positive maps {\it Phys. Rev. Lett.} {\bf 102} 100402; {\it ibid} {\bf 116} 049901(E) (2016)



\bibitem{Brodutch13} Brodutch A, Datta A, Modi K,  Rivas A and Rodr{\'i}guez-Rosario C A 2013 Vanishing quantum discord is not necessary for completely positive maps {\it Phys. Rev. A} {\bf 87} 042301 

\bibitem{Buscemi14} Buscemi F 2014 Complete positivity, Markovianity, and the quantum data-processing inequality, in the presence of initial system-environment correlations {\it Phys. Rev. Lett.} {\bf 113} 140502 

\bibitem{Dominy16} Dominy J M, Shabani A and  Lidar D A 2016 A general framework for complete positivity {\it Quantum Inf. Proc.} {\bf 15} 465

\bibitem{Breuer_RMP} Breuer H-P,  Laine E-M, Piilo J and Vacchini B 2016 Colloquium: Non-Markovian dynamics in open quantum systems {\it Rev. Mod. Phys.} {\bf 88} 021002
 
 \bibitem{werlang1}  Werlang T, Souza S, Fanchini F F and Villas Boas C J 2009 Robustness of quantum discord to sudden death \emph{Phys. Rev. A} {\bf 80} 024103

 \bibitem{decay1} Maziero J,  C\'{e}leri L C, Serra R M and Vedral V 2009 Classical and quantum correlations under decoherence \emph{Phys. Rev. A} {\bf 80} 044102
 
 \bibitem{decay2} Maziero J, Werlang T, Fanchini F F, C\'{e}leri L C and Serra R M 2010 System-reservoir dynamics of quantum and classical correlations \emph{Phys. Rev. A} {\bf 81} 022116
 
 \bibitem{decay6} Wang B, Xu Z-Y, Chen Z-Q and Feng M 2010 Non-Markovian effect on the quantum discord
 \emph{Phys. Rev. A} {\bf 81} 014101
 
 \bibitem{decay7} Altintas F and Eryigit R 2010 Quantum correlations in non-Markovian environments \emph{Phys. Lett. A} {\bf 374} 4283
 
 \bibitem{mazzola1} Mazzola L, Piilo J and Maniscalco S 2010 Sudden transition between classical and quantum decoherence \emph{Phys. Rev. Lett.} {\bf 104} 200401

\bibitem{freezing-exp1} Xu J-S, Xu X-Y, Li C-F , Zhang C-J , Zou X-B and Guo G-C 2010 Experimental investigation of classical and quantum correlations under decoherence \emph{Nat. Commun.} {\bf 1} 7

\bibitem{Geo_QD_decoherence5}
Karpat G  and  Gedik  Z  2011 Correlation Dynamics of Qubit-Qutrit Systems in a Classical Dephasing Environment  \emph{Phys. Lett. A} {\bf  375}  4166 

 \bibitem{decay3} Berrada K, Eleuch H and Hassouni Y 2011 Asymptotic dynamics of quantum discord in open quantum systems \emph{J. Phys. B: At. Mol. Opt. Phys.} {\bf 44} 145503

 \bibitem{decay9} Bellomo B, Compagno G, Franco R L, Ridolfo A and Savasta S 2011 Dynamics and extraction of quantum discord in a multipartite open system \emph{Int. J. Quantum Inf.} {\bf 09} 1665 
 
 \bibitem{decay10} Xi Z, Lu X-M, Sun Z and Li Y 2011 Dynamics of quantum discord in a quantum critical environment \emph{J. Phys. B: At. Mol. Opt. Phys.} {\bf 44} 215501
 
 \bibitem{decay13} He Q-L, Xu J-B, Yao D-X and Zhang Y-Q 2011 Sudden transition between classical and quantum decoherence in dissipative cavity QED and stationary quantum discord \emph{Phys. Rev. A} {\bf 84} 022312

 \bibitem{decay4} Pal A K and Bose I 2012 Markovian evolution of classical and quantum correlations in transverse-field XY model \emph{Eur. Phys. J. B} {\bf 85} 277 

\bibitem{decay12} Daoud M and Laamara R A 2012 Quantum discord of Bell cat states under amplitude damping \emph{J. Phys. A: Math. Theor.} {\bf 45} 325302

\bibitem{decay5} Pinto J P G, Karpat G and Fanchini F F 2013 Sudden change of quantum discord for a system of two qubits \emph{Phys. Rev. A} {\bf 88} 034304
 
 \bibitem{decay11} Franco R L, Bellomo B, Maniscalco S and Compagno G 2013 Dynamics of quantum correlations  in two-qubit systems within non-Markovian environments \emph{Int. J. Mod. Phys. B} {\bf 27} 1345053
  
 \bibitem{decay15} L\"{u} Y-Q, An J-H, Chen X-M, Luo H-G and Oh C H 2013 Frozen Gaussian quantum discord in photonic crystal cavity array system \emph{Phys. Rev. A} {\bf 88} 012129
 
\bibitem{decay16} Karpat G and Gedik Z 2013 Invariant quantum discord in qubit-qutrit systems under local dephasing \emph{Phys. Scr.} {\bf T153} 014036

\bibitem{decay17} Guo J-L, Li H and Long G-L 2013 Decoherent dynamics of quantum correlations in qubit-qutrit systems \emph{Quantum Inf. Process} {\bf 12} 3421

\bibitem{haikka-frz} Haikka P, Johnson T H and Maniscalco S 2013 Non-Markovianity of local dephasing channels and time-invariant discord \emph{Phys. Rev. A} {\bf 87} 010103(R)

\bibitem{sarandy-dec} Montealegre J D, Paula F M, Saguia A and Sarandy M S 2013 One-norm geometric quantum discord under decoherence \emph{Phys. Rev. A} {\bf 87} 042115
 
\bibitem{frz-dekhechi} Paula F M, Silva I A, Montealegre J D, Souza A M, deAzevedo E R, Sarthour R S, Saguia A, Oliveira I S, Soares-Pinto D O, Adesso G and Sarandy M S 2013 Observation of Environment-Induced Double Sudden Transitions in Geometric Quantum Correlations \emph{Phys. Rev. Lett.} {\bf 111} 250401 

 
\bibitem{decay18} Rong X, Jin F, Wang Z, Geng J, Ju C, Wang Y, Zhang R, Duan C, Shi M and Du J 2013 Experimental protection and revival of quantum correlation in open solid systems \emph{Phys. Rev. B} {\bf 88} 054419

 
\bibitem{open_extra_new11}
Galve F,  Plastina F, Paris M G A and  Zambrini R 2013   Discording Power of Quantum Evolutions \emph{Phys. Rev. Lett.} {\bf  110}  010501


\bibitem{open_extra_new12}
Buscemi F and Bordone P 2013 Time evolution of tripartite quantum discord and entanglement under local and nonlocal random telegraph noise  
\emph{Phys. Rev. A} {\bf 87} 042310

\bibitem{open_extra_new13}
Benedetti C,  Buscemi F,  Bordone P  and  Paris M G A 2013 Dynamics of quantum correlations in colored-noise environments   \emph{Phys. Rev. A}
{\bf  87} 052328

\bibitem{death-abar1}
 Zou H-M and Fang M-F 2015 Discord and entanglement in non-Markovian environments at finite temperatures \emph{Chin. Phys. B} {\bf 25} 090302 




\bibitem{mark} You B and Cen L-X 2012 Necessary and sufficient conditions for the freezing phenomena of quantum discord under phase damping \emph{Phys. Rev. A} {\bf 86} 012102 


\bibitem{mark1} Mazzola L, Piilo J and Maniscalco S 2011 frozen discord in non-markovian dephasing channels \emph{Int. J. Quantum Inf.} {\bf 09} 981 
 
\bibitem{franco-frz} Franco R L, Bellomo B, Andersson E and Compagno G 2012 Revival of quantum correlations without system-environment back-action \emph{Phys. Rev. A} {\bf 85} 032318

 \bibitem{mannone-frz} Mannone M, Franco R L and Compagno G 2013 Comparison of non-Markovianity criteria in a qubit system under random external fields \emph{Phys. Scr.} {\bf T153} 014047


\bibitem{xu-exp} Xu J-S, Sun K, Li C-F, Xu X-Y, Guo G-C, Andersson E, Franco R L and Compagno G 2013 Experimental recovery of quantum correlations in absence of system-environment back-action \emph{Nat. Com.} {\bf 4}  2851

\bibitem{yan-dis} Yan X-Q and Zhang B-Y 2014 Collapse-revival of quantum discord and entanglement \emph{Ann. Phys.} {\bf 349} 350
  

\bibitem{song-frz} Song W and Cao Z -L Conditions for the freezing phenomena of geometric measure
of quantum discord for arbitrary two-qubit X-states under non-dissipative dephasing noises 2014 \emph{Int. J. Theor. Phys.} {\bf 53} 519

\bibitem{freezing_new_new1}
Yao Y, Li M W, Yin Z Q and  Han Z F 2012 Geometric interpretation of the geometric discord \emph{Phys. Lett. A} {\bf 376} 358


\bibitem{karmakar-local} Karmakar S, Sen A, Bhar A  and Sarkar D 2015 Effect of local filtering on freezing phenomena of quantum correlation \emph{Quantum Inf. Proc.} {\bf 14} 7

\bibitem{Jianwei-Xu1}
Xu J 2013 Analytical expressions of global quantum discord for two classes
of multi-qubit states \emph{Phys. Lett. A} {\bf 377} 238

\bibitem{Aaronson-frz} Aaronson B, Franco R L and Adesso G 2013 Comparative investigation of the freezing phenomena for quantum correlations under nondissipative decoherence \emph{Phys. Rev. A} {\bf 88} 012120 

\bibitem{cianciaruso-frz} Cianciaruso M, Bromley T R, Roga W, Franco R L and Adesso G 2015 Universal freezing of quantum correlations within the geometric approach \emph{Sci. Rep.} {\bf 5} 10177


 \bibitem{DD1} Fanchini F F, de Lima E F and Castelano L K 2012 Shielding quantum discord through continuous dynamical decoupling \emph{Phys. Rev. A} {\bf 86} 052310

\bibitem{DD2} Addis C, Karpat G and Maniscalco S 2015 Time-invariant discord in dynamically decoupled systems \emph{Phys. Rev. A} {\bf 92} 062109

\bibitem{DD3} Singh H, Arvind and Dora K 2016 Experimentally freezing quantum discord in a dissipative environment using dynamical decoupling arXiv:1610.02755
 
 \bibitem{SMMdBSC11}  Soares-Pinto D O,  Moussa M H Y,  Maziero J,  deAzevedo E R,  Bonagamba T J, Serra R M and  C\'eleri L C 2011 Equivalence
between Redfield- and master-equation approaches for a time-dependent quantum system and coherence control {\it Phys. Rev. A} {\bf 83} 062336 

\bibitem{RWD14}  Ren B-C,  Wei H-R and  Deng F-G 2014 Correlation dynamics of a two-qubit system in a Bell-diagonal state under non-identical local noises {\it Quantum Inf. Process.} {\bf 13} 1175


\bibitem{LSZ10}  Liu B-Q, Shao B and Zou J 2010 Quantum discord for a central two-qubit system coupled to an XY-spin-chain environment {\it Phys. Rev. A} {\bf 82} 062119

\bibitem{SLS10}  Sun Z,  Lu X-M and  Song L 2010 Quantum discord induced by a spin chain with quantum phase transition {\it J. Phys. B: At. Mol. Opt. Phys.} {\bf 43} 215504



%\bibitem{QWS07LMG}  Quan H T,  Wang Z D and  Sun C P 2007 Quantum critical dynamics of a qubit coupled to an isotropic Lipkin-Meshkov-Glick bath {\it Phys. Rev. A} {\bf 76} 012104 






%%%%%  bunch 4  %%%%

\bibitem{Hu10DM3}  Hu M-L 2010 Disentanglement dynamics of interacting two qubits and two qutrits in an XY spin-chain environment with the Dzyaloshinsky-Moriya interaction {\it Phys. Lett. A} {\bf 374}  3520

\bibitem{Hao10}  Hao X,  Ma C-L and  Sha J 2010 Decoherence of quantum discord in an asymmetric-anisotropy spin system {\it J. Phys. A: Math. Theor.} {\bf 43} 425302

\bibitem{YQY11LMG}  Yan Y-Y,  Qin L-G and Tian L-J 2011 Dynamics of quantum correlations for central two-qubit coupled to an isotropic Lipkin-Meshkov-Glick bath arXiv:1112.2285

\bibitem{YTQ11DM}  Yan Y-Y,  Qin L-G,   Tian L-J 2012 Decoherence from a spin chain with Dzyaloshinskii—Moriya interaction {\it Chin. Phys. B} {\bf 21} 100304 

\bibitem{TZQ12threesite}  Tian L-J,  Zhang C-Y,  Qin L-G 2013 Sudden Transition in Quantum Discord Dynamics: Role of Three-Site Interaction {\it Chin. Phys. Lett.} {\bf 30} 050303

\bibitem{AZCM15DM2}  Abdel-Aty A,  Zakaria N,  Cheong L and  Metwally N 2014 Effect of Spin-Orbit Interaction
(Heisenberg XYZ Model) On partial entangled Quantum Network {\it Quantum Inf. Sci.} {\bf 4} 1 

\bibitem{Guo16threesite}  Guo J L and  Zhang X Z 2016 Quantum correlation dynamics subjected to critical spin environment with short-range anisotropic interaction {\it Sci. Rep.} {\bf 6}  32634 

\bibitem{VL98} Viola L and  Lloyd S 1998 Dynamical suppression of decoherence in two-state quantum systems {\it Phys. Rev. A} {\bf 58} 2733

\bibitem{Rossini07} Rossini D,  Calarco T, Giovannetti V, Montangero S and  Fazio R 2007 Decoherence induced by interacting quantum spin baths {\it Phys. Rev. A } {\bf 75} 032333 

\bibitem{YL08} Yang W and  Liu R-B 2008 Quantum many-body theory of qubit decoherence in a finite-size spin bath {\it Phys. Rev. B} {\bf 78} 085315 %

\bibitem{XuXu11} Xu H-S and Xu J-B 2011 Enhancement of quantum correlations for the system of cavity QED by applying bang-bang pulses {\it Europhys. Lett.} {\bf 95}  60003 %

 \bibitem{Francica10} Francica F, Plastina F and Maniscalco S 2010 Quantum Zeno and anti-Zeno effects on quantum and classical correlations {\it Phys. Rev. A} {\bf 82} 052118
 
 \bibitem{LLXY11}  Luo D-W,  Lin H-Q,  Xu J-B and  Yao D-X 2011 Pulse control of sudden transition for two qubits in XY spin baths and quantum phase transition {\it Phys. Rev. A} {\bf 84} 062112 

\bibitem{Ge10}  Ge R-C, Gong M,  Li C-F,  Xu J-S and  Guo G-C 2010 Quantum correlation and classical correlation dynamics in the spin-boson model {\it Phys. Rev. A} {\bf 81} 064103 

\bibitem{Man11} Man Z-X,  Xia Y-J and An N-B 2011 The transfer dynamics of quantum correlation between systems and reservoirs
 {\it J. Phys. B: At. Mol. Opt. Phys.} {\bf 44} 095504

\bibitem{open_extra_new5} 
Wang C and Chen Q H 2013 Exact dynamics of quantum correlations of two qubits coupled to bosonic baths \emph{New J. Phys.} {\bf 15} 103020
 
  \bibitem{Wall17} Wall M L, Safavi-Naini A and Rey A M 2017 Boson-mediated quantum spin simulators in transverse fields: XY model and spin-boson entanglement {\it Phys. Rev. A} {\bf 95} 013602

\bibitem{Manzano13}  Manzano G, Galve F and Zambrini R 2013 Avoiding dissipation in a system of three quantum harmonic oscillators {\it Phys. Rev. A} {\bf 87} 032114 % –  QD in an open quantum system composed by three coupled and detuned harmonic oscillators in the presence of a common heat bath

\bibitem{Lorenzo15} Lorenzo S, Farace A, Ciccarello F,  Palma G M and Giovannetti V 2015 Heat flux and quantum correlations in dissipative cascaded systems {\it Phys. Rev. A} {\bf 91} 022121 

\bibitem{Zhang11}  Zhang Y-J, Zou X-B,  Xia Y-J and Guo G-C  2011 Quantum discord dynamics in the presence of initial system–cavity correlations {\it  J. Phys. B: At. Mol. Opt. Phys.} {\bf 44} 035503


\bibitem{McEndoo13} McEndoo S, Haikka P, De Chiara G M,  Palma and Maniscalco S 2013  Entanglement control via reservoir engineering in ultracold atomic gases {\it Europhys Lett.} {\bf 101} 60005

\bibitem{Vasile10} Vasile R, Giorda P, Olivares S,  Paris M G A and Maniscalco S 2010 Nonclassical correlations in non-Markovian continuous-variable systems {\it Phys. Rev. A} {\bf 82} 012313 

\bibitem{Isar11} Isar A 2011 Quantum Entanglement and Quantum Discord of Two-Mode Gaussian States in a Thermal Environment {\it Open Sys. Inf. Dynamics} {\bf 18} 175

\bibitem{Isar12} Isar A 2012 Entanglement and discord in two-mode Gaussian open quantum systems 
{\it Phys. Scr.} { \bf T147} 014015

\bibitem{Isar13} Isar A 2013 Quantum correlations of two-mode Gaussian systems in a thermal environmet {\it Phys. Scr.} { \bf T153} 014035

\bibitem{Isar13_1} Isar A 2013 Quantum Discord of Two Bosonic Modes in Two-Reservoir Model {\it Open Sys. Inf. Dynamics} {\bf 20} 1340003

\bibitem{Suciu15} Suciu S and Isar A 2015 Gaussian geometric discord in terms of Hellinger distance {\it AIP Conf. Proc.} {\bf 1694} 020013 

\bibitem{Xu10PRA} Xu J-S,  Li C-F,  Zhang C-J, Xu X-Y, Zhang Y-S and Guo G-C 2010 Experimental investigation of the non-Markovian dynamics of classical and quantum correlations {\it Phys. Rev. A} {\bf 82} 042328

\bibitem{Benedetti13} Benedetti C, Shurupov A P, Paris M G A, Brida G and Genovese M 2013 Experimental estimation of quantum discord for a polarization qubit and the use of fidelity to assess quantum correlations {\it Phys. Rev. A} {\bf 87} 052136

\bibitem{Cialdi14} Cialdi S, Smirne A, Paris M G A, Olivares S and Vacchini B 2014 Two-step procedure to discriminate discordant from classical correlated or factorized states {\it Phys. Rev. A} {\bf 90} 050301(R) % local detection expt photon 2

\bibitem{TangGessner15}   Tang J S,  Wang Y T,  Chen G,  Zou Y,  Li C F,  Guo G C,  Yu Y,  Li M-F, Zha G-W,  Ni H-Q,  Niu Z C,  Gessner M and Breuer H-P 2015 Experimental detection of polarization-frequency quantum correlations in a photonic quantum channel by local operations {\it Optica} {\bf 2} 1014

\bibitem{Gessner13NP} Gessner M, Ramm M,  Pruttivarasin T, Buchleitner A,Breuer H-P and H\''{a}ffner 2013 Local detection of quantum correlations with a single trapped ion {\it Nat. Phys.} {\bf 10} 105 % local detection expt ion trap



\bibitem{Soares10NMR} Soares-Pinto D-O, C\'eleri L-C, Auccaise R, Fanchini F-F, deAzevedo E-R, Maziero J, Bonagamba T J and Serra R M 2010 Nonclassical correlation in NMR quadrupolar systems {\it Phys. Rev. A} {\bf 81} 062118


\bibitem{Rong13} 
Rong X, Jin F, Wang Z, Geng J, Ju C, Wang Y, Zhang R, Duan C, Shi M and Du J 2013 Experimental protection and revival of quantum correlation in open solid systems {\it Phys. Rev. B} {\bf 88} 054419 %– Published 26 August 2013


\bibitem{Freitas12} Freitas J N and Paz J P 2012 Dynamics of Gaussian discord between two oscillators interacting with a common environment {\it Phys. Rev. A} {\bf 85} 032118 


\bibitem{Correa12} Correa L A, Valido A A and Alonso D 2012 Asymptotic discord and entanglement of nonresonant harmonic oscillators under weak and strong dissipation {\it Phys. Rev. A} {\bf 86} 012110
% – Published 20 July 2012



\bibitem{Marian15} Marian P, Ghiu L and Marian T A 2015 Decay of Gaussian correlations in local thermal reservoirs {\it Phys. Scr.} {\bf 90} 074041

\bibitem{Cazzaniga13} Cazzaniga A, Maniscalco S, Olivares S and Paris M G A 2013 Dynamical paths and universality in continuous-variable open systems {\it Phys. Rev.  A} {\bf 88} 032121 % – Published 30 September 2013

\bibitem{Qars15} Qars J E,  Daoud M and Laamara A 2015 Entanglement versus Gaussian quantum discord in a double-cavity opto-mechanical system {\it Int. J. Quantum Inf.} {\bf 13} 1550041 


%expt open qd {Gessner13NP, TangGessner15, Cialdi14, freezing-exp1


\bibitem{Blandino12} Blandino R, Genoni M G, Etesse J,  Barbieri M, Paris M G A, Grangier P and Tualle-Brouri R 2012 Homodyne estimation of gaussian quantum discord {\it Phys. Rev. Lett.} {\bf 109} 180402 %– Published 2 November 2012

\bibitem{Buono12} Buono D, Nocerino G, Porzio A and  Solimeno S 2012 Experimental analysis of decoherence in continuous-variable bipartite systems {\it Phys. Rev. A} {\bf 86} 042308

\bibitem{Vogl13} Vogl U, Glasser R T, Glorieux Q,  Clark J B, Corzo N V and Lett P D 2013 Experimental characterization of Gaussian quantum discord generated by four-wave mixing {\it Phys. Rev. A} {\bf 87} 010101(R)

\bibitem{Hosseini14} Hosseini S,  Rahimi-Keshari S, Haw J Y, Assad S M, Chrzanowski H M, Janousek J, Symul T,  Ralph T C and  Lam P K 2014 Experimental verification of quantum discord in continuous-variable states {\it J. Phys. B: At. Mol. Opt. Phys.} {\bf 47} 025503

\bibitem{Valente15} Valente P, Auyuanet A, Barreiro S, Failache H and Lezama A 2015 Experimental characterization of the Gaussian state of squeezed light obtained via single passage through an atomic vapor {\it Phys. Rev. A} {\bf 91} 053848 %– Published 26 May 2015

\bibitem{Laura16} Laura T K, Christian T S, Osvaldo J F, Stephen P W and Miguel A L 2016 Entanglement-breaking channels and entanglement sudden death {\it Phys. Rev. A} {\bf 94} 012345



\bibitem{GQD_QD}
Okrasa M and Walczak Z 2002 On two-qubit states ordering with quantum discords  \emph{Europhys. Lett.} {\bf  98} 40003
 
\bibitem{Geo_QD_decoherence3}
Bellomo B,  Franco R L and  Compagno G 2012 Dynamics of geometric and entropic quantifiers of correlations in open quantum systems \emph{Phys. Rev. A} {\bf 86} 012312 
 
\bibitem{Geo_QD_decoherence4}
Ramzan M 2013  Decoherence dynamics of discord for multipartite quantum systems \emph {Eur. Phys. J. D} {\bf 67} 170


\bibitem{Geo_QD_decoherence7}
Ma Z H,  Chen Z H and  Fanchini F F 2013   Multipartite Quantum Correlations in Open Quantum Systems \emph{New J. Phys.} {\bf  15} 043023

\bibitem{Geo_QD_decoherence8}
Silva I A,  Girolami D,  Auccaise R, Sarthour R S,  Oliveira I S, Bonagamba T J,  deAzevedo E R, SoaresPinto D O  and  Adesso G  2013  Measuring Bipartite Quantum Correlations of an Unknown State
 \emph{Phys. Rev. Lett.} {\bf 110} 140501

\bibitem{Geo_QD_decoherence9}
Mohamed A-B A 2013 Quantum discord and its geometric measure with death entanglement in correlated dephasing two qubits system  \emph{Quantum Inf. Rev.} {\bf  1} 1 

\bibitem{Geo_QD_decoherence10}
Roszak K,  Mazurek P and  Horodecki P 2013  Anomalous decay of quantum correlations of quantum-dot qubits  \emph{Phys. Rev. A} {\bf 87} 062308 
 
\bibitem{Geo_QD_decoherence11}
Hu M L and  Tian D P 2014 Preservation of the geometric quantum discord in noisy environments \emph{Ann. Phys.} {\bf 343} 132 


\bibitem{Hill_EOF} Hill S and Wootters W K 1997 Entanglement of a Pair of Quantum Bits \emph{Phys. Rev. Lett.} {\bf 78} 5022 

\bibitem{Wooters} Wootters W K 1998 Entanglement of formation of an arbitrary state of two qubits \emph{Phys. Rev. Lett.} {\bf 80} 2245 

\bibitem{woot-eof}
Wootters W K 2001 Entanglement of formation and concurrence \emph{Quantum Inf. Comput.} {\bf 1} 27

\bibitem{Hayden_eof} Hayden P M,  Horodecki M and Terhal B M 2001 The asymptotic entanglement cost of preparing a quantum state \emph{J. Phys. A: Math. Gen} {\bf 34} 6891 

\bibitem{Bennett_monogamy}  
Bennett C H, Bernstein H J, Popescu S and Schumacher B 1996 Concentrating partial entanglement by local operations
\emph{Phys. Rev. A} {\bf 53} 2046

\bibitem{Osborne}  
Osborne  T and  Verstraete F 2006 General monogamy inequality for bipartite qubit entanglement \emph{Phys. Rev. Lett.}  {\bf 96} 220503 

\bibitem{Giorgi} 
Giorgi G L 2011 Monogamy properties of quantum and classical correlations  \emph{Phys. Rev. A} {\bf 84} 054301

\bibitem{Bai_discordmono} Bai Y-K, Zhang N, Ye M-Y and Wang Z D 2013 Exploring multipartite quantum correlations with the square of quantum discord \emph{Phys. Rev. A} {\bf 88} 012123 

\bibitem{QICgroup_monogamyreview} 
Dhar H S, Pal A K, Rakshit D, Sen(De) A, Sen U 2016 Monogamy of quantum correlations - a review 
	arXiv:1610.01069 

\bibitem{Barry_monoreview} Kim J S,  Gour G and Sanders B C 2012 Limitations to Sharing
Entanglement \emph{Contemp. Phys.} {\bf 53} 417 


\bibitem{Aditi_monogamydef} 
Sen(De) A and  Sen U 2012 Locally accessible information of multisite quantum ensembles violates entanglement monogamy 
\emph{Phys. Rev. A} {\bf 85} 052103 

\bibitem{Prabhu_state_discrimi} 
Prabhu R,  Pati A K, Sen(De) A and  Sen U 2012
Conditions for monogamy of quantum correlations: Greenberger-Horne-Zeilinger versus W states 
\emph{Phys. Rev. A} {\bf 85} 040102(R) 

\bibitem{manab_mono} 
Bera M N, Prabhu R, Sen(De) A and Sen U 2012 Characterization of tripartite quantum states with vanishing monogamy score \emph{Phys. Rev. A} {\bf 86} 012319 

\bibitem{Prabhu_lightcone}  
Prabhu R, Pati A K, Sen(De) A and Sen U 2012 Relating monogamy of quantum correlations and multisite entanglement
\emph{Phys. Rev. A} {\bf 86} 052337 


\bibitem{Winter_squashed} 
Christandl M and Winter A 2004 ``Squashed entanglement": An additive entanglement measure {\it J. Math. Phys.} {\bf 45} 829 

\bibitem{onew_disent} Bennett C H,  Brassard G,  Popescu S, Schumacher B, Smolin J A and Wootters W K 1996 Purification of noisy entanglement and faithful teleportation via noisy channels \emph{Phys. Rev. Lett.} {\bf 76} 722;  \emph{ibid.} {\bf 78} 2031(E) (1997)

\bibitem{Ou_Fan}  
Ou Y C and Fan H 2007 Monogamy inequality in terms of negativity for three-qubit states \emph{Phys. Rev. A} {\bf 75} 062308 

\bibitem{Oliveira}  
de Oliveira T R, Cornelio M F and  Fanchini F F 2014   Monogamy of entanglement of formation \emph{Phys. Rev. A} {\bf 89} 034303



\bibitem{nonmono_EOF} Fanchini F, de Oliveira M, Castelano L and Cornelio M 2013  Why entanglement of formation is not generally monogamous \emph{Phys. Rev. A} {\bf 87} 032317 

\bibitem{Bai_monogamy} 
Bai Y-K,  Xu Y-F and  Wang Z D 2014 General monogamy relation for the entanglement of formation in multiqubit systems
\emph{Phys. Rev. Lett.} {\bf 113} 100503 

\bibitem{Gerardo_contangle}  
Adesso G and  Illuminati F 2006 Continuous variable tangle, monogamy inequality, and entanglement sharing in Gaussian states of continuous variable systems \emph{New J. Phys.} {\bf 8} 15 

\bibitem{Gaussian_tangle}  
Hiroshima T, Adesso G and Illuminati F 2007 Monogamy Inequality for Distributed Gaussian Entanglement \emph{Phys. Rev. Lett.} {\bf 98} 050503 

\bibitem{Addeso_CVent} 
Adesso G and Illuminati F 2007 Strong Monogamy of Bipartite and Genuine Multipartite Entanglement: The Gaussian Case \emph{Phys. Rev. Lett.} {\bf 99} 150501

\bibitem{Addeso_CVent2} 
Adesso G and Illuminati F 2008 Genuine multipartite entanglement of symmetric Gaussian states: Strong monogamy, unitary localization, scaling behavior, and molecular sharing structure \emph{Phys. Rev. A} {\bf 78} 042310 

\bibitem{KimJ_monogamy} 
Kim J, Das A and Sanders B 2009 Entanglement monogamy of multipartite higher-dimensional quantum systems using convex-roof extended negativity  \emph{Phys. Rev. A} {\bf 79} 012329 


\bibitem{asu_monogamy} Kumar A, Prabhu R, Sen(De) A and  Sen U 2015 Effect of a large number of parties on the monogamy of quantum correlations \emph{Phys. Rev. A} {\bf 91} 012341 


\bibitem{Song_Renyi} 
Song W, Bai Y-K, Yang M, Yang M and Cao Z-L 2016 General monogamy relation of multiqubit systems in terms of squared R{\'e}nyi-$\alpha$  entanglement \emph{Phys. Rev. A} {\bf 93} 022306
 
 \bibitem{Yuan_Tsalli}  
Yuan G-M,  Song W, Yang M, Li D-C, Zhao J-L and Cao Z-L 2016 Monogamy relation of multi-qubit systems for squared Tsallis-q entanglement \emph{Sci. Rep.} {\bf 6} 28719


 
\bibitem{Streltsov} 
Streltsov A,  Adesso G,  Piani M and  Bru{\ss} D 2012 Are General Quantum Correlations Monogamous? \emph{Phys. Rev. Lett.} {\bf 109} 050503 
 

\bibitem{salini} 
Salini K, Prabhu R, Sen(De) A and  Sen U 2014 Monotonically increasing functions of any quantum correlation can make all multiparty states monogamous \emph{Ann. Phys.}  {\bf 348} 297 


 \bibitem{GQD_monogamy1}
    Daoud M, Laamara R A,  Essaber R and  Kaydi W 2014     Global quantum correlations in tripartite nonorthogonal states and monogamy properties \emph{Phys. Scr.} {\bf  89} 065004 
    
    
 \bibitem{GQD_monogamy2}
Cheng S and  Hall M J W  2017  Anisotropic invariance and the distribution of quantum correlations  \emph{Phys. Rev. Lett.} {\bf 118} 010401

\bibitem{HJW_theorem} Hughston L P, Jozsa R and Wootters W K 1993 A complete classification of quantum ensembles having a given density matrix \emph{Phys. Lett. A} {\bf 183} 14




%%%% CONNECTION OF ENT WITH DISCORD MONOGAMY %%%%%%
%%%%%===================================%%%%%%

\bibitem{GGM}  
Sen(De) A and Sen U 2010 Channel capacities versus entanglement measures in multiparty quantum states \emph{Phys. Rev. A} {\bf 81} 012308

\bibitem{GGM2} Sen(De) A and Sen U 2010 Bound genuine multisite entanglement: Detector of gapless-gapped quantum transitions in frustrated systems arXiv:1002.1253

 \bibitem{GGM_anindya} 
 Biswas A,  Prabhu R, Sen(De) A and Sen U 2014 Genuine multipartite entanglement trends in gapless-gapped transitions of quantum spin systems \emph{Phys. Rev. A} {\bf 90} 032301 

 %GGM branch ended
 
% GM bunch started 

\bibitem{GM_shimony}  Shimony A 1995 Degree of Entanglement \emph{Ann. N. Y. Acad. Sci.} {\bf 755} 675 

\bibitem{GM_Barnum} Barnum H and Linden N 2001 Monotones and invariants for multi-particle quantum states  \emph{J. Phys. A: Math. Gen.} {\bf 34} 6787 

\bibitem{GM_Wei_Goldbart} Wei  T-C and Goldbart P M 2003 Geometric measure of entanglement and applications to bipartite and multipartite quantum states \emph{Phys. Rev. A}  {\bf 68} 042307  
 



\bibitem{GHZ}  
Greenberger D M, Horne M A  and  Zeilinger A 1989 \emph{Bell's
Theorem, Quantum Theory, and Conceptions of the Universe} ed. M. Kafatos (Kluwer Academic, Dordrecht, The Netherlands)

\bibitem{Dur1}   D\" ur W,  Vidal G and  Cirac J I 2000 Three qubits can be entangled in two inequivalent ways  \emph{Phys. Rev. A} {\bf 62} 062314 

\bibitem{Asu_global_local}  
Kumar A, Singha Roy S, Pal A K,  Prabhu R, Sen(De) A and  Sen U 2016 Conclusive identification of quantum channels via monogamy of quantum correlations \emph{Phys. Lett. A} {\bf 80} 3588


\bibitem{distri_DC} 
Bru{\ss} D, D'Ariano G M, Lewenstein M, Macchiavello C, Sen(De) A and  Sen U  2004 Distributed Quantum Dense Coding \emph{Phys. Rev. Lett.} {\bf 93} 210501

\bibitem{Bruss_mDC} Bru{\ss} D, Lewenstein M,  Sen(De) A, Sen U,  D'Ariano G M and  Macchiavello C 2006 Dense coding with multipartite quantum states \emph{Int. J. Quantum Inf.} {\bf 4} 415 

\bibitem{Nepal} Nepal R,  Prabhu R, Sen(De) A and Sen U 2013  Maximally-dense-coding-capable quantum states \emph{Phys. Rev. A} {\bf 87} 032336 

\bibitem{Horodecki_DCadv}  Horodecki M and  Piani M 2012  On quantum advantage in dense coding  \emph{J. Phys. A: Math. Theor.} {\bf 45} 105306

\bibitem{TamoghnaDC}  Das T, Prabhu R, Sen(De) A and  Sen U 2014  Multipartite dense coding versus quantum correlation: Noise inverts relative capability of information transfer
\emph{Phys. Rev. A} {\bf 90} 022319

\bibitem{Shadman} Shadman Z, Kampermann H,  Macchiavello C and  Bru{\ss} D 2010 Optimal super dense coding over noisy quantum channels \emph{New J. Phys.} {\bf 12} 073042 

\bibitem{Shadman2} Shadman Z, Kampermann H, Bru{\ss} D and Macchiavello C 2011 Optimal superdense coding over memory channels
\emph{Phys. Rev. A} {\bf 84} 042309

\bibitem{Shadman3} Shadman Z, Kampermann H, Bru{\ss} D and Macchiavello C 2012 Distributed superdense coding over noisy channels
\emph{Phys. Rev. A} {\bf 85} 052306 

\bibitem{Shadman4} Shadman Z, Kampermann H, Bru{\ss} D and Macchiavello C 2013 A review on super dense coding over covariant noisy channels {\it Quantum Measur. Quantum Metro.} {\bf 1} 21

\bibitem{Holevo_covariant} Holevo  A S  2005 Additivity  conjecture and covariant channels  \emph{Int. J. Quantum. Inf.} {\bf 3} 41

\bibitem{Pauli_channel} Fivel D I 1995 Remarkable Phase Oscillations Appearing in the lattice dynamics of Einstein-Podolsky-Rosen states \emph{Phys. Rev. Lett.} {\bf 74} 835



\bibitem{prabhu_exclusion} Prabhu R, Pati A K, Sen(De) A and Sen U 2013 Exclusion principle for quantum dense coding
\emph{Phys. Rev. A} {\bf 87} 052319

\bibitem{Liang}  Qiu L, Tang G,  Yang X-Q  and  Wang A-M 2014 Relating tripartite quantum discord with multisite entanglement and their performance in the one-dimensional anisotropic XXZ model \emph{Europhys. Lett.} {\bf 105} 30005


\bibitem{Koteswara}  Rao K R K,  Katiyar H, Mahesh T S, Sen(De) A,  Sen U and  Kumar A 2013 Multipartite quantum correlations reveal frustration in a quantum Ising spin system \emph{Phys. Rev. A} {\bf 88} 022312 

\bibitem{Allegra} Allegra M, Giorda P and  Montorsi A 2011 Quantum discord and classical correlations in the bond-charge Hubbard model: Quantum phase transitions, off-diagonal long-range order, and violation of the monogamy property for discord \emph{Phys. Rev. B} {\bf 84} 245133

\bibitem{Sarovar} Sarovar M, Ishizaki A,   Fleming G R and Whaley K B 2010
 Quantum entanglement in photosynthetic light harvesting complexes \emph{Nat. Phys.}  {\bf 6} 462 

\bibitem{QB_magrec} Bandyopadhyay J N, Paterek T and Kaszlikowski D 2012 Quantum coherence and sensitivity of avian magnetoreception \emph{Phys. Rev. Lett.} {\bf 109} 110502 

\bibitem{Lambert_QB} Lambert N, Chen Y-N, Cheng Y-C, Li C-M, Chen G-Y and Nori F 2013 Quantum biology \emph{Nat. Phys.} {\bf 9} 10  


\bibitem{Coherence_measure} Girolami D 2014 Observable Measure of Quantum Coherence in Finite Dimensional Systems \emph{Phys. Rev. Lett.} {\bf 113} 170401

\bibitem{Coherence_measure2} Baumgratz T, Cramer M and Plenio M B 2014 Quantifying Coherence 
\emph{Phys. Rev. Lett.} {\bf 113} 140401 

\bibitem{Legget_garg} Leggett A J and Garg A 1985 Quantum mechanics versus macroscopic realism: Is the flux there when nobody looks? \emph{Phys. Rev. Lett.} {\bf 54} 857

\bibitem{Legget_Garg_ROPP}  Emary C, Lambert N and Nori F 2014 Leggett-Garg inequalities
\emph{Rep. Prog. Phys.} {\bf 77} 016001 

 \bibitem{Engel_FMO}  Engel G S, Calhoun T R, Read E L, Ahn T-K,   Man{\v c}al T,  Cheng Y-C,  Blankenship R E and   Fleming G R 2007 Evidence for wavelike energy transfer through quantum coherence in photosynthetic systems \emph{Nature} {\bf 446} 782 

\bibitem{Coherence_FMO}  Olaya-Castro A,  Lee C F, Olsen F F and Johnson N F 2008 Efficiency of energy transfer in a light-harvesting system under quantum coherence
\emph{Phys. Rev. B} {\bf 78} 085115  

\bibitem{Coherence_FMO2} Plenio M B and Huelga S F 2008 Dephasing-assisted transport: quantum networks and biomolecules  \emph{New. J. Phys.} {\bf 10} 113019

\bibitem{Coherence_FMO3} Rebentrost P, Mohseni M, Kassal I, Lloyd S and Guzik A A 2009 Environment-assisted quantum transport  \emph{New J. Phys.} {\bf 11} 033003 

\bibitem{Wilde_Leggett}  Wilde M M,  McCracken J M and Mizel A 2010 Could light harvesting complexes exhibit non-classical effects at room temperature? \emph{Proc. R. Soc. A: Math. Phys. Eng. Sc.} {\bf 466} 1347

\bibitem{bell-original}
Bell J S 1964 On the Einstein Podolosky Rosen paradox \emph{Physics} {\bf 1} 195

\bibitem{Gisin_Bellv}  Gisin N 1991 Bell's inequality holds for all non-product states \emph{Phys. Lett. A} {\bf 154 }  201
 
 \bibitem{Popescu_Bellv} Popescu S and Rohrlich D 1992 Generic quantum nonlocality
\emph{Phys. Lett. A} {\bf 166} 293

\bibitem{Multiparty_BV} {\.Z}ukowski M,  Brukner {\v C}, Laskowski W and  Wi{\'e}sniak M 2002 Do All Pure Entangled States Violate Bell’s Inequalities for Correlation Functions? \emph{Phys. Rev. Lett.} {\bf 88} 210402 
 
\bibitem{Multiparty_BV2}  Sen(De) A, Sen U and {\.Z}ukowski M 2002 Functional Bell inequalities can serve as a stronger entanglement witness than conventional Bell inequalities
 \emph{Phys. Rev. A} {\bf 66} 062318 
 
 \bibitem{Yu_entnoBV}  Yu S, Chen Q, Zhang C, Lai C H and Oh C H 2012 All entangled pure states violate a single Bell's inequality \emph{Phys. Rev. Lett.} {\bf 109} 120402 

\bibitem{Liang_entnoBV} Liang Y-C, Masanes L and Rosset D 2012 All entangled states display some hidden nonlocality
\emph{Phys. Rev. A} {\bf 86} 052115 

 \bibitem{Horo_Bell} Horodecki R,  Horodecki P and  Horodecki M 1995 Violating Bell inequality by mixed spin-$\frac{1}{2}$ states: necessary and sufficient condition \emph{Phys. Lett. A} {\bf 200} 340 

\bibitem{Bell_monogamy}  Kurzy\'nski P,  Paterek T,  Ramanathan R,  Laskowski W and  Kaszlikowski D 2011 Correlation Complementarity Yields Bell Monogamy Relations
\emph{Phys. Rev. Lett.} {\bf 106} 180402 

\bibitem{Kunal} Sharma K, Das T, Sen(De) A and  Sen U 2016 
Distribution of Bell-inequality violation versus multiparty-quantum-correlation measures
\emph{Phys. Rev. A} {\bf 93} 062344 

\bibitem{Avijit_Bell} Pandya P, Misra A and Chakrabarty I 2016 Complementarity between tripartite quantum correlation and bipartite Bell-inequality violation in three-qubit states \emph{Phys. Rev. A} {\bf 94} 052126


\bibitem{GQD_CHSH2}
Yao Y,  Li H W,  Li M,  Yin Z Q,  Chen W and  Han Z F 2012  Bell violation versus geometric measure of quantum discord and their dynamical behavior  \emph{Eur. Phys. J. D} {\bf  66} 295  

\bibitem{modi-vedral1}
Modi K and Vedral V 2011 Unification of quantum and classical correlations and quantumness measures \emph{AIP Conf. Proc.} {\bf 1384} 69 

\bibitem{Dissension} Chakrabarty I,  Agrawal P and  Pati A K 2011 Quantum dissension: Generalizing quantum discord for three-qubit states \emph{Eur. Phys. J. D} {\bf 65} 605 

\bibitem{xu-ggqd-dekho}
Xu J 2012 Geometric global quantum discord \emph{J. Phys. A: Math. Theor.} {\bf 45} 405304


\bibitem{CHI-GLOBAL}
Chi D P, Kim J S and Lee K 2013
Generalized entropy and global quantum discord in multiparty quantum systems \emph{Phys. Rev. A} {\bf 87} 062339


\bibitem{coto-discord}
Coto R and Orszag M 2014 Determination of the maximum global quantum discord via measurements of excitations in a cavity QED network \emph{J. Phys. B: At. Mol. Opt. Phys.} {\bf 47} 095501

\bibitem{qiang-discord}
Qiang W-C, Zhang H-P and Zhang L 2015 Geometric global quantum discord of two-qubit X states \emph{Int. J.  Theor. Phys.} {\bf 55} 1833



\bibitem{Wei_rel} Wei T-C, Ericsson M, Goldbart P M and Munro W J 2004 Connections between relative entropy of entanglement and geometric measure of entanglement \emph{Quantum Inf. Comput.} {\bf 4} 252

\bibitem{Briegel_cluster} Briegel H J and Raussendorf R 2001 Persistent entanglement in arrays of interacting particles \emph{Phys. Rev. Lett.} {\bf 86} 910 

\bibitem{Nielsen_cluster} Nielsen M A 2006 Cluster state quantum computation \emph{Rep. Math. Phys.} {\bf 57} 147 


\bibitem{Dissonance_Roa}  Roa L,  Retamal J C and  Alid-Vaccarezza M 2011  Dissonance is required for assisted optimal state discrimination \emph{Phys. Rev. Lett.}  {\bf 107} 080401

% \bibitem{GQD_multiparty} Daoud M, Laamara R A and Seddik S 2015 Hilbert-Schmidt measure of Pairwise Quantum Discord for Three-Qubit X States   \emph{Rep. Math. Phys.} {\bf 76} 207 

\bibitem{Newcomb}  Newcomb S 1981 Note on the frequency of use of the different digits in natural numbers \emph{Am. J. Math.} {\bf 4} 39 


\bibitem{Benford} Benford F 1938 The law of anomalous numbers \emph{Proc. Am. Philos. Soc.} {\bf 78}  551

\bibitem{Benford_earthquake} Sambridge M, Tkal{\v c}i{\'c}  H and Jackson A 2010 Benford’s law in the natural sciences  \emph{Geophys. Res. Lett.} {\bf 37} L22301

\bibitem{Aditi_Benford_law} 
Sen(De) A and Sen U 2011 Benford's law detects quantum phase transitions similarly as earthquakes \emph{Europhys. Lett.} {\bf  95} 50008

\bibitem{Bhattacharya_metric} Bhattacharya  A 1943 On A Measure of Divergence Between Two Statistical Populations Defined by their Probability Distributions \emph{Bulletin of Cal. Math. Soc.} {\bf 35} 99 


\bibitem{Utkarsh_Benford} Rane A D, Mishra U, Biswas A, Sen(De) A and Sen U 2014 Benford's law gives better scaling exponents in phase transitions of quantum XY
 models \emph{Phys. Rev. E} {\bf 90} 022144

\bibitem{Titas_benford} Chanda T,  Das T,  Sadhukhan D, Pal A K, Sen(De) A and  Sen U 2016 Statistics of leading digits leads to unification of quantum correlations \emph{Europhys. Lett.} {\bf 114} 30004

\bibitem{EUR_review1}
 Wehner S and  Winter A 2010 Entropic uncertainty relations-a survey \emph{New J. Phys.} {\bf 12} 025009 
 
\bibitem{EUR_review2} Coles P J, Berta M,  Tomamichel M and  Wehner S 2017 
Entropic cuncertainty relations and their applications
\emph{Rev. Mod. Phys.} \textbf{89} 015002 
 
 \bibitem{EUR1}
  Deutsch D 1983 Uncertainty in Quantum Measurements \emph{Phys. Rev. Lett.} {\bf 50} 631 
 
 \bibitem{EUR5}
Maassen H and  Uffink J B M  1988 Generalized entropic uncertainty relations \emph{Phys. Rev. Lett.} {\bf  60} 1103 
 

 \bibitem{EUR_barta}
Berta M, Christandl M,	 Colbeck R,	 Renes J M  and  Renner R	 2010  The uncertainty principle in the presence of quantum memory \emph{Nat. Phys.} {\bf 6}    659

\bibitem{QKD_pati}
Pati A K,  Wilde M  M,  Usha Devi A R, Rajagopal A K and Sudha 2012 Quantum discord and classical correlation can tighten the uncertainty principle in the presence of quantum memory
\emph{Phys. Rev. A} {\bf  86} 042105  

\bibitem{EUR2_EUR3_compare}
Yao C, Chen Z, Ma Z, Severini S and Serafini A 2014  Entanglement and discord assisted entropic uncertainty relations under decoherence  \emph{Sci. China Phys. Mech. Astron.} {\bf 57} 1703


\bibitem{EUR_discord_upper_bound}
Hu M L and  Fan H 2013 Upper bound and shareability of quantum discord based on entropic uncertainty relations \emph{Phys. Rev. A} {\bf 88} 014105

\bibitem{tighther_pati_EUR}
Ma Z H,  Yao C M,  Chen Z H,  Severini S and  Serafini A 2013 A universal, memory-assisted entropic uncertainty relation arXiv:1302.1011

\bibitem{bera-comp}
Bera A, Kumar A, Rakshit D, Prabhu R, Sen(De) A and Sen U 2016 Information complementarity in multipartite quantum states and security in cryptography \emph{Phys. Rev. A} {\bf 93} 032338 

\bibitem{B92} 
Bennett C H 1992 Quantum cryptography using any two nonorthogonal states \emph{Phys. Rev. Lett.} {\bf 68} 3121



%\bibitem{cakmak-discord}
%Çakmak B and Gedik Z 2013 Quantum discord of SU(2) invariant states \emph{J. Phys. A: Math. Theor.} {\bf 46} 465302

%\bibitem{zurek-comple}
%Zwolak M and Zurek W H 2013 Complementarity of quantum discord and classically accessible information \emph{Sc. Rep.} {\bf 3} 1729

%\bibitem{xu-lazy}
%Xu J 2014 Lazy states, discordant states and entangled states for 2-qubit systems arXiv:1401.4260

%\bibitem{demon-braga}
%Braga H C, Rulli C C, de Oliveira T R
%and Sarandy M S 2014 Maxwell’s demons in multipartite quantum correlated systems \emph{Phys. Rev. A} {\bf 90} 042338

%\bibitem{Jara-power}
%Jara-Figueroa C, Klimov A B and Roa L 2014 Discording power of Hamiltonian interactions \emph{The Euro. Phys. J. D} {\bf 68} 51

%\bibitem{dan-discord}
%Li-Dan G, Xiao-Qian W, Yu-Mei X and Yuan-Yuan S 2014 Quantum Discord in Two-Qubit System Constructed from the Yang-Baxter Equation
%\emph{Comm. Theor. Phys.} {\bf 61} 349 

% \bibitem{Wang_DC} Wang X, Qiu L, Li S, Zhang C and Ye B 2015 Relating quantum discord with the quantum dense coding capacity \emph{J. Exp. Theor. Phys.} {\bf 120} 9

%\bibitem{Fedorov-discord}
%Fedorov A K, Kiktenko E O, Man'ko O V and  Man'ko V I 2015 Tomographic discord for a system of two coupled nanoelectric circuits \emph{J. Phys. A: Math. Theor.} {\bf 90} 055101



%\bibitem{kanno-discord}
%Kanno S, Shock J P and Soda J 2016 Quantum discord in de Sitter space \emph{Phys. Rev. D} {\bf 94} 125014





\bibitem{LN_Lee} Lee J, Kim M S, Park Y J and  Lee S 2000 Partial teleportation of entanglement in a noisy environment \emph{J. Mod. Opt.} {\bf 47} 2151 

\bibitem{LN_Plenio}  Plenio M B 2005 Logarithmic Negativity: A Full Entanglement Monotone That is not Convex
\emph{Phys. Rev. Lett.}  {\bf 95} 090503 

\bibitem{VedralRMP02} Vedral V 2002 The role of relative entropy in quantum information theory
{\it Rev. Mod. Phys.} {\bf 74} 197 


\bibitem{HHH00} Horodecki M,  Horodecki P and  Horodecki R 2000 Limits for Entanglement Measures
{\it Phys. Rev. Lett.} {\bf 84} 2014 

\bibitem{Plenio01} Plenio M B and Vedral V 2001 Bounds on relative entropy of entanglement for multi-party systems \emph{J. Phys. A: Math. Gen.} {\bf 34} 6997

\bibitem{Anandan_fubini}  Anandan J and  Aharonov Y 1990 Geometry of quantum evolution
\emph{Phys. Rev. Lett.} {\bf 65} 1697  
  
\end{thebibliography}
\end{document}